\documentclass[adp,fleqn]{w-art}

\usepackage{times}
\usepackage{w-thm}

\theoremstyle{plain}

\theoremstyle{definition}

\usepackage[]{graphicx}
\chardef\bslash=`\\ 

\usepackage{graphicx}
\usepackage[usenames]{color}
\usepackage{amsmath}
\usepackage{amssymb}
\usepackage{subfigure}

\usepackage{dcolumn}
\usepackage{multirow}
\usepackage{float}

\usepackage{color}

\def\dn{\downarrow}
\def\up{\uparrow}

\font\bbf=cmssi10
\font\msbm=msbm10
\font\smallMsbm=msbm6
\def\vek#1{\mbox{{\bbf #1}}\,}
\def\Z{\mbox{{\msbm Z}}}
\def\smallZ{\mbox{{\smallMsbm Z}}}
\def\op#1{\widehat{#1}}
\def\wt#1{\widetilde{#1}}

\def\unitVec#1{\mbox{{\bbf e}}_{#1}}

\def\d{{\mathrm d}}

\def\ve{\varepsilon}
\def\vp{\varphi}

\def\vt{\vartheta}

\def\bra#1{\langle#1|}
\def\braket#1#2{\langle #1|#2\rangle}
\def\ket#1{|#1\rangle}

\def\unit#1{\ {\rm #1}}   

\def\pd#1#2{\frac{\partial #1}{\partial #2}}
\def\pder#1#2{\pd{#1}{#2}}

\def\Ft{{\cal F}} 

\definecolor{crimson}{rgb}{0.3,0,0}
\definecolor{darkgreen}{rgb}{0,0.6,0.2}
\definecolor{darkblue}{rgb}{0,0.2,0.6}
\definecolor{darkyellow}{rgb}{0.7,0.7,0}
\definecolor{orange}{rgb}{0.7,.35,0}

\def\note#1{{\color{orange} #1}}
\def\NOTE#1{{\bf\color{red} #1}}

\def\subsubsubsection#1{\par\bigskip{\bf\Large #1}\par\smallskip}

\newcommand{\captionfonts}{\small}

\makeatletter  
\long\def\@makecaption#1#2{%
  \vskip\abovecaptionskip
  \sbox\@tempboxa{{\captionfonts #1: #2}}%
  \ifdim \wd\@tempboxa >\hsize
    {\captionfonts #1: #2\par}
  \else
    \hbox to\hsize{\hfil\box\@tempboxa\hfil}%
  \fi
  \vskip\belowcaptionskip}
\makeatother   

\def\ot{{1/3}}
\def\tt{{2/3}}
\def\tf{{2/5}}

\def\op#1{\relax#1}

\def\gsI{GS$_{\infty}${}}

\def\cunit{\ (e^2/\ve\ell_0)}   

\def\subsubsubsection#1{\par\bigskip{\sl #1}\par\smallskip}

\def\ttfont#1{$\mathtt{#1}$}

\font\bbf=cmssi10   
\font\msbm=msbm12   
\font\smallMsbm=msbm7   

\def\eff{\mathit{eff}}

\def\kr{k^r}
\def\krv{\vek{k}^r}
\def\krvd{\wt{\vek{k}}{}^r}

\begin{document}

\def\note#1{\relax}
\def\NOTE#1{\relax}

\def\footlabel#1{\relax}

\DOIsuffix{theDOIsuffix}
\Volume{12}
\Issue{1}
\Copyrightissue{01}
\Month{01}
\Year{2003}
\pagespan{1}{}



\title{Spin in fractional quantum Hall systems}

\author[K. V\'yborn\'y]{Karel V\'yborn\'y\footnote{e-mail: 
{\sf vybornyk@fzu.cz}, Phone: +420\,220\,318\,459,
     Fax:  +420\,233\,343\,184}\inst{1}\inst{2}} 

\address[\inst{1}]{1. Institut f\"ur Theoretische Physik,
  Universit\"at Hamburg, Jungiusstr. 9, 20355 Hamburg, Germany}
\address[\inst{2}]{Institute of Physics,
  Academy of Sciences of the Czech Republic, Cukrovarnick\'a 10, 
  16253 Praha 6, Czech Republic}

\begin{abstract}
A system at filling factor $\tt$ could be a candidate for
a quantum Hall ferromagnet at integer filling factor of composite
fermions. Using exact diagonalization with electrons on a
torus we study the transition from the singlet ground state to the 
polarized ground
state at this filling and look for signatures of quantum Hall
ferromagnetism. Differences between the fractional and corresponding
integer systems are emphasised. Most interestingly,
we find around the transition a low excited half-polarized
state  which might become the ground state in the
thermodynamical limit. We study its structure and compare it to the
singlet and polarized ground states. A new interpretation of the singlet state
is suggested and comparison of the filling factors $\tt$ and
$\tf$ is presented.
Adding magnetic inhomogeneities into the system we investigate the
stability of all the three involved states and the 
tendency to build up domains like in conventional ferromagnets.

\end{abstract}

\maketitle                   

\vspace*{-1.0cm}

\tableofcontents

Similar to superconductivity, the {\em fractional quantum Hall effect}
\cite{chakraborty:1995,yoshioka:2002} 
is a unique field, where correlations
between electrons give rise to macroscopically well observable ground
states which we would not expect on the level of a Hartree-Fock
approximation. The correlations are introduced by interelectronic
interaction and, contrary to atomic physics for instance, the 
quantization of single-electron 
energy levels is a consequence of the strong magnetic field (Landau
levels) and of the suppressed motion in the direction of the field
(quasi two-dimensional systems).
The latter phenomenon leads to another unusual feature of the fractional
quantum Hall systems: the many-electron states in a non-interacting
system are highly (macroscopically) degenerate, since all electrons
within one Landau level have the same energy. In particular, 
for filling factors below one, where it is useful to be restricted to
the lowest Landau level, {\em all} many-electron states have the same
energy. Now, the effect of interactions between electrons 
cannot be investigated by perturbation
theory, as there is no single ground state to start
with or, in other words, there is no small parameter in which we could
expand the perturbation series. Since energy spacing between the
many-body states is zero, the interaction is never a small perturbation,
regardless of how weak it is. This fact renders the fractional quantum Hall
systems unique from the theoretical point of view and makes completely
novel types of quantum-mechanical ground states possible. The best
known of these are the incompressible quantum liquids.

{\em Quantum Hall ferromagnetism} was one of the companions of the {\em
integer} quantum Hall effect (Subsect. \ref{pos-ch02-16}). The
observed long-range spin order can be explained by exchange energy
gain in the ferromagnetic state and hence Hartree-Fock models are
sufficient to describe the ongoing physics. However, new experimental
publications appeared in late nineties.
Phenomena reminiscent of ferromagnetism have also
been observed in the {\em fractional} quantum Hall regime, being most
noticeable at filling factors $\tt$ and $\tf$
(\cite{kronmuller:09:1998,cho:09:1998}). In this situation,
the Hartree-Fock approximation is no longer acceptable, the spin-ordered
states are highly correlated. This area is not very well
explored. Instead of a lattice of spins which are all pointing in the
same direction, here, we are dealing with itinerant electrons which
are either in a fully polarized or in a spin singlet state
(Subsect. \ref{pos-ch03-17}). Although both states are
incompressible, their structure is quite different~\cite{chakraborty:07:2000}.

How far can we extend the analogy between an Ising spin-lattice
ferromagnet and fractional quantum Hall systems where two ground
states with different spin order compete with each other? This was the
central question of this thesis.
There are several fundamental differences between these two systems.
The latter one is itinerant and the liquid-like ground state is
stable only owing to correlations while, in a spin-lattice, the
electrons are spatially fixed and the ferromagnetism occurs also in
classical systems with suitable site-to-site coupling.  By observing
e.g. hysteresis in magnetotransport, experimentators have provided a lot of
evidence that the two phenomena are indeed very closely related
\cite{smet:03:2001,smet:01:2002,kraus:12:2002}, 
on the other hand, observations without an
analogy to usual Ising systems have been reported too
\cite{kukushkin:05:1999}. 

Before we start the theoretical introduction, let us summarize the
basic experimental facts. At filling factors $\tt$ and $\tf$ two
different ground states may appear. Depending on the ratio
between the Zeeman to Coulomb energy, $E_Z/E_C$, it is the
fully spin polarized ($E_Z/E_C\to \infty$) or the
spin-singlet one (${E_Z/E_C\to 0}$) \cite{kukushkin:05:1999}. 
This transition can be
accomplished (a) by varying the electron density at a fixed filling factor
\cite{hashimoto:04:2002}, (b) by tilting the
magnetic field \cite{clark:03:1989,eisenstein:03:1989} or
(c) by applying hydrostatic pressure which modifies the bulk $g$-factor
\cite{leadley:11:1997}. 

When the two ground states are brought to degeneracy, transport
experiments show hysteresis, time-dependent resistance with
Barkhausen jumps \cite{bertotti:1998} (see Refs. above) and 
huge longitudinal magnetoresistance \cite{kronmuller:09:1998} which
is related to the spin polarization of the ion lattice of the hosting GaAs
(NMR experiments \cite{kronmuller:05:1999,dietsche:??:2001}). 
These could be related to formation of
spatial domains of the two ferromagnetic ground states, even though
surface acoustic wave experiments could not confirm this
\cite{dunford:07:2002}. On the other hand, optical experiments
\cite{kukushkin:05:1999,freytag:09:2001}, suggest that a
half-polarized ground state occurs near the transition.

\setcounter{footnote}{0}
\section{Theoretical basics}
\label{pos-ch02-00}

\subsection{One electron in magnetic field}
\label{pos-ch02-07}


When the mutual interactions are left aside, electrons 
in a plane
subject to 
homogeneous perpendicular magnetic field $B$ fill the macroscopically 
degenerate equidistant Landau levels (LLs) with energies
$E=(n+\frac{1}{2})\hbar\omega$, $n=0,1,2,\ldots$ The degeneracy of all the
levels is the same, and it increases proportionally to the magnetic
field. Therefore, occupancy of the Landau levels, the {\em filling factor},
depends both on the number of electrons $N_e$ (per area $L^2$) and
on $B$:
\begin{equation}\label{eq-ch02-38}
   \nu =  \frac{N_e/L^2}{eB/h} = \frac{N_e}{L^2/(2\pi\ell_0^2)} 
   = \frac{N_e}{(BL^2)/(h/e)} = \frac{N_e}{\Phi/\Phi_0} =
   \frac{N_e}{N_m}\,.
\end{equation}
Note that this $\nu$, i.e. number of Landau levels occupied in
the ground state
is equal to the {\em inverse} number of magnetic
flux quanta $\Phi_0=h/e$ per electron in the system (the second last
expression in (\ref{eq-ch02-38}).

These facts can easily be obtained by solving the single-electron
Schr\"odinger equation with Hamiltonian
\begin{equation}\label{eq-ch02-16}
   H_0 = \frac{1}{2m} (\vek p + e\vek A)^2\,,\qquad
   \nabla\times\vek A= (0,0,B)\,,
\end{equation}
a particularly nice and understandable example of this calculation is
given by Murthy and Shankar \cite{murthy:10:2003}. Suitable energy and
length units  are the cyclotron energy
$\hbar\omega=\hbar eB/m$ and the magnetic length $\sqrt{\hbar/eB}$
denoted by $\ell_0$.

Let us now focus on the lowest Landau level. There are $eB/h\cdot$ states
per unit area having the same energy $\hbar\omega/2$ and
infinitely many possibilities of constructing a basis of this space.
Choosing the Landau (symmetric) gauge in (\ref{eq-ch02-16}) which
is translationally symmetric in one direction (rotationally symmetric
around the origin) we are lead to the following bases
\begin{eqnarray}\label{eq-ch02-28}
    \mbox{Landau:}\, \vek{A}=(0,Bx,0)\,,&\ & 
    \psi_{k_y}(x,y) = \exp(-ik_y y) \exp[-(x+k_y)^2/2\ell_0^2]\,,\\
    \nonumber && \hskip5cm k_y/(2\pi /L)=0,1,2,\ldots \\
    \label{eq-ch02-25}\mbox{symmetric:}\,
    \vek{A}=\frac{1}{2}(y,-x,0)\,,&\ &
    \psi_m(z) = z^m \exp(-|z|^2/4\ell_0^2)\,,\quad
    m=0,1,2,\ldots
\end{eqnarray}
Especially for the latter basis, formulae are often more transparent
if we use a complex variable $z=x+iy$ rather than $x,y$ separately to
address the points in the plane.

\subsubsection{Magnetic translations}

A plane with perpendicular homogeneous magnetic field is obviously
translationally invariant. However, spatial translations applied to the
Hamiltonian may alter the gauge even though they leave
the magnetic field unchanged. Operators which
conserve also the gauge (and which therefore commute with $H_0$) are
the {\em magnetic translations} 
\cite{zak:06:1964,zak:06-2:1964}. These operators will thus
replace the ordinary translations applicable to systems without
magnetic field.

Magnetic translation operators depend on the choice of the gauge, in
particular for the Landau gauge (\ref{eq-ch02-28})
\begin{equation}\label{eq-ch02-27}
  \vek u=(u_1,u_2)\,:\qquad T(\vek u) = \exp(-iu_x y/\ell_0^2) t(\vek u)\,,
\end{equation}
where $t(\vek u)$ is the ordinary translation operator 
$\exp(i\vek u\cdot\vek p/\hbar)$. General explicit formula for any
gauge can be found e.g. in the article of 
Haldane and Rezayi \cite{haldane:02:1985}.

Note, that ordinary and magnetic translations (\ref{eq-ch02-27}) 
coincide for $u_x=0$, exactly as a wavefunction in the form
(\ref{eq-ch02-28}) remains unchanged up to a constant phase under the
replacement $y\to y+u_y$.
For $\vek u=(u_x,0)$ this is not the case and that is
why we must resort to {\em magnetic} translations.

\subsection{What to do when Coulomb interaction comes into play}
\label{pos-ch02-08}


The quantum mechanical solution of one electron -- or many
non-interacting electrons in a plane subject to a perpendicular
magnetic field is at the root of the integer quantum Hall effect. The
basic fact is that for integer filling factors, {\em any}, even
arbitrarily small excitation costs at least the energy
$\hbar\omega$. This gap renders the ground state incompressible. The
fractional quantum Hall effect cannot be explained in this picture.
For instance at filling factor $\nu=\ot$, a non-interacting system has
a manyfold degenerate ground state, or, some excitations cost zero
energy (those which involve only rearrangement of electrons within the
lowest LL), and the ground state should be compressible.  Today it is
well established that the effect is due to electron-electron
interactions which select among those states one special ground state
and separate it by a gap from the excitations.


Now, the Hamiltonian of the many-electron system consists of two terms:
the kinetic energy (leading to Landau level quantization) and the
electron-electron interaction. 
%
\begin{equation}\label{eq-ch02-29}
  H = \sum_{i=1}^{N_e} \frac{\vek{p}_i^2}{2m} + 
      \frac{e^2}{4\pi\ve }\frac{1}{2}\sum_{i\not= j} 
                       \frac{1}{|\vek r_i-\vek r_j|}
\end{equation}
Consider some particular filling factor, $\nu=\ot$ for example, and
let us vary the magnetic field. Since $\nu=n/(2\pi
\ell_0^2)=n/(eB/\hbar)$, this implies changing the electron density $n$
simultaneously.  
The kinetic energy will change proportionally to $\hbar\omega\propto B$.
The interaction energy on the other hand scales with $1/a\propto \sqrt{B}$, 
as the mean electron-electron distance $a$ is proportional to the
magnetic length $a=\sqrt{1/n}=\sqrt{1/(\nu eB/\hbar)}\propto \ell_0$
(for a more thorough discussion see Yoshioka \cite{yoshioka:2002}, Chap. 4)

In the high field limit we can therefore expect the Coulomb
interaction to be a small perturbation which lifts the degeneracy of
Landau levels. Looking for a (high-$B$) ground state at some
particular $\nu<1$ we can therefore omit the higher Landau
levels and study only states {\em within the lowest Landau level}. This
model gives qualitatively correct predictions (for high $B$) and
the inclusion of the Landau level mixing leads only to quantitative
corrections (e.g. in the ground state energy, see Chakraborty and
Pietil\"ainen \cite{chakraborty:1995}).

\subsubsection{Ground states: analytical many-body wavefunctions}

It is very surprising that even though we now handle a {\em
many-body} Hamiltonian (\ref{eq-ch02-29}), there are still {\em
analytic} (correlated) wavefunctions which describe the ground state at
some special filling factors. The best known example was suggested by
R. B. Laughlin \cite{laughlin:05:1983} earning him the Nobel Prize:
\begin{equation}\label{eq-ch02-09}
\Psi_L(z_1,\ldots,z_n) = 
       \exp\big(-(|z_1|^2+\ldots+|z_n|^2)/4\ell_0^2\big)
       \prod\nolimits_{i<j} (z_i-z_j)^3\,.
\end{equation}

There are several beautiful physical arguments why this wavefunction
must be the ground state at $\nu=\ot$ (in fact, an excellent approximation
to it, see Subsec. \ref{pos-ch02-01}). These are explained
in other more detailed publications
\cite{yoshioka:2002} (Chap. 4), \cite{chakraborty:1995,haldane:08:1983}.
Let us mention here only two basic ideas about the interpretation of $\Psi_L$.

First, the $(z_i-z_j)^3$ term makes the Laughlin wavefunction isotropic and
translationally invariant. More detailed studies (density-density
correlations, see Subsec. \ref{pos-ch03-17}) suggest that it resembles
a liquid. Second, $\Psi_L$ resembles $\Psi_1$, the wavefunction of completely
occupied lowest Landau level which is the same as
(\ref{eq-ch02-09}), only with $(z_i-z_j)^3$ replaced by $(z_i-z_j)$. In
fact, 
$$
  \mbox{$\nu=1$, GS:}\ \Psi_1 \propto \prod_{i<j}(z_i-z_j) 
  \stackrel{\times \Pi_{i<j}(z_i-z_j)^2}{\longrightarrow} 
  \mbox{$\nu=\frac{1}{3}$, GS:}\ \Psi_L
  \propto \prod_{i<j}(z_i-z_j)^3\,.
$$
Now consider a single electron in a state $\psi_m(z)\propto z^m$ 
(\ref{eq-ch02-25}). If we pierce the system by an infinitely
thin solenoid at $z_0$ and pass two magnetic flux quanta adiabatically through
it, this state will evolve into $\psi_m(z) (z-z_0)^2$. This leads to
the following interpretation. The Laughlin state is just the
completely filled lowest Landau level, but the constituent particles are
electrons with two attached magnetic flux quanta rather than bare electrons.

So far we have only spoken about the filling factor $\nu=\ot$ and
fully spin-polarized electrons. Generalizations of these concepts are
possible also to systems where electrons are not fully spin
polarized. Halperin proposed the following WFs \cite{halperin:??:1983}
\begin{equation}\label{eq-ch02-56}
\hskip-5mm\Phi_{mm'n}[z] = \prod_{i<j\le N_\up} (z_i-z_j)^m
			\prod_{k<l\le N_\dn} (z'_k-z'_l)^{m'}
			\prod_{{i\le N_\up}\atop{k\le N_\dn}}
			     (z_i-z'_k)^n
			\prod_{i,j} 
			\exp(-|z_i|^2/4\ell_0^2)\exp(-|z'_j|^2/4\ell_0^2)\
\end{equation}
with $m,m'$ odd. The state assumes $N_\up$ ($N_\dn$) particles with
spin up (down) and $z_i$ ($z'_j$) describe their positions.
The filling factors of the two components are
$$
	\nu_\up = \frac{m'-n}{mm'-n^2}\,,\qquad
	\nu_\dn = \frac{m-n}{mm'-n^2}\,.
$$
Thus, they describe a state at filling
$\nu=\nu_\up+\nu_\dn$ and polarization $p=(\nu_\up-\nu_\dn)/\nu$. For
example, the choice $m=m'=3$ and $n=2$ leads to the total filling factor
$\tf$ and zero spin polarization ($\nu_\up=\nu_dn$).

These analytical results will always be a good starting point for
investigations going to regions where only numerical methods are
possible. Before continuing, however, it must be emphasised, that
wavefunctions in (\ref{eq-ch02-09},\ref{eq-ch02-56}) are not the only
analytical trial wavefunctions known in the lowest Landau level. A
more detailed review can be found in \cite{sarma:1997} (MacDonald and
Girvin).

\subsection{Other types of electron-electron interactions}



%
%
%
%

\label{pos-ch02-01}

A model of {\em short-range interaction} (SRI) in fractional quantum
Hall systems is the main issue of this section. We will explain
how a general interaction $V(r)$ between two particles within the
lowest Landau level (LL) can be represented by a set of {\em
Haldane pseudopotentials} $\{V_m\}$ \cite{haldane:08:1983} and show
how this concept makes it easier to study different classes of
interaction. In particular, this discussion 
will unveil under what conditions the
Laughlin wavefunction (\ref{eq-ch02-09}) becomes the exact
many-body ground state.

\subsubsection{Two particles, magnetic field and a general isotropic
  interaction}

Let us consider two negatively charged particles in a plane subjected to
a perpendicular magnetic field $B$. Assume that their interaction is
described by a potential (energy) $V(r)$ which depends only on their
mutual distance. Classically, when starting from rest, the particles
would move along a straight line towards or away from each other were it
not for the magnetic field. The Lorentz force bends their trajectories
and makes them orbit around their centre-of-mass on a circular
trajectory. In quantum mechanics, this circular motion is quantized
just as in case of an electron orbiting around a hydrogen
nucleus. {\em Roughly speaking}, only discrete separations $r_m$
between the two particles are allowed.  Interaction energies
$V(r_m)=V_m$ rather than the full form $V(r)$, $r\in(0;\infty)$ fully
determine the spectrum of a many-body system of particles interacting
via $V(r)$.

\begin{figure}
\begin{center}
\leavevmode\raise2.5cm\hbox{\parbox{2cm}{
${E_{rel}=}$
\hbox{$\hskip.7cm {\color{darkgreen}\hbar\omega(i+\frac{1}{2})}$}
\hbox{$\hskip.7cm -{\color{blue}m\hbar\omega}$}}}
\hskip-5mm
\includegraphics[scale=0.4]{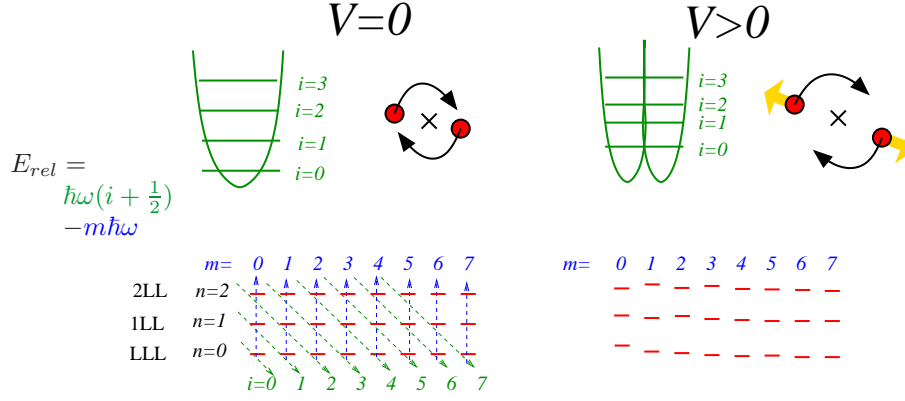}
\end{center}
\caption{Spectrum of one particle confined to a plane subject to
a perpendicular magnetic field or, equivalently, the spectrum
corresponding to the relative motion of two particles
($H_{rel}$). {\em Left:} without particle-particle interaction, the
two terms, 'harmonic oscillator' (quantum number $i$) and 'angular
momentum' (quantum number $m$) combine into degenerate Landau levels (quantum
number $n$). {\em Right:} interaction lifts the degeneracy. 
If the interaction potential is small
compared to the harmonic oscillator term (i.e. $\langle
V\rangle \ll \hbar\omega$), Landau levels are roughly preserved. The
energy levels $V_m$ {\em within} the lowest Landau level (sorted
according to $m=\langle L_z/\hbar\rangle$) are then the
Haldane pseudopotentials.}\label{fig-ch02-02}
\end{figure}

Let us now follow this idea in more detail and derive the precise claims.
The Hamiltonian for two particles reads
\begin{equation}\label{eq-ch02-04}
  H = \frac{1}{2m}(\vek{p}_1+|e|\vek{A}_1)^2 +
      \frac{1}{2m}(\vek{p}_2+|e|\vek{A}_2)^2 + V(|\vek{r}_1-\vek{r}_2|)\,.
\end{equation}
Following Laughlin \cite{laughlin:03:1983}, we
write it as a sum of the centre-of-mass (CM) and relative 
parts [$\vek{r}_{CM} = (\vek r_1 + \vek r_2)/2$, 
$\vek{r}_{rel}= (\vek r_1 - \vek r_2)/\sqrt{2}$]. 
Without $V(r_{rel})$, both parts will be equivalent to a single
particle in a plane in magnetic field. Hence, a Landau level index
($N=n_{CM}$, $n=n_{rel}$) and an angular momentum ($M=m_{CM}$, $m=m_{rel}$)
will be attributed to both parts. Out of these, $N$ is fixed to zero
(two-particle state within the lowest LL) and $M$ is unimportant as
it is merely tantamount to fixing the CM to some particular position 
in the plane.
Eigenstates of the relative part will be sorted as shown in
Fig. \ref{fig-ch02-02} on the left.

With $V(r_{rel})$ included, the relative part reads
\begin{equation}\label{eq-ch02-03}
  H_{rel} = \underbrace{\frac{p_{rel}^2}{2m} + 
                       \frac{1}{8}\hbar\omega (r_{rel}/\ell_0)^2 -
		       \frac{1}{2}\omega L^z_{rel}}_{H_{rel, kin}} +
                       V(|\vek{r}_{rel}|)\,
\end{equation}
with $L^z_{rel}$ denoting the ($z$-component of relative) angular momentum.
This is the well known Fock-Darwin form, the standard usage of which 
is to describe one particle in a magnetic field and confining
potential $V$. However, we will use it in a different way here: 
we consider $V(r)$ to be weak, and for
example a repulsive $\propto 1/r$ potential, compared to the parabolic
term in $H_{rel, kin}$. It is only our initial assumption that the
interaction is much weaker than the cyclotron energy, $E_C\ll \hbar\omega$
(Subsec. \ref{pos-ch02-08}). 

Owing to $[H_{rel},L^z_{rel}]=0$, the eigenstates of $H_{rel}$ can
still be classified by angular momentum. Moreover, assuming the states of
the lowest LL to have no admixtures from higher LLs 
($E_C\ll \hbar\omega$), the eigenstates 
\begin{equation}\label{eq-ch02-05}
\psi_{rel}^m(r,\vp)=\exp(-i m\vp)\, r^m \exp(-r^2/4\ell_0^2)\,. 
\end{equation}
will not depend on $V(r_{rel})$ at all. 
This is a combined effect of the $L_z$ symmetry and the requirement of
analyticity (confinement to the lowest LL): angular part of 
the form $\exp(-i m\vp)$ implies $\psi^m(z)\propto z^m$.

On the other hand, energies of these states will shift differently for
different $m$'s (Fig. \ref{fig-ch02-02}, right). 
The energy shift due to the interaction between particles is
$V_m=\bra{\psi_{rel}^m}V\ket{\psi_{rel}^m}$ in a state with relative
angular momentum $m$. Since $r_m=\bra{\psi_{rel}^m}
r\ket{\psi_{rel}^m}=\ell_0\sqrt{2m+1}$ we can {\em roughly} estimate
$V_m$ to be $V(r_m)$.

The operator of any weak ($E_C\ll \hbar\omega$) 
interaction (in the lowest Landau level) can
be then written in terms of its spectral decomposition
$$
  V(r_{rel})  =  \sum_{m=0}^\infty \ket{\psi_{rel}^m}V_m\bra{\psi_{rel}^m}\,.
$$
The spectrum and eigenstates of a {\em many-body} system confined to the
lowest Landau level and {\em interacting}
by $V(r_{rel})$ is thus completely determined by the discrete set of numbers
$\{V_m\}$.

The quantities $\{V_m\}$ are called {\em Haldane pseudopotentials}. They
were first introduced in \cite{haldane:08:1983} in the context of
interacting electrons on a sphere.
Finally, we add two remarks.

{\em Fermions and bosons.} A careful reader may have noticed that we
    have spoken just about two {\em particles} so far. An additional
    constraint that e.g. (spatial part of the) wavefunction should be
    antisymmetric implies
    $$\psi_{rel}(\vek r_1,\vek r_2) = -\psi_{rel}(\vek r_2,\vek r_1)\qquad
      \Rightarrow\qquad \psi_{rel}(r,\vp) = - \psi_{rel}(r,-\vp)\,.
    $$
    Therefore, only the states with $m$ odd (\ref{eq-ch02-05})
    are allowed in the case of two electrons with the same spin (where
    a symmetric spinor part implies an antisymmetric orbital part of
    the wavefunction).
    In other words: only
    the values of $V_1,V_3,\ldots$ are needed when we describe motion
    of fully spin polarized electrons.

{\em Uniqueness.} If $V(r)$ is given, the pseudopotentials $V_m$
    are determined uniquely. However, the opposite is not true: 
    knowing only the values of $V_m$, we cannot reconstruct the full
    form of $V(r)$.

\subsubsection{Particular values of Haldane pseudopotentials
for the Coulomb interaction}

The ideas above are not valid exclusively for the lowest Landau level.

\begin{figure}
\begin{center}
\begin{tabular}{cccc}
(a) & (b) & (c) & (d)\\
\includegraphics[scale=0.55]{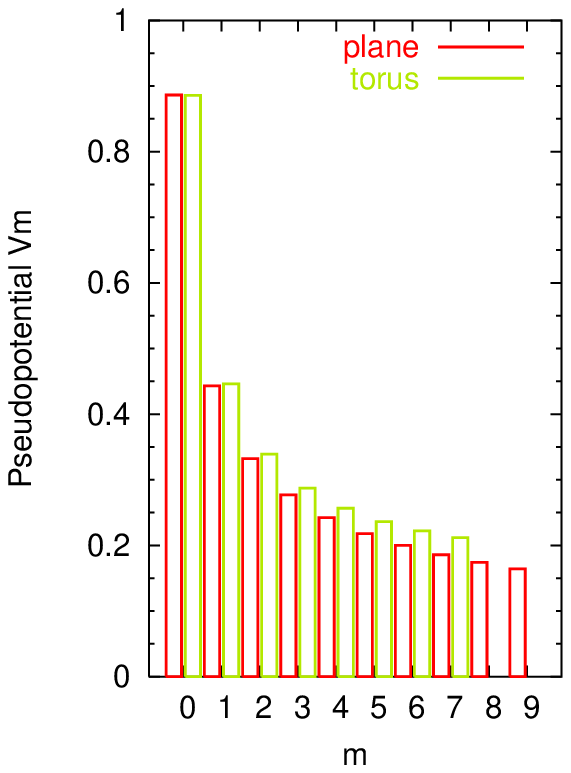} &
\hskip-3mm\includegraphics[scale=0.55]{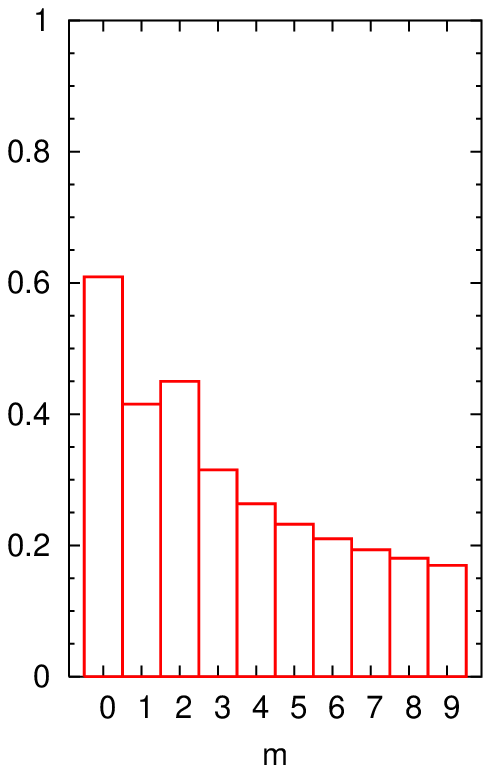} &
\hskip-3mm\includegraphics[scale=0.55]{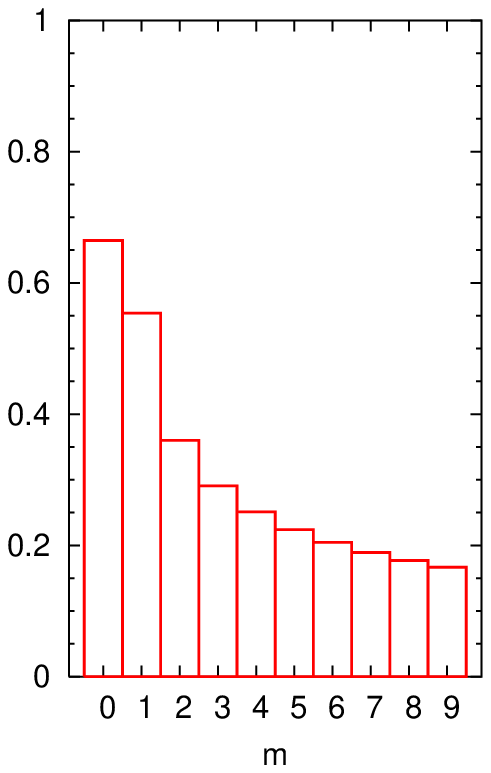} &
\hskip-3mm\includegraphics[scale=0.55]{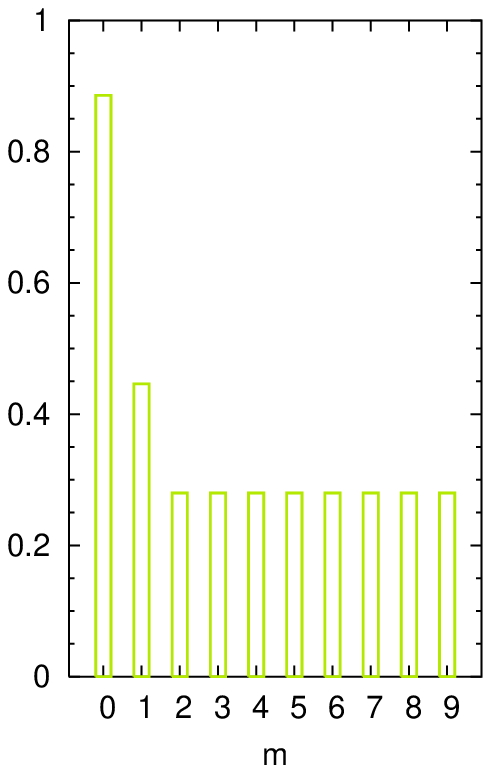} 
\end{tabular}
\end{center}
\caption{Values of Haldane pseudopotentials $V_m$ for
Coulomb-interacting electrons (a-c):
between two particles (a) both in the lowest Landau
level, $(n_1,n_2)=(0,0)$, (b) both in the first Landau level,
$(n_1,n_2)=(1,1)$, and (c) one in the lowest and the second in the first
Landau level, $(n_1,n_2)=(1,0)$. The pseudopotentials referring to
electrons in a plane (a,b,c) are 'universal', those related to
electrons on a torus (a,d) depend on its size (here $N_m=30$,
Subsec.~\ref{pos-ch02-05}). Values of
$\{V_m\}$ chosen for as a model for short-range interaction in this
work are shown in (d).}\label{fig-ch02-04}
\end{figure}

Let us consider a numerical example
for electrons in a plane, one of them 
located in {\em arbitrary} Landau level $n_1$ and another in
$n_2$. Their relative angular momentum be $m$.
Given these three numbers, the state is uniquely defined, up to the
center-of-mass part of the wavefunction, as we have already stated. Assuming
interaction of the form $V(q)$ (in the Fourier space), their
interaction energy can be written as \cite{prange:1987}
\begin{equation}\label{eq-ch02-13}
V_m^{n_1,n_2} = \int_0^\infty q\d q V(q) L_{n_1}(q^2/2) L_{n_2}(q^2/2)
                L_m(q^2) \exp(-q^2)\,.
\end{equation}
The {\em Laguerre polynomials} are defined by $L_n(x)=(1/n!) [x^n
e^{-x}]^{(n)} e^x\,.$ 
For the case of Coulomb interaction, $V(q)= \alpha/|q|$, the integrals in
(\ref{eq-ch02-13}) can be evaluated (easily and)
analytically. Figure \ref{fig-ch02-04} shows their values for the
cases (a) both particles in the Lowest Landau level ($n=0$), (b) both
particles in the first Landau level ($n=1$) and (c) one in the lowest
and one in the first Landau level.

For $n_1=n_2=0$ the
coefficients $V_m$ decay
monotonically with increasing $m$, exactly as the Coulomb energy does
with increasing distance. The non-monotonic structure of $V_m$ for the
case of particles in the first Landau level is due to the additional
structure of wavefunctions in higher Landau levels (e.g. a node
at $r=0$ for $n=1$).

\subsubsection{Model interactions: hard core, hollow core}
\label{pos-ch02-09}

Why is a hard-core interaction ({\em short-range
interaction}) important for the physics of the lowest Landau level?

There are three reasons: 
(i) it is the strongest part of the Coulomb interaction, (ii) the
Laughlin wavefunction is an exact (zero energy) 
gapped ground state for this interaction
and (iii) the ground state changes only little if the other terms of
the Coulomb interaction are considered.

Let us discuss this in more detail. For the purposes of this paper, 
the short-range interaction (SRI) 
for {\em spin polarized electrons} is defined by the 
Haldane pseudopotentials
\begin{equation}\label{eq-ch02-14}
  \mbox{short-range int. (spin polarized electrons): }
  \{V_1,V_3,V_5,\dots \} = 
  \{1,0,0,\dots \}\,.
\end{equation}
For the {\em Coulomb} interaction, Fig. \ref{fig-ch02-04}a, 
$V_1$ is indeed the strongest pseudopotential.

On the other hand, considering the Laughlin wavefunction of $N$
particles (\ref{eq-ch02-09}), 
any pair of electrons in it is in a state with relative
angular momentum $m=3$: owing to factors $(z_i-z_j)^3$. As there are 
no pairs with angular momentum $m=1$, the total energy of this state
will be $V_1\cdot 0+V_3\cdot N(N-1)/2=0$ for SRI. Also, this state 
is rigid: any
excitation of this state must remove the triple zero from some of the
electrons (leaving only a single zero required by antisymmetry) thereby
creating some pairs with $m=1$. This implies a finite excitation gap.

These are analytical results. A surprising numerical result is that
the many-body ground state changes only slightly if other
pseudopotentials $V_3,V_5,\dots$ are 'turned on' up to their Coulomb
values (Fig. \ref{fig-ch02-04}a). This has been confirmed by Haldane and Rezayi
\cite{haldane:01:1985} (later also by others,
e.g. \cite{fano:08:1986}) by calculating the overlap between the real
ground state and the Laughlin state for different sets of $V_m$. It is
also shown in \cite{haldane:01:1985} that if $V_1$ is lowered beyond
some critical value (while keeping other pseudopotentials on their
Coulomb values), the gap collapses rendering the ground state
compressible. These observations have been systematised by W\'ojs and
Quinn \cite{wojs:02:2000}: they argued that both Coulomb and
short-range (\ref{eq-ch02-14}) interactions belong to the
same class of {\em superharmonic} pseudopotentials where particles try
to avoid low $m$ pair states, implying the Laughlin ground state. A
drastic change in the ground state occurs first when we leave the
mentioned class of interactions. The 'superharmonicity' means roughly
that $V_m$ decays fast enough with growing $m$ \cite{wojs:02:2000}. 
The aforementioned collapse of the gap is then no longer surprising as
the decreased value of $V_1$ will eventually violate potentials
superharmonicity between $m=1$ and $3$.

An advantageous property of the interaction (\ref{eq-ch02-14}) is that
it is effectively non-parametric, the only present parameter $V_1$
determines only the
overall scaling of the energy scale within the lowest Landau level.

These results can be summarized by stating that the short-range
interaction is the component of a realistic interaction which
determines almost completely the properties of the $\nu=1/m$
spin-polarized ground states.

Obviously, for non-fully spin polarized systems it is not possible to
keep $V_1\not= 0$ only. Electrons with equal spin are still  closest in
the state $m=1$ (with energy $V_1$), electrons of unlike spin
however are closest in the state $m=0$ with energy $V_0$. Such a model
is obviously not as elegant as in the former case, it contains two
parameters $V_0$, $V_1$ whose ratio cannot be factored out of the
Hamiltonian. An alternative might be the potential
\begin{equation}\label{eq-ch02-15}
  \{V_0,V_1,V_2,\dots \} = 
  \{\infty , 1, 0, 0, 0, \ldots \}\,.
\end{equation}
%

Another typical model potential presented by Haldane and Rezayi
\cite{haldane:03:1988} was inspired by the low value of $V_0$ in the first
Landau level compared to the lowest Landau level
(Fig. \ref{fig-ch02-04}). They suggested the hollow-core potential 
%
$$
  \mbox{hollow-core interaction: }\{V_0,V_1,V_2,\dots \} = 
  \{0 , 1, 0, 0, 0, \ldots \}\,.
$$
%
and tried to explain the even-denominator fractional quantum Hall
effect at $\nu=\frac{5}{2}$ using this interaction.

 
\subsubsection{An alternative definition of Haldane pseudopotentials}

Haldane introduced the quantities $V_m$ originally for interacting
electrons on a sphere \cite{haldane:08:1983}. In that case, or for
electrons in a plane, $m$ can be
identified with the relative angular momentum of the electron pair,
having in mind that $m$ is closely related to the average separation
between the particles increases. In contrast to that, rotational
symmetry of the configuration space is lost on a torus and angular
momentum is no longer a good quantum number. Here we will introduce an
alternative definition of Haldane pseudopotentials which is applicable
also for particles on a torus
\cite{trugman:04:1985}.

First, recall that matrix elements of the {\em Coulomb} interaction
on a torus can be conveniently evaluated in Fourier space
(Subsec. \ref{pos-ch02-10}) where 
$$
  \quad V(\vek r) = e^2/|\vek r|\qquad \Rightarrow
  \qquad V(\vek q) = {e^2}/{|\vek q|}\,.
$$
Second, consider a general (radial, bounded) interaction with its Fourier
transforms $V=V(|\vek q|)$ and expand $V(|\vek q|)$ into a Taylor
series. Owing to $V(\vek r)=V(-\vek r)=V(|\vek r|)$, the series will
be free of odd powers $q^{2k+1}$. Now, go back to the direct
space and use $\Ft [f(r)]^{(k)} = (iq)^k \Ft f(r)$
\begin{equation} \label{eq-ch02-06}
  V(q) = \wt{v}_0 + \wt{v}_2 q^2 + \wt{v}_4 q^4 + \ldots \qquad
  \Rightarrow \qquad
  V(r) = \wt{v}_0 \delta(r)- \wt{v}_2 \nabla^2 \delta(r) 
         + \wt{v}_4 \nabla^4 \delta(r) - \ldots\,.
\end{equation}
The coefficients $\wt{v}_i$ now fully characterize the particle-particle
interaction. In an extension to the article of Trugman {\em et al.}
\cite{trugman:04:1985}, let us show how to translate them into
$V_m$'s, i.e. interaction energy of the two-particle state in a plane
(or on a sphere) with relative angular momentum $m$.

For the evaluation of $V_m = \bra{\psi_m} V(r) \ket{\psi_m}$, 
let us take the functions $\psi^m_{rel}$ from the planar
system, (\ref{eq-ch02-05}) plus normalization. If ${V(q)=q^{2k}}$ then 
\begin{equation}\label{eq-ch02-07}
  V_m = (-1)^k\int \d r^2 \psi_m \psi_m^* \nabla^{2k} \delta(\vek r) =
  \frac{(-1)^k}{2^m m!} \left[
  \left(\frac{1}{r} \frac{\d}{\d r}r\frac{\d}{\d r}\right)^k r^{2m}
  e^{-r^2/2} \right]_{r=0}\hskip-.1cm
\end{equation}
This is a unique prescription of how an interaction of the type
$V(q)=q^{2k}$ can be transcribed into the terms of $V_m$. Table
\ref{tab-ch02-01} contains these coefficients for
several lowest powers of $q$. Note that $V_m=0$ for $m>k$.

\begin{table}
\begin{center}
\begin{tabular}{l||rrrrrr}
           & $V_0$       & $V_1$       &$V_2$&$V_3$&$V_4$&$V_5$ \\ \hline\hline
$q^0$ $(k=0)$& $1\cdot 2^0$ &   0         &  0         & 0          & 0&0\\
$q^2$ $(k=1)$& $1\cdot 2^1$& $-1\cdot 2^1$&  0         & 0          & 0	   &0\\
$q^4$ $(k=2)$& $2\cdot 2^2$ & $-4\cdot 2^2$& $2\cdot 2^2$  & 0          &0&0\\
$q^6$ $(k=3)$& $6\cdot 2^3$& $-18\cdot 2^3$&$18\cdot 2^3$ &$-6\cdot 2^3$&0&0\\
$q^8$ $(k=4)$& $24\cdot 2^4$&$-96\cdot 2^4$& $144\cdot 2^4$&$-96\cdot2^4$& 
                                                             $24\cdot 2^4$&0\\
\end{tabular}
\end{center}
\caption{Values of Haldane pseudopotentials corresponding to
particle-particle interactions of the type $V(q)=q^{2k}$. These
values are additive, e.g. $V(q)=-\frac{1}{2}q^2+1$ corresponds to
the 'hollow core interaction': $\{V_m\}=\{0,1,0,0,0,\ldots
\}$.}\label{tab-ch02-01}
\end{table}

In conclusion, an interaction potential defined by some particular set of
values of Haldane pseudopotentials $V_m$ can be recalculated into the
coefficients $\wt{v}_i$ in (\ref{eq-ch02-06}) (Taylor series of
$V(q)$) using Table
\ref{tab-ch02-01} or, more generally using (\ref{eq-ch02-07}).

Again, several remarks should be made.

(1) The expansion in (\ref{eq-ch02-06}), being first suggested by Trugman
    and Kivelson \cite{trugman:04:1985}, looks unusual. In the
    distributional sense, we say that a non-zero ranged potential
    $V(r)$ can be written as a sum of terms with 'zero range'.

    Instead of a $\delta$-function imagine rather a sharp peaked
    function $\delta_b$, a Lorentzian of width $b$, for
    instance. Functions $\nabla^{2k} \delta_b(\vek r)$ will then have
    'the longer the range the higher the $k$ is': 
    it is instructive to draw a sketch of the first few functions
    $(\delta_b)^{(2k)}$ and consider as 'range' the position of the
    local extreme which is the most distant from the origin.
    In this sense,
    (\ref{eq-ch02-06}) is an expansion of $V(q)$ in terms of
    'increasing ranges'. 

(2) When calculating the Coulomb matrix elements for particles on a
    torus (Subsec. \ref{pos-ch02-10}), we do not use the full
    function $V(q)$ but only its values in discrete 'lattice' points $\vek
    q$. This is obviously due to the periodic boundary conditions.
    In particular, $\vek q=0$ is missing among these points. 

    Thus, we need not worry about the long-rangedness of the Coulomb
    potential, $V(q\to 0)\to\infty $ which renders it unexpandable
    into a power series of $q$. Instead of $1/q$ we may
    imagine to have considered any other polynomial in $q$ which
    matches the values of $1/q$ at the 'lattice' points. Both
    interactions must lead to the same results.

(3) {\em Example:} consider two electrons  
    in the lowest Landau level interacting via
    $V(q)=\alpha q^2$. Eigenstates sorted according the to increasing value of
    the particle-particle distance $\langle r\rangle$
    may be indexed by an integer, say $m$. The state $m=0$ will
    have an energy of $-\alpha$, the state $m=1$ will have an energy of
    $\alpha$ and all other states (with larger interparticle
    separation) will have zero energy.

    The state with $m=0$ will have a symmetric wavefunction and will
    be thus prohibited for electrons with equal spins. Thus there will be
    only one state with non-zero energy for this case and it is the
    state with the lowest interparticle separation. The potential 
    $V(q)=\alpha q^2$ defines therefore a hard-core interaction.

\subsubsection{Short-range interaction on a torus}
\label{pos-ch02-11}

The decomposition of the Coulomb interaction (in the lowest LL) into
the set of Haldane pseudopotentials has already been shown in
Fig. \ref{fig-ch02-04}. This is also the spectrum of two
Coulomb-interacting particles on a sphere.

Let us now consider a pair of particles on a torus,
Fig. \ref{fig-ch02-04}a. The index $m$ is no longer the angular momentum
of the pair as this is not a good quantum number. The wavefunctions
$\psi_{rel}^m(z)\propto z^m\exp(-|z|^2/4\ell_0^2)$ in
(\ref{eq-ch02-05}) must be modified, in order to
comply with the periodic boundary conditions.

\begin{figure}
\begin{center}
\begin{tabular}{p{3.5cm}c}
\hskip2cm (a) & (b) \\
\leavevmode\raise5cm\hbox{\unitlength=1mm\includegraphics[scale=0.3,angle=-90]%
{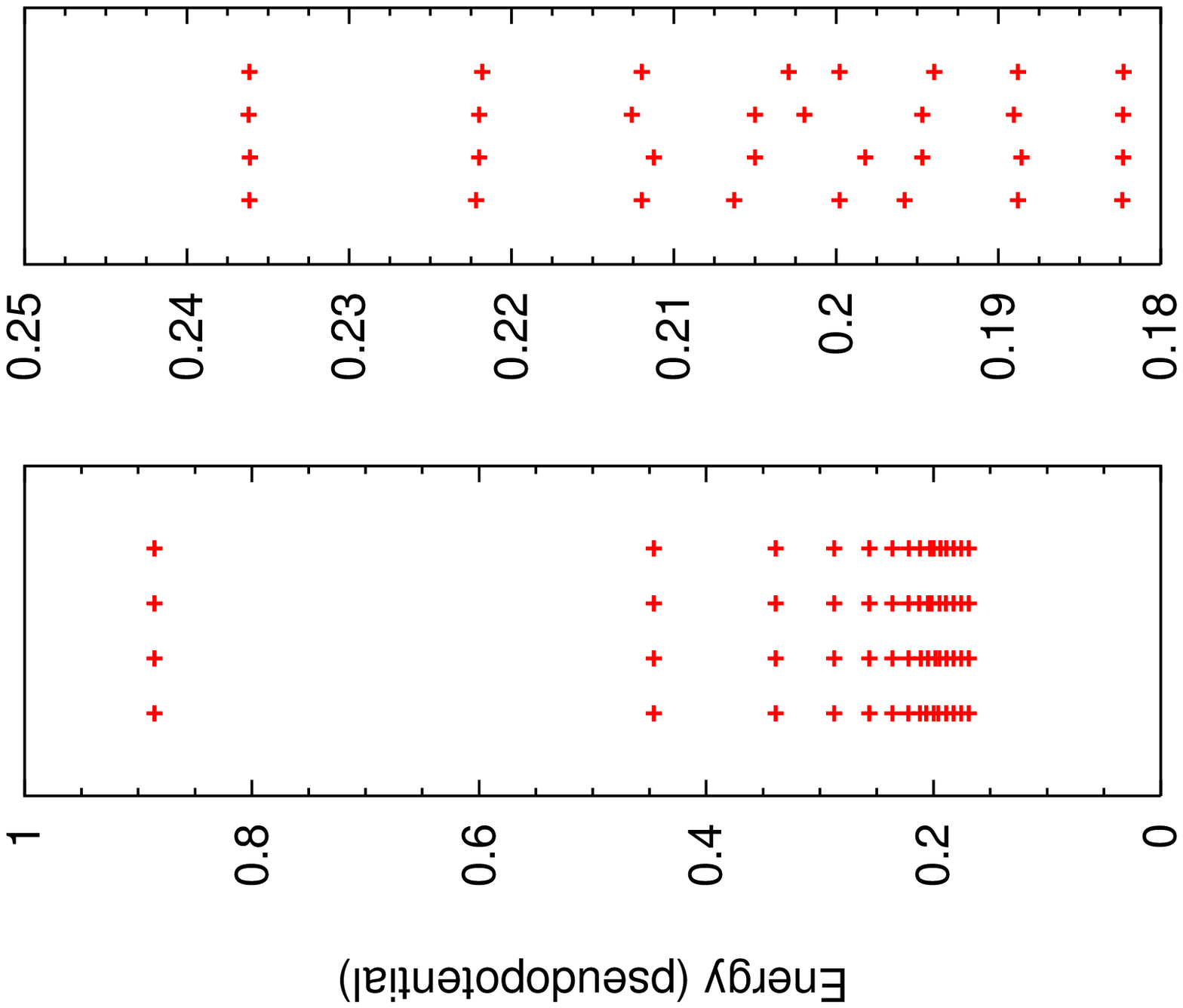}} 
\put(-182,110){$\uparrow$ {\ttfont st01} }
\put(-182,65){$\downarrow$ {\ttfont st03} }
\put(-182,15){$\uparrow$ {\ttfont st28,} }
\put(-182,5){\phantom{$\uparrow$} {\ttfont st29} }
\put(-121,55){{\small $\downarrow$\ttfont st17} }
\put(-124,21){{\small $\nearrow$\ttfont st21} }
&
\includegraphics[scale=0.6]{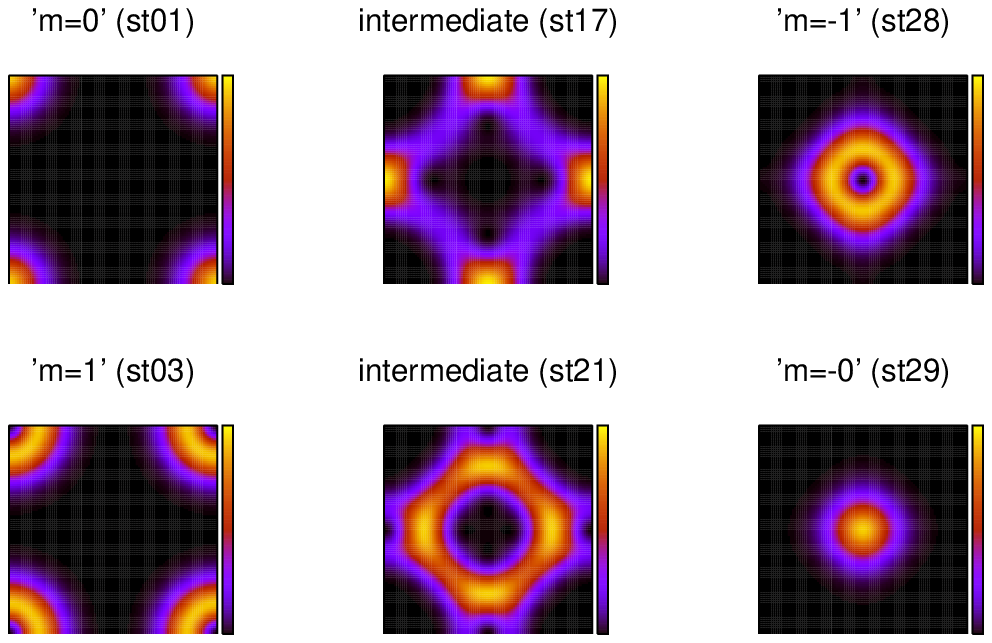} 
\end{tabular}
\end{center}
\caption{Two-particle eigenstates of Coulomb-interacting particles
in the lowest LL confined to a torus. States are shown
irrespective of the symmetry or antisymmetry of the wavefunction. {\em
Left:} spectrum (horizontal axis has no meaning: levels are 'randomly'
distributed into four groups in order to show degeneracies).  
{\em Right:} relative part of the wavefunction, 
$|\psi^m(\vek{r}_{rel})|^2$, for several states. Here, $\vek{r}_{rel}=0$
corresponds to the corner of the square (the four corners are identical due
to periodic boundary conditions). Some of these states resemble the
eigenstates of relative angular momentum, see text.}\label{fig-ch02-06}
\end{figure}

In Fig. \ref{fig-ch02-06} we show some of the
wavefunctions corresponding to the relative motion on a torus of size
$N_m=30$ ($=ab/2\pi\ell_0^2$, Subsec. \ref{pos-ch02-05}).
We will denote them simply by $\psi^m$, 
$m=0,1,\ldots,m_{max}$ and skip all other indices which would be
appropriate, e.g. to indicate
that they depend on the size of the torus ($a\times a$).
Even though these states are more complicated than
those in (\ref{eq-ch02-05}), they can still be sorted according to
growing values of $r_m=\bra{\psi^m} r\ket{\psi^m}$. It is not
surprising that the states $\psi^m$ for low $m$ ($\ll N_m$),
Fig.~\ref{fig-ch02-06}b, look very
similar to the eigenstates of angular momentum $m$ for infinite
systems (\ref{eq-ch02-05}). First when $r_m$ becomes comparable to the
system size, deviations from the circular form of $|\psi|^2$ occur
(middle column of Fig. \ref{fig-ch02-06}b). It is an
intriguing property of the periodic boundary conditions that the
states with very high $m$ look very similar to those with very low
$m$. If we fix one electron to $\vek r=(0,0)$, then the second
electron orbits around $(0,0)$ at a distance $r_m$
in the state $\psi^m$, whereas in the state $\psi^{m_{max}-m}$ 
it orbits around $(a/2,a/2)$ at the same distance.
This can be seen by comparing the left and 
right columns of Fig.~\ref{fig-ch02-06}b.

Now, with $\psi^m$, as a substitute for the relative angular momentum
eigenstates, we can define Haldane pseudopotentials on a torus by
$V_m=\bra{\psi^m} V(r_{rel})\ket{\psi^m}$. Their values
(Fig. \ref{fig-ch02-06}a) are almost equal to $V_m$ in a plane, as
long as $\psi^m$ is not affected by the periodic boundary conditions
(Fig. \ref{fig-ch02-04}a) i.e. for small values~of~$m$.

A reasonable model mimicking the short-range interaction keeps
the first two energies of the spectrum in Fig. \ref{fig-ch02-04},
i.e. the pseudopotentials $V_0$, $V_1$ at their 'Coulomb' values while
setting the other ones to zero. Table \ref{tab-ch02-01} gives a
prescription how to encode such an interaction into $V(q)$. We thus
arrive at an interaction potential defined by $V(q)=0.34q^2 -1.51$ which is
used throughout this work to model a short-range interaction unless
something else is explicitly stated.

\subsection{Composite fermion theories}


\label{pos-ch02-17}

Let us recall the observation from Subsec. \ref{pos-ch02-08}:
three zeroes are bound to each electron
in the Laughlin state $\Psi_L$. A good way to see this is to 
fix the positions of $z_2,\ldots,z_n$
and use the last 'free' coordinate $z_1$ to inspect (the zeros of) the
wavefunction. One zero is required 
by the Pauli principle (when $z_1=z_2$, the wavefunction must vanish),
the others are 'voluntary'. 

Whenever an electron goes once around a zero in $\Psi_L$, the wavefunction
acquires a phase equal to the Aharonov-Bohm phase corresponding to
one magnetic flux quantum. From this point of view, the Laughlin state
can be interpreted as the $\nu=1$ state where two magnetic flux 
quanta are attached to each electron. These objects (electron dressed
by two flux quanta) are called {\em composite fermions}
(CF).
Note however that the precise definition of a composite fermion may vary in
different theories as will be explained below.

Intuitively, this concept explains the existence of a gapped ground
state at filling factor $\nu=\ot$. Originally, there are {\em three} flux
quanta per electron (\ref{eq-ch02-38}) and the huge Hilbert space
of many-electron states in the lowest Landau level is completely degenerate
without interaction. In other words, we expect no gap without interaction.
If we now assume, that the Coulomb interaction leads to the formation of
composite objects, an electron and {\em two} flux quanta, then there remains
only {\em one} free flux quantum per CF. This 
in turn implies the filling factor of 
$\nu_{CF}=1$ for CF (\ref{eq-ch02-38}).
We know that in this case the ground state of particles obeys Fermi
statistics (see comment \cite{comm:ch02-03})
and is gapped. If a Landau level is completely filled, then
any, even infinitesimal, excitation requires promoting at least 
one CF into a higher CF Landau level which costs the finite 
energy $\ge\hbar\omega_{CF}$. Now let us describe some of the
current composite fermion theories in a little more detail.

\subsubsection{Chern-Simons transformation}

Looking at the Laughlin wavefunction in the way sketched above, 
we might find it reasonable to incorporate the flux attachment
into the Hamiltonian. 

The Chern-Simons (CS) transformation is just a gauge transformation of
the magnetic field
\begin{equation}\label{eq-ch02-58}
   \vek{a}_{CS}(\vek{r}) = \alpha\Phi_0 \int \d^2 r_1
   \frac{\unitVec{z}\times (\vek{r}-\vek{r}_1)}{|\vek{r}-\vek{r}_1|^2}
   \Psi^\dag(\vek{r}_1)\Psi(\vek{r}_1)\,.
\end{equation}
It does not change the magnetic field ('gauge
transformation') felt by the 
electrons only owing to the fact that two electrons
cannot be simultaneously on the same place (see comment \cite{comm:ch02-04}). 
The price for this is that the
transformation is singular, $\vek{a}_{CS}$ diverges for $\vek
r=\vek{r}_i$. The objects $\Psi^\dag(\vek r)$ are the one-electron field
operators and $\alpha$ is the number of attached magnetic fluxes.

After this transformation the full Hamiltonian 
%
\begin{equation}\label{eq-ch02-39}
    H = \frac{1}{2m}\int \d^2 \vek{r} \Psi^\dag(\vek{r})
        \left[-i\hbar \nabla_r + e\vek{A}(\vek{r}) -
	       e\vek{a}_{CS}(\vek{r})\right]^2 \Psi(\vek{r})
\end{equation}
contains -- apart from one-particle terms -- two-particle terms
(those containing $\vek{a}_{CS}$) and also three-particle terms 
$\Psi^\dag(\vek{r})\Psi(\vek{r})\Psi^\dag(\vek{r}_1)
\Psi(\vek{r}_1)\Psi^\dag(\vek{r}_2)\Psi(\vek{r}_2)$ 
(they originate from $\vek{a}_{CS}^2$). The CS transformation alone 
thus does not really simplify the original Hamiltonian.

A mean field approximation can be made at this point where the density
operator $\Psi^\dag(\vek{r}_1)\Psi(\vek{r}_1)$ in $\vek{a}_{CS}$ is
replaced by the mean value $n_S$. We arrive at a single particle
problem with an effective magnetic field $B_{CF}=B-\alpha\phi_0n_S$.
In illustrative terms:
\begin{center}
  \begin{tabular} {p{3cm}cp{3.8cm}cp{3.3cm}}
                      & \small CS transf.         & 
   & \small mean field&  \\
  a many-body system at $\nu=\ot$ & $\longrightarrow$ 
  & a very complicated many-body problem at $\nu=\ot$ &  $\longrightarrow$ 
  & a simple one-particle problem at $\nu=1$
  \end{tabular}
\end{center}
The final one-particle problem at $\nu=1$ has a
non-degenerate ground state, the lowest Landau
level fully occupied by CF. 
By means of this procedure we thus circumvented the original
problem that the Coulomb interaction must select the ground state out
of the vast number of degenerate many-body $\nu=\ot$ states within the
lowest LL.

A mean field approximation is not the only possible treatment
of the Hamiltonian (\ref{eq-ch02-39}). 
However, theories beyond the mean field
i.e. those treating fluctuations of the gauge field, are very
complex \cite{mariani:12:2002}.

Using the CS transformation we attach $2s=\alpha$ vortices
(not zeroes) to each electron. In the mean field approximation 
the problem is equivalent to non-interacting particles in reduced
magnetic field $B_{CF}$ which then corresponds to a filling factor
$\nu_{CF}$. It turns out that many of the experimentally observed fractions
$\nu$ (exceptions see in \cite{pan:01:2003})
correspond to integer $\nu_{CF}$. Let us conclude with an overview
of relations between quantities referring to electrons and to
CF, cf. (\ref{eq-ch02-38}).
\begin{eqnarray}\label{eq-ch02-08}
  B_{CF} = B(1-2s\nu) = B - 2s n_S \Phi_0\,,\qquad
  \ell^\ast\equiv \ell_{CF} = \frac{\ell_0}{\sqrt{1-2s\nu}}\,,\qquad\\
  \frac{1}{\nu_{CF}} = \frac{1}{\nu} - 2s\,, \qquad
  \nonumber
  \hskip4cm\quad 
  \nu = \frac{p}{2sp+1}\,,\mbox{ ($p,s$ integer).}
\end{eqnarray}

\subsubsection{Composite fermions {\em \`a la} Jain}
\label{pos-ch02-18}

Compared to the Chern-Simons transformation, Jain chooses to go in
some sense the same path but in the opposite direction 
\cite{jain:07:1989,jain:11:1994}. It
starts with a wavefunction of particles (fermions) at integer filling
$\nu_{CF}=p$, attaches $s$ zeroes ({\em not} vortices) to each particle
and, after projection into the lowest Landau level, it presents the
result as a trial wavefunction for the ground state at filling
$\nu=p/(2sp+1)$ (\ref{eq-ch02-08}). This procedure reproduces exactly the
Laughlin wavefunction and at other fractions 
it gives wavefunctions with very high overlap with ground states
calculated numerically by exact diagonalization.

There are two central reasons why this approach is very popular. On one
hand, it gives a simple single-particle picture of what is
going on in the highly correlated many-body problem. On the other hand,
it offers explicit formulae to work with since it is easy to write down
a wavefunction of $p$ full Landau levels. A very pleasant feature of
this approach is that it allows to incorporate the spin of electrons
easily \cite{wu:07:1993}. 
Take $p_\up$ of full Landau levels with spin up and $p_\dn$ of
full Landau levels with spin down.
These Landau levels are then called {\em composite fermion Landau
levels}. The magnetic field felt by the CF, i.e. the field
corresponding to filling factor $\nu_{CF}=p$ is called {\em effective
magnetic field} $B_{eff}$. It is weaker than magnetic field $B$
corresponding to the electronic state at $\nu$ (\ref{eq-ch02-08}).

Note, that the filling factors in (\ref{eq-ch02-08}) are all in range
$\nu<\frac{1}{2}$. For $\frac{1}{2}<\nu<1$, Jain {\em et al.}
\cite{wu:07:1993} suggest the idea of {\em antiparallel flux
attachment}: the effective field $B_{eff}$ is
antiparallel to the real field $B$, however, the additional flux quanta are
added in parallel to $B$ i.e. antiparallel to $B_{eff}$.
In terms of (\ref{eq-ch02-08}) this means $p\to -p$ or
$\nu=p/(2sp-1)$.

An example of candidates for ground states and their polarization
provided by Jain's composite fermion theory is given in
Tab. \ref{tab-ch02-02} (see Chakraborty \cite{chakraborty:07:2000} for 
a review regarding ground states with various spins).

\begin{table}
$$
\begin{array}{c|ccccc|cccc}
 p_{\up} & 1 & 2 & 3 & -2 & \ldots & 1 & 2 & 2 & -1 \\
 p_{\dn} & 0 & 0 & 0 &  0 &        & 1 & 1 & 2 & -1 \\ \hline
 p=p_{\up}+p_{\dn}       
         & 1 & 2 & 3 & -2 &        & 2 & 3 & 4 & -2 \\ \hline\hline
 \phantom{\displaystyle\int}\nu=p/(2sp+1)     
         & \frac{1}{3} 
             & \frac{2}{5}
	         & \frac{3}{7}
		     & \frac{2}{3}
		                  &&\frac{2}{5}
		                  & \frac{3}{7}
				      & \frac{4}{9} 
				          & \frac{2}{3} \\ \hline
 S/\frac{n}{2}&1&1&1 & 1 &        & 0 &\frac{1}{3}
                                      &\frac{1}{2} 
                                          & 0 \\
\end{array}
$$
\caption{The scheme of construction of 
 Jain's wavefunctions for CF with two flux
 quanta attached: examples of composite fermion filling factors
 ($p_{\up}$, $p_\dn$ are numbers of fully occupied spin up and spin
 down CF-Landau levels) and corresponding electronic filling
 factors. 
}
\label{tab-ch02-02}
\end{table}

\subsubsection{Composite fermions {\em \`a la} Shankar and Murthy
  (Hamiltonian theory)}

The Hamiltonian theory of FQHE (Shankar and Murthy \cite{murthy:10:2003}) 
builds on previous works of Jain and those concerning the CS transformation,
quoting words of its authors, it combines the strengths of the both theories.

It provides a projected Hamiltonian of the lowest Landau level
which scales only with the Coulomb interaction. In addition to each electron a
new independent object is introduced: a pseudovortex. Its
definition on the level of commutation relations (Eq. 129 in
  \cite{murthy:10:2003})
assures, that if an electron goes around a pseudovortex, 
it picks up the phase of
$2\pi\, 2s$ i.e. it has the same effect as an insertion of $2s$ 
flux quanta. Note however that it
is {\em not} a zero of the wavefunction. 
The projected Hamiltonian is written in coordinates which are a
combination of the electron and pseudovortex position 
(Eq. 138 in \cite{murthy:10:2003}). This combination is then
called {\em composite fermion coordinate}.

For this Hamiltonian an ansatz for a ground state can be written
down. At filling $\nu=p/(2sp+1)$, it is $p$ Landau levels filled
with CF. It is then possible to evaluate their Hartree-Fock energies. 

The first substantial success of this theory is that it produces
the correct scaling of spectra within the lowest Landau level ($\propto
\sqrt{B}$). Compared to Jain's theory, it keeps track of the fact that
the two fluxes (which sit exactly at each electron in the Laughlin
state) can be only loosely bound to electrons. This is owing to
the dynamical degree of freedom given to the pseudovortices. 
On the other hand,
the electronic coordinates are actually the only really independent
ones, for instance in the Laughlin wavefunction, all the zeroes
$(z_i-z_j)$ are expressed in terms of electronic coordinates.
Thus, the price we must pay for
the extension of the Hilbert space is that we must perform a
projection to the space of physical states at the end.

Nevertheless, this does not seem to be a substantial problem and thus the
Hamiltonian theory is probably the most advanced achievement in an
effort to understand the many-body physics in the FQHE.

\subsection{How to test the CF theory?}


\label{pos-ch02-14}

The concept of flux attachment (Sec. \ref{pos-ch02-17}) provides a
well-understandable model of the FQHE. However transparent it seems 
at the first look, predictions based on it must be tested against a model
which contains less approximations. Exact diagonalization (ED) is a
good choice for this purpose. In exchange for extensive numerics to
perform, the only substantial approximation of the method is to take a
finite instead of an infinite system.

The main part of this Section concerns the exact diagonalization (ED)
\cite{chakraborty:1995}.  It is definitely not the only numerical
method used in the context of the FQHE.  Some numerics is at the end
of nearly any method as soon as many-body problems are concerned, be
it a Hartree-Fock treatment of CFs or Monte Carlo simulations of the
Laughlin state mapped onto a one-component plasma. O

We take the complete many-body Schr\"odinger equation but confine 
the interacting electrons moving actually in an infinite plane onto a
compact (i.e. finite-sized) surface, possibly without edges. The standard
choices are a sphere \cite{haldane:08:1983},
a torus (square with periodic boundary conditions) 
\cite{yoshioka:04:1983} and a disc \cite{laughlin:05:1983}
(see Yoshioka \cite{yoshioka:2002} for an overview). Although these
manifolds are locally flat and therefore with growing system size a
convergence towards infinite-plane results can be expected, they all
break some of the symmetries of the infinite plane. For instance, the
sphere keeps the angular momentum while the torus retains the
translational symmetry. In any case, the hope is that effects
inflicted by the finite size can be separated from those generic to a
two-dimensional electron gas. Another usual yet not necessary
approximation is to neglect Landau level mixing, i.e. restriction to
the lowest Landau level only. Also note, that there is a long way from
an ideal 2D system which study here, to the experimental reality
(impurities, effective mass approximation, finite thickness of the 2D
electron gas etc.).

\subsubsection{Torus boundary conditions}
\label{pos-ch02-05}

One possibility to model an infinite plane by a finite manifold
without edges is a rectangle with area $a\cdot b$ 
with periodic boundary
conditions (PBC). Topologically, this is the same as a torus, although
it is better to stay with the former picture for the sake of twisted PBC,
even if we sometimes use the word 'torus' as a shortcut for this model.

What are the single particle states of the lowest Landau level in
this case? Recall (\ref{eq-ch02-28}) where single-particle states
complying with translational symmetry along $y$ are given
%
$$
    \psi_{0,k_y'}(x',y') = \exp(-ik_y' y') \exp[-(x'+k_y')^2/2]\,,
$$
%
primed variables are in units of magnetic length, $x'=x/\ell_0$,
$k'=k\ell_0$. Periodic boundary conditions along $y$
admit only discrete values of
$k_y'=(2\pi\ell_0/b)j$ with $j$ integer. The wavefunction is centered
in the $x$-direction around $X_j=k_y\ell_0^2$ and if we require $X_j$ to
lie within $[0;a)$, we have $0\le-k_y'<a/\ell_0$. Thus, up to a sign,
\begin{equation}\label{eq-ch02-59}
0\le j<\frac{ab}{2\pi\ell_0^2}\equiv m\,.
\end{equation}
Equation (\ref{eq-ch02-38}) 
with $L^2=ab$ implies that $ab/2\pi\ell_0^2$ is equal
to the number of magnetic flux quanta ($\Phi/\Phi_0$) which pass through
the rectangle and by virtue of (\ref{eq-ch02-59}) 
it must be an integer. This brings us to the
central insight that {\em there is only a finite number $m=N_m$ of states
in a square with periodic boundary conditions} (subject to magnetic
field and discarding all but the lowest Landau level) and that the 
size of the torus (area in units of $\ell_0^2$) 
can be measured by the number of magnetic flux quanta $N_m$ penetrating the
torus: $ab=2\pi\ell_0^2 N_m$.

States $\psi_{0,k_y'}(x',y')$ shown above are not periodic in the $x$
direction and this
can be accomplished by periodic continuation: $\psi(x,y) \to
\psi(x,y) + \psi(x+a,y) + \ldots$ . The (non-normalized) single particle
states we will be dealing with are thus
\cite{yoshioka:04:1983,yoshioka:06:1984,yoshioka:xx:1984}
\begin{eqnarray}\nonumber
\vp_j(x',y') &=& 
\displaystyle
    \sum_{k=-\infty}^{\infty} \exp\left[iy'(\frac{j}{m}+k)\zeta 
    - \frac{1}{2}\left(x'- (\frac{j}{m}+k)\zeta \right)^2\right]\,,\quad
    \zeta = \sqrt{\frac{a}{b}\cdot 2\pi m}\,, \\
  && \label{eq-ch02-40}
    \hskip5cm j=0,1,2,\ldots, m-1\,.
\end{eqnarray}
These states constitute the single-particle basis of the lowest
Landau level.

\subsubsubsection{Twisted boundary conditions} 

Consider what happens if we
require the modulus of $\psi$ rather than $\psi$ itself to be periodic,
similar to Bloch's theorem. Thus,
the wavefunction may acquire a non-trivial phase when going once
around the torus. Mathematically, this can be described using the
magnetic translation operators (\ref{eq-ch02-27})
\begin{equation}\label{eq-ch02-57}
  T(a\unitVec{x}) \psi = \exp(i\phi_x) \psi\,,\qquad
  T(b\unitVec{y}) \psi = \exp(i\phi_y) \psi\,.
\end{equation}
Fixing phases $\phi_x$, $\phi_y$, the correct (non-normalized) 
periodic single particle states are
\begin{eqnarray}\nonumber
  \vp_j(x,y) &=& \displaystyle
  \sum_{k=-\infty}^\infty \underbrace{\exp(ik\phi_x)
  t(ka\unitVec{x})}_{T(ka\unitVec{x})}
  \exp(-iX_j y/\ell_0^2+i\phi_y y/b) \exp[-(x-X_j)^2/2\ell_0^2]\,,\\
  && \label{eq-ch02-42}
  \hskip4cm X_j=\frac{j}{m}a\,,\qquad j=0,1,\ldots, m-1\,,
\end{eqnarray}
where $t(\xi\unitVec{x})$ is an ordinary translation, i.e. 
an operator transforming $\psi(x,y)$ into $\psi(x+\xi,y)$.
For $\phi_x,\phi_y=0$ the original result (\ref{eq-ch02-40}) is
recovered. This choice of $\phi_x,\phi_y$ is also used throughout this work.

{\em Interpretation of $\phi_x,\phi_y$.} By imposing the PBC 
we arrived at the statement that wavefunctions must be
centered (along $x$) at $X_j=(a/m)\cdot j$, $j=0,1,\ldots$ There is no
{\em a priori} reason for the point $x=X_0=0$ to be more important
than $x=X_{0.5}=(a/m)\cdot 0.5$ which is not among the just mentioned $X_j$'s. 
By varying $\phi_x$, the set $\{X_0,X_1,\ldots\}=(a/m)\{0,1,\ldots\}$ is
transformed into $(a/m)\{0+\phi_x/2\pi, 1+\phi_x/2\pi, \ldots
\}$. Thus, sweeping $\phi_x$ from $0$ to $2\pi$, we probe all points
between $0$ and $a$ in the $x$
direction. Independently on this, we may sweep through all $k_y$
points in the interval $[0;2\pi/b]$ by changing $\phi_y$. 
Thus, $\phi_x$ and $\phi_y$ are analogous to lattice wavevectors
within the first Brillouin zone in an ordinary periodic system
defined by ordinary rather than magnetic translations.

In summary, by considering only a finite system, we have only $m$
states to probe the whole plane (i.e. $[0,a]$ in $x$ and $[0,2\pi/b]$
in $y$). Sweeping $\phi_x$, $\phi_y$ from $0$ to $2\pi$ we can access
an arbitrary point in the plane.

Another interpretation of $\phi_x,\phi_y$ was given by Tao and Haldane
\cite{tao:03:1986} in terms of additional magnetic
fluxes. These come from two ideal anuloids (closed solenoids):
one goes inside the torus and another around the torus outside.
$(h/e)(\phi_{x,y}/2\pi)$. It was also shown
\cite{haldane:02:1985,helias:2003} that
$\phi_x$ increasing linearly in time acts as a homogeneous electric
field in $x$ direction.

\subsubsubsection{General basis of single-particle states on a torus: 
complex coordinates}

A precise discussion of one-particle states on a torus including the
phases $\phi_x,\phi_y$ was first given by Haldane and Rezayi
\cite{haldane:02:1985}. They showed that the most
arbitrary state is
\begin{equation}\label{eq-ch02-41}
  \psi(x,y) = \exp(-{\textstyle \frac{1}{2}x^2}) \cdot \underbrace{\exp(ikz)
  \prod_{l=1}^m \vt_1\left(\pi(z-z_l)/b| i \right)}_{\mbox{\small analytic}}
  \,, \qquad
  z=x +i y
\end{equation}
where $\vt_1(u|\tau)$ is an elliptic theta function
(\cite{gradshteyn:1980}, p. 921), $k$ is a real number in the range $|k|<
\pi m/b$ and $z_i$ are some fixed complex numbers within the rectangle
$[0,a]\times [0,b]$. In the terminology of (\ref{eq-ch02-42}), these
states correspond to any $j$ and any $\phi_x$, $\phi_y$. 
The most important things to know about the theta functions are that it
is analytic, that 
$\vt_1(z-z_l|i)\propto z-z_l$ for $|z-z_l|\to 0$ and that $z_l$ is
its only zero in the rectangle. In this form, it
is also clear that $m$ is equal to the number of flux quanta in the
elementary cell (the rectangle). Going once around the rectangle, the
wavefunction gathers a phase of $2\pi\times $ number of zero points
inside. That number is just $m$, each factor in (\ref{eq-ch02-41})
contributes by a single zero.

\begin{figure}
\begin{center}
\subfigure[Modulus.]{\includegraphics[scale=.8]{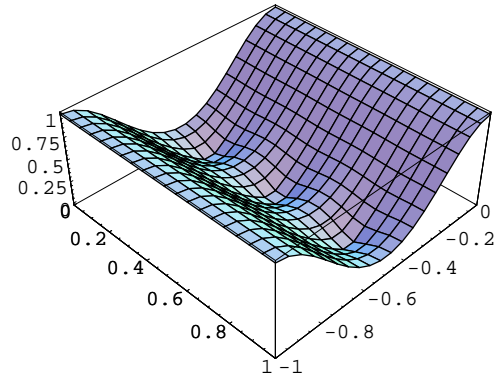}}
\hskip2cm
\subfigure[Phase.]{\includegraphics[scale=.35]{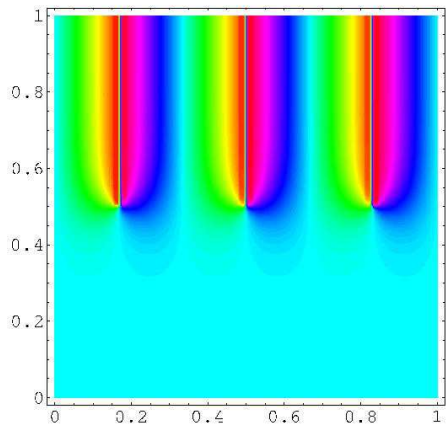}}
\end{center}
\caption{One possible one-particle state on a torus pierced by three
flux quanta (i.e. $m=3$).}
\label{fig-ch02-10}
\end{figure}

By choosing fixed $\phi_x,\phi_y$, there arise  $m$ possible choices
for the values of $k$ and $z_0=\sum_l z_l$, say $j=0,1,\ldots,
m-1$. For each pair $(k,z_0)$ we can construct one function of the
form (\ref{eq-ch02-41}) and the resulting $m$ functions will
constitute a basis of the lowest Landau
level, just as the basis in (\ref{eq-ch02-42}). There is naturally a
large freedom in choosing one particular basis. This happens by
choosing some particular position of the zero points $z_l$'s while
observing the constraint on $z_0$. The basis in (\ref{eq-ch02-42})
can be obtained from (\ref{eq-ch02-41}) by putting the zeroes on a
line, $z_l=i\cdot b l/m+j/ma$ and choosing $k=(2\pi/b)j$ for the state
$\vp_j$ with $\phi_x=\phi_y=0$. Even though it is by far not obvious
in (\ref{eq-ch02-42}), Fig. \ref{fig-ch02-10} shows a 2D plot
of one of such functions.

In principle, the wavefunctions in (\ref{eq-ch02-41}) are very similar
to those obtained in the circular gauge (\ref{eq-ch02-25}) except for
substituting
$z$ by $\vt_1(z|i)$. This is a manifestation of the fact, that even
on a torus, circular symmetry is approximately preserved at short
distances and deviations occur first when $\vt_1(z|i)$ deviates from
$z$ at larger distances. One could say, $\vt_1(z|i)$ is the
function $f(z)=z$ adapted to the torus i.e. deformed to comply with
periodic boundary conditions.

On the other hand contrary to the infinite plane, each single-electron
wavefunction on a torus has as many zeroes as there are flux quanta
passing through the torus.

\subsubsection{Many-body symmetries on a torus}
\label{pos-ch02-06}

\subsubsubsection{Center-of-mass}

What changes if we consider $n$-body states instead of
single-particle ones \cite{haldane:02:1985}? 
Given the considered Hamiltonian (\ref{eq-ch02-48}),
the most obvious symmetry is the separation of center-of-mass and
relative part of the wavefunction
\begin{equation}\label{eq-ch02-11}
  \Psi(z_1,\ldots,z_n) = \Psi_{CM}(Z) \psi_{rel}\,,\qquad
  Z=z_1+\ldots+z_n\,.
\end{equation}
The center-of-mass part is just a one-particle wavefunction. 
Hence it must have the form shown in
(\ref{eq-ch02-41}). Haldane and Rezayi \cite{haldane:02:1985} showed
that it has $q$ zeroes in the region $[0;a]\times[0;b]$ for filling
factor $\nu=N_e/(qN_e)$. Again as for
single-particle states, there are $q$ basis states for
$\Psi_{CM}$. Since the energy does not depend on the center-of-mass position
in a homogeneous system, these three states will lead to degenerate
many-body states, provided $\psi_{rel}$ remains the same.

\begin{figure}
  \subfigure[A basis leading to more 
inhomogeneous densities.]{\label{fig-ch02-11a} 
  \hskip-2cm\begin{tabular}{cccccc}
  \multicolumn{2}{c}{$(\alpha)$} & \multicolumn{2}{c}{$(\beta)$} & 
  \multicolumn{2}{c}{$(\gamma)$} \\[3mm]
    $|\Psi_{CM}(x,y)|$ & Density $n(x,y)$ &
    $|\Psi_{CM}(x,y)|$ & Density $n(x,y)$ &
    $|\Psi_{CM}(x,y)|$ & Density $n(x,y)$ \\
  \raise.7cm\hbox{\includegraphics[scale=0.33]{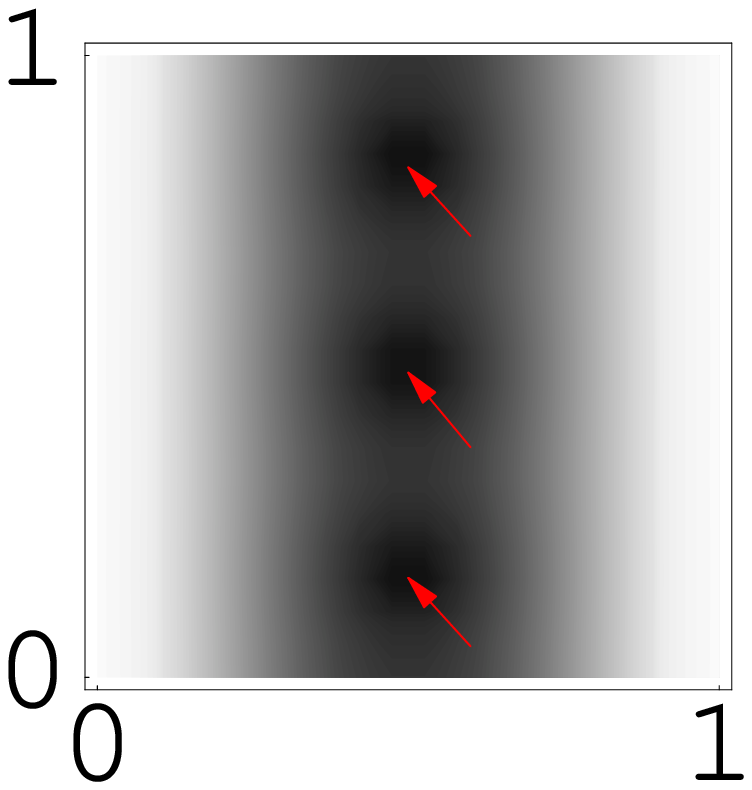}} & 
  \hbox{\hskip-3.2cm%
  \includegraphics[scale=0.5]{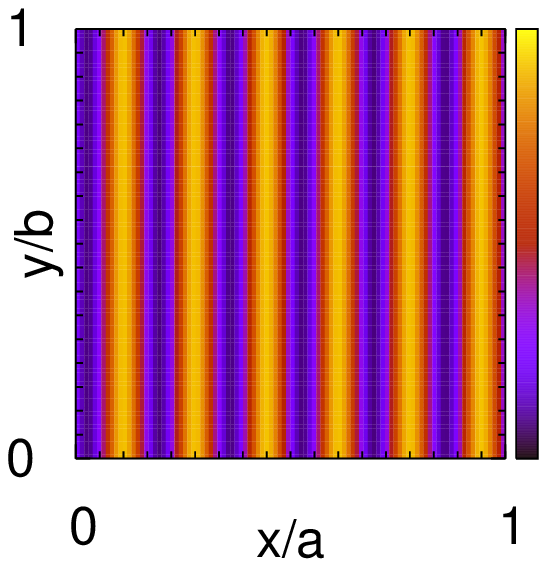}\hskip-2.8cm}&
  \raise.7cm\hbox{\includegraphics[scale=0.33]{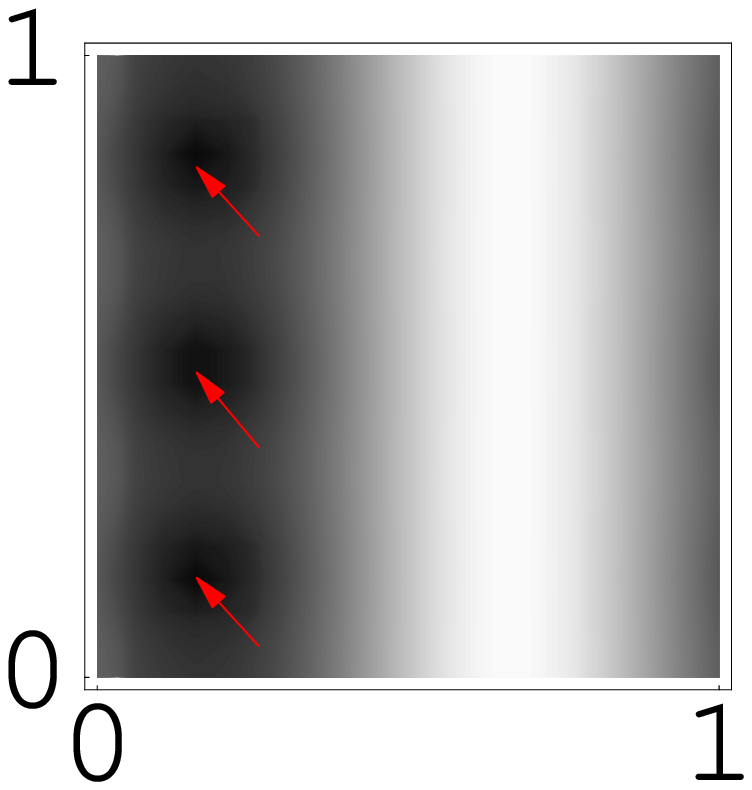}} & 
  \hbox{\hskip-3.2cm%
  \includegraphics[scale=0.5]{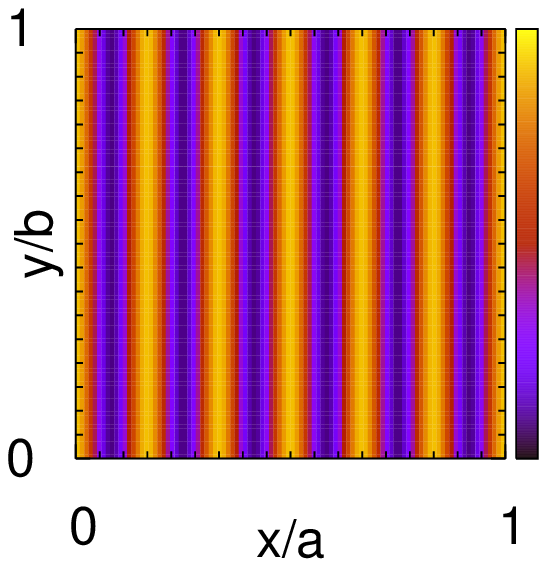}\hskip-2.8cm}&
  \raise.7cm\hbox{\includegraphics[scale=0.33]{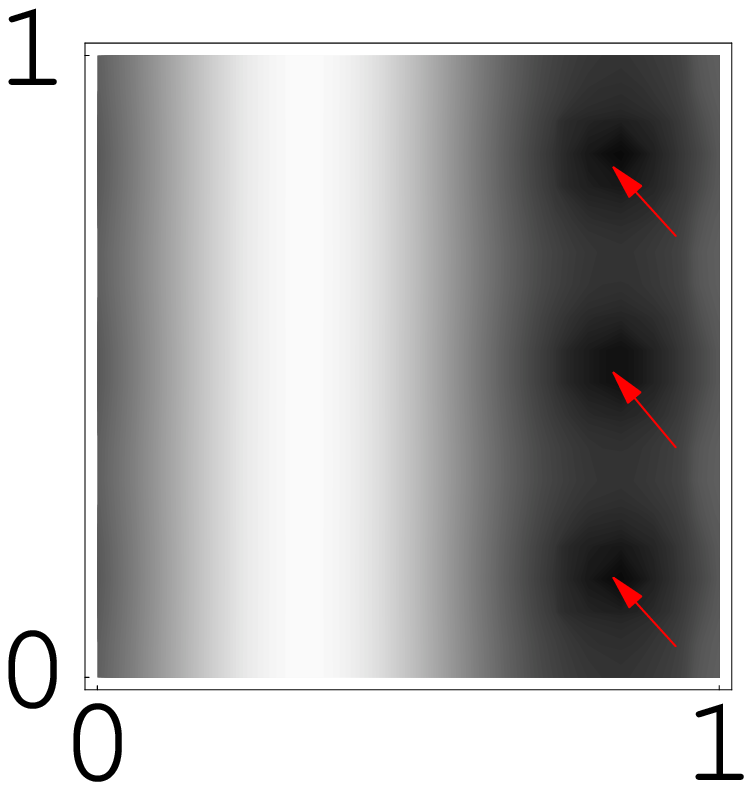}} & 
  \hbox{\hskip-3.2cm%
  \includegraphics[scale=0.5]{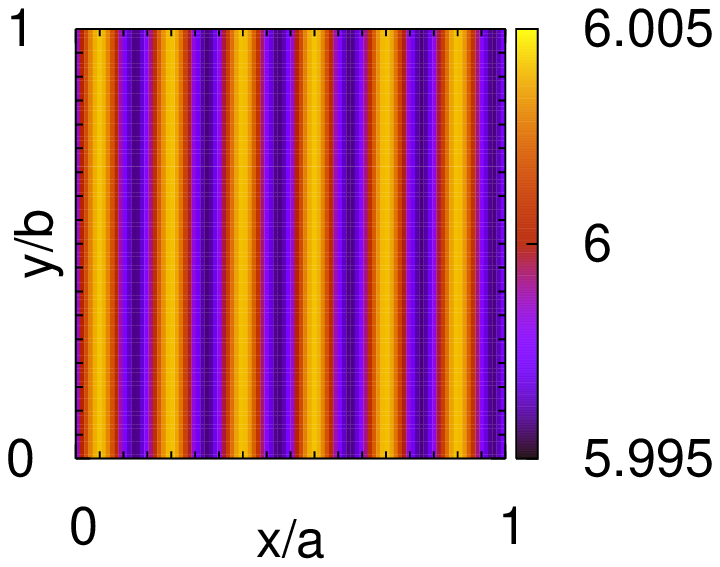}\hskip-2.8cm} 
  \\[-5mm]
  \multicolumn{2}{c}{\includegraphics[scale=0.4]{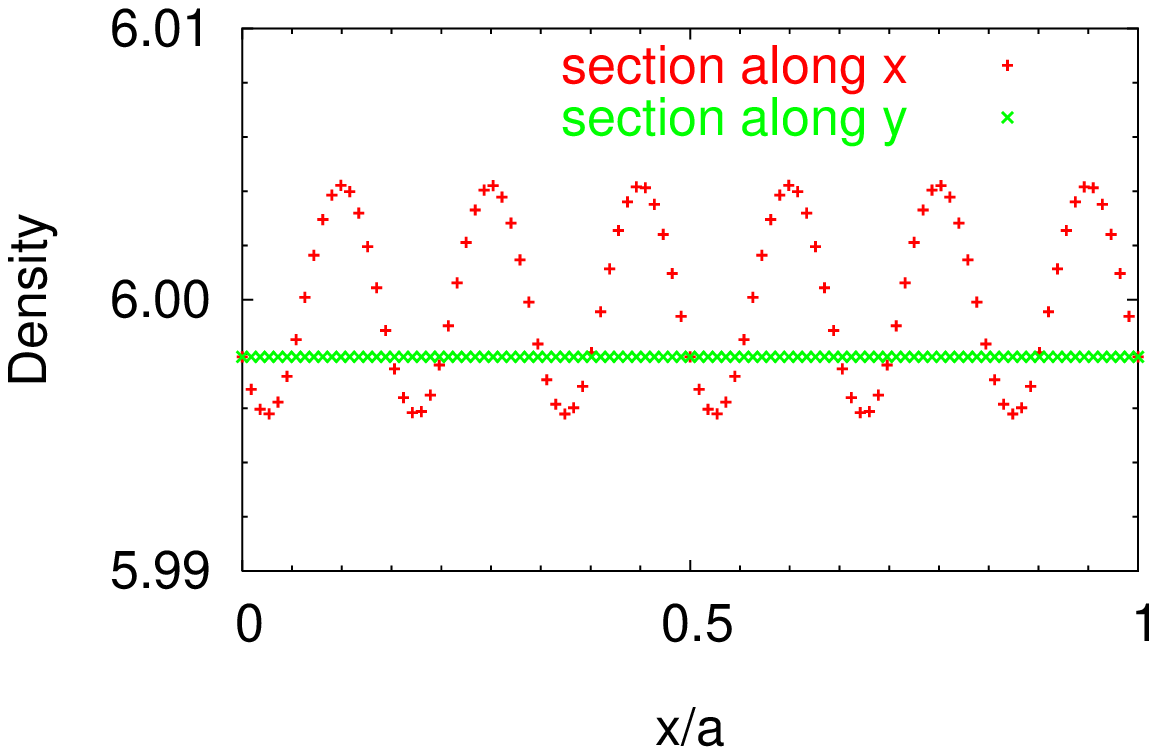}}&
  \multicolumn{2}{c}{\includegraphics[scale=0.4]{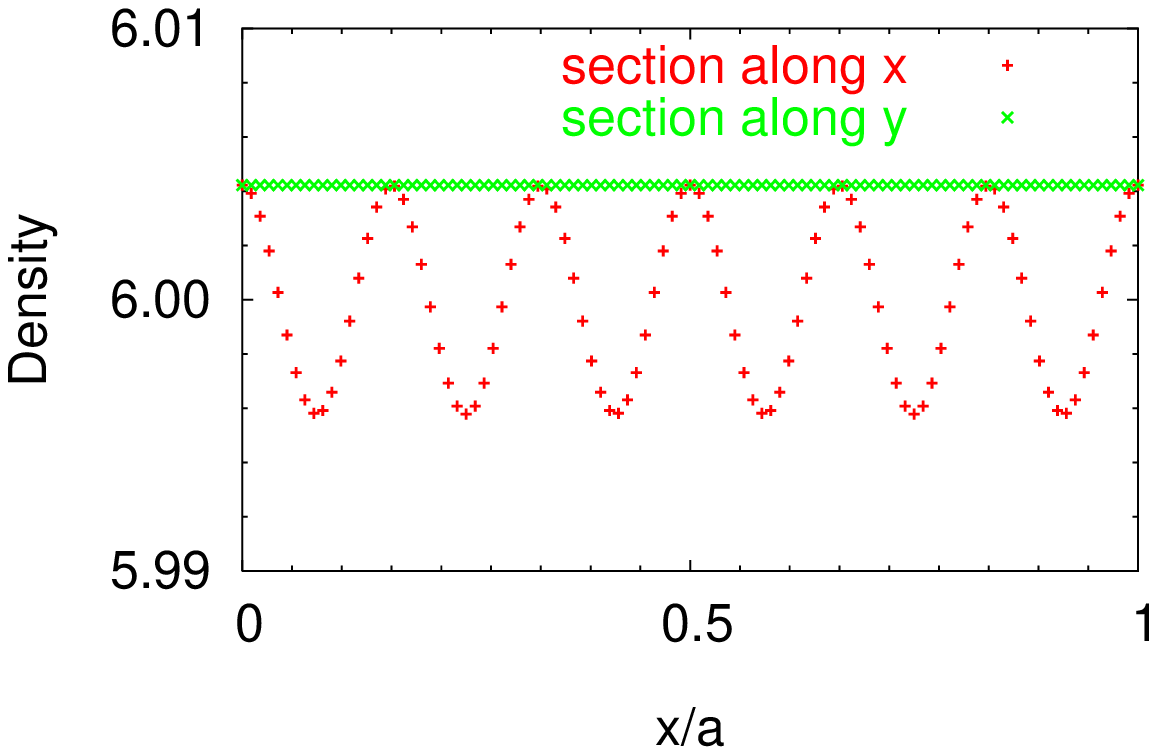}}&
  \multicolumn{2}{c}{\includegraphics[scale=0.4]{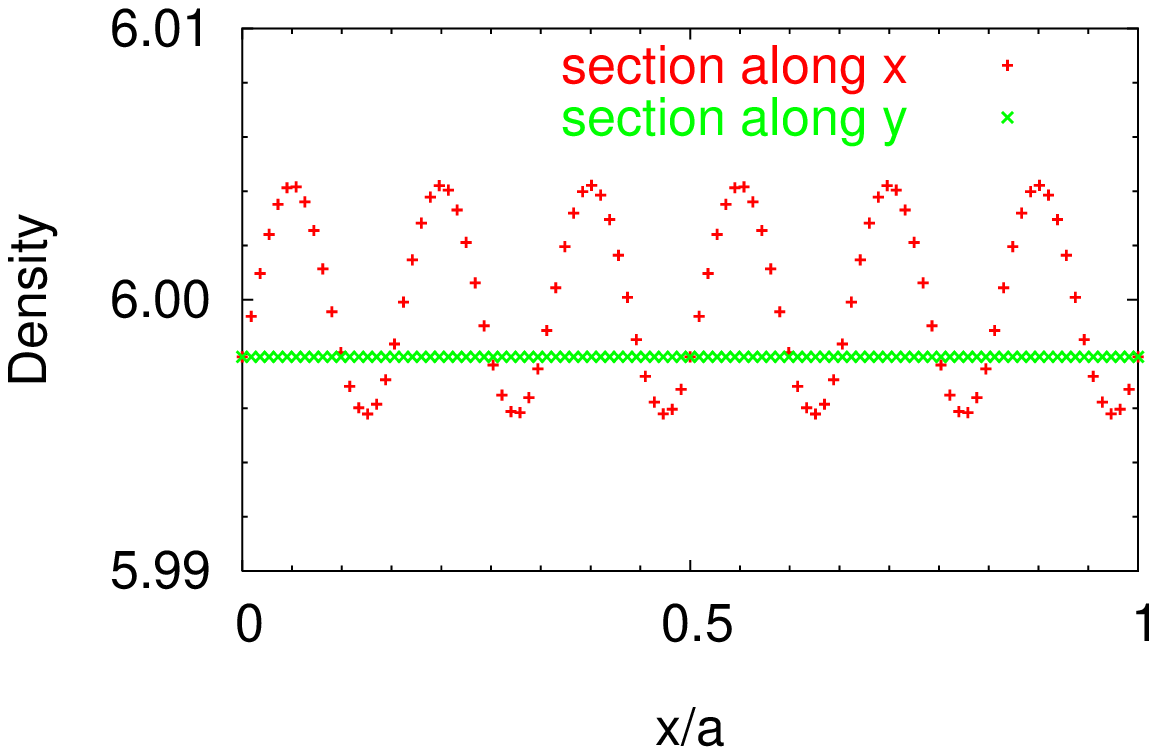}}
  \end{tabular}
}

  \subfigure[A basis leading to less inhomogeneous densities.]{%
  \hskip-2cm\begin{tabular}{cccccc}
  \multicolumn{2}{c}{$(\alpha)$} & \multicolumn{2}{c}{$(\beta)$} & 
  \multicolumn{2}{c}{$(\gamma)$} \\[3mm]
    $|\Psi_{CM}(x,y)|$ & Density $n(x,y)$ &
    $|\Psi_{CM}(x,y)|$ & Density $n(x,y)$ &
    $|\Psi_{CM}(x,y)|$ & Density $n(x,y)$ \\
  \raise.7cm\hbox{\includegraphics[scale=0.33]{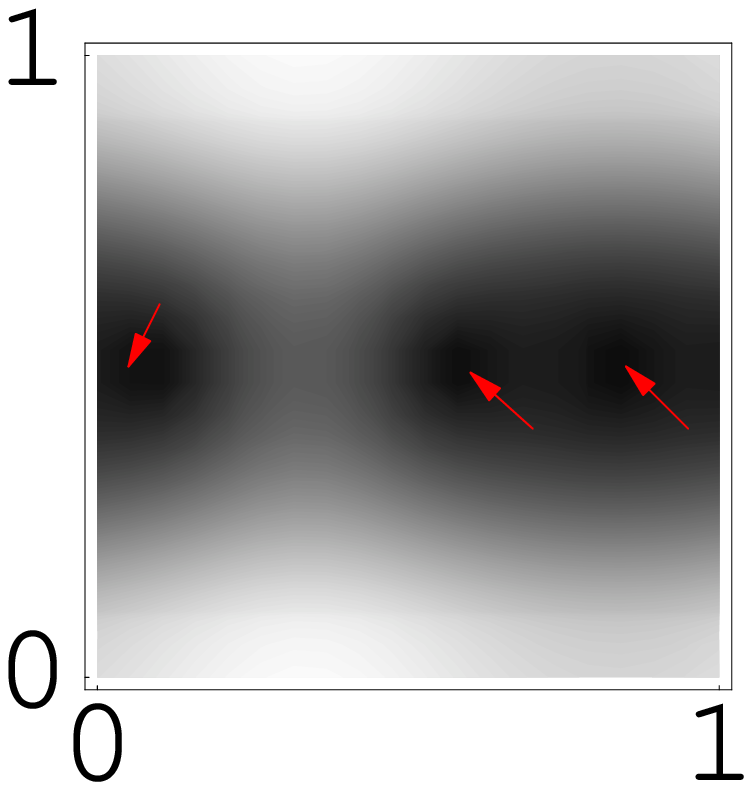}} & 
  \hbox{\hskip-3.2cm%
  \includegraphics[scale=0.5]{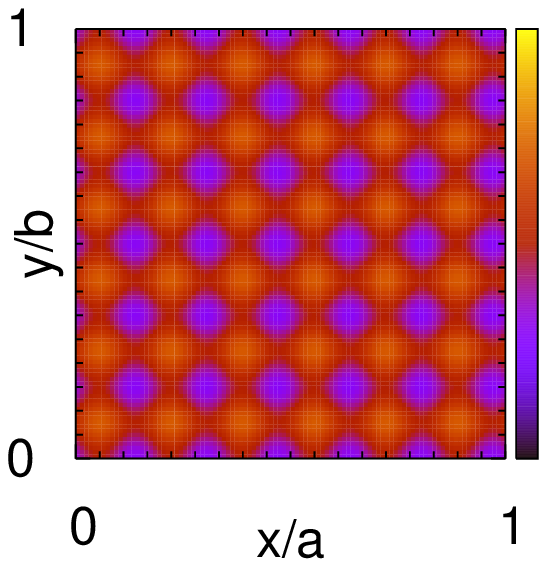}\hskip-2.8cm}&
  \raise.7cm\hbox{\includegraphics[scale=0.33]{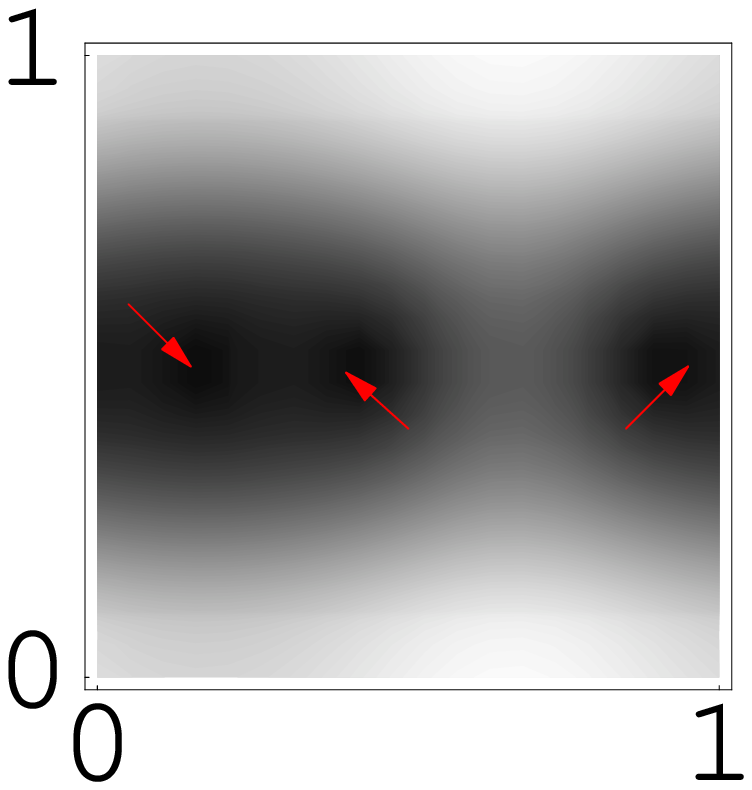}} & 
  \hbox{\hskip-3.2cm%
  \includegraphics[scale=0.5]{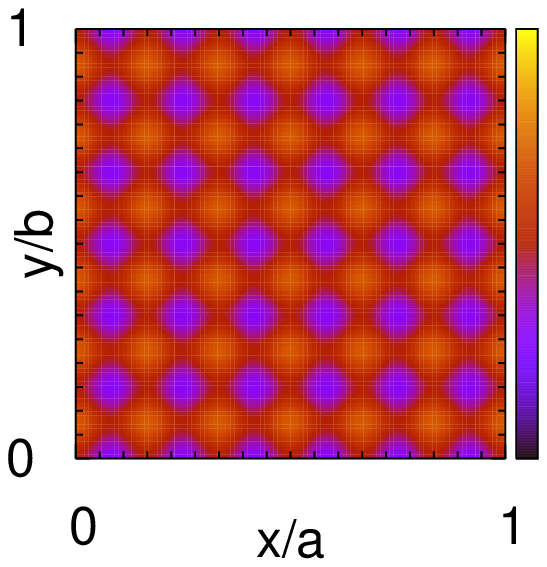}\hskip-2.8cm}&
  \raise.7cm\hbox{\includegraphics[scale=0.33]{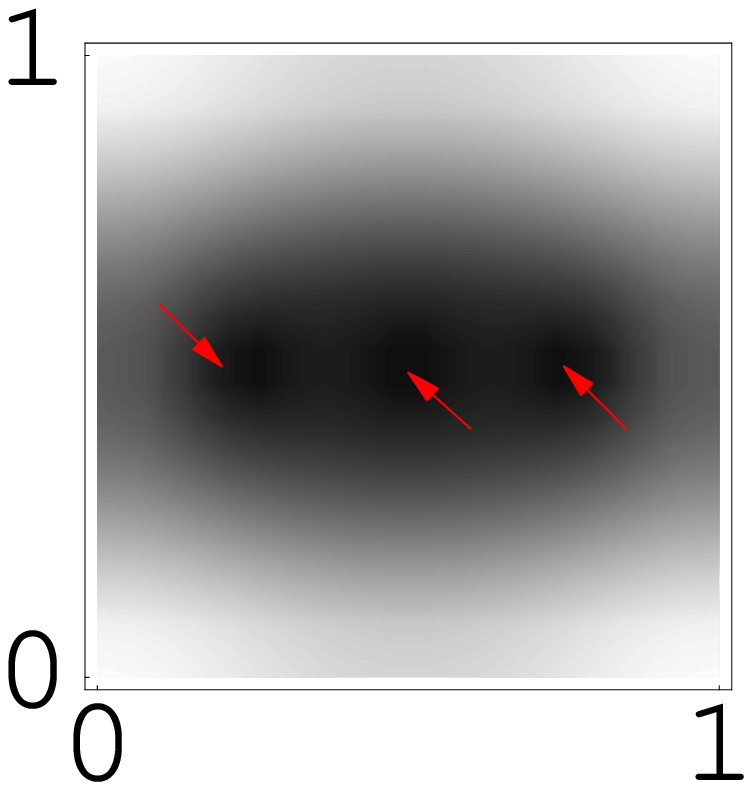}} & 
  \hbox{\hskip-3.2cm%
  \includegraphics[scale=0.5]{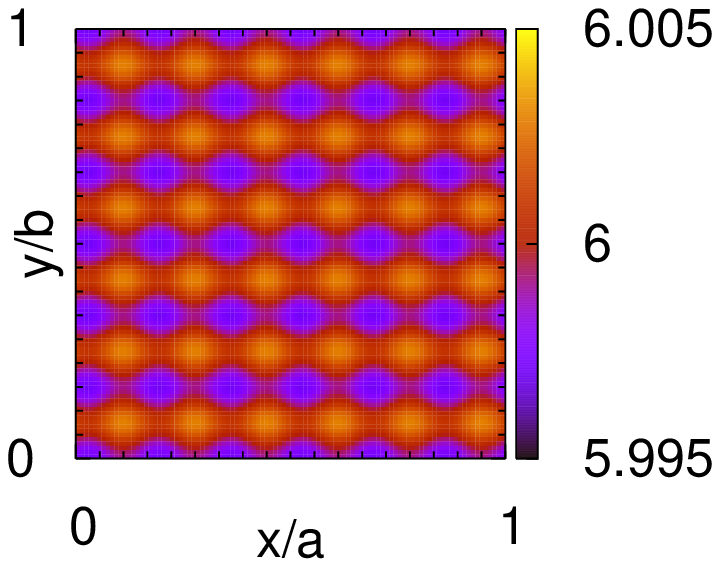}\hskip-2.8cm} 
  \\[-5mm]
  \multicolumn{2}{c}{\includegraphics[scale=0.4]{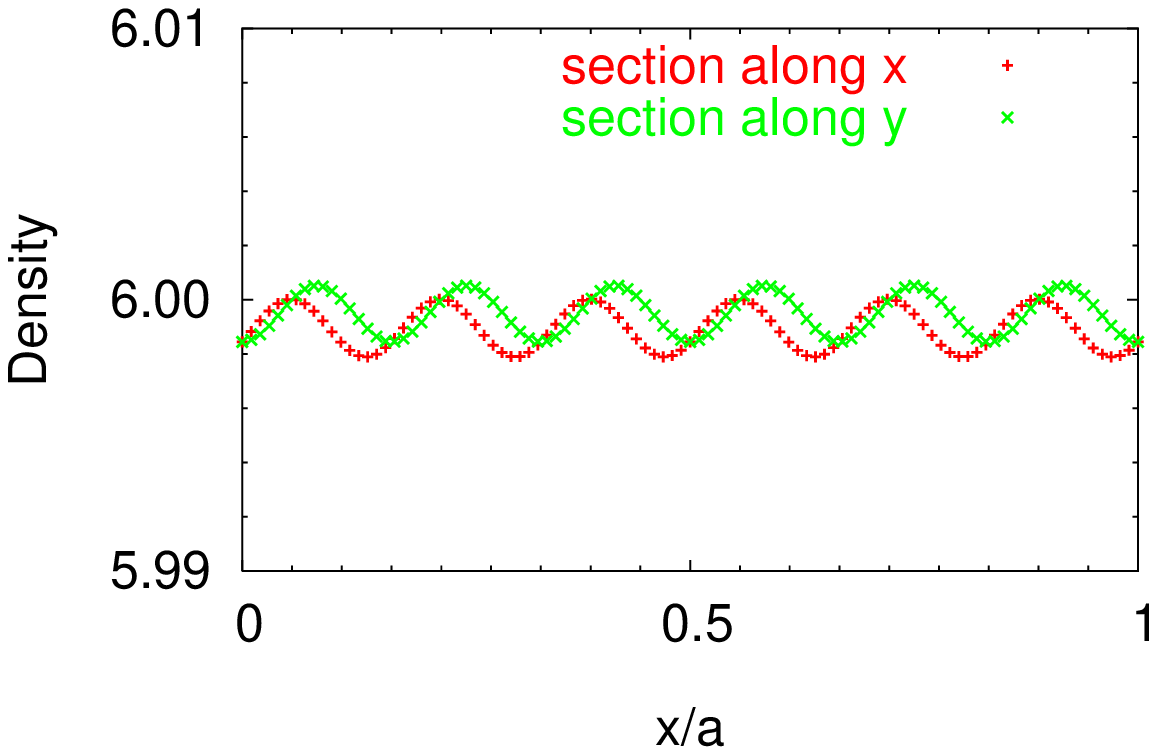}}&
  \multicolumn{2}{c}{\includegraphics[scale=0.4]{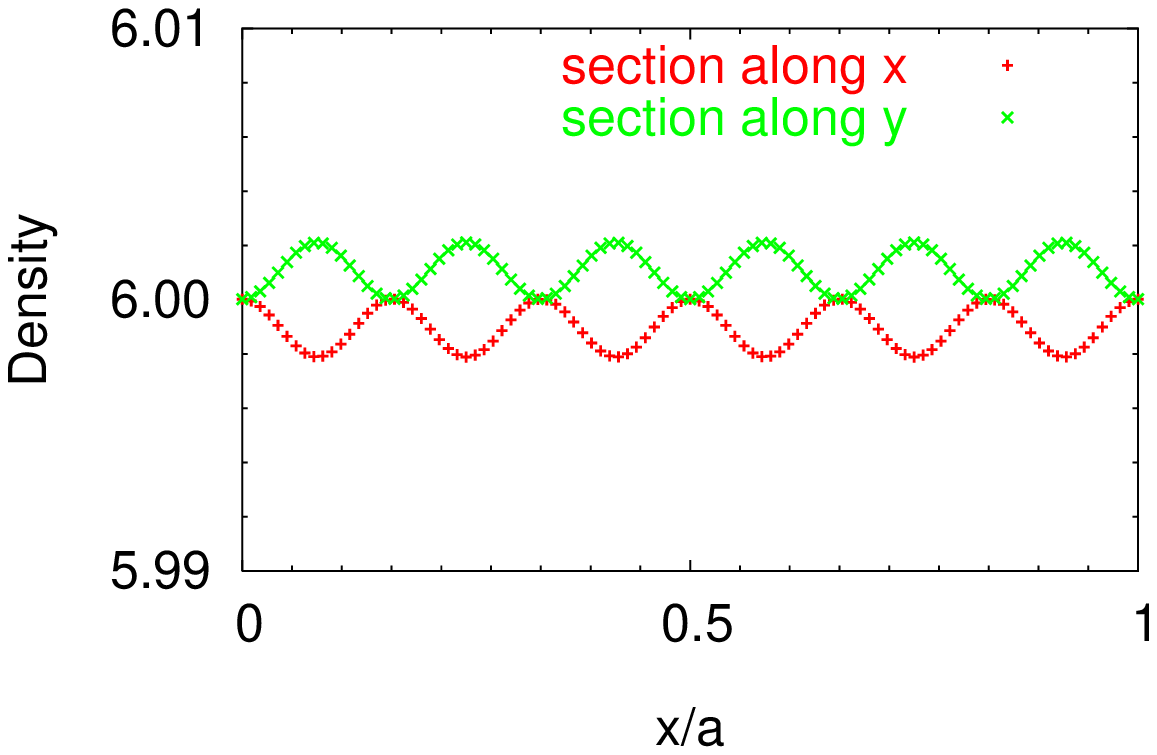}}&
  \multicolumn{2}{c}{\includegraphics[scale=0.4]{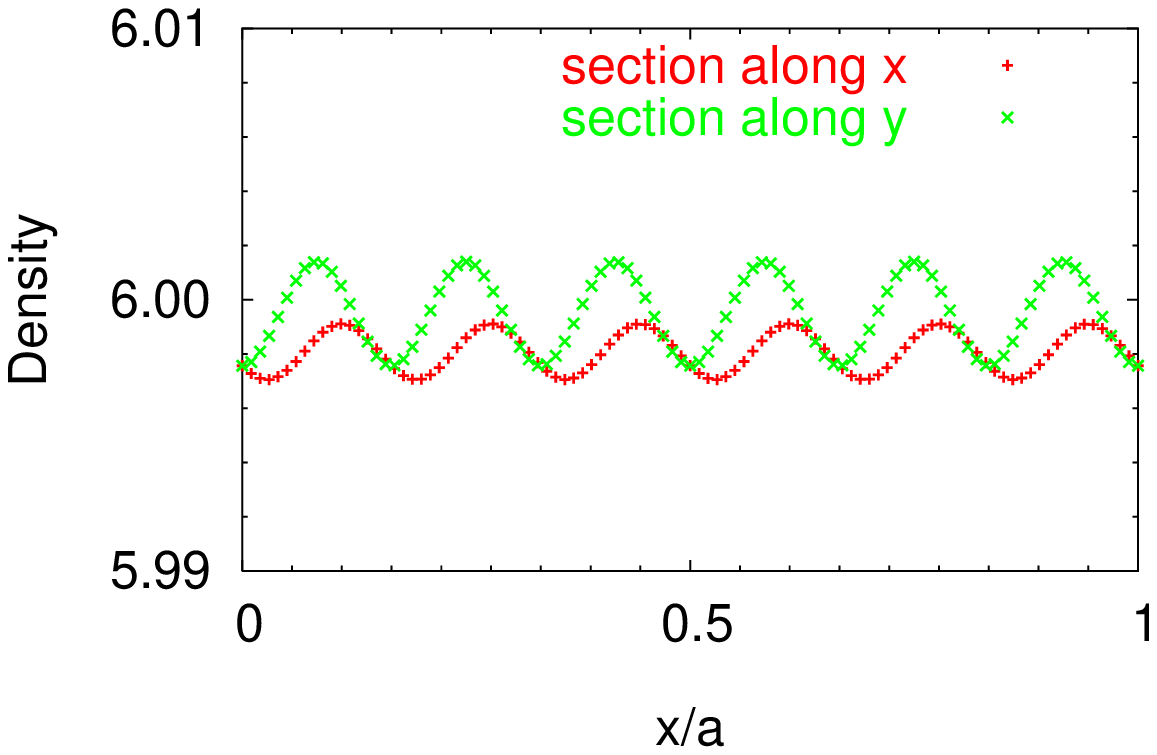}}
  \end{tabular}
\label{fig-ch02-11b}}
\caption{Two different bases for $\Psi_{CM}$, (a) and (b). 
  At filling $\nu=\ot$
  there are three allowed CM states on a torus. They are labeled
  $\alpha,\beta,\gamma$ in this figure. For each element of each basis we show
  the modulus of $\Psi_{CM}$, the density of the corresponding
  Laughlin state with six electrons, 
  i.e. the state $\Psi_{CM}\Psi_L$, and section of the density
  along $x$ and along $y$. Note the positions of the three 
  zeroes in different $\Psi_{CM}$'s (marked by the red arrows).}
\label{fig-ch02-11}
\end{figure}

This introduces a delicate topic. The electron density in a
given state depends on the center-of-mass part of the
wavefunction. Different choices of bases in the
$q$-fold i.e. threefold for $\nu=\ot$,  degenerate space of
center-of-mass wavefunctions may lead to a $q$-tuple of states with
practically homogeneous density in some cases or with quite strongly
varying density in other cases (Fig. \ref{fig-ch02-11}). 
This is true in spite of that we always
describe the same ground state subspace. Even worse, in
homogeneous systems we often want to
study only the relative part of the wavefunction, which must be the
same in all cases. If it is for example the Laughlin wavefunction, we know
that it leads to a homogeneous density. The central trouble is then that
the Hamiltonian eigenstates obtained by exact diagonalization contain
$\Psi_{CM}$.

\subsubsubsection{Relative-motion part of the wavefunction}

The discussion in the previous paragraph is based on (magnetic) translations
of the center-of-mass $T_{CM}(\vek u)$. In an $n$-body state, 
the translation of a single ($i$-th)
particle (Subsect. 7.2 in \cite{chakraborty:1995}), 
$t_i(\vek v)$, can be split into a translation of the center
of mass $T_{CM}(\frac{1}{n}\vek v)$ 
and a relative translation $T_{rel,i}(\vek v- \frac{1}{n}\vek v)$.
Owing to the indistinguishability of particles, the effect of
the relative translation $t_i(\vek v)$ on a particular many-particle 
state is the same for any $i$.
We may thus omit the index and imagine $i=1$, for instance.

Again, as in Bloch's theorem, wavevector $\krv$ can be
attributed to these relative translations
\cite{haldane:11:1985}
\begin{equation}\label{eq-ch02-44}
  T_{rel}(\vek v)\psi = \exp(i\krv\cdot \vek v) \psi\,.
\end{equation}
Since $T_{rel}(\vek v)$
commutes with the Hamiltonian
(\ref{eq-ch02-48}), the Hamiltonian eigenstates can be sorted
according to values of $\krv$. In Bloch's theorem, the
allowed translations are given by an arbitrary lattice vector $\vek
v$. Not all of them are allowed for $T_{rel}$ though
\cite{haldane:11:1985}.

This concept is very similar to a single particle in a periodic
potential. However, there is no real periodic potential in an infinite
plane and we introduced one particular period artificially.
The largest period
possible within our model is the size of the rectangle.

\begin{figure}
\input{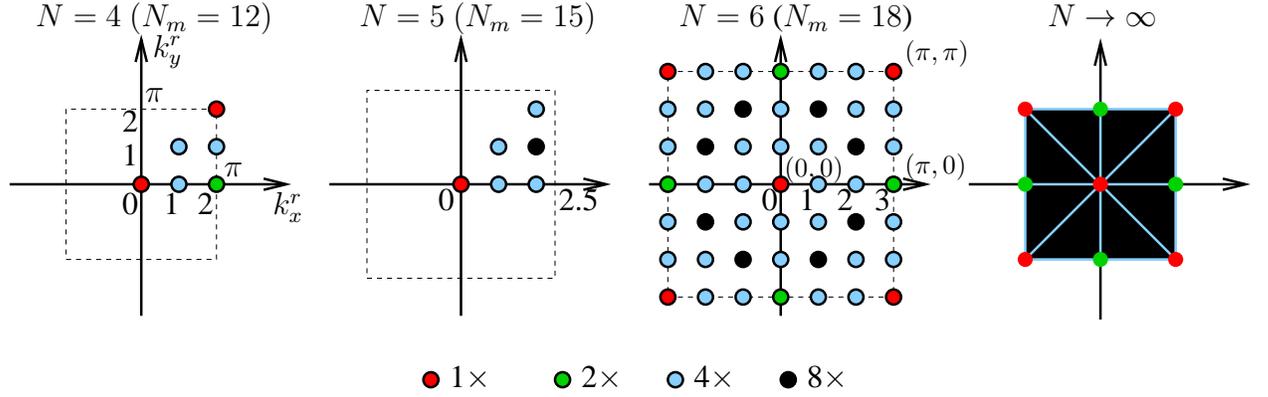}
\caption{The first Brillouin zone for relative translations on
the torus (square with periodic boundary conditions). Its size depends on
the number of particles in the system, cf. (\ref{eq-ch02-45}). 
At filling factor $\nu=p/q$
with $N_e=Np$ particles (and $N_m=Nq$ fluxes), number of allowed
$\krv$-points is $N^2$ and the upper right corner has
$\krv=\gamma(\pi,\pi)$, $\gamma=\sqrt{N/2\pi q}/\ell_0$. Different colours
indicate points of different symmetry (or degeneracy number of a state
with this $\krv$ in a homogeneous system), the rightmost figure shows
the limit of large $N$.} 
\label{fig-ch02-12}
\end{figure}

The Brillouin zone for $\krv$ is rectangular (Fig. \ref{fig-ch02-12}) 
and its {\em size} grows with
the size of the elementary cell. For filling factor $\nu=p/q$ ($p,q$
with no common divisor $>1$) and number of flux quanta per cell
$N_s=Nq$, the allowed values of $\krv$ are
\begin{equation}\label{eq-ch02-45}
 \krv \ell_0 = \sqrt{\frac{2\pi}{N_s\lambda}} (s,t)\,,\qquad
 |s|,|t|\le N/2\mbox{ and integer.} 
\end{equation}
The quantity $\lambda$ is the aspect ratio. For the sake of comparison between
systems of different sizes we will sometimes use size-independent
units for $\krv$, where $\krvd=(\pi,\pi)$ 
means the upper right corner of the Brillouin zone, i.e. $s=t=N/2$.

It can be verified (\cite{chakraborty:1995}, p. 169), that application
of the operator
\begin{equation}\label{eq-ch02-43}
 \mbox{CDW:}\ \sum_j \exp(i\vek q\cdot \vek{r}_i)\,, \mbox{ or }
 \qquad
 \mbox{SDW:}\ \sum_j S_j^+\exp(i\vek q\cdot \vek{r}_i)\,
\end{equation}
to an arbitrary state corresponding to $\krv$
increases its wavevector $\krv$ by $\vek q$. On the other hand, the operator
(\ref{eq-ch02-43}) generates a charge-density wave (spin-density wave) 
with wavevector
$\vek q$, as can be best verified by the simple example of the Fermi gas.
Isotropic states are supposed to have $\krv=0$. 

The wavevector $\krv$ for states on a torus 
is also related to the angular momentum of the corresponding states on
a sphere or on a disc,
$|\krv|=(|\vek L|/\hbar)/R$, where $R$ is radius of the
sphere \cite{haldane:08:1983,chakraborty:1995}. 
The direction of $\vek L$ (or alternatively $L_z$, for
instance) is related to the direction of $\krv$: for example a plane wave
going around the equator specified by the wavevector $\krv$ 
will have $\vek L$ pointing to the pole and $L_z=\hbar|\krv| R$.
This correspondece allows to directly compare spectra for FQH
states obtained for different boundary conditions 
\cite{haldane:01:1985} and this, in turn, helps to sort out
the finite size effects.

\subsubsubsection{Momentum}

So far, we have introduced two sorts of translational symmetries of
states on a torus. One of the center-of-mass part of the wavefunction
and another of the relative part. Since the corresponding magnetic
translation operators commute with the homogeneous Hamiltonian, it
would, in principle, be possible to split the basis of the whole lowest
Landau level into several smaller bases and diagonalize in the
subspaces separately. Each basis would be characterized by a particular
value of $\vek{k}_{CM}$ and $\krv$.

This procedure can help to treat larger systems but it costs some
extra effort to implement it and moreover it is only possible in
homogeneous systems. We will now discuss another of Hamiltonian's
symmetries, described by a new quantum number $J$, 
which is a combination of the previous two. This symmetry is
preserved with a certain class of inhomogeneities and
it can be implemented straightforwardly. When constructing the
basis for a particular value of $J$, we only have to select the matching
Slater determinants (\ref{eq-ch02-52}) rather than to construct
linear combinations of them. 

The homogeneous Hamiltonian in the Landau gauge ($i$ is the
particle index, $V_{int}$ is the Coulomb interaction between particles)
$$
   H=V_{int}+\sum_i H_0^i\,,\qquad
   H_0 = \frac{1}{2}\hbar\omega \left[ -\pder{^2}{x'^2} + 
          \left(-i\pder{}{y'} + x'\right)^2\right]\,,\qquad
	  (x',y')=(x/\ell_0,y/\ell_0)
$$
conserves the total momentum in $y$ direction. Due to the periodic boundary
conditions allowed values of $k_y$ are $(2\pi/b)j$, $j=0,1,\ldots,m-1$.
In an $n$-body state constructed as a
Slater determinant of single-electron states  $\vp_{j_i}$
(\ref{eq-ch02-40}), the total momentum along $y$ is 
\begin{equation}\label{eq-ch02-47}
  (b/2\pi) K_y = (b/2\pi) \sum_{i=1}^n k_y^i = j_1+\ldots +j_n 
  (\mathrm{mod}\ m) \equiv J\,.
\end{equation}
Values of $J$ thus range for instance from $0$ to $m-1$.
It is useful to keep in mind, that $j_i$ is (up to the factor) the
point in $x$-direction at which $\vp_{j_i}$ is centered,
$X_{j_i}=(j_i/m)a$. Thus, $J$ can also be interpreted
as the $x$-coordinate of the center-of-mass of the $n$-electron state.

Without proof, let us now present the precise connection between $J$
and the wavevectors following from $T_{CM}$ and $T_{rel}$
(i.e. $\vek{k}_{CM}, \krv$).
Let $\nu=p/q$ ($p,q$ with no common divisor $>1$) and $m=Nq$ the number of
flux quanta per cell. An arbitrary $J$ can be decomposed into
two parts
\begin{equation}\label{eq-ch02-54}
  J=J_{CM}\cdot N + J_{rel}\,, \quad
  |J_{rel}|\le N/2\,,\ J_{CM}\mbox{ integer,}
\end{equation}
i.e. $J_{rel}$ is $J$ modulo $N$ and $J_{CM}$ is $J$ divided by $N$.
The quantity $J_{rel}$ is directly the $y$-component of $\krv$, more precisely,
$J_{rel}=t$ or $J_{rel}=N/2-t$ in (\ref{eq-ch02-45}), the former for
$pq(n-1)$ even, the latter for $pq(n-1)$ odd \cite{chakraborty:1995}.

$J_{CM}$ distinguishes states which differ {\em only} in the
center-of-mass coordinate. By a successive application of $T_{CM}$ to
one state $\Psi$ we can go through all possible values of
$J_{CM}=0,1,\ldots,\allowbreak {q-1}$. 

Each subspace with definite $J$ contains states of all different $\kr_x$.
Since $T_{rel}$ by allowed translation vectors
commute with the total momentum along $y$, it is in principle possible to
split a basis corresponding to a particular $J$ into subspaces with 
$\kr_x\ell_0 \sqrt{N/2\pi}=-N/2,\ldots N/2$. 
However, the basis state will no longer have the simple form of
antisymmetrized product states of $\vp_j$ (\ref{eq-ch02-40}).

\subsubsection{Exact diagonalization}
\label{pos-ch02-10}

Many interacting electrons in a rectangle with periodic boundary
conditions can be described in the following way. 

\begin{itemize}
\item{} Choose the number of flux quanta penetrating the rectangle
($m$). All allowed single-particle states $\vp_j$ are those written
in (\ref{eq-ch02-40}) or (\ref{eq-ch02-42}) for nontrivial
boundary-condition phases $\phi_x,\phi_y$. Their number is $m$.

\item{} Construct all possible linearly independent 
$n$-particle states (for the given
number of flux quanta $m$). Most conveniently, these can be
antisymmetrized products  (Slater determinants) of $n$ states
$\vp_{j_i}$, denote them by
\begin{equation}\label{eq-ch02-52}
  \ket{j_1\ldots j_n} = a_{j_1}^\dag \dots a_{j_n}^\dag \ket{0}\,.
\end{equation}

\item{} The filling factor is then $\nu=n/m$, cf. (\ref{eq-ch02-38}).

\item{} Take an arbitrary 
many-body Hamilton operator and 
calculate its matrix elements in the basis $\ket{j_1\ldots
j_n}_k$, $k=1,\ldots, N$. 
The dimension of the matrix is $N={n\choose m}$.

\item{} Diagonalize the Hamilton matrix. Eigenvalues are the total 
energies $E_i$,
eigenvectors $\vek{v}_i=(v_i^1,\ldots, v_i^N)$ are related to the
many-body eigenstates by
\begin{equation}\label{eq-ch02-53}
  H\ket{\psi_i} = E_i\ket{\psi_i}\,,\qquad
  \ket{\psi_i}  = \sum_{k=1}^N v_i^k \ket{(j_1\ldots j_n)_k}\,.
\end{equation}
\end{itemize}

This procedure is {\em exact} if we consider a
system where electrons in the lowest Landau level form a periodic system. The
approximation rests therefore in representing an infinite system by a
periodic repetition of a representative finite cell, a procedure
which has been very successfully applied in condensed matter theory.
Formulated in other words: the Hamiltonian is
exact and all approximations are implemented by the choice of the basis.
The dimension of the matrix is finite by construction, 
no cutoff for one-particle states is needed.

In the rest of this Subsection we present the particular form of
the Coulomb matrix elements (\cite{yoshioka:04:1983} or 
\cite{chakraborty:1995}, Sect. 5.1).

The exact Hamilton operator in first and in second quantization is
\begin{eqnarray}\label{eq-ch02-48}
  H &=& \frac{e^2}{4\pi\ve}\sum_{i<j} V(|\vek{r}_i-\vek{r}_j|) \\
  \nonumber
  H &=& \sum_{j} {\cal W} a_j^\dag a_j +
        \sum_{{j_1,j_2 \atop j_3,j_4}} 
        {\cal A}_{j_1,j_2,j_3,j_4} 
        a_{j_1}^\dag a_{j_2}^\dag a_{j_3} a_{j_4}\,,
\end{eqnarray}
where $a_{j}^\dag$ create single-electron states.
The latter expression 
assumes already periodic boundary conditions. The first sum
is the Madelung-type energy of the electron interacting with its own
periodic images \cite{bonsall:02:1977}
\begin{equation}\label{eq-ch02-10}
{\cal W} = -\frac{e^2}{4\pi\ve\ell_0} \frac{1}{\sqrt{2\pi m}} \left[2 - 
\sum_{l_1,l_1 \atop (l_1,l_2)\not=(0,0)} \vp_{-\frac{1}{2}} 
\left(\pi (l_1^2 \lambda + l_2^2/\lambda)\right)\right]\,,\qquad
\vp_n(z) \equiv \int_1^\infty \d t\, e^{-z t}t^n\,.
\end{equation}
If only the electrons were considered, this
energy would diverge at least as $\sum_n 1/n$. To keep it finite,
a neutralizing positive background must be considered \cite{bonsall:02:1977}.

Choosing the single-electron basis according to (\ref{eq-ch02-40})
or (\ref{eq-ch02-42}), the interaction matrix elements are 
\begin{eqnarray} \nonumber
\hskip-1cm{\cal A}_{j_1,j_2,j_3,j_4} &=& 
 \frac{1}{2} \int \d \vek{r}_1 \d \vek{r}_2
 \vp_{j_1}^\ast (\vek{r}_1) \vp_{j_2}^\ast (\vek{r}_2) 
 V(|\vek{r}_1-\vek{r}_2|)
 \vp_{j_3} (\vek{r}_2) \vp_{j_4} (\vek{r}_1) = \\ \label{eq-ch02-49}
 &=& \frac{e^2}{4\pi\ve\ell_0} \frac{1}{2 m} 
     \sum_{{q_x = (2\pi/a) s\atop q_y = (2\pi/b) t} \atop 
            {s,t\in \smallZ \atop (s,t)\not= (0,0)}}
     \delta'_{j_1+j_2, j_3+j_4} \delta'_{s,j_1-j_4} \frac{V(\vek q)}{\ell_0} 
     \exp\left[-\frac{1}{2}\vek{q}^2 \ell_0^2\right] \times \\[-5mm]
 &&  \nonumber \hskip4cm \times
     \exp\left[-2\pi i t (j_1-j_3)/m\right] \times 
     \alpha(j_1+j_2-j_3-j_4, \phi_y)\,.
\end{eqnarray}
with both integrals taken over the rectangle
$[0;a]\times[0;b]$. Primed Kronecker $\delta$ compares the two
arguments {\em modulo} $m$. The
last factor $\alpha$ is solely due to the boundary condition phase
$\phi_y$
$$
   \alpha(\Delta J,\phi_y) = \delta_{J,0} + 
   \delta_{\Delta J,m}\exp(i\phi_y) + \delta_{\Delta J,-m}\exp(-i\phi_y)\,,
$$
and the matrix elements do not depend on $\phi_x$, only the basis
vectors $\ket{j_1\ldots j_n}_k$ do.

The periodic continuation of the Coulomb interaction in two dimensions
is given by
\begin{equation}\label{eq-ch02-50}
\frac{4\pi\ve}{e^2} V(\vek r) = \left. \frac{1}{|\vek r|}\right|_{per}
=\frac{1}{ab} \sum_{\vek q} \frac{2\pi}{|\vek q|}
\exp(i\vek q\cdot \vek r)\,,\qquad
\vek q = \left(\frac{2\pi}{a}s, \frac{2\pi}{b} t\right)\,,\ 
s,t\in\Z\,,
\end{equation}
hence the Fourier series of $V(\vek r)$ used in (\ref{eq-ch02-49})
has  $V(\vek q)=1/|\vek q|$.

The Hamiltonian (\ref{eq-ch02-48}) assumes spin-polarized particles. Its
extension to particles which may have different spin is
straightforward, since the Coulomb interaction conserves spin
\cite{zhang:12:1984},
\begin{equation}\label{eq-ch02-51}
  H = \sum_{j} {\cal W} a_j^\dag a_j +
        \sum_{{j_1,j_2 \atop j_3,j_4}\atop \sigma,\sigma'} 
        {\cal A}_{j_1,j_2,j_3,j_4} 
        a_{j_1\sigma}^\dag a_{j_2\sigma'}^\dag a_{j_3\sigma'} a_{j_4\sigma}\,.
\end{equation}
operators $a_{j\sigma}^\dag$ must be extended appropriately. They
create a particle in state $\vp_j$ either with spin up or spin down.

\subsubsection{Symmetries and choices of bases}
\label{pos-ch02-03}

Regarding the structure of the basis of our choice (\ref{eq-ch02-52})
there are two Hamiltonian symmetries which are easy to use:
conservation of $J$, total momentum along $y$
(\ref{eq-ch02-47}) and conservation of the $z$-component of the
total spin $S_z$.

'Easy to use' means here that the basis of the whole lowest Landau level in
the form of Slater determinants $\ket{(j_1\sigma_1\ldots j_n\sigma_n)_k}$ 
can be sorted into groups corresponding to particular values of $J$
and $S_z$. 

Sorting according to $J$ splits the basis into $m$ subspaces of
approximately the same size $\approx ({m \atop n})/m$. Utilisation of
$S_z$ brings a smaller profit, since the $S_z=0$ subspace is larger than
the $S_z=n/2$ subspace by a substantial factor of about $({n\atop n/2})\approx
2^{n-1}/\sqrt{2\pi n}$. The size of the largest group is then not simply
the number of all states divided by the number of subspaces.

Other symmetries of the homogeneous Hamiltonian would correspond to
conservation of the total spin $S^2$ and conservation of $\kr_x$
(Subsect. \ref{pos-ch02-06}). The eigenstates of these operators, however,
are generally not of the simple product form (\ref{eq-ch02-52}),
but they are linear combinations of such states. More importantly, these
symmetries are gone if inhomogeneous systems are considered. However, suitably
chosen inhomogeneities can preserve the 'easy-to-use' symmetries mentioned
previously (Subsect. \ref{pos-ch04-02}) while still lowering the
total symmetry of the Hamiltonian.

If the aim is to choose $n$ as high as possible, then the largest accessible
systems have about ten electrons. At filling $\nu=\ot$ with $J$
symmetry employed and $S_z=n/2$, the basis counts 1\,001\, 603
elements
for $n=10$ and $J=5$. The largest bases used in this work contained $5\times
10^6$ elements, extremely elaborate programs can handle bases up to
sizes about an order of magnitude larger \cite{morf:08:2002}. An
alternative to the classical exact diagonalization is presented in
Subsect. \ref{pos-ch02-21}

\subsubsubsection{Particle-hole symmetry}

Particle-hole symmetry provides a mapping between systems at fillings
$\nu$ and $1-\nu$ (spinless electrons) or $\nu$ and $2-\nu$ (spinful
electrons). The mapping is exact provided Landau level mixing is
absent. As an illustrative example consider fully polarized
electrons i.e. only the  
lowest LL with spin up is relevant, lowest LL
spin down and all higher LLs are so far in energy that they
can be neglected.
The spectra of a $\nu=\ot$ and $\tt$ systems are identical up to a
constant shift and the corresponding wavefunctions are related by a
simple transformation.

Think of the lowest Landau level as of a 1D chain. The Landau gauge is
particularly illustrative for this as the one-electron states
(\ref{eq-ch02-28}) are localised along $x$. The basic idea of the
particle-hole symmetry is that two electrons at positions $i$ and $j$
feel the same repulsive force as two holes at the same positions,
i.e. when the whole 1D chain is full and only at $i$ and $j$ electrons are
missing.

Let us put this into mathematical terms. Let $a_j^\dag$ be a creation
operator of a single-electron state with momentum $k_y=2\pi j/b$,
being therefore localised in $x$-direction around $X_j=k_y\ell_0^2$
(\ref{eq-ch02-28}). Assuming that $j$ can take values
$0,\ldots, m-1$, particle-hole conjugated $n$-body states are
\begin{equation}\label{eq-ch02-34}
   a_{j_1}^\dag \ldots a_{j_n}^\dag \ket{0} \mbox{ (particles) } 
   \quad\longleftrightarrow\quad
   a_{j_1} \ldots a_{j_n} \ket{1} \mbox{ (holes),} 
\end{equation}
where $\ket{0}$ is an empty Landau level (vacuum) while 
$\ket{1} \equiv a_0^\dag
\ldots a_{m-1}^\dag \ket{0}$ is a completely filled Landau level. 
For example $a_0^\dag a_2^\dag\ket{0}\equiv
\ket{\bullet\cdot\bullet\cdot\cdot\cdot}\longleftrightarrow 
\ket{\cdot\bullet\cdot\bullet\bullet\bullet}\equiv a_0a_2\ket{1}=
a_1^\dag a_3^\dag a_4^\dag a_5^\dag\ket{0}$.

A straightforward calculation shows that 
matrices of a translationally invariant two-body
operator $\hat{A}$ are the same (up to a
multiple of identity matrix and complex conjugation) in an arbitrary
$n$-particle basis and its conjugated $(m-n)$-electron basis. The only
approximation we must concede is to neglect the Landau level mixing.

Result of the calculation is the following. The diagonal terms of an
operator $A$ in the particle basis and in the hole basis fulfil  
\begin{equation}\label{eq-ch02-32}
  \bra{1}a_{j_1}^\dag \ldots a_{j_n}^\dag\ {A}\
  a_{j_n}\ldots a_{j_1} \ket{1}  = 
  \frac{m-2n}{m} \bra{1} \ {A}\ \ket{1}
  +   \bra{0}a_{j_1} \ldots a_{j_n}\ {A}\
  a_{j_n}^\dag \ldots a_{j_1}^\dag \ket{0}
\end{equation}
and the off-diagonal terms remain the same up to the complex
conjugation.

Two cases are worth of special attention:

The {\em spectra} of (fully polarized) systems at
  $\nu=n/m$ and $\nu=(m-n)/m$ are the same up to a shift
  \begin{equation}\label{eq-ch02-33}
    E_{\nu}^i = E_{1-\nu}^i + E_f\, (m-2n)/m \,,
  \end{equation}
  where $E_f$ is the energy of a completely filled (lowest) Landau
  level.  This result does not depend on the
  form of the interaction $V(r)$.
  A nice demonstration of this formula is 
  shown in Fig. \ref{fig-ch03-50}b (see the comment
  \cite{comm:ch02-05}).

  Conjugated states (\ref{eq-ch02-34}) may have
  different values of $J$ (\ref{eq-ch02-54}). 
  For instance: for $m=4$, consider a
  three-electron state $\ket{j_1j_2j_3}=\ket{013}$ 
  and its particle-hole conjugate
  $\ket{j_1}=\ket{2}$. The former has $J=0$ while the latter has 
  $J=2$.

  For the {\em density-density correlation function} 
  $g_\Psi(\vek r)=\langle \sum_{i<j}\delta(\vek r
  -\vek{r}_i+\vek{r}_j)\rangle_\Psi $ we get 
  \begin{equation}\label{eq-ch02-35}
  g_\Psi(\vek r) = \frac{m-2n}{m} \big( 1-\exp(-r^2/2\ell_0^2) \big) +
                   g_{\Psi'}(\vek r)\,,
  \end{equation}
where $\Psi$ and $\Psi'$ are arbitrary particle-hole conjugated
states. Note that $g_{\Psi'}$ refers to {\em
electrons} in the 'hole' state. Correlations between holes in
$\Psi'$ are the same as those between electrons in $\Psi$.

  Note, that $g(\vek r)$ in (\ref{eq-ch02-35})
  is {\em not} defined in the normalized form
  $\delta(\vek r-\vek{r}_i+\vek{r}_j )/(n(n-1))$. 
  Also $g(\vek r)$ of a full
  Landau level may depend on system (finite-size) parameters,
  e.g. in a rectangle with periodic boundary conditions, it depends on
  aspect ratio.

Let us mention that densities of particle-hole conjugated states are
related by $n_\Psi(\vek r) = m - n_{\Psi'}(\vek r)$, exactly as we
expect from the picture of a hole as a missing particle. 
The plus sign in (\ref{eq-ch02-35}) might look
puzzling. At the second glance, however, $g_\Psi=n\cdot n$
(schematically) and therefore
$g_{\Psi'}=(1-n)\cdot (1-n) = 1-2n+g_\Psi$.

\subsubsection{Density matrix renormalization group}
\label{pos-ch02-21}

Exact diagonalization as it has just been presented, boasts of taking
the complete basis of the lowest Landau level on a torus. As long as
the low-energy states are considered, many of the basis states will be
almost absent in the product-state expansion (\ref{eq-ch02-53}).
Especially those which place many electrons close to each other and
thus contribute with a large Coulomb energy.
Leaving out such states from the basis will not
affect the calculated ground state noticeably while it 
reduces the matrix sizes.

Density matrix renormalization group (DMRG) is a systematic method to leave out
irrelevant basis states. Roughly, its basic idea is to successively enlarge the
considered system and to use only the most important $n$-particle
states for calculating the $(n+1)$-particle ground state.

The idea was used originally for one-dimensional systems (a review in
\cite{schollwock:09:2004,schollwoeck:04:2005}). Shibata and
Yoshioka
\cite{shibata:06:2001,shibata:03:2003,shibata:08:2003,shibata:03:2004,yoshioka:xx:2002}
noticed that the single-electron basis of the lowest Landau level
{\em is} in principle one-dimensional (\ref{eq-ch02-42}) and
adapted this method as an extension of the exact diagonalization for
studies of the lowest Landau level. They were thus able to study
systems with up to about 20 particles at fillings close to $\nu=\ot$.

\subsection{Quantum Hall Ferromagnets}


\label{pos-ch02-16}

Consider the situation $\nu=1$ and vanishing
Zeeman energy \cite{girvin:07:1999}. What is the ground state? 

In the absence of Zeeman splitting, the lowest Landau levels ($n=0$) for
spin up and for spin down have the same energy, thus, without
interaction, there are $2eB/h$ single-electron states available with energy
$\hbar\omega/2$, which is the lowest energy an electron can
have in the presence of a magnetic field $B$. Filling factor one means that
only $eB/h$ states (per unit area) are occupied. Hence there is a vast
number of degenerate many-electron ground states without interaction. 

One of these states has the form
$$
  \Psi_{H} = \Phi(z_1,\ldots,z_n) \ket{\!\up\up\ldots \up}\,.
$$
Antisymmetry of $\Psi_{H}$ implies antisymmetry of $\Phi$, or in other
words $\Phi$ vanishes when any $z_i$ approaches any
$z_j$. Each particle is surrounded by a correlation hole,
cf. (\ref{eq-ch03-01}). Moreover, the state $\Psi_H$ is the only
one (with $\nu=1$ within the lowest Landau level) 
whose orbital part is fully antisymmetric, up to $SU(2)$ spin
rotations. If we do {\em not} neglect the repulsive interaction
between electrons, the 'optimal correlation hole' of the state
$\Psi_H$ will make its Coulomb energy lower than for any other $\nu=1$
state and $\Psi_H$ becomes the absolute ground state even at zero
Zeeman energy. 
The long-range order in spins (all are pointing
in the same direction) which are not localised at fixed positions,
e.g. as it is in a spin lattice, renders $\Psi_H$ an itinerant
ferromagnetic state.
In the absence of the Zeeman splitting the $\nu=1$ quantum Hall system
constitutes an example of a Heisenberg ferromagnet.

\begin{figure}
  \subfigure[The $\nu=1$ QHF: $(0,\up)$ and $(0,\dn)$ levels are active,
  the rest is empty. A Heisenberg ferromagnet. The $SU(2)$ spin symmetry
  is manifest in the degeneracy of all possible $S_z$ states for one
  given $S$.]{%
    \label{fig-ch02-09a}
\includegraphics[scale=1.2]{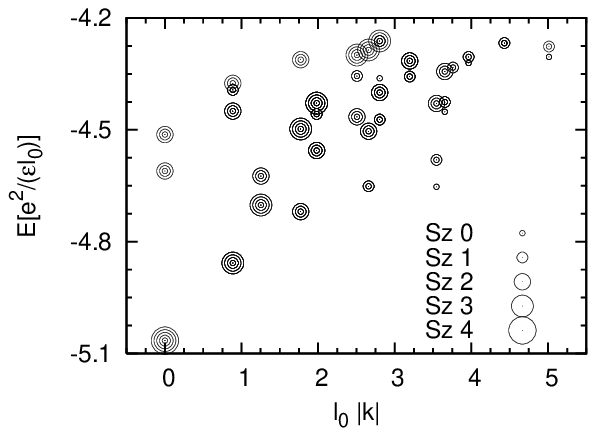}}
  \subfigure[The $\nu=2$ QHF: $(0,\up)$ and $(1,\dn)$ levels are
             active, $(0,\dn)$ is full and treated as inert, the rest is
             empty. An Ising ferromagnet. The $Z_2$ symmetry implies
             that only $S_z$ and $-S_z$ levels are degenerate; $S$
             is no good quantum number. Ground state energy was
             shifted to zero.]{%
    \label{fig-ch02-09b}
\includegraphics[scale=1.2]{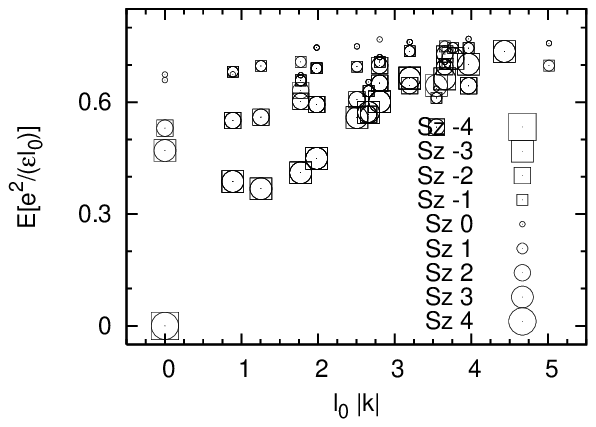}}
\caption{Spectra of two examples of quantum Hall ferromagnets (exact
    diagonalization, Coulomb interaction, eight electrons). The
    ferromagnetic ground state occurs in both cases at $\krv=0$ which
    is a necessary condition for the state to be non-degenerate (in
    orbital degrees of freedom). See Fig. \ref{fig-ch02-13} for
    an explanation of the choice of Landau level indices and spins of
    involved levels $(n,\sigma)$. }
\label{fig-ch02-09}
\end{figure}

For Coulomb interaction, the energy cost of a single electron
flip,
which implies a violation of the antisymmetry of $\Phi$, can be
evaluated analytically: $E=(e^2/\ve \ell_0) \sqrt{\pi/8}$
\cite{girvin:07:1999}. Quantitatively, this number is comparable to
the cyclotron energy $\hbar\omega$ at magnetic fields in the range of
few tesla in GaAs.  The fully polarized state then becomes the ground
state stabilized by the huge gain in exchange energy. A spectrum
obtained by the exact diagonalization in a small system is shown in
Fig. \ref{fig-ch02-09}a. In agreement with the argumentation above,
the ground state has $S=n/2$ and it is well separated from excited
states.

For the Pauli principle to apply ($\Phi$ vanishes as $z_i\to z_j$), it is
only important that all spins have the {\em same} direction, not that
they are all pointing upwards. 
Thus, the $\nu=1$ ground state is characterized by full spin
polarization ($S=n/2$) and arbitrary $S_z$. All states $(S^-)^k
\Psi_{H}$, $k=0,1,\ldots n$ are degenerate ground states. A finite
Zeeman energy will lift this degeneracy and the $\nu=1$ system will
then have a nondegenerate ground state $\Psi_H$, e.g. $S_z=n/2$ for
$B$ pointing in the $z$-direction.

Other types of integer quantum Hall ferromagnets are possible, but
they all share the common scheme: two degenerated Landau levels which
provide $2eB/h$ 'free places' and only $eB/h$ of them should be
occupied. Depending on {\em which} two Landau levels are degenerate,
different types of ferromagnets can follow. A classification of
possible cases was given by Jungwirth and MacDonald 
\cite{jungwirth:12:2000}.

\begin{figure}
\parbox{.5\textwidth}{\input{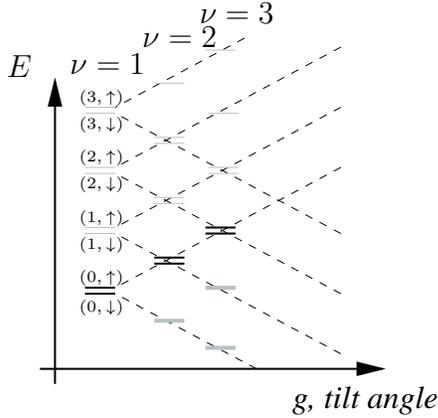}}
\parbox{.5\textwidth}{%
\caption{Integer quantum Hall ferromagnets occur when two crossing
Landau levels should be only half-filled ($2eB/h$ free states, $eB/h$
electrons to occupy them). Thick (thin) grey levels indicate completely
filled (empty) Landau levels, the pair of black levels are the
'active' ones. Despite the large number of possibilities of how to
distribute electrons in the two levels, the Coulomb interaction
selects (up to degeneracy in $S_z$) one state. Depending on
which Landau level crossing is active, different types of ferromagnets
occur: Heisenberg type for $\nu=1$, Ising type for $\nu=2,3$.
}
\label{fig-ch02-13}}
\end{figure}

Let us introduce one more example, the $\nu=2$ QHF which turns out to be
an Ising type ferromagnet (see also Jungwirth {\em et al.}
\cite{jungwirth:11:2001}). By changing the ratio between Zeeman and
cyclotron energy, 
$(n,\sigma)=(0,\up)$ and $(1,\dn)$ Landau levels can be brought
to coincidence (Fig. \ref{fig-ch02-13}). Experimentally, this can be
accomplished either by changing the $g$-factor (it decreases with
pressure \cite{cho:09:1998} or by tilting the magnetic field
\cite{dePoortere:??:2000} (cyclotron energy depends
only on the component perpendicular to the 2DEG plane, Zeeman energy
depends on the total field).
The low lying $(0,\dn)$ Landau level is fully occupied
($eB/h$ states) and can be taken as inert. The remaining $eB/h$ states
(giving in total $\nu=2$) can be distributed among the $2eB/h$
available places of the {\em two} crossing Landau levels 
(Fig. \ref{fig-ch02-13}). Contrary to
the $\nu=1$ QHF, there are only two ground states now: either
$(0,\up)$ is full or $(1,\dn)$ is full
(Fig.~\ref{fig-ch02-09}b). To obtain this result we should use the exact
diagonalization because of the large degeneracy present when interaction is
switched off. However, the fact that distributing the electrons
between the $(0,\up)$ level and the $(1,\dn)$ level costs extra energy
(compared to placing all electrons into one of the levels),
is probably a consequence of the
fact that spin up orbitals are not the same as spin down orbitals
\cite{jungwirth:12:2000} as they lie in different Landau levels. 

For a more detailed discussion of spectra of a Heisenberg and an Ising QHF
(Fig. \ref{fig-ch02-09}), see in Subsec. \ref{pos-ch03-06}.

Quantum Hall ferromagnets which occur at integer filling factor have
the advantage that they can often be well described by Hartree-Fock
models, at least as far as the ground state is considered. Even here, exact
diagonalization studies can sometimes unveil unexpected ground states,
as shown by Nomura \cite{nomura:2003} in bilayer systems (spin degree
of freedom is substituted by pseudospin which refers to the two
layers).

The principial question which is addressed in this thesis is, whether
quantum Hall ferromagnetism can also occur at fractional filling
factors. Naively, one may expect that phenomena occuring for electrons
(integer $\nu$)
would also occur for the CF (fractional $\nu$). The Coulomb 
energy (of CF cyclotron energy) would take over
the role of the cyclotron energy within the integer QHF. A pleasing
fact is that now the 
ratio of the CF cyclotron energy and the Zeeman energy is
$B$-dependent, so that the coincidence of the CF LL can be induced
just by adjusting $B$ at a given filling factor.
Experimentally, there are strong hints on the existence of
ferromagnetism even at fractional fillings
\cite{smet:01:2002,eom:09:2000} and this work should contribute to
the understanding of these phenomena from the side of theory.

\setcounter{footnote}{0}
\section{Structure of the incompressible states and of the
half--polarized states}

\label{pos-ch03-00}

\subsection{Basic characteristics of the incompressible ground states}









\label{pos-ch03-17}

Being interested in phenomena occurring at the transition between two
incompressible ground states, the spin-polarized and the singlet one,
it is reasonable to get acquainted with these two ground states first.

\begin{figure}
\subfigure[Systems at filling factors $\nu=\ot,\tt$ and $\tf$ correspond
to $\nu_{CF}=1,2$ and $2$ within the non-interacting CF picture. The
composite fermions (CF) are electrons with two magnetic flux quanta
attached parallel (for $\nu=\ot,\tf$) or antiparallel (for $\nu=\tt$)
to the effective magnetic field $B_{\eff}$ (but always parallel to the
real external field $B$). When Zeeman splitting is
increased, crossings between CF Landau levels occur and spin
polarization of the ground state changes.]{
\includegraphics[scale=0.38]{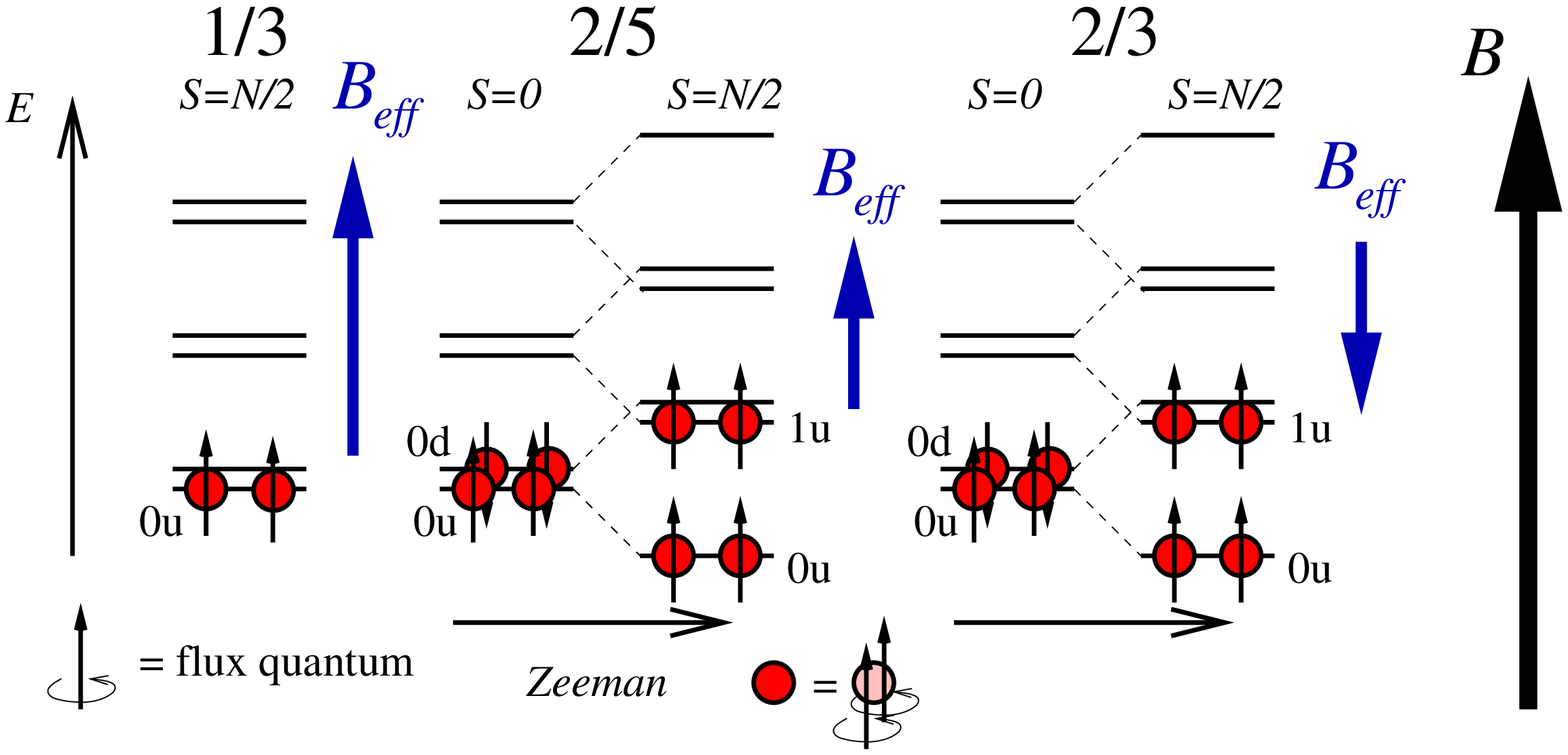}
\label{fig-ch03-02}}
\subfigure[Energy levels of 8 particles on a torus at filling factors
$\tt$ and $\tf$ without Zeeman splitting (short-range interaction,
see Subsec. \ref{pos-ch02-01}). Note the large excitation energies
(related to the incompressibility gaps) for the
ground states at $S=0$ and $S=4$, as compared to other inter-level
separations.]{
\includegraphics[scale=0.3]{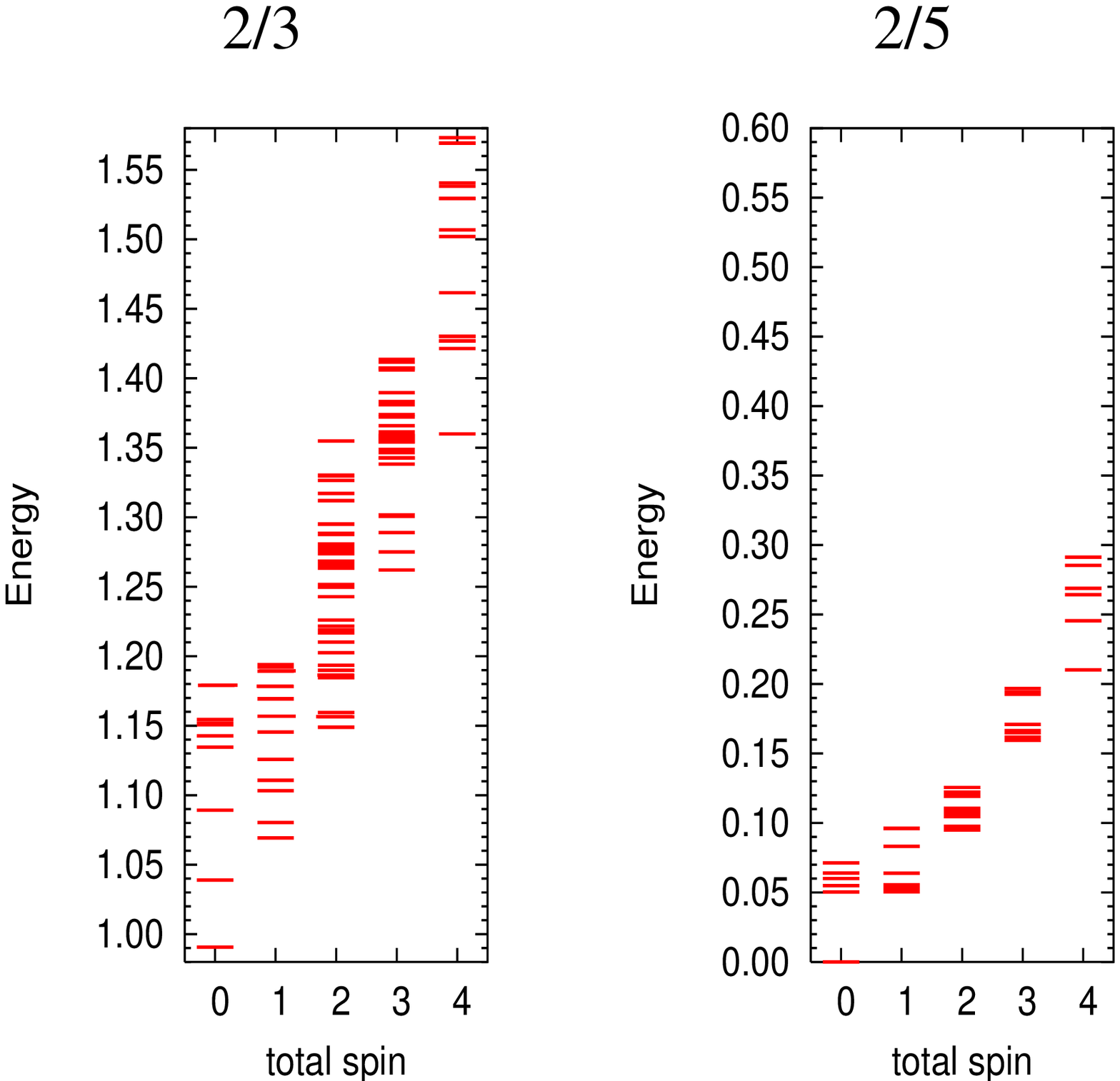}\label{fig-ch03-01}}
\caption{Ground states at filling 
factors $\ot$, $\tf$ and $\tt$ from the point of view
of a composite fermion picture and the exact diagonalization.}
\label{fig-ch03-01and02}
\end{figure}

In the very illustrative model of non-interacting composite fermions
(NICF), introduced in Subsect. \ref{pos-ch02-18},
the ground state at electronic filling factor $\nu=\tt=2/(2\cdot
2-1)$ corresponds to two completely filled composite fermion 
Landau levels (LL). If, in
some particular situation, the CF cyclotron energy is smaller than the
Zeeman splitting, these will be the $n=0,\up$, $n=1,\up$ CF Landau
levels and the ground state will be fully spin polarized,
Fig. \ref{fig-ch03-01and02}a. If the ratio between Zeeman and CF cyclotron
energies is reversed, the ground state has $n=0,\up$, $n=0,\dn$ CF
Landau levels filled and is therefore a spin singlet, cf. comment
\cite{comm:ch03-09}.  Here, the CFs are electrons
with two flux quanta attached {\em antiparallel} to the effective
magnetic field $B_{\eff}$
\cite{wu:07:1993},
which leads to a minus sign in the denominator of
the CF filling factor (\ref{eq-ch02-08}).

A similar situation, i.e. occurrence of two incompressible ground
states, the singlet and the polarized one, occurs also at filling
factor $\nu=\tf$. Here, the ground state can be interpreted as two
filled CF LLs where the two flux quanta were attached {\em parallel}
to $B_{\eff}$. Thus, these ground states should be completely
equivalent to the ground states at $\nu=\tt$ within the NICF
approximation.

Let us compare this picture of an infinite
two-dimensional system with a finite system
treated exactly.  Looking at the exact spectra of a $\nu=\tt$
and a $\nu=\tf$ finite system, Fig. \ref{fig-ch03-01and02}b, 
we readily recognize
ground states in the $S=0$ and the $S=N/2$ sector which are well
separated from excited states, as compared to the typical level
separation within the excitation spectrum or in subspaces with other
values of the total spin. Also, as the NICF model predicts, the spin
singlet ground state ($n=0,\up$, $n=0,\dn$) has a lower energy
$E(S=0)$ than the polarized one ($n=0,\up$, $n=1,\up$), $E(S=N/2)$ if
the Zeeman energy is set to zero. Both ground states have $\krv=(0,0)$
which corresponds to $\vek{L}=0$ in a system with circular symmetry,
Subsect. \ref{pos-ch02-06}.  Angular momentum
equal to zero is in turn a property inevitable in any state with no
partially filled Landau levels, corresponding argumentation is
analogous to the comment \cite{comm:ch03-09}. 

It should be
emphasised at this place that however strong support for the NICF model these
findings provide, they cannot be taken as a proof of its complete
correctness. The interacting electrons can{\em not} be exactly mapped to {\em
non-interacting} CFs and even the quality of the approximation is
hard to control. Although the NICF model gives correct answers to questions 
indicated above, there is no guarantee of correct answers in other
cases, especially at other filling factors.
In the following, we will continue discussing properties of both
incompressible states at $\nu=\tt$ and of those at $\nu=\tf$ as
calculated by exact
diagonalization and we will occasionally mention links to composite
fermion theories.

\subsubsection{Densities and correlation functions}
\label{pos-ch03-15}

Having computed a many-particle wavefunction numerically usually does
not automatically mean that we can say much about the nature of the
state it describes. Very often, the only statement to be made is that the
state is highly correlated, or entangled. By this we mean that the
state cannot be written as a single Slater determinant
\cite{comm:ch03-01}, not even approximately, and thus its description
goes far beyond any Hartree-Fock model.

In order to learn more about the state it is apt to evaluate expectation values
of observables such as density or density-density correlation
functions. In the first quantization formalism these are the following
operators 
\begin{eqnarray}\label{eq-ch03-12}
  \op{n}(\vek{r}) = \displaystyle 
                      \sum_i \delta(\vek r - \vek{r}_i)\,, &&
  \op{g}(\vek{r}) = \displaystyle
                    \frac{1}{N_e(N_e-1)}\sum_{i\not= j} 
                    \delta\big(\vek r - (\vek{r}_i-\vek{r}_j)\big)\,,
\end{eqnarray}
summations running over all particles in the system. For inhomogeneous
systems it is also useful to consider an 
unaveraged density-density correlation operator
$$
  \op{g}(\vek{r}'',\vek{r}') =
                    \frac{1}{N_e(N_e-1)}\sum_{i\not= j} 
                    \delta(\vek{r}'-\vek{r}_i)
                    \delta(\vek{r}''-\vek{r}_j)\big)\,,
$$
where an average over $\vek{r}''$ gives (\ref{eq-ch03-12}).
This is the probability density of finding a particle at place
$\vek{r'}$ provided there is a particle at place $\vek{r}''$. The
function $\op{g}(\vek{r})$ is just $\op{g}(\vek{r}+\vek{r}',\vek{r}')$
averaged over all $\vek{r}'$, hence $g(\vek{r})\propto
g(\vek{r}+\vek{r}',\vek{r}')$ for homogeneous systems, 
i.e. both quantities are the same up to a proportionality constant.

For not fully spin polarized states it is also useful to watch
quantities $\op{n}_\up(\vek r)$ or $\op{g}_{\up\dn}(\vek{r})$
and its analogues with other spin indices. For example
\begin{equation}\label{eq-ch03-13}
  \op{g}_{\up\dn}(\vek{r}) =
                    \frac{1}{N_e(N_e-1)}\sum_{i\not= j}
		     \delta_{\sigma_i\up}\delta_{\sigma_j\dn}
		     \delta\big(\vek r - (\vek{r}_i-\vek{r}_j)\big)\,.
\end{equation}
The normalization of density and density-density correlation functions we
chose in  (\ref{eq-ch03-12},\ref{eq-ch03-13}) 
is the following:
\begin{equation}\label{eq-ch03-14}
  \int \d \vek r n(\vek r) = N_e\,,\qquad
  \int \d \vek r g(\vek r) = 1\,,\qquad
  \int \d \vek r g_{\sigma\sigma}(\vek r) = 
                  \frac{N_\sigma(N_{\sigma}-1)}{N_e(N_e-1)}\,,\ 
		  \sigma\in\{\up,\dn\}\,,
\end{equation}
where integrals are taken over the whole system i.e. elementary cell.

As long as homogeneous systems are concerned we naturally expect
density and also polarization to remain
constant. For the incompressible states this is true  only up to finite
size effects. The density shows a slight modulation which decays rapidly as
the system size is increased. Discussion of these effects which have no
relevance for the real infinite 2D system will be presented later,
Subsec. \ref{pos-ch03-01}.

In the following, by $g(r)$ we mean $g(\vek r)$ with $r=|\vek r|$ for
isotropic and homogeneous systems. Also, whenever we will speak about
'correlation functions' we mean equal time density-density
correlation functions.

\subsubsubsection{Fully occupied Landau levels}

The density-density correlation function can be analytically evaluated for
a state with $\nu=n$ fully occupied Landau levels
\cite{kamilla:11:1997}. This is the ground state of non-interacting
electrons at integer filling factor. For the spin polarized case,
\begin{equation}\label{eq-ch03-01}
  g(r) = 1- \frac{1}{n^2} \exp(-[(rk_F)^2/4n]) 
         \left[L_{n-1}^1 \left(\frac{(rk_F)^2}{4n}\right)\right]^2\,.
\end{equation}
Here $L_n^\alpha(x)$ are the associated Laguerre
polynomials \cite{comm:ch03-03,gradshteyn:1980}. In particular,
\begin{equation} \label{eq-ch03-08}
\hskip-.5cm \nu=1\,:\quad g_{\nu=1}(r)=1-\exp(-r^2/2\ell_0^2)\,,\qquad
\nu=2\,:\quad g_{\nu=2}(r)=1-\exp(-r^2/4\ell_0^2)\cdot
                             \frac{1}{4}[2-r^2/4\ell_0^2]^2\,.
\end{equation}

\begin{figure}
\includegraphics[scale=0.5,angle=0]{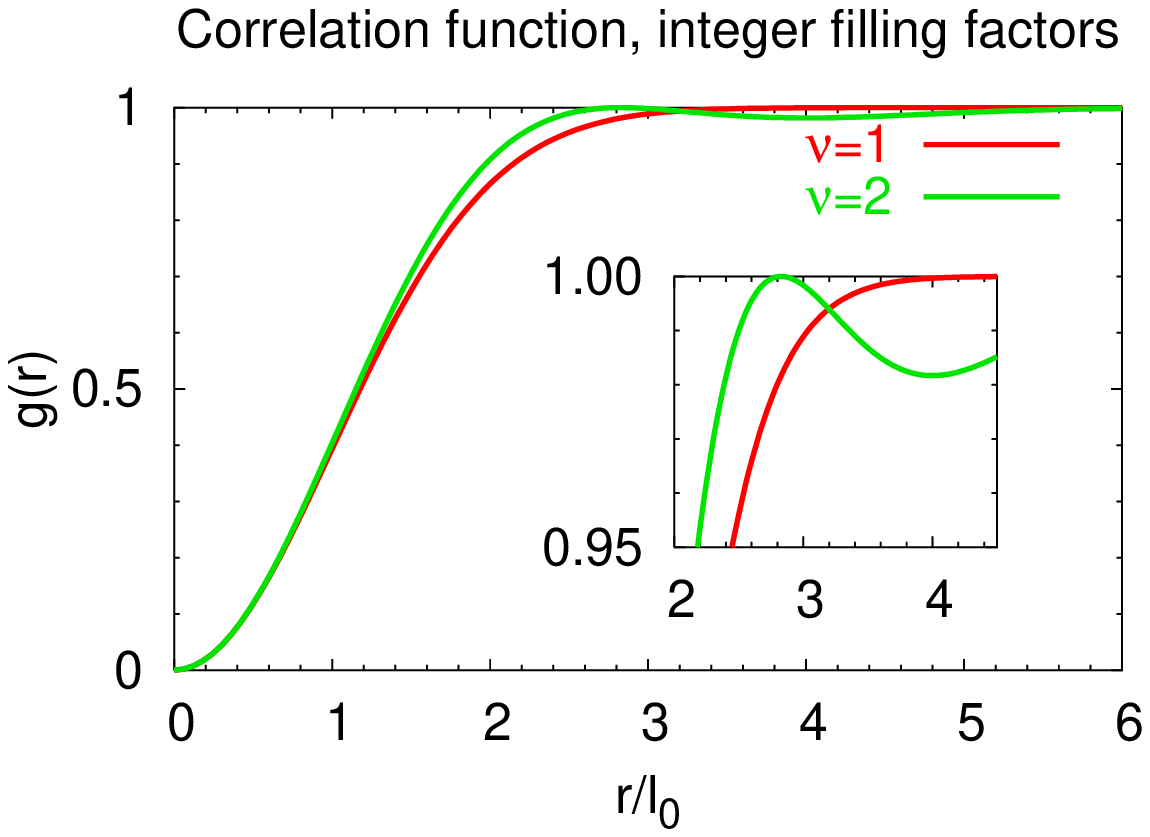}\hfill
\includegraphics[scale=0.5,angle=0]{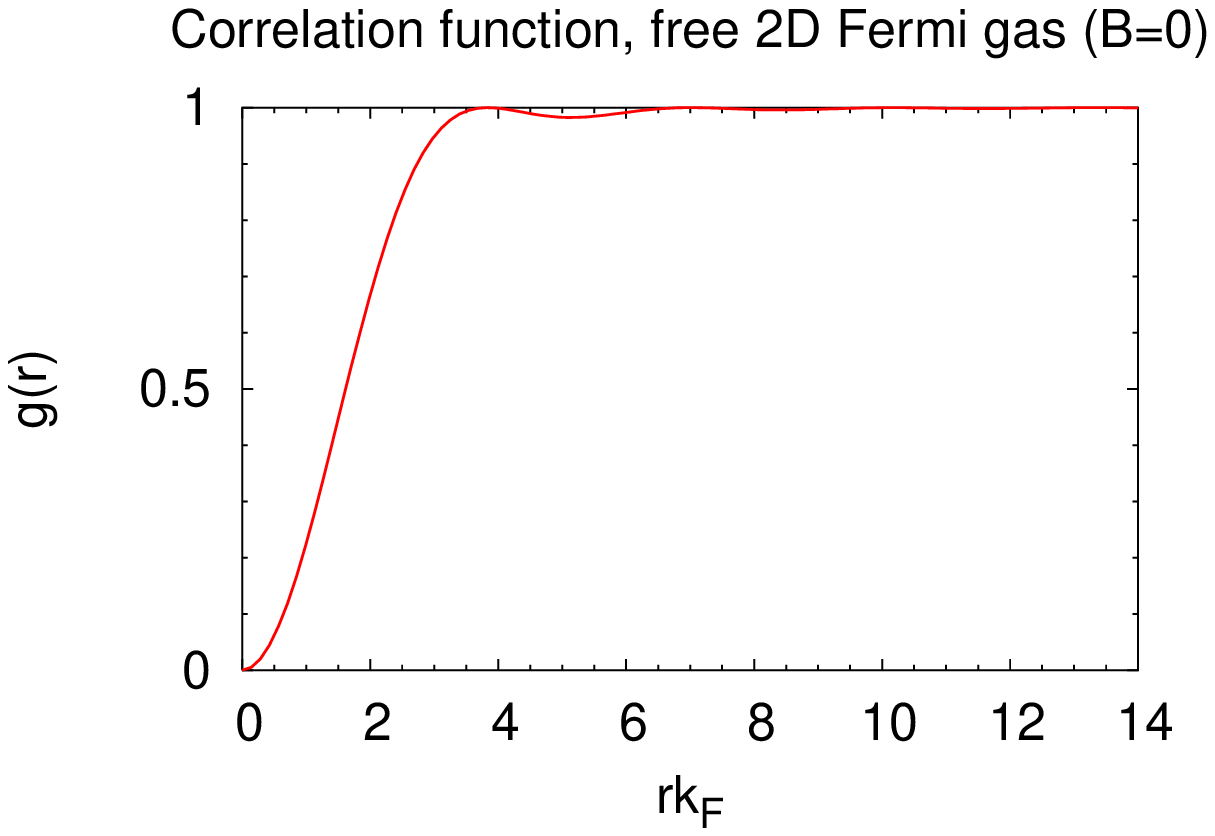}\hskip-2cm
\caption{Density-density correlation in a free 2D electron gas and in
magnetic field at integer filling factors (spin-polarized electrons).}
\label{fig-ch03-56}
\end{figure}

The Fermi wavevector $k_F$ for a system subjected to a perpendicular
magnetic field is defined as $k_F$ in exactly the same system
(i.e. the same areal density of electrons) just with magnetic field
switched off. In this scheme
\begin{eqnarray}\label{eq-ch03-03}
(k_F\ell_0)^2=2\nu\,,\mbox{ or}&& k_F= \sqrt{2\nu} \ell_0^{-1}\,.
\end{eqnarray}
It is a pleasant news that by taking the limit
$\nu=n\to\infty$ in (\ref{eq-ch03-01}) we obtain
\begin{eqnarray}\label{eq-ch03-02}
g_{FS}(r) &=& 1-\left[\frac{2}{k_Fr}J_1(k_Fr)\right]^2\,,
\end{eqnarray}

which is the correlation function of free electrons in two
dimensions (Fermi sea). It should not be anything else because
$\nu\to\infty$ with $k_F$ kept constant means that $B$ is decreased to
zero at a given areal density of electrons.

\subsubsubsection{Filling factor $\nu=\ot$}

Provided Landau level mixing is absent and considering only the
short-range interaction between particles (Sec.
\ref{pos-ch02-01}), the ground state at filling factor $\nu=\ot$ is
described by the Laughlin wavefunction $\Psi_L(z_1,\ldots,z_n)$,
(\ref{eq-ch02-09}). Up to my knowledge, no {\em closed}
\cite{comm:ch03-05} analytical expression 
of the correlation function in this state is available. Only
the short range behaviour can be determined 
analytically. For $(z_1-z_2)\to 0$, $|\Psi_L|^2$ vanishes proportional to
$(z_1^*-z_2^*)^3(z_1-z_2)^3=|z_1-z_2|^6$, hence
$g(r)=cr^6+o(r^6)$ for $r\to 0$.

\begin{figure}
\begin{tabular}{ccc}
\includegraphics[scale=0.2]{figs/ch03/ch03-fig03a.epsi} &
\raise.6cm\hbox{\includegraphics[scale=0.235]{figs/ch03/ch03-fig03b.epsi}} &
\raise-.2cm\hbox{\includegraphics[scale=0.32]{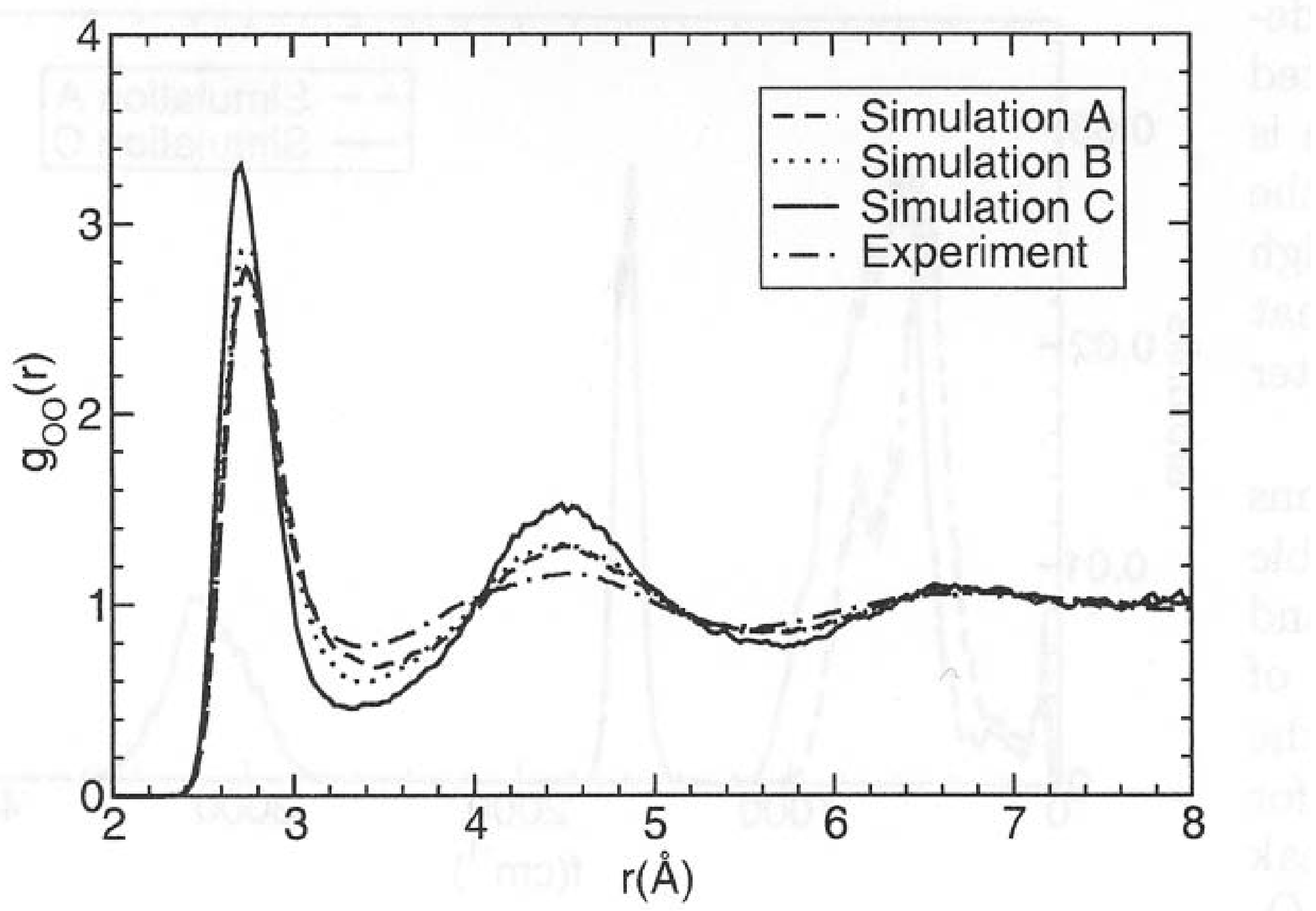}} \\
$\nu=\frac{1}{3}$ & $\nu=\frac{2}{5}$ & Water
\end{tabular}
\caption{{\em Left and middle:} Correlation functions of the ground states
of 50-60 particles at filling factors $\ot$ and $\tf$ of the principal 
Jain's sequence, $\nu=p/(2p+1)$ (cf. Subsect. \ref{pos-ch02-18}). 
The wavefunctions (WF) predicted by composite fermion theory were taken
(for $\nu=1/3$ this is identical with the Laughlin WF) and $g(r)$ was
calculated by a Monte Carlo method. Taken from
Ref. \cite{kamilla:11:1997}.
{\em Right:} correlation function between
oxygen atoms in liquid water as an example of a density-density
correlation function in a well-known liquid 
(see text on p. \pageref{pag-ch03-01}). Results of both numerical
simulation and experiments are shown, see the original paper by
Allesch {\em et al.} \cite{allesch:01:2004} for details.
}\label{fig-ch03-03}
\end{figure}

\begin{figure}
\subfigure[$N_e=10$.
Due to the absence of circular symmetry on the torus, $g(\vek r)$ is in
general not only a function of $r=|\vek r|$. For $|\vek r|\ll a$,
$g(\vek r)$ is however quite isotropic, Fig. \ref{fig-ch03-04+05}b inset.
The first electron is sitting at the corner, the four corners are
identical owing to the periodicity. The lower plot differs from the
upper one only by a finer $z$-scale which highlights the structures
in $g(\vek r)$ at larger distances.]{\raise7cm\hbox{\hskip1.2cm%
\includegraphics[scale=0.4,angle=-90]%
{figs/ch03/ch03-fig05.compr.epsi}\hskip1.2cm}\label{fig-ch03-05}%
}%
\subfigure[Section of $g(\vek r)$ for $N_e=5,8,9,10$-electron ground
states along
$\vek r=(x,x)$; the perfectness of match to $g(r)$ in
Fig. \ref{fig-ch03-03} gives us a feeling how little the ground state
is affected by the finiteness of the system. It is noteworthy that
$g(r)$ can be astonishingly well fitted by the 
$\noexpand{[}g_{FS}(r)\noexpand{]}^3$
(\ref{eq-ch03-02}) up to distances beyond the first maximum (up to vertical
scaling, only $k_F$ must be fitted, see the text). {\em Inset:} 
sections along diagonal and side of the square, i.e. 
$g(x/\sqrt{2},x/\sqrt{2})$ and $g(x,0)$   for the $N_e=10$ system.
The good match of the two curves as far as well beyond the first
maximum ($\approx 6\ell_0$) indicates that the isotropy on length scales
$<6\ell_0$ is not much affected by the rectangular geometry (periodic
boundary conditions). ]{\label{fig-ch03-04}%
\raise1cm\hbox{\includegraphics[scale=0.7,angle=0]{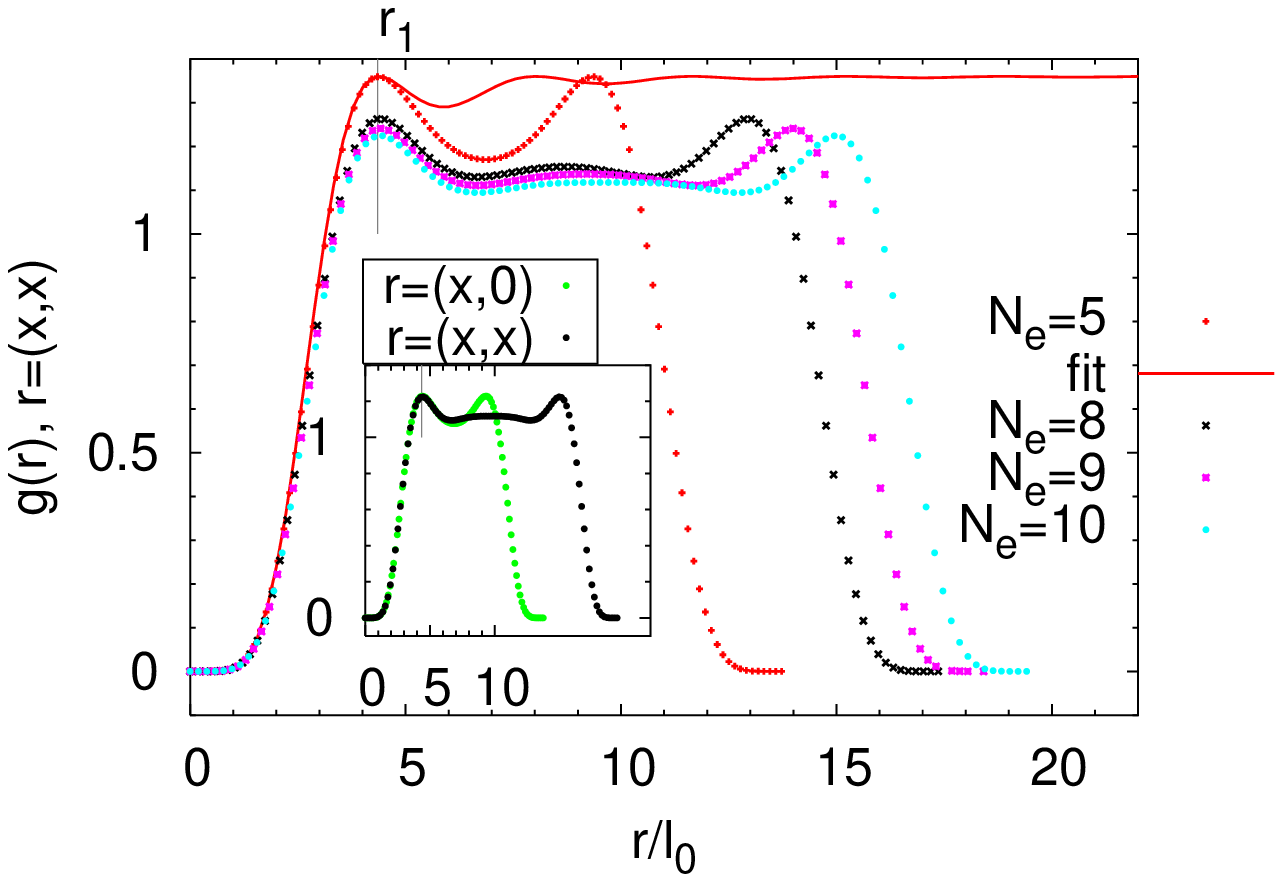}
}}
%
\caption{Correlation functions in the ground state of $N_e$ electrons
on a torus (square, length of sides $a=b$, with periodic boundary
conditions) at filling factor $\nu=\ot$.  The function $g(\vek r)$
gives the probability of finding 
an electron at position $\vek r=(x,y)$ provided there is an electron
sitting at $\vek{r}'=(0,0)$. For a homogeneous system the choice of
$\vek{r}'$ does not influence the probability distribution of finding
the second electron.}\label{fig-ch03-04+05}
\end{figure}

Numerically, $\bra{\Psi_L} g(r)\ket{\Psi_L}$ can be evaluated by
various Monte Carlo techniques,
Fig. \ref{fig-ch03-03}. These results are closer to the
thermodynamic limit, referring to larger numbers of particles,
than $g(r)$ which can be obtained from exact
diagonalization, Fig. \ref{fig-ch03-04+05}. This
is however only because we know an analytic WF of the GS for
arbitrarily large systems in this case,
$\Psi_L$. Exact diagonalization can be performed only for systems
with $N_e\lesssim 10$ electrons, but it is not necessary to know
anything about the ground state in advance apart of that it lies in
the lowest Landau level. Therefore, exact diagonalization provides us
a way to confirm that $\Psi_L$ is indeed the ground state or a good
approximation to it, e.g. for Coulomb-interacting electrons. Note
also that Figs. \ref{fig-ch03-04+05} refer to electrons on torus
whereas Fig. \ref{fig-ch03-03} refers to the disc geometry.
Indeed, the correlation functions are
very similar in both geometries, compare Fig. \ref{fig-ch03-04} and
Fig. \ref{fig-ch03-03}. This fact supports the hypothesis that the
corresponding states, $\Psi_L$ on a disc and those on tori, 
are universal and hence basically the same as the
ground state in an infinite 2D system.

The correlation
function $g(\vek r)$ in Fig. \ref{fig-ch03-04+05} is rather isotropic,
at least on distances smaller than $a/2$. This distinguishes the
Laughlin state from a Wigner crystal (Subsect. \ref{pos-ch03-09}) or a
unidirectional charge density wave in which some special
directions exist, 
Subsect. \ref{pos-ch03-11}. This fact motivates also the incompressible
{\em liquid} terminology. 
Regarding the distinction between liquids and gases, the difference is
the strength of interparticle interaction. Whereas negligible in gases, the
  interaction in liquids is strong compared to kinetic energy. In the
  lowest Landau level, kinetic energy is zero, or better a constant
  $\hbar\omega/2$.

The first maximum in $g(r)$ occurs at $r_1\approx 4.4\ell_0$,
Fig. \ref{fig-ch03-04+05}b, and this separation can be taken as a typical
interparticle distance in the Laughlin 
state. This distance lies close to the mean interparticle
  distance determined by the filling factor,
  $r_{mean}/\ell_0=\sqrt{2\pi/\nu}\approx 4.35$
  (\ref{eq-ch02-38}).

After $r_1$,
oscillations in $g(r)$ decay rapidly. The overall form of $g(r)$ in
the Laughlin state clearly differs from the correlation function of a
free 2D Fermi gas (\ref{eq-ch03-02}). We will emphasise three aspects.

(i) Laughlin state, Fig. \ref{fig-ch03-04+05}b: the first peak of $g(r)$ is
relatively high, measured for
instance by ratio $g(r_1)/g(a/\sqrt{2})\gtrsim 1.1$. Here
$a/\sqrt{2}\approx 10\ell_0$ is the maximum interparticle distance in
the considered finite system.

2D Fermi gas, Fig. \ref{fig-ch03-56}: the first structure
of $g(r)$ is about ten times weaker. Here, it is more appropriate to
watch the depth of the first minimum, see (ii).

(ii) 2D Fermi gas: all maxima (at $r_{FS}^i$) of $g(r)$ have the same value,
$g(r_{FS}^i)=1$. Laughlin state: the first maximum $g(r_1)\approx 1.1$
(for $N_e\to\infty$) is much higher than other maxima.

(iii) 2D Fermi gas: $g(r)\propto r^2$ for $r\to 0$. This is purely
the effect of Pauli exclusion principle. Mathematically, it comes from
the antisymmetry of the wavefunction $\Psi$, in other words, $\Psi$
is a Slater determinant. Laughlin state: $g(r)\propto
r^6$. This is a manifestation of correlations in the state, i.e. of
the fact that $\Psi_L$ cannot be written as a single Slater
determinant. $g(r)\propto r^6$ also means that any two electrons avoid
being close to each other very efficiently and this helps to minimize
the Coulomb energy which is high at short inter-particle distances
\cite{haldane:01:1985},
Subsect. \ref{pos-ch02-09}. 

Just as an illustration, a correlation function $g(r)$ of liquid water is shown
in Fig. \ref{fig-ch03-03}, right. Of course, it is not possible to directly
compare water and a 2D electron gas in the fractional quantum Hall
regime. \label{pag-ch03-01} 
Nevertheless, the pronounced structures in $g(r)$ beyond the
correlation hole in
the Laughlin state, Fig. \ref{fig-ch03-03} left, are definitely 
more similar to $g(r)$ of {\em liquid} 
water, Fig.~\ref{fig-ch03-03}, right, rather than to $g(r)$ of a
2D Fermi {\em gas}, Fig. \ref{fig-ch03-56}.

The Laughlin state, Fig. \ref{fig-ch03-04+05}b, also differs
from integer filling factor states apparently, Fig.
\ref{fig-ch03-56}. The latter ones ($i=1,2,\ldots$) namely have
always $g_{\nu=i}(r)\propto r^2$ at $r\to 0$. Also $g_{\nu=i}(r)$ has
exactly $i-1$ maxima, i.e. $g_{\nu=1}(r)$ is free of maxima. 

This demonstrates the fact, that in the $\nu_{CF}=1$
composite fermion (CF) state, which is the model of the $\nu=\ot$
electronic ground state (Sect. \ref{pos-ch02-17}), the {\em
electron-electron} correlations are different to those in a $\nu=1$
electronic state. This is a bit counterintuitive, since the CFs were
created by adding two zeroes to electrons {\em in the
$\nu=1$ state} and we could have therefore expected that the electrons
'remained at their original positions' 
under this transformation. Figures \ref{fig-ch03-04+05}b and
\ref{fig-ch03-56} however show that even though
the CF density equals the electronic one
the electron-electron {\em correlations} are different in both states.

On 'intermediate length scales' ($1$ to $5$ magnetic lengths), 
the correlation function
of the Laughlin state $g(r)$  in Fig. \ref{fig-ch03-04+05}b can be
strikingly well fitted by 
\begin{equation}\label{eq-ch03-09}
c\cdot [g_{FS}(r)]^3\,,
\end{equation}
where $g_{FS}(r)$ is the correlation function of a free 2D Fermi gas,
(\ref{eq-ch03-02}). Herefore, we put $k_F\approx 0.874
\ell_0^{-1}$ which is only by about $7\%$ more than what we would
expect for filling factor $\nu=\ot$, (\ref{eq-ch03-03}). 

The quality of the match relies on the choice of $m=3$ for the exponent in
Expr. \ref{eq-ch03-09}
(for $r\to 0$) and on the fitting constants $c$ and $k_F$ 
(around $r\approx r_1$). The surprising fact is therefore only the
good match {\em between} $r=0$ and $r=r_1$. Also note that long-range
($r\gg r_1$) behaviour of expression (\ref{eq-ch03-09}) and of $g(r)$ of the
Laughlin state are different. This again emphasises the differences between
the Laughlin state and the Fermi gas. Expression (\ref{eq-ch03-09}) provides
therefore only another representation of the exchange hole, parallel
to approximate formulae given e.g. by Girvin \cite{girvin:07:1984}.

In conclusion, we have seen that the correlation function of the
correlated $\nu=\ot$ ground state (Fig. \ref{fig-ch03-03}) has a strong first
maximum (near to $4.4\ell_0$) and an unusual exchange hole
$g(r)\propto r^6$. These features distinguish the $\ot$ state from
both free 2D Fermi gas and completely filled Landau levels and
indicate the {\em liquid-like} and 
{\em correlated} nature of the Laughlin state.

\subsubsubsection{Filling factor $\nu=\tt$}

Provided the Landau level mixing is absent, the particle-hole
symmetry in one Landau level gives a direct relation (isomorphism)
between Hilbert subspaces of fully polarized states at $\nu=\tt=1-\ot$
and $\nu=\ot$, Subsect. \ref{pos-ch02-03}. Owing to this relation
eigenvectors of any
radial two-particle interaction are exactly the same\footnote{In the
  following sense: Take an eigenvector for $\nu=\ot$. This is a linear
  combination of Slater determinants from the $\nu=\ot$ space. Replace
  each of them by its particle-hole counterpart and the resulting
  state from the $\nu=\tt$ space is an eigenstate.} 
in both spaces and corresponding eigenvalues are identical up to
a constant shift.

\begin{figure}
\subfigure[The polarized state.]{%
\includegraphics[angle=0,scale=0.6]{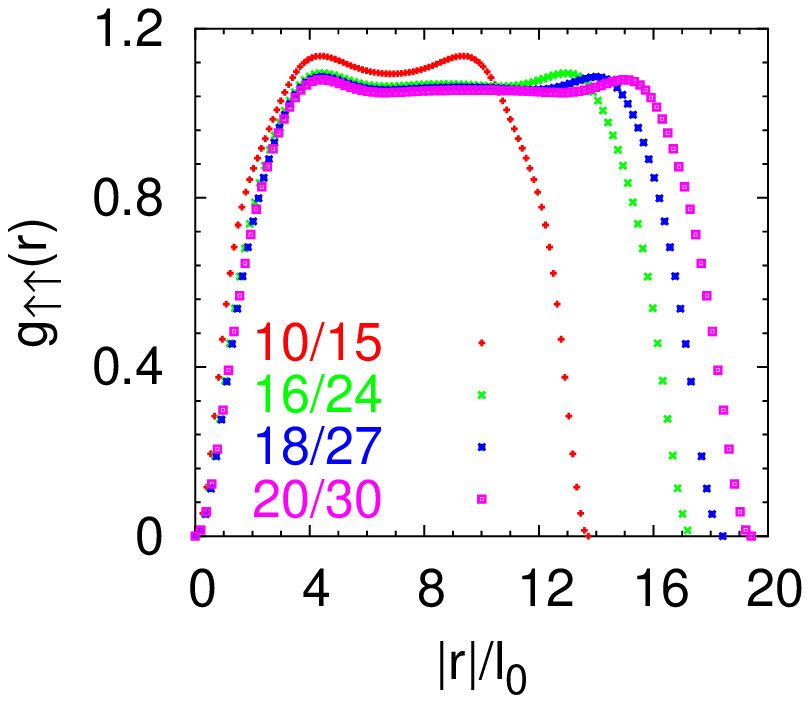}%
\label{fig-ch03-06}}
\subfigure[The spin-singlet state: correlation between unlike spins
({\em left}) and like spins ({\em right}).]{%
\includegraphics[angle=0,scale=0.52]{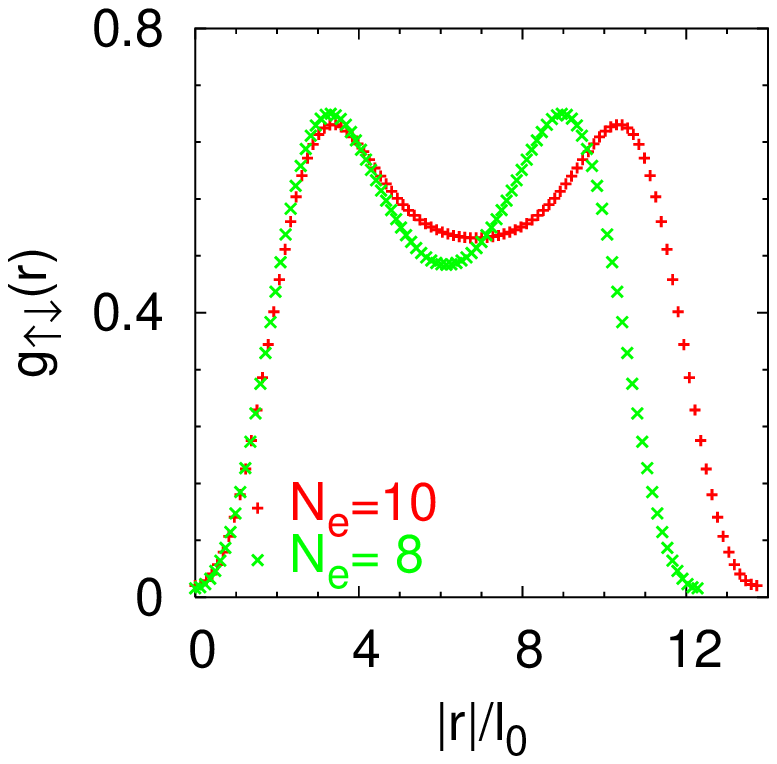}%
\includegraphics[angle=0,scale=0.52]{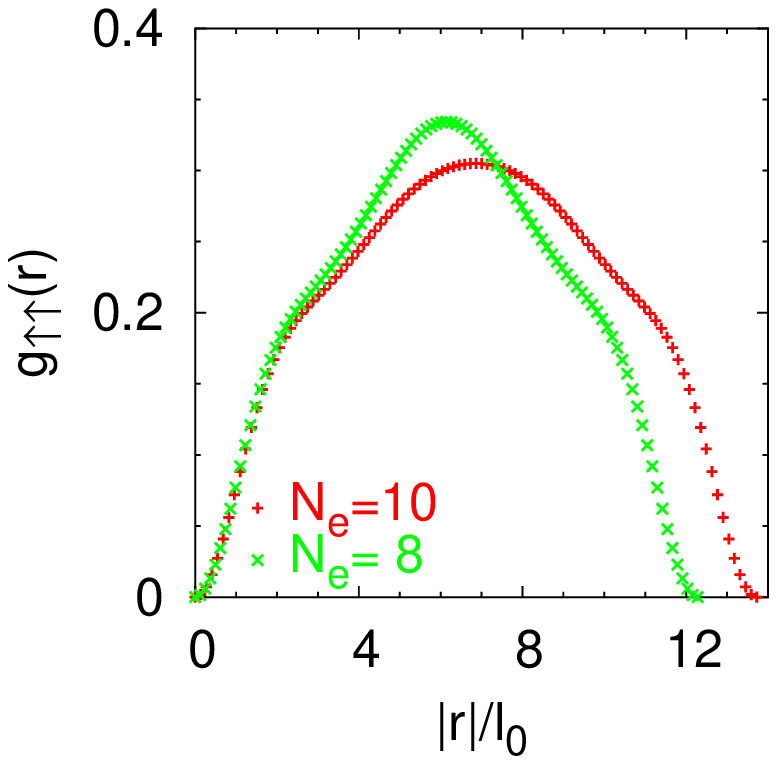}}
\caption{The $\nu=\tt$ ground states shown in their correlation
functions. Section along the diagonal of the square elementary cell is
shown, i.e. $g(r)=g(x/\sqrt{2},x/\sqrt{2})$ 
and systems of different sizes are compared.}\label{fig-ch03-07}
\end{figure}

The correlation function in the {\em fully polarized} $\nu=\tt$
ground state, Fig. \ref{fig-ch03-07}a, is thus linked to the one of the
Laughlin WF by an analytical formula (\ref{eq-ch02-35}).
For a system with
$N_m$ flux quanta, i.e. having an area of $2\pi\ell_0^2 N_m$,
it reads
\begin{equation}\label{eq-ch03-04}
  \tt N_m(\tt N_m -1)g_{\nu=\tt}(\vek r) = 
  \ot N_m(\ot N_m -1)g_{\ot}(\vek r) + \ot N_m^2 g_{\nu=1}(\vek r)\,.
\end{equation}
The $g(r)\propto r^6$ short range behaviour is thus obscured by the second
term.

\begin{figure}
\subfigure[The polarized state.]{%
\includegraphics[angle=0,scale=0.52]{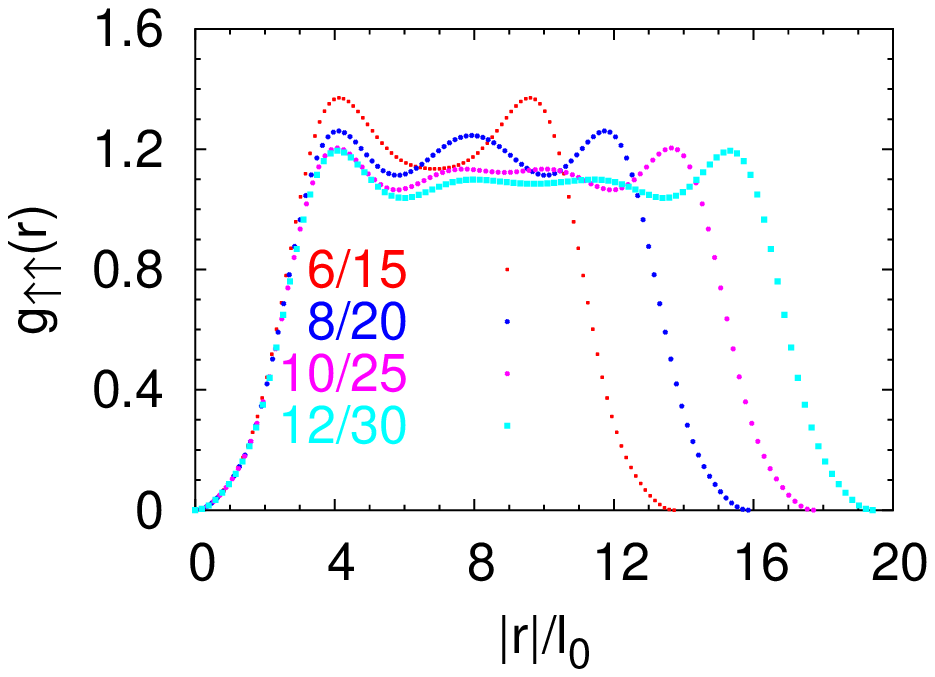}%
\label{fig-ch03-10}}
\subfigure[The spin-singlet state: correlation between unlike spins
({\em left}) and like spins ({\em right}).]{%
\includegraphics[angle=0,scale=0.52]{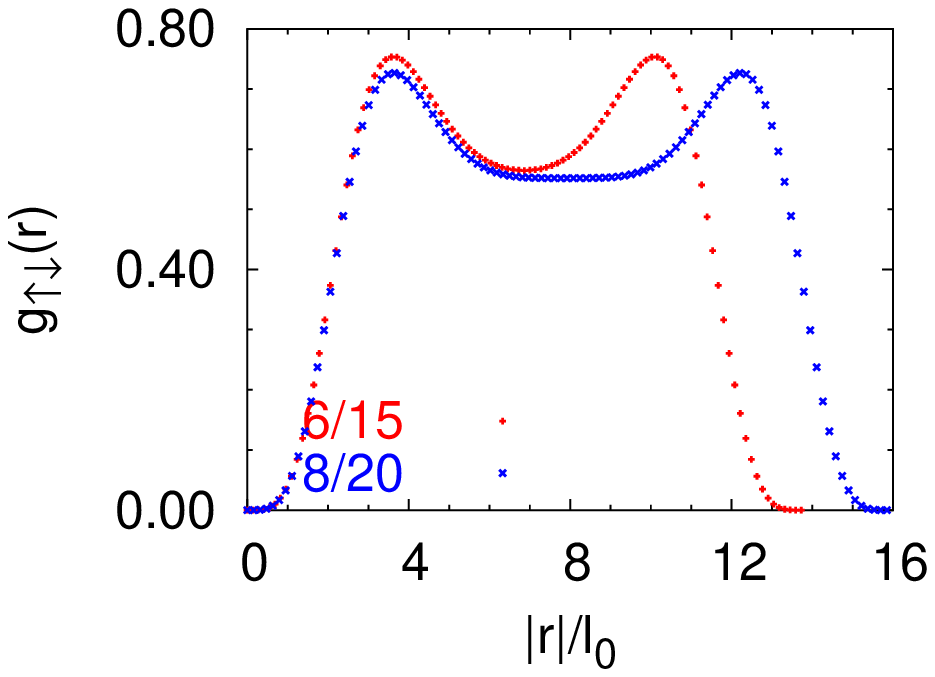}%
\includegraphics[angle=0,scale=0.52]{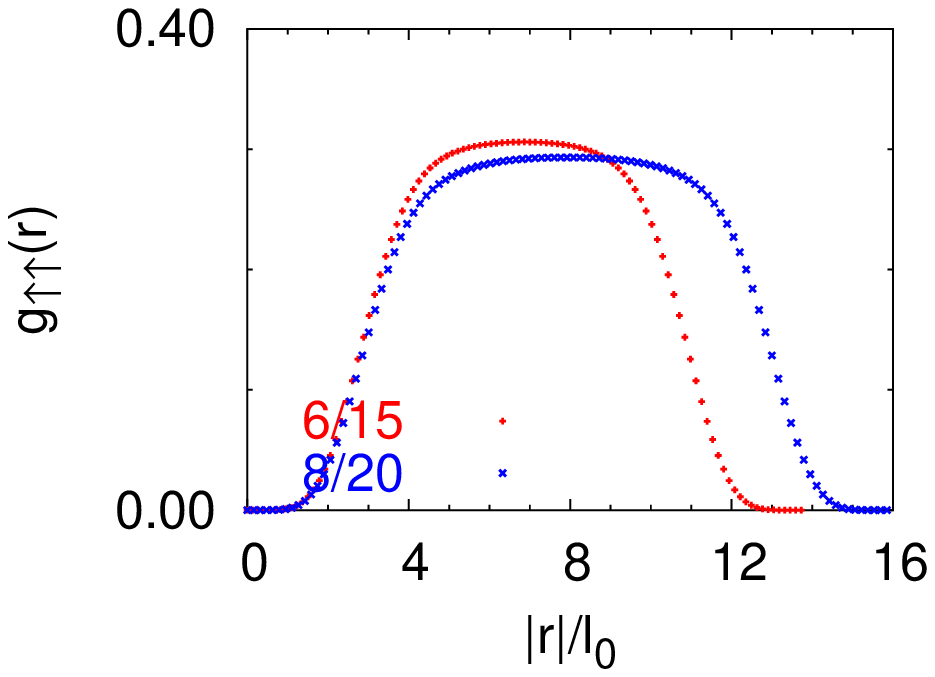}}
\caption{The $\nu=\tf$ ground states, correlation functions
$g(r)=g(x/\sqrt{2},x/\sqrt{2})$. Note the good match of peak positions
in systems of different sizes.}\label{fig-ch03-09}
\end{figure}

The {\em spin singlet} ground state at $\nu=\tt$ has a different
character. Here, we can distinguish between correlation functions for
electrons of like spin, $g_{\up\up}(\vek r)$, and for electrons of
opposite spin, $g_{\up\dn}(\vek r)$, Fig. \ref{fig-ch03-07}. 
Neither of them bears any apparent resemblance to either the $\nu=\tt$
or $\nu=\ot$ polarized ground states. We should like to point out some
of their particular features.

(i)    The ring-like form of $g_{\up\dn}(\vek r)$
suggests that the state consists of pairs of particles with opposite
spin with average separation $r_{\up\dn}\approx 3.3\ell_0$. 

(ii) There is a deep hole in $g_{\up\dn}(r)$ around zero. 
This can{\em not} be due to Pauli exclusion principle which applies
only to electrons of like spin, but rather solely due to Coulomb
repulsion. As a check (not presented here), 
a comparison between $g_{\up\dn}(r)$ in
Fig. \ref{fig-ch03-07} and the 'lowest LL Pauli hole' $g_{\nu=1}(r)$
(\ref{eq-ch03-08}) reveals that their forms are indeed different.
Also note that the
value of $g_{\up\dn}(0)$ is not exactly zero, it is several percent of
the maximal value of $g_{\up\dn}(r)$, Subsect. \ref{pos-ch03-12}.

\begin{figure}
\vspace*{-4cm}
\subfigure[Filling factor $\tt$. The shoulder at
$r\approx 2\ell_0$ is apparently caused by a term proportional to
$1-\exp(-r^2/2\ell_0^2)$, i.e. $g_{\nu=1}(r)$, cf. (\ref{eq-ch03-08}),
which contributes to the total $g_{\up\up}(r)$. After this term was
subtracted, a local power analysis of $g_{\up\up}(r)$ has been performed.
]{\label{fig-ch03-11a}
\hbox{\includegraphics[angle=0,scale=0.38]{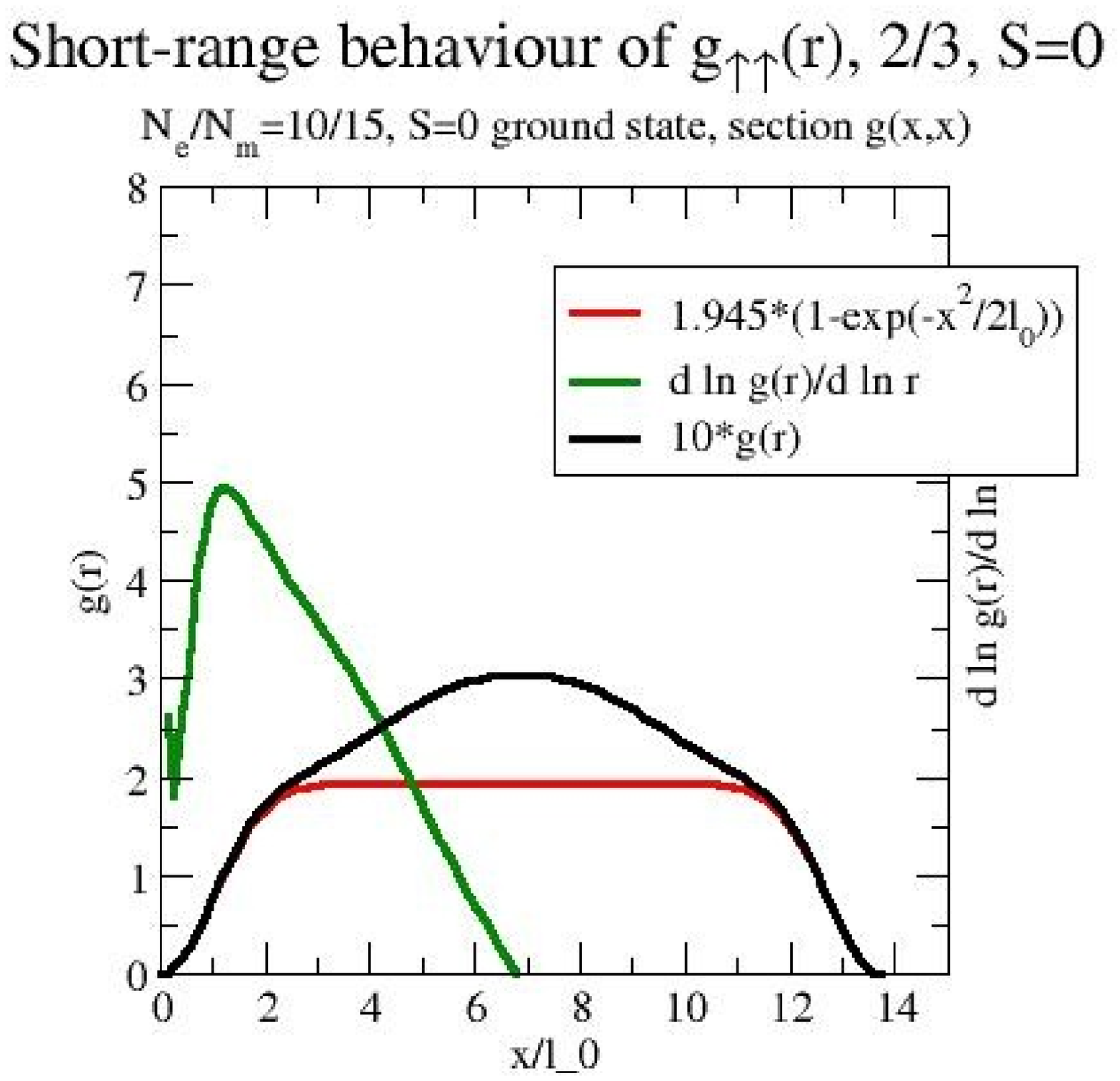}}}
\subfigure[Filling factor $\tf$. Local power analysis near 
to $r=0$. Noise at very small distances is purely due to numerical
inaccuracies: values of $g(r)$ are already very small there.]{\raise10cm\hbox{%
\includegraphics[angle=-90,scale=0.45]{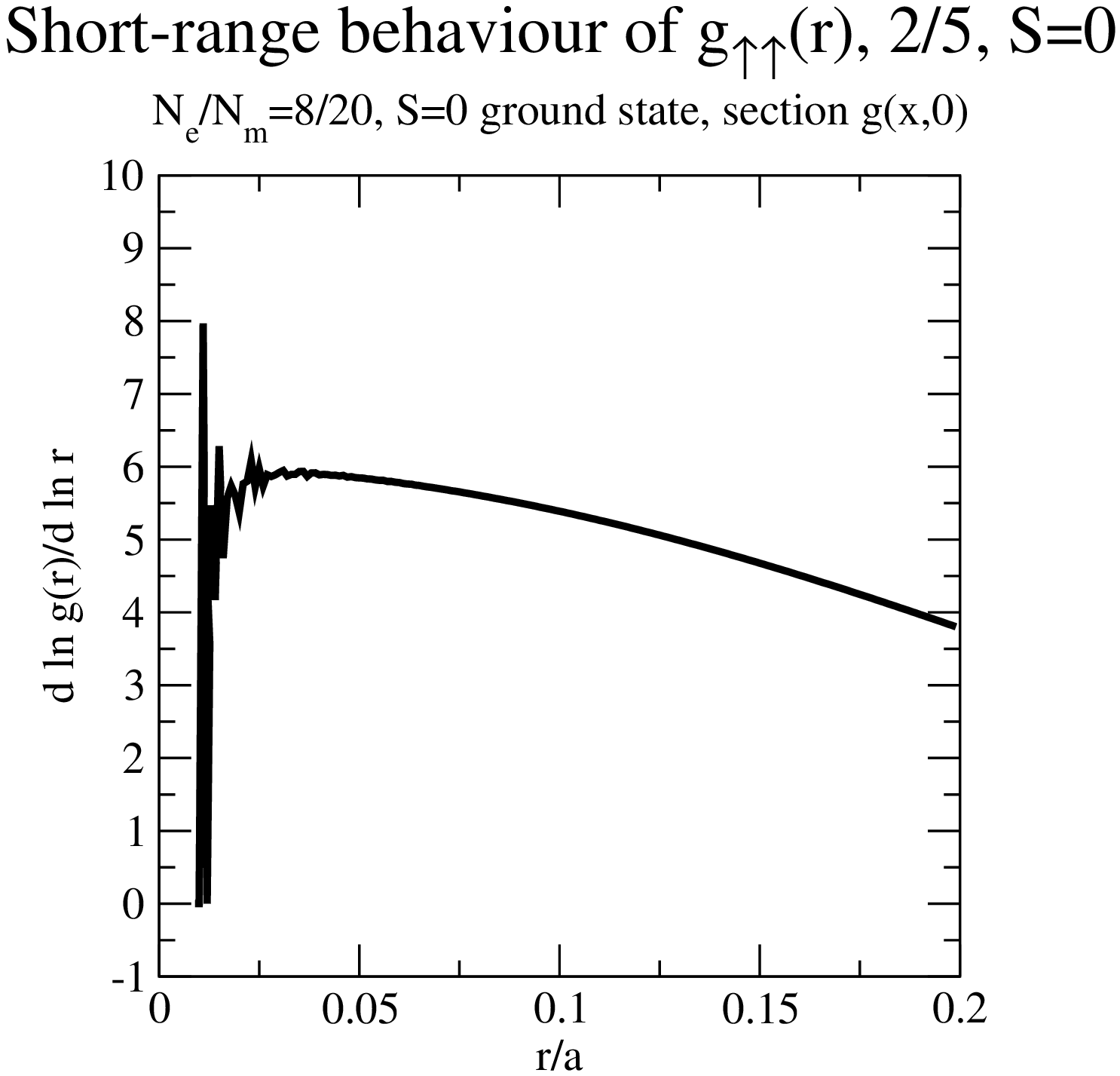}}%
\hskip-5cm\label{fig-ch03-11b}}
\caption{Correlation of like spins, $g_{\up\up}(\vek r)$, of the
singlet ground states at filling factors $\tt$ and $\tf$. Local power
analysis (\ref{eq-ch03-10}) 
shows, that both correlation functions are $\propto r^6$ for
$r\to 0$; however, the shoulder in the state at filling factor $\tt$
has to be subtracted first.}
\label{fig-ch03-11}
\end{figure}

(iii) There is a well pronounced shoulder in $g_{\up\up}(r)$
around $r\approx 2\ell_0$. It is very suggestive, how well this
shoulder can be fitted by the correlation function of a full lowest
LL, $g_{\nu=1}(r)$, i.e. the lowest LL exchange hole
(\ref{eq-ch03-08}). This is shown in Fig. \ref{fig-ch03-11}a. 

This feature reminds of the relation between $\ot$ and $1-\ot$ systems
(\ref{eq-ch03-04}). This is also supported by the fact, that after
the shoulder is subtracted [$g_{\nu=1}(r)$ times a constant], 
the remaining part of $g_{\up\up}(r)$ is
$\wt{g}(r)\propto r^6$ at short distances (Fig. \ref{fig-ch03-11}a), 
just as it is the case in the $\ot$ Laughlin state. 
However, particle-hole conjugation between filling factors $\ot$ and $\tt$ is
applicable only for spin-polarized states.

\begin{SCfigure}[20]
\vspace*{-2cm}

\leavevmode\raise3cm\hbox{%
\includegraphics[angle=-90,width=0.6\textwidth]{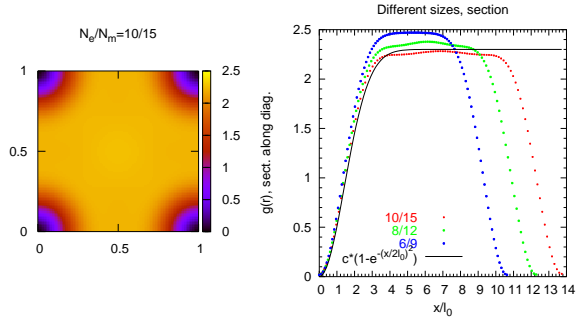}}
\caption{Different density-density correlation functions in the $\nu=\tt$
singlet state seem to be related to each other: $g_{\up\dn}(\vek
r)+g_{\up\up}(\vek r)$ (properly normalized, see text) is very similar to the
function $1-\exp(-r^2/4\ell_0^2)$, the density-density correlation in
a full Landau level with $\ell_0\sqrt{2}$ in the place 
of~$\ell_0$. }
\label{fig-ch03-08}
\end{SCfigure}

(iv) The sum of $g_{\up\up}(r)$ and $g_{\up\dn}(r)$
properly scaled for $N_e\to\infty$ lies very close to $g_{\nu=1}(\vek r)$ 
with $\ell_0$ substituted by $\ell_0\sqrt{2}$,
Fig. \ref{fig-ch03-08}. Proper scaling means that $g_{\up\up}(r)$ and
$g_{\up\dn}(r)$ should  have the same norm, e.g. equal to one, in sense of
  (\ref{eq-ch03-14}). With the current notation
  (\ref{eq-ch03-13}) this is true only for $N_e\to\infty$.

Therefore, if spin is disregarded, the
singlet ground state at $\nu=\tt$, created by magnetic field $B$,
strongly resembles the state of a completely filled lowest LL at
magnetic field $B/2$.

{\em Summary:} the {\em polarized} ground state at $\nu=\tt$ is the
particle-hole conjugate of the Laughlin state at
$\nu=\ot$. The electronic correlation function of the $\tt$ state reproduces
the liquid-like maximum at $r_1\approx 4.4\ell_0$ but the $\nu=\ot$ broad
exchange hole with $g(r)\propto r^6$ is hidden behind the lowest
LL exchange hole, $g_{\nu=1}(r)$.

The {\em singlet} GS seems to consist of pairs of spin up and spin
down electrons with characteristic size of $3.3\ell_0$. Together with the sum
rule, point (iv) above, this could be interpreted as
that $N_e$ electrons in the singlet GS form $N_e/2$ pairs, each with total
$S_z=0$ and these pairs form the same state as $N_e/2$ fermions at
$\nu=1$ in the ground state.

In particular, it should be emphasised that the singlet state  can{\em not}
be described as a mixture of two mutually uncorrelated $\nu=\ot$
Laughlin liquids, one with spin up, another with spin down, as we
could wrongly infer from the picture of non-interacting composite
fermions, see comment \cite{comm:ch03-06}.


\subsubsubsection{Filling factor $\nu=\tf$}

This filling factor should be the counterpart to $\nu=\tt$ within the CF
picture. The two magnetic fluxes are attached parallel rather than
antiparallel to the effective magnetic field and in both cases the CF
filling is two (Sect. \ref{pos-ch02-17}).  In spite of this relation the
density-density correlations between electrons show significant
differences.

The correlation hole of the {\em polarized} ground state
(Fig. \ref{fig-ch03-09}a or Fig. \ref{fig-ch03-03}, middle) is much
broader for $\nu=\tf$. The first maximum occurs in both systems ($\tf$
and $\tt$) at about the same distance $\approx 4.1\ell_0$, it is
however much better pronounced in the $\tf$ system and also more
structure is present beyond the first maximum here. Around $r=0$ both
systems follow $g(r)\propto r^2$. However, whereas
$g(r)$ for $\nu=\tt$ is dominated by the 'exchange hole',
i.e. $g_{\nu=1}(r)$, cf. (\ref{eq-ch03-04}), the $\tf$ state has a
much broader minimum around $r=0$. 

These findings are not unexpected. Consider two systems of the same
area $2\pi\ell_0 N_m$, one at fillings $\tt$ and $\tf$,
respectively. The latter will be more diluted ('emptier'), 
since it contains only
$\tf N_m$ electrons, compared to $\tt N_m$ in the $\nu=\tt$ system
(\ref{eq-ch02-38}). Therefore, the correlation hole in $g(r)$ can
be broader in the $\tf$ system. This conclusion is not a controversy
of the CF picture, rather, it is a warning. By far not all claims
which are true for electronic Landau levels (e.g. $\nu=2$ state remains the
same regardless of the direction of the magnetic field) 
are true for composite fermions as well (it matters whether
$B_{\eff}$ is pointing parallel and antiparallel to the attached flux
quanta).

There is also a close relation between the polarized
$\tf$ GS and the Laughlin $\ot$ state according to the CF picture.
The latter one corresponds to filling factor one, the former one to
filling factor two of composite fermions. Comparing these two states,
we find a bit stronger structures in the density-density correlation
of the $\nu=\tf$ GS and also the first maximum shifts to smaller
distances ($4.4\ell_0$ at $\nu=\ot$ and $4.1\ell_0$ at
$\nu=\tf$). Both effects are quite similar to what happens when going
from $\nu=1$ to $\nu=2$, cf. Figure next to
(\ref{eq-ch03-08}). Comparing the $\nu=2$ and $\tf$ systems,
we again (cf. $\nu=1$ and $\ot$)
find much stronger structures of $g(r)$ in the latter case, just as 
we expect for a liquid state.




Some marked differences occur also in the {\em singlet ground states}
at both filling factors. At $\tf$, correlation functions
$g_{\up\up}(\vek r)$ as well as $g_{\up\dn}(\vek r)$ seem to be quite
flat beyond $r_m\approx 6\ell_0$. We may speculate that the same is true
for the filling $\tt$, Fig. \ref{fig-ch03-09} (the shoulder in
$g_{\up\up}$ would probably have to be subtracted first), but then the plateau
would occur first beyond some larger distance $r_m$ 
which is not accessible by exact diagonalization.

It is remarkable that after subtracting the shoulder from
$g_{\up\up}(r)$ of the $\tt$ singlet state (point (iii) in the
discussion of $\tt$), the rest $\wt{g}(r)$ is $\propto r^6$ near to
$r=0$. This is the same behaviour as we find in $g_{\up\up}(r)$ of the
$\tf$ singlet state, see Fig. \ref{fig-ch03-11}.

Correlations of unlike spins exhibit one clear maximum which is, as
compared to $\tt$, slightly but perceptibly shifted to a bit larger
$r_{\up\dn}\approx 3.7\ell_0$. This agrees with the above
argument that $\tf$ systems are more diluted than the $\tt$ ones,
but quantitatively this shift is too small. It is only $\approx 30\%$
of what we would naively expect from comparing the areal electron
densities.

Finally, the $r\to 0$ behaviour of the $\tf$ singlet state,
$g_{\up\dn}(r)\propto r^4$ and $g_{\up\up}(r)\propto r^6$, matches the
behaviour of the $\{3,3,2\}$-Halperin wavefunction
(\ref{eq-ch02-56}) and this $\Phi_{332}[z]$ is in turn
identical
with the ground state wavefunction proposed by Jain's theory,
Subsect. \ref{pos-ch02-18}. This is because $\Phi_{nn'm}[z]$ lies completely in
  the lowest LL and thus the last step of Jain's procedure, namely the
  projection to the LLL, is out of effect. Seen from the opposite
  direction: the singlet $\tf$ state corresponds to filling only the
  lowest CF LL spin up and spin down.

{\em Summary:} From the viewpoint of composite fermion theories, the
{\em polarized} $\tf$ state ($p=2$), Tab. \ref{tab-ch02-02}, is
related both to the $\ot$ Laughlin state ($p=1$) and $\tt$ polarized
ground state ($p=-2$). The electron-electron correlations in exactly
diagonalized systems clearly support the former relation, the latter one
($\tf$ with $\tt$) is however far from being obvious in this way. 

Neither  is the analogy between
$\tf$ and $\tt$ apparent for the {\em singlet} ground state. 
Although similarities exist, perhaps most importantly pairing between
electrons of unlike spin, short range behaviour of correlation
functions is very different.

\subsubsection{Ground state for Coulomb interaction and for a
short-range interaction}
\label{pos-ch03-12}

Short-range interactions as they were introduced in Section
\ref{pos-ch02-01} have a special significance for the FQHE. It has been
repeatedly emphasised that the Laughlin WF is on one hand
an {\em extremely good} approximation of the ground state of a
Coulomb-interacting (CI) system while on the other hand, it is the
{\em exact} ground state of electrons feeling only a short-range
mutual interaction (SRI) as it was defined in section
\ref{pos-ch02-01}.  Consequently, it is very popular to say that a
short-range interaction Hamiltonian captures the essential physics
of the FQHE by inducing the correct correlations
in the ground state. By the correct
  correlations we mean the $\Psi\propto (z_i-z_j)^3$ behaviour when
  two particles approach each other.

The SRI was used in most of the calculations presented in this
work. This choice has been made for two reasons. It brings better
chances in finding analytical results like the Laughlin WF. Moreover
we may hope that the results in finite systems converge faster to the
thermodynamical limit ($N\to\infty$) because the electrons see only
as far as their interaction reaches and thus -- sooner than for a
long-range interaction -- they will not 'realize' anymore that they
live on a torus and not in an infinite plane. Aim of the following
section is to show and discuss how the ground states at $\nu=\tt$
change if the character of the interaction changes.

The ground state energies for CI and SRI are naturally quite
different. This is however for the largest part only an unessential
shift, a part of it is the missing Madelung constant
(\ref{eq-ch02-10}). Under SRI an electron of course cannot
interact with its own image in the neighbouring primitive
cell. More importantly, the gap energies are quite
similar in both cases, Subsect. \ref{pos-ch04-04}.

\begin{figure}
\includegraphics[scale=0.6,angle=-90]{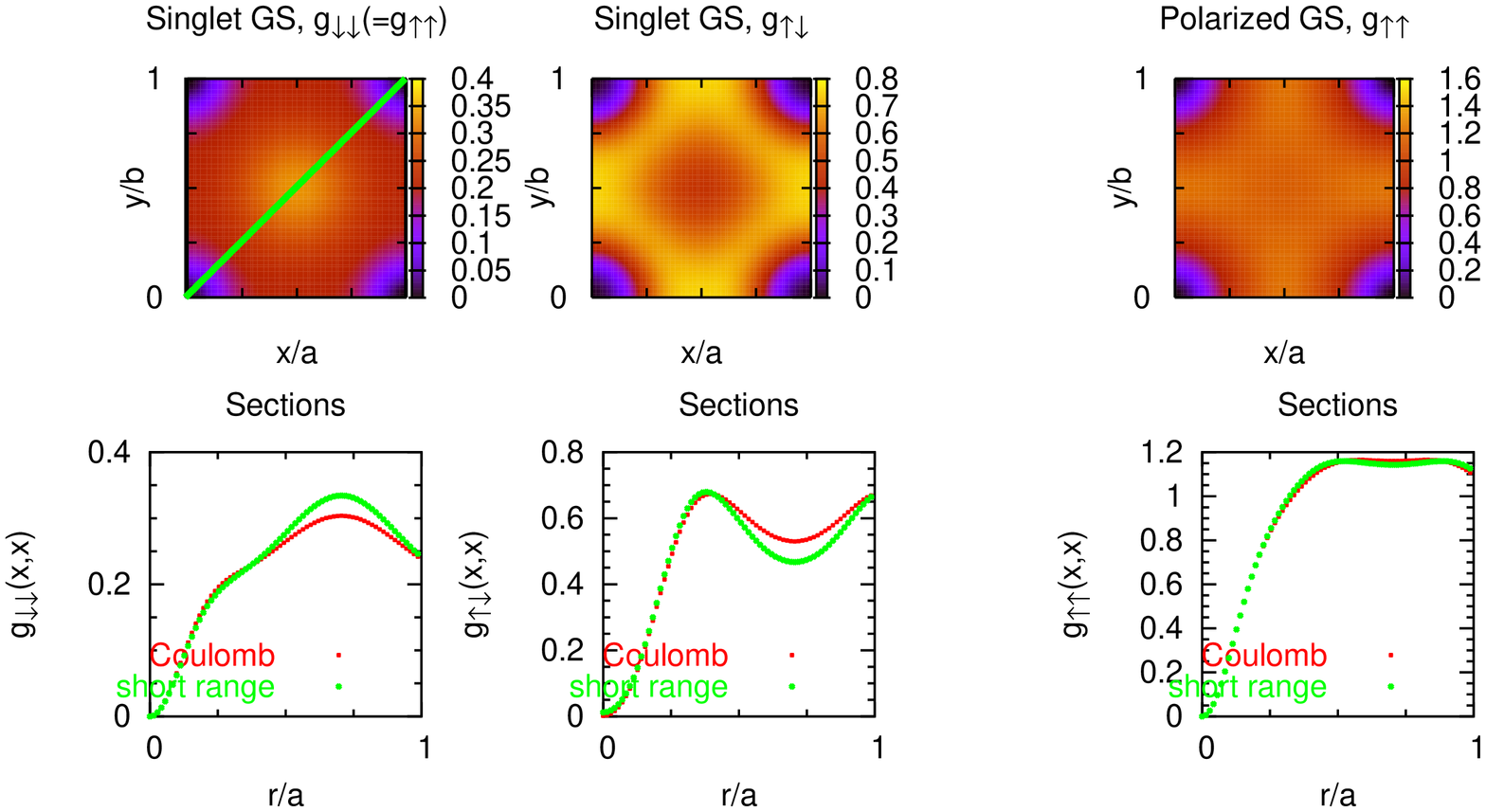}
\vskip-1.2cm

\caption{Correlation functions of the singlet and polarized
ground states at $\nu=\tt$: comparison between the Coulomb
and short-range interaction. The curves are identical for
small $r$ and slight deviations occur at longer scales. This is
another way to demonstrate that it is sufficient to consider
short-range interaction in order to get (almost) correct ground
states under FQHE conditions.}\label{fig-ch03-12}
\end{figure}

Since the density of the incompressible ground states should always be
constant, up to finite size effects to be discussed later, let us now
focus on correlation functions. The three plots in
Fig. \ref{fig-ch03-12} show $g_{\up\up}(r)$ and $g_{\up\dn}(r)$ of the
singlet state and $g(r)$ of the polarized state. In all three cases,
the correlation functions of the CI state and the SRI state are quite similar.
Most apparent differences appear at large distances. On a torus,
the largest possible separation between two electrons is $r=a/\sqrt{2}$.
On the other hand, the correlation functions 
are very precisely identical for small $r$. This shows that,
e.g. in the polarized GS, the wavefunction contains 
the factor $(z_i-z_j)^3$ for CI as well as
for SRI. In other words,
the Laughlin state (as the GS for SRI) describes {\em exactly} the
short-range behaviour of an incompressible state of even long-range
interacting electrons. Fig.  \ref{fig-ch03-12} demonstrates that this
is true (at least in a very good approximation) also for other ground 
states where the analytical wavefunction is not available 
(e.g. the singlet GS).

In fact, for the singlet GS there is a tiny but perceptible difference
in $g_{\up\dn}(0)$ for the two types of interaction. Since
$g_{\up\dn}(0)$ is almost zero, this observation suggests that a yet
modified interaction might lead to analytical results, 
$\{V_0,V_1,\ldots\}=\{\infty,\alpha,0,0,\ldots\}$ in terms of
pseudopotentials, Sect. \ref{pos-ch02-01}. Such an interaction
enforces $g_{\up\dn}(0)=0$, which is anyway almost fulfilled for the
current SRI, and on the other hand it retains the pleasant property of
SRI in polarized systems, i.e. it is one-parametric.

There is yet another significant difference between SRI and CI which
is not obvious in Fig. \ref{fig-ch03-12} at first glance. 
The difference concerns 
the placement of zeroes in the wavefunction and we will concentrate
on the $\nu=\ot$ ground state now (see Sec. \ref{pos-ch02-01}).

\begin{figure}
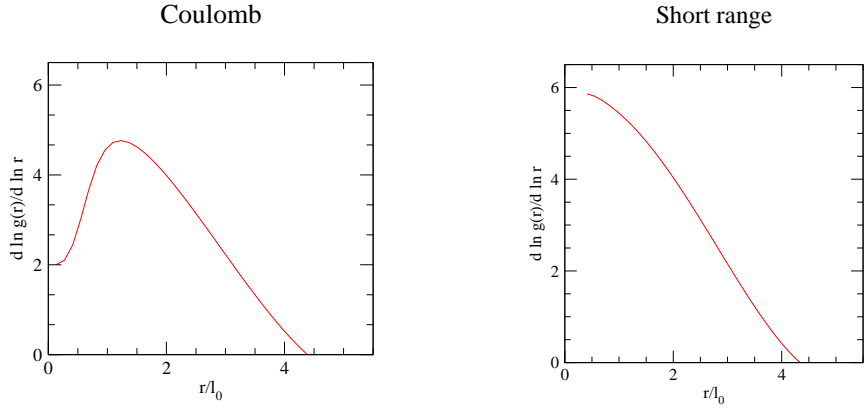

\begin{center}
\includegraphics[scale=0.4]{figs/ch03/ch03-fig13c.eps}
\hskip2cm
\includegraphics[scale=0.4]{figs/ch03/ch03-fig13d.eps}
\end{center}
\caption{The incompressible ground state at $\nu=\ot$ (with 
ten electrons), Coulomb
interaction (left) and a short-range interaction (right). Section
through the density-density correlation function $g(\vek r)$ along
$\vek r=(x,x)$ is taken and the 'local degree' of the polynomial
behaviour is determined (see the text).
While the local behaviour around $r=0$ is
$g(r)\propto r^6$ for the SRI, indicating that there is exactly a
{\bf triple} ($6=2\cdot {\mathbf 3}$) zero of the wavefunction 
on on each electron, we can
clearly see only {\bf one} zero at each electron's position for the
Coulomb interaction, $g(r)\propto r^2$ and $2=2\cdot {\mathbf
1}$. However, going away from $r=0$, the 'local degree' grows and
beyond $\approx 1.5\ell_0$ it approaches the curve of the SRI state.
The conclusion is that one zero (the obligatory Pauli exclusion 
principle zero) is fixed to each electron ($r=0$) in the Coulomb
state and the other two zeroes are only loosely bound to the
electron. First from distances $\gtrsim 1.5\ell_0$ this compound object
looks like an electron with two attached flux quanta.}\label{fig-ch03-13}
\end{figure}

In a general fermionic state, there must always be a zero bound to
each electron in order to fulfil the Pauli exclusion principle: two
electrons (of the same spin) cannot be at the same point in space
simultaneously, {\em ergo} if $z_1=z_i$ then the wavefunction must
vanish. Factors $(z_i-z_j)^3$ in the Laughlin state mean that there are
two extra zeroes exactly at the position of each electron. That is
why $g(r)\propto r^6$ for small $r$'s, Fig. \ref{fig-ch03-13},
right and (\ref{eq-ch03-10}). 
For CI, the Laughlin WF is only an {\em approximation} to the
ground state. In the real ground state, the one obligatory zero is
still sitting on each electron and the two others are only near
rather than exactly on the top of the electron. In
Fig. \ref{fig-ch03-13} we can even see how far they are on
average. These two extra zeroes are now mobile and their position
depends on the position of all other electrons. 
Note that this distance depends on the system
size \cite{mueller:2005}.

\subsubsubsection{Local power analysis}

A comment is due
on the way how the plots in Figs. \ref{fig-ch03-13},\ref{fig-ch03-11}
were obtained. It is basically a section of
$g(\vek r)$ along one straight line going through $\vek r=0$. 
This function was then transformed by 
\begin{equation}\label{eq-ch03-10}
   g(r) \ \longrightarrow  \frac{\d \ln g(r)}{\d \ln r}
\end{equation}
which gives a local degree of the polynomial behaviour. Let us give an
two examples. If $g(r)$ were $\alpha r^n$ then $\d \ln
g(r)/\d \ln r = n$. If $g(r)\propto (r-r_0)^n$ then $\d \ln
g(r)/\d \ln r = n r/(r-r_0)\to n$ for $r\gg r_0$. In other words, if
there is a dominant $r^n$ term in $g(r)$, the quantity plotted in
Fig. \ref{fig-ch03-04} gives the exponent. Of course, it is only
approximate except for the case $g(r)=\alpha r^n$ but it is quite easy
to evaluate and moreover it gives a global property of the
wavefunction as compared to fixing electron positions $z_2,\ldots,
z_n$ and examining the WF as a function of $z_1$ where results depend
on where we fix the electrons $z_2,\ldots, z_n$.

\subsubsection{Some excited states}
\label{pos-ch03-11}

There is a rich variety of excitations to the incompressible
FQH states.
For instance quasiholes, excitons (quasihole-quasielectron pairs),
charge density waves (CDW) or spin density waves (SDW), all of them
can be described analytically (at least to some extent), and then of
course all the rest of excitations 
which has not been understood up to now. Following
the introduction given around (\ref{eq-ch02-43}), we will now
demonstrate how to identify some of these excitations in spectra 
obtained by exact diagonalization at the example of $\nu=\ot$.

\subsubsubsection{Charge density waves}

CDWs can be excited for example in the liquid GS at $\nu=\ot$. 
Disregarding the possibility of spin flips
(as it may be reasonable when Zeeman energy is too high), it turns out
that these are the lowest excitations. 

\begin{SCfigure}
\includegraphics[scale=0.6,angle=0]{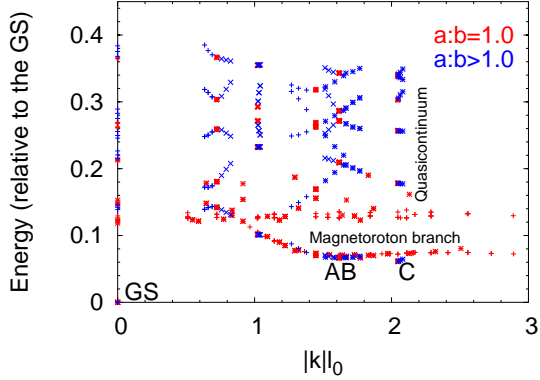}
\caption{The ground state (at $k=0$) and low excitations in SRI
fully polarized $\nu=\ot$ systems of different sizes (4-10 electrons).
Energy plotted against the $|\krv|$, see
Sec. \ref{pos-ch02-06}. To be able to study more points in $k$ space,
systems of different sizes are compared and also systems with aspect
ratios slightly varying from one, see text. Note the well pronounced
magnetoroton branch. Note that energies calculated in systems of
{\em different} sizes lie on the {\em same} branch indicating that
these states are not much system-size-dependent (and therefore relevant
even in infinite systems).
The correlation functions for three states lying on this branch
(A,B,C) are depicted in 
Fig. \ref{fig-ch03-55}.
}\label{fig-ch03-31}
\end{SCfigure}

In Fig. \ref{fig-ch03-31} spectra of several short-range-interacting
$\nu=\ot$ systems (tori of different sizes) are presented. The horizontal
axis is modulus of $\krv$, i.e. the 'crystallographic $k$-vector'
described in Subsec. \ref{pos-ch02-06}. The Laughlin state has
$\krv=0$ and a CDW of wavevector $\vek Q$ excited from this state has
$\krv=\vek Q$. Beware however, that not every state which has
$\krv\not= 0$ must be a charge density wave! Apart from other possible periodic
excitations, there are also basically nonperiodic excitations
(e.g. quasiholes) and such states are forced into periodicity only
'artificially' by the periodic boundary conditions imposed in our
exact diagonalization model.

The lowest excitations in Fig. \ref{fig-ch03-31} form a well developed
branch $E(\krv)$, which is usually called {\em magnetoroton branch},
and other excited states form a quasicontinuum. 
The dispersion of the magnetoroton branch can be
calculated analytically in the single mode approximation. The original
calculation by Girvin {\em et al.} \cite{girvin:02:1985} for {\em
Coulomb } interacting systems at $\nu=\ot$ showed a well pronounced
minimum in $E(|\krv|)$ of the magnetoroton branch at
$\krv\ell_0\approx 1.4$. In a short-range interacting system, shown
in Fig. \ref{fig-ch03-31}, the situation is slightly different. Having
reached its minimum value, $E(|\krv|)$ remains constant beyond
$\krv\ell_0\approx 1.4$.

A point worth of emphasis is that the magnetoroton branch in
Fig. \ref{fig-ch03-31} contains points (energies) from exactly diagonalized
systems of {\em different} sizes. This confirms our hope that these
states are not bound to some particular geometry of the elementary
cell and that they appear also in an infinite system. 

Dealing with finite systems, we will always have only a finite, and
usually quite small, number of allowed values for $\krv$
(\ref{eq-ch02-45}). On the other hand, the more points in
$\krv$-space we can access, the better we can recognise modes in
exact diagonalization spectra, just like the magnetoroton branch in
Fig. \ref{fig-ch03-31}. Note also the large space between $\krv=0$ and
the next smallest $|\krv|\approx 0.5\ell_0^{-1}$ in
Fig. \ref{fig-ch03-31} which corresponds to the longest wavelength
compatible with the periodic boundary conditions.

The traditional way to improve these limits (few $\krv$-points, too
large smallest $|\krv|>0$) is to study larger systems. This is however
prohibitively difficult with exact diagonalization. An alternative
approach may be to study systems with aspect ratios $\lambda=a:b$ slightly
deviating from one. This allows us to deform the lattice
of allowed $\krv$-points continuously (\ref{eq-ch02-45} contains
$\lambda$), and on the other hand, we can expect that the states will
not suffer from the {\em slight} asymmetry in $a:b$ in
line with the argument that these states are not bound to any
particular geometry of the elementary cell. This method is
demonstrated in Fig. \ref{fig-ch03-31} by the blue points.
The aspect ratio was varied from one up to 1.3. Since the energies of the
CDW states still lie well on the magnetoroton branch, we can conclude
that this variation was still only a small perturbation, 
i.e. acceptable for studying this branch. A more reliable
critierion would be to check overlaps of wavefunctions at $a:b=1$ and $a:b>1$.

\begin{figure}
\hskip1cm\begin{tabular}{cccc}
  \hskip-2.5cm\includegraphics[scale=0.45]{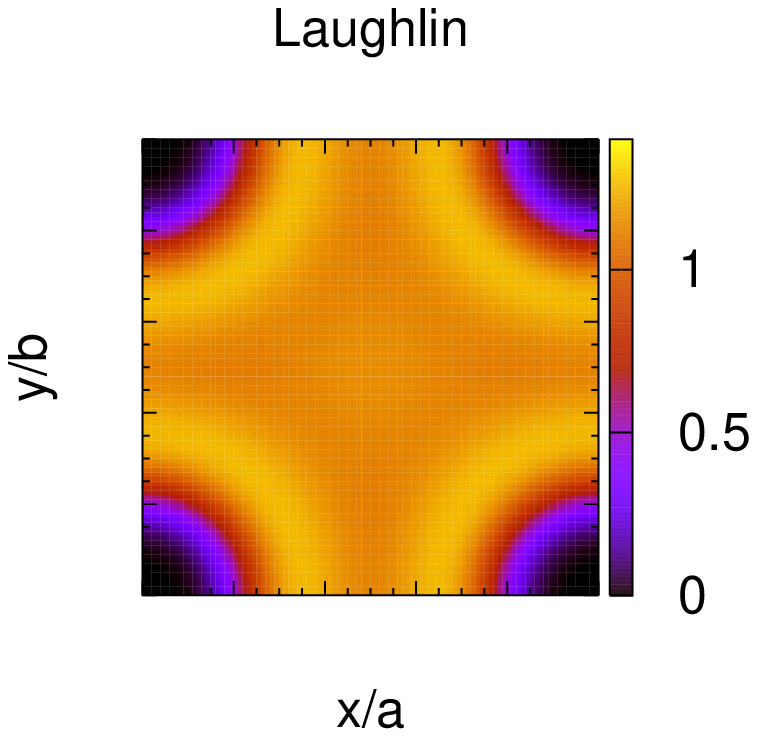} &
  \hskip-2.5cm\includegraphics[scale=0.45]{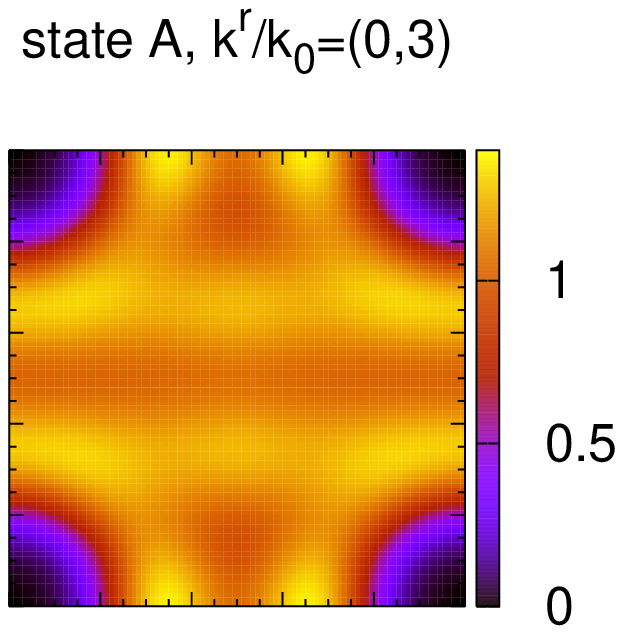} &
  \hskip-2.5cm\includegraphics[scale=0.45]{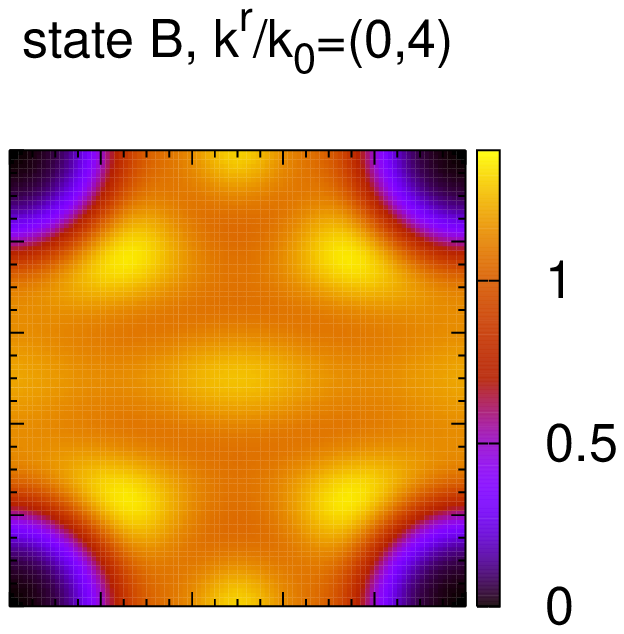} &
  \hskip-2.5cm\includegraphics[scale=0.45]{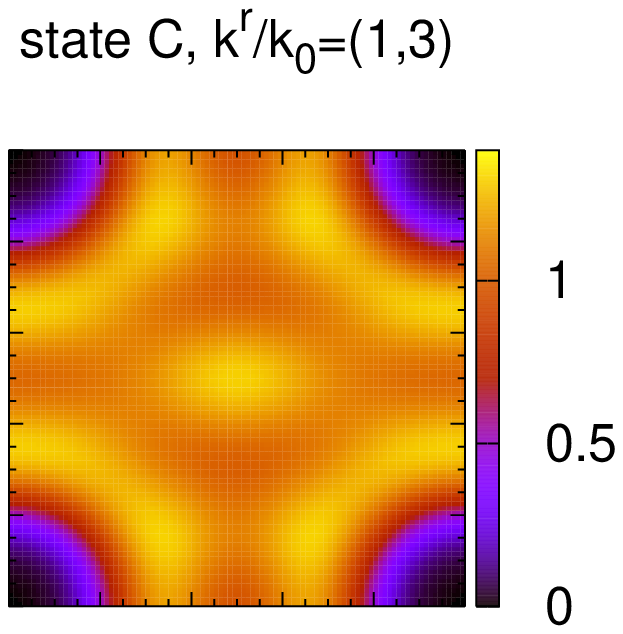} \\
  &
  \hskip-2.5cm\includegraphics[scale=0.45]{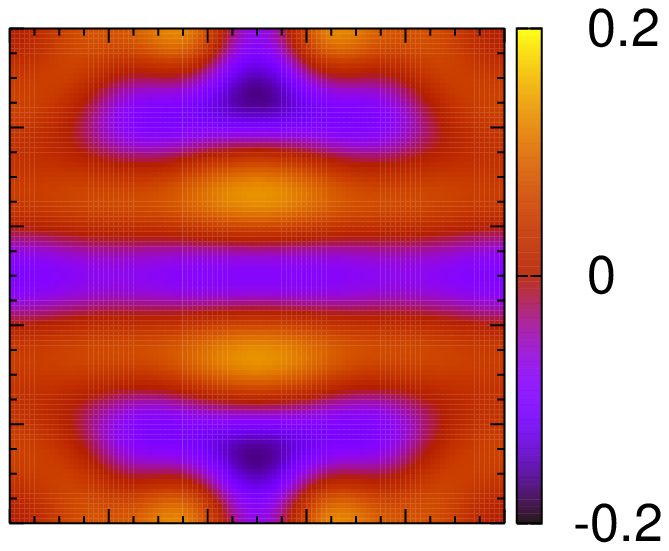} &
  \hskip-2.5cm\includegraphics[scale=0.45]{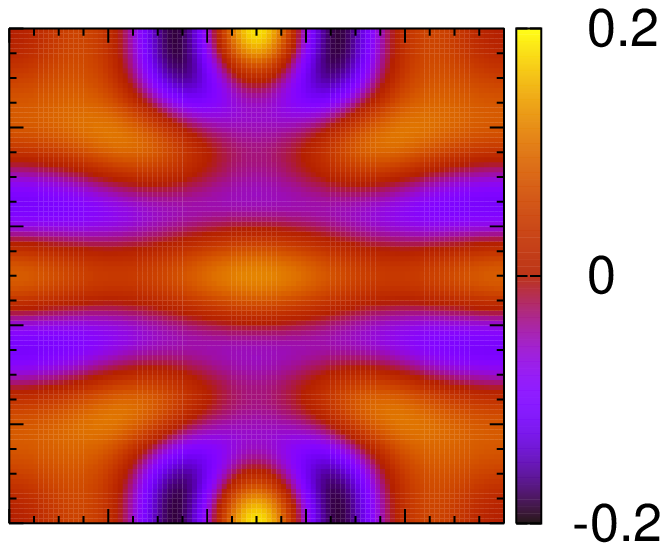} &
  \hskip-2.5cm\includegraphics[scale=0.45]{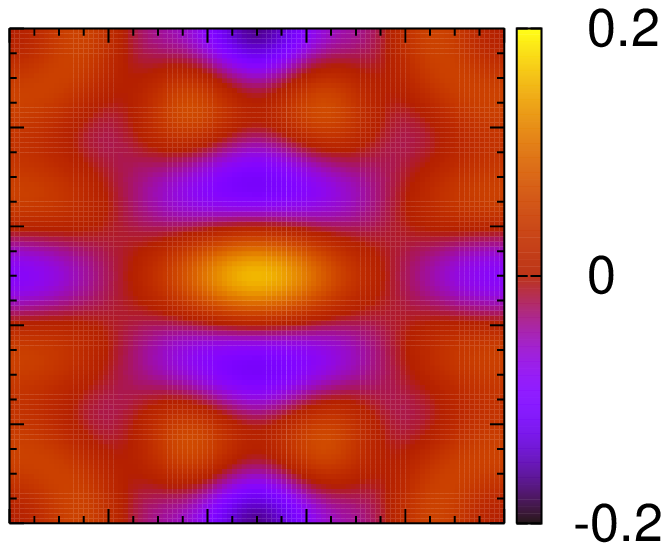} 
\end{tabular}
\vskip-.5cm
\caption{Correlation functions of the ground state (Laughlin) and 
  several CDW states which lie on the
  magnetoroton branch of a $\nu=\ot$ system with short-range
  interaction (eight electrons). 
  {\em Upper row:} correlation functions $g(\vek r)$,
  {\em lower row:} $g(\vek r)$ of the CDW states from which the
  Laughlin state $g(\vek r)$ has been subtracted.}
\label{fig-ch03-55}
\end{figure}

Correlation functions of several states in the magnetoroton branch
(Fig. \ref{fig-ch03-31}) are shown in Fig. \ref{fig-ch03-55}. The first
look at $g(\vek r)$ (upper row in Fig. \ref{fig-ch03-55}) may be
sometimes not enough to distinguish their charge density wave
nature. The CDW is superimposed on the structure of the mother
Laughlin state, which these states are an excitation of. The periodic
structure of $g(\vek r)$ is thus more clear if we subtract the
corresponding correlation function 
of the Laughlin state first, Fig. \ref{fig-ch03-55} lower
row. We can find three periods in $y$ direction (horizontal
waves at $x=0$, $0.3$ and $0.6$) in the state A or 4
periods in $y$ and one period in $x$ in the state C, in agreement with
their values of $\krvd/k_0=\krvd/(N_m\pi/6)$. Note, that
it is harder to distinguish the
periodic structure in the $\krvd=(0,\pi)$
state (B), which may be partly because this is a point of high
symmetry in  $\krvd$-space, Fig.~\ref{fig-ch02-12}.

In conclusion, we have shown how (the best known type of) charge
density wave states on a torus can be identified in the exact diagonalization
spectra and in correlation functions. Generally, we can expect that
charge density waves excited from incompressible liquid states will
form branches in $E(|\krv|)$, provided of course that their energy is
not hidden in a quasicontinuum of other excited states. Correlation
functions show indeed the expected periodicity of a CDW superimposed
on the structure of the ground state. 


\subsubsection{Finite size effects}
\label{pos-ch03-01}

\begin{figure}
\begin{center}
\subfigure[Inhomogeneous density in an $N_e=4$ system.]{%
\includegraphics[angle=0,scale=0.5]{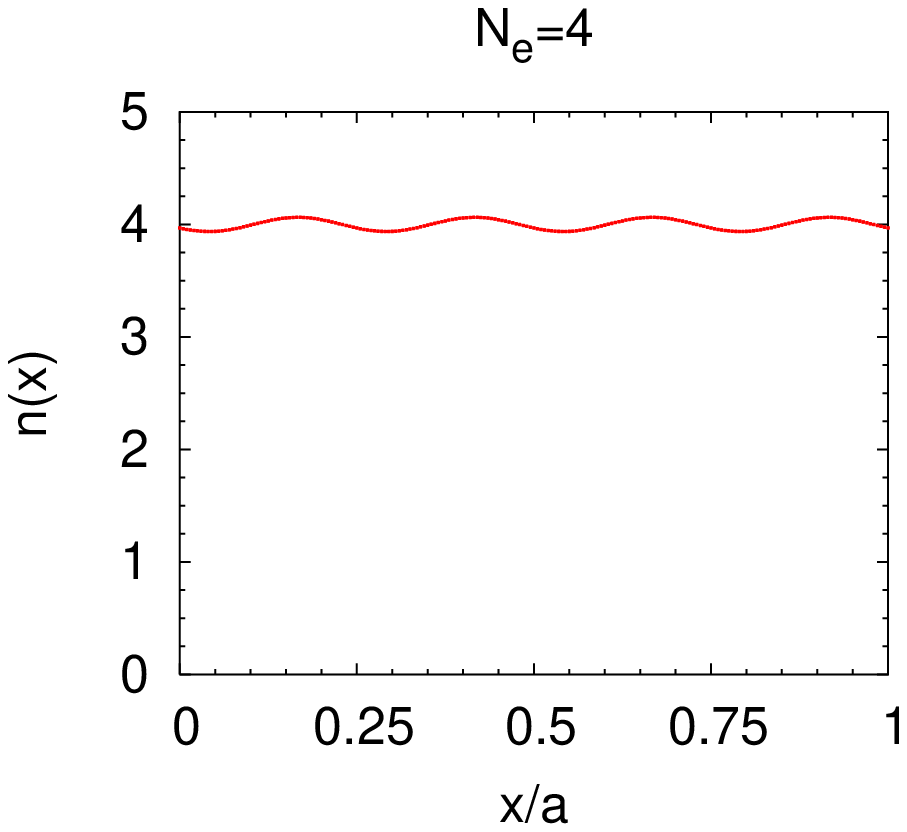}
\label{fig-ch03-15a}}
\subfigure[Comparison between systems of different sizes. Note the
scaling of the $x$ axis.]{%
\includegraphics[angle=0,scale=0.5]{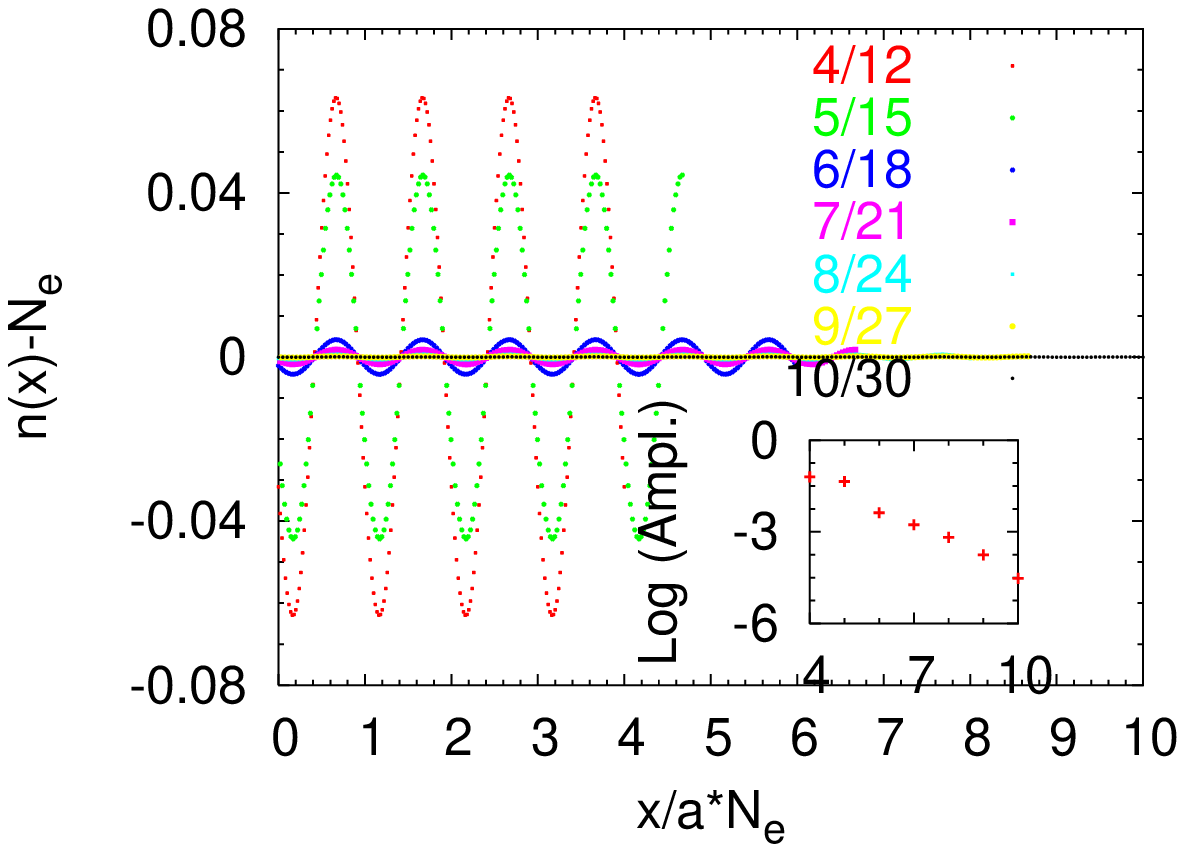}
\label{fig-ch03-15b}}
\end{center}
\caption{Density of the $\nu=\ot$ incompressible ground state as
obtained in torus geometry for different system sizes. The
oscillations can be traced back to the centre-of-mass part of the
wavefunction. As far as this
effect is considered, differences between Coulomb interaction and
short-range interaction are small.}\label{fig-ch03-15}
\end{figure}

Consider a $\nu=\ot$ system with its exact
GS written as $\Psi_L$, the Laughlin wavefunction (WF), see (\ref{eq-ch02-09}).
Particle density in the state $\Psi_L$ is
very precisely constant provided we stay within the disc of radius
$\ell_0\sqrt{2\pi\cdot 3N}$.
The first striking observation is that the density of the ground state obtained
from exact diagonalization varies quite strongly, Fig. 
\ref{fig-ch03-15}. At the same time we notice that the ground state, 
which is claimed to be incompressible, hence non-degenerate, 
is actually triply degenerate.

Fortunately, this does not mean that finite size calculations are
completely wrong. Both facts can be attributed to the
centre-of-mass part of the wavefunction (CMWF) which is {\em not}
present in $\Psi_L$ but {\em is} present in numerical calculations,
Subsec. \ref{pos-ch02-06}. As far as isotropic states are
considered, this seems to be the most serious effect coming from the
finite-sizedness of the system and in the following we will discuss
its origin and how it can be eliminated.

\subsubsubsection{Centre-of-mass part of the wavefunction}

\begin{figure}
\hbox{\hskip1cm (a)\hskip3.2cm (b)\hskip3.2cm (c)\hskip3.2cm (d)\hskip3.5cm }
\hbox{\includegraphics[angle=-90,scale=0.26]{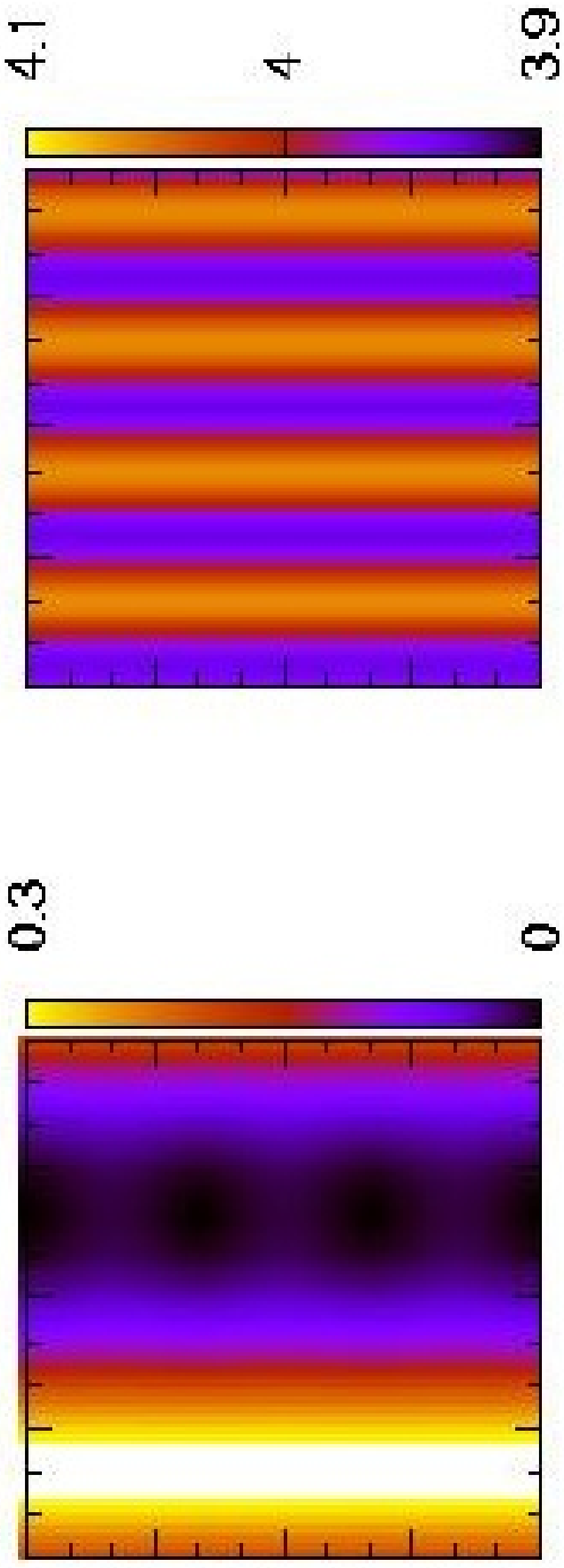}%
\includegraphics[angle=-90,scale=0.29]{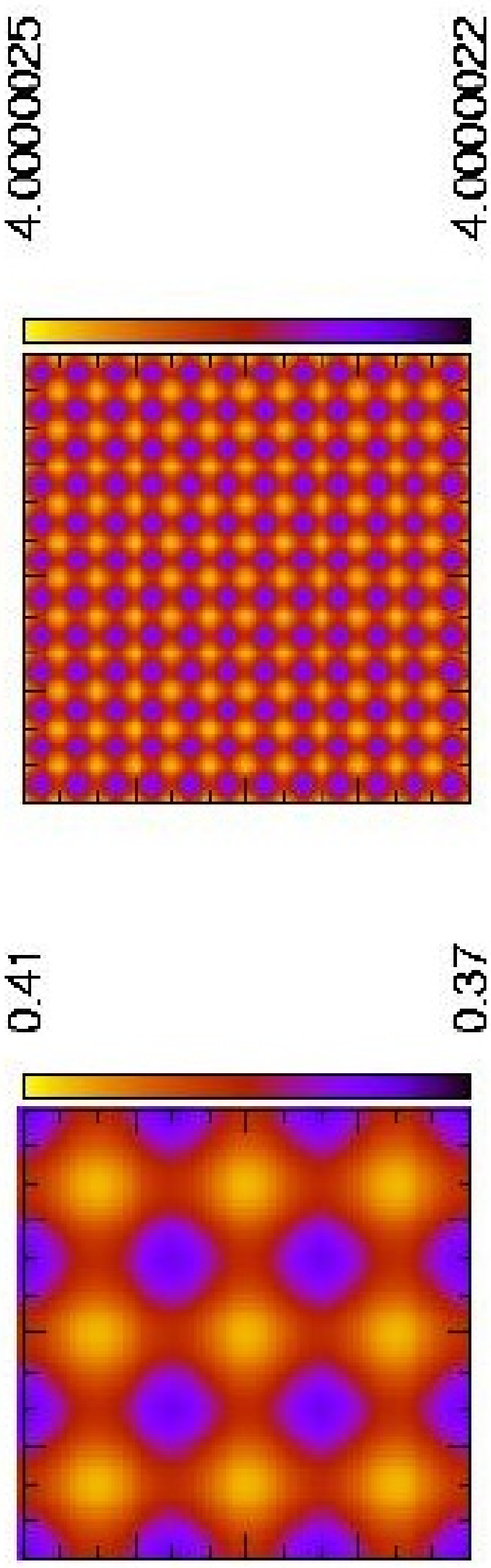}}
\caption{{\em Left to right:} (1) $|\Psi_{CM}(Z)|^2$,
$Z=z_1+z_2+z_3+z_4$ for one of the three degenerate ground states in a
4 particle $\nu=\ot$ system and (2) the density $n(\vek r)$ of this
state. (3) $|\Psi_{CM}^1(Z)|^2+|\Psi_{CM}^2(Z)|^2+|\Psi_{CM}^3(Z)|^2$
of those three states and (4) the sum of their densities (divided by
three). Note that the last density is nearly constant (as it should be
for the Laughlin state) and thus by adding up densities of the three
states differing only in the CM part, we eliminated the effect of the
CM part of WF.}\label{fig-ch03-17}
\end{figure}

The complete WF of the Laughlin state at $\nu=\ot$ (for $n$ particles)
in the disc geometry 
(Subsec. \ref{pos-ch02-08}, \ref{pos-ch02-05})  
might be
\begin{equation}\label{eq-ch03-05}
   \Psi_{1/3}(z_1,\ldots, z_n) = 
       \underbrace{F(Z)\exp(-|Z|^2/2\ell_0^2)}_{\Psi_{CM}(Z)}\times
       \underbrace{\exp\big(-(|z_1|^2+\ldots+|z_n|^2)/4\ell_0^2\big)
       \prod_{i<j} (z_i-z_j)^3}_{\Psi_L(z_1,\ldots,z_n)}
\end{equation}
with $Z=z_1+\ldots+z_n$ and for example $F(Z)=Z^3$
or any other analytic function with three zeroes $Z_1,Z_2,Z_3$. The
CMWF $\Psi_{CM}$ has the form (Sec. \ref{pos-ch02-05}) of a WF for one
particle somewhere in the lowest Landau level to which a single
variable $Z$ is attributed. In torus geometry, the WF must be changed
in order to comply with periodic boundary conditions (PBC)
which amounts to replacing $(Z-Z_i)$ terms by theta functions of the
same argument, Subsec. \ref{pos-ch02-06}. Example of
$\Psi_{CM}$ obtained from the numerically calculated
ground state $\Psi_{GS}$ in a system with four
particles is shown in Fig. \ref{fig-ch03-17}a.
The CM part was extracted from the complete WF by
  the scheme
  $\Psi_{CM}(4\Delta)=
  \Psi(z_1+\Delta,\ldots,z_4+\Delta)/\Psi(z_1,\ldots,z_4) 
  \Psi_{CM}(0)$.
 Note that this result
fully matches what we expect from analytic considerations,
$\alpha$ in Fig. \ref{fig-ch02-11}a.

If we calculate quantities
like the density or correlation function in the state $\Psi=\Psi_r\Psi_{CM}$,
we evaluate integrals of the type
\begin{equation}\label{eq-ch03-17}
  n_{\Psi_{r}\Psi_{CM}}(z) = \int \d z_1\dots \d z_n
  |\Psi_r(z_1,\ldots, z_n)|^2 |\Psi_{CM}(z_1+\ldots+z_n)|^2
  \delta(z_1-z)\,.
\end{equation}
Recast in CM and relative variables, this integral is a
multidimensional convolution of $\Psi_{CM}$ and $\Psi_r$.
Assuming that $\Psi_r$ is isotropic but non-constant which is true
for the Laughlin WF, the function $n(z/4)$ can be thus shown to have the
same periodicity
as $\Psi_{CM}(z)$. Less exactly but in more illustrative terms:
  $n(z)$ is basically a smeared $|\Psi_{CM}(4z)|^2$. Note that this
  explains why $n(z)$ varies much stronger along $x$ than along $y$.

These considerations can be summarized in the following way. 
Even though e.g. the Laughlin state is translationally
invariant, the CM part of the wavefunction which is always present in the
exact diagonalization studies, will cause the density to be
inhomogeneous. Consider eigenstates of
$J$. This determines the form of $\Psi_{CM}$ to be as in
  Fig. \ref{fig-ch03-17}a, cf (\ref{eq-ch02-54})
the density $n(z)$ of an $N_e$-electron state
(a) will be $1/N_e$ periodic along $x$, Fig. \ref{fig-ch03-15} or
Fig. \ref{fig-ch03-17}b,
(b) will be $1/3N_e$ periodic along $y$, Fig. \ref{fig-ch03-17}d,
(c) will be modulated much stronger along $x$ than along $y$ (compare
scales in Fig. \ref{fig-ch03-17}b and \ref{fig-ch03-17}d) and
(d) will rapidly converge to a constant for $N_e\to\infty$.

Similar ideas have first been presented by  Haldane and Rezayi
\cite{haldane:02:1985}.

\subsubsubsection{How to suppress the effect of the CM part of the WF}

To suppress the effect of the CMWF it would be ideal to calculate the
density as
$$
  n_{\Psi_r}(z) = \int \d z_1\dots \d z_n |\Psi_r(z_1,\ldots,
  z_n)|^2\delta(z_1-z)
$$
rather than by (\ref{eq-ch03-17}).
In other words, it would be nice if we could 
replace $\Psi_{CM}(z_1+\ldots+z_n)$ by a
constant in the numerically calculated wavefunction $\Psi_r\Psi_{CM}$.

Even though we could numerically calculate $\Psi_{CM}$ and then
calculate the density in the state $\Psi/\Psi_{CM}$, this is technically
quite labourious and requires numerical evaluation of ($n-1$)-fold
integrals. Instead we can make a trick. Consider again
the example of the $\nu=\ot$ GS. The state is triply degenerated in the
CM part and the three different $\Psi_{CM}^{1,2,3}$ (as they come
from ED in subspaces with sharp $J$) have the pleasant property that
the sum of their squared moduli is nearly constant, or in a more
restrained (and honest) terminology, its variations are much weaker
than those of individual $|\Psi_{CM}^i|^2$, Fig. \ref{fig-ch03-17}.

With this in mind we expect that the sum
$n_{\Psi_{CM}^1\Psi_r}(z)+n_{\Psi_{CM}^2\Psi_r}(z)+n_{\Psi_{CM}^3\Psi_r}(z)$
will be a good approximation to $n_{\Psi_r}(z)$. The reader may check
with Fig. \ref{fig-ch03-17} how well this is fulfilled.

\subsubsubsection{Other finite size effects}

Here, we will try to abstract from the effects due to the CM part of
the calculated wavefunctions. Since the operator for density-density
correlation depends only on relative coordinates we expect that
$g(\vek r)$ will be free of the finite size effects described in
previous paragraphs. Since the curves for $g(\vek r)$ obtained from
the $\nu=\ot$ ground state in systems of different sizes
(Fig. \ref{fig-ch03-04+05}b) match very well for $|r|$ going at least to
one third of the elementary cell we may have good confidence in these
results even within the scope of infinite systems. In clear terms, we
may believe that $g(r)$ of the infinite system is nearly the same as
$g(r)$ obtained in a finite system (with $a:b=1$) as far as up to
$r\approx 0.35a$.

Another type of finite size effects which are 'finer' than those originating
from the CM part of the WF is shown in Fig. \ref{fig-ch03-17}, the 
rightmost plot. The density plotted should be constant after
averaging over the three states degenerated in the CM part in the
infinite system. The weak $(1/N_m)$-periodic
structure ($N_m=12$ in Fig. \ref{fig-ch03-17}) which we still observe 
reflects the quantization of one particle momenta by the PBC. 
One particle can be localized only around one of $N_m$ 
discrete set of points in the $x$-direction. This
effect is the same along $x$ and $y$, since we have lost the quantum
number $J$ (\ref{eq-ch02-54}), 
by averaging over the three states, belonging to
$J=2,6,10$ in the present case.
Note, how extremely small this finite size effect is.

\subsubsection{Conclusion: yet another comparison to composite fermion models}
\label{pos-ch03-10}

For a large part we were concerned with the $\nu=\ot$, $\tt$ and $\tf$
incompressible ground states in this section. All these states,
including their possible spin polarizations, can
be described in terms of Landau levels (LL) filled with composite
fermions (CF), Fig. \ref{fig-ch03-01and02}a and Sec. \ref{pos-ch02-17}. In
particular, wavefunctions suggested by Jain,
Subsec. \ref{pos-ch02-18}, are very close to the many-electron ground
states calculated by exact diagonalization, as it is demonstrated by
comparing the wavefunctions calculated by the two approaches in terms
of overlaps which approach unity \cite{wu:07:1993}
or of correlation functions shown in this Section,
Figs. \ref{fig-ch03-03}, \ref{fig-ch03-04+05} and
\ref{fig-ch03-09}a.  

However, we have seen in this Section that this picture is not as
intuitive as someone may believe. Correlation functions of states with
$p$ filled CF LLs are quite different from those of states with $p$
filled electronic LLs. Changing orientation of the effective magnetic
field following from the CF LL 'quantization' alters the
correlation functions drastically. It is hard to establish
a relation between the ground states at $\nu=\tt$ and $\tf$ on the
level of comparing the {\em electronic} 
correlation functions. We should also mention a
discrepancy in the CF model for the $\nu=\tt$ polarized state. It is
both a particle-hole conjugate to the $\nu=\ot$ Laughlin state and a
state with two filled CF Landau levels and effective magnetic field
antiparallel to the real magnetic field or attached flux quanta. As
Wu, Dev and Jain \cite{wu:07:1993} noted already in their original
work about antiparallel flux attachment, these two approaches give two
non-equivalent 
microscopic wavefunctions. Surprisingly enough, both wavefunctions
have high overlaps ($\approx 0.99$) with the polarized ground state obtained
by exact diagonalization \cite{wu:07:1993}. Thus, either 
both models are in fact indeed equivalent or this result shows
that even such high overlaps may be not enough to prove the
correctness of a trial many-body wavefunction. 

\begin{SCfigure}
\includegraphics[scale=0.7,angle=0]{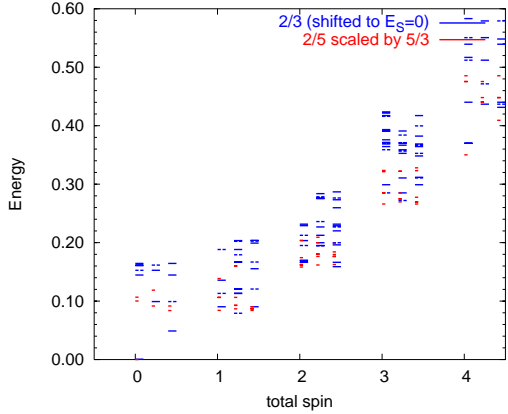}
\caption{Spectra of the $\nu=\tt$ and $\tf$ systems with eight electrons,
zero Zeeman energy (SRI). Energies have been shifted to set the singlet GS
in both systems to zero energy.
In the picture of
non-interacting composite fermions, the energies of the polarized
($S=4$) ground states (the $\tt$ and $\tf$ ones) should then be the same
when $B_{\eff}$ is correctly rescaled (the indicated scaling of
$5/3$). For each spin, the levels are sorted according to the orbital
quantum number $J$ (\ref{eq-ch02-54}), 
which can take on three non-equivalent values (left to right): $0,1,2$. }
\label{fig-ch03-26}
\end{SCfigure}

Another point worth of notice
is that the 'CF cyclotron energies' (sometimes denoted by 
$\hbar\omega_{CF}$)
extracted from exact diagonalization with electrons 
are not quite the same in $\tt$ and $\tf$ systems, 
Fig. \ref{fig-ch03-26},  the scaling factor $5:3$ makes  $B_{\eff}$
equal in both systems. In the picture of
non-interacting CFs, only the direction of the effective field $B_{\eff}$ is
reversed. Thus, if $B_{\eff}$ has the same modulus in both cases and
Zeeman energy vanishes then $E_p(N_e=8)-E_u(N_e=8)$, i.e. the difference
of energies of the polarized and singlet GSs for 8-electron systems,
should be equal to four times the CF cyclotron energy in the both
systems (cf. Fig. \ref{fig-ch03-01and02}b). 
In the exact diagonalization spectra of 
$N_e=8$ systems are regarded, the difference
of $\hbar\omega_{CF}$ for $\nu=\tt$ and $\tf$ is small,
about 5\%, Fig. \ref{fig-ch03-26}.
However, the quantitative agreement becomes worse when we attempt to
extrapolate the energies to larger systems.

Also comparing $\tf$ to $\tt$, differences in the lowest excitations from the
polarized and singlet ground states (energies, quantum numbers) 
are quite apparent, Fig. \ref{fig-ch03-26}. 

All these facts demonstrate that it can be misleading to think of the
$\tt$ and $\tf$ states as of an exact copy of Landau levels completely
filled with electrons. Composite fermion models must be taken
seriously since they provide us with many very good predictions
(explicite forms of wavefunctions, e.g.) but apart of that they are
not exact, they fail to describe some phenomena like e.g. position of
zeroes in Coulomb interacting states, Subsect. \ref{pos-ch03-12}, the
analogy between electronic and CF Landau levels is sometimes
weak. One of the inherent problems not mentioned so far 
  is the question of mixing
  between CF Landau levels: whereas LL mixing can be neglected for
  electrons in the limit $B\to\infty$, there is no such case for CFs.

The nature of many incompressible FQH states is therefore still not
completely clear, for example the ground states at $\nu=\tt$ and
$\tf$. Results in this Section indicate that the singlet states at
these filling factors comprise of pairs of spin up and spin down
electrons which we would not expect from the CF analogy -- at least
not at first glance. Furthermore, in the $\nu=\tt$ singlet with
electron density $n$, the
$\up-\dn$ pairs seem to form a state which could be constructed by
taking a system with the lowest LL completely filled with electrons
of density $n/2$
and then replacing each electron by an $\up-\dn$ pair. This behaviour
is not observed in the $\nu=\tf$ singlet. We may again conclude, that even
though the $\tf$ and $\tt$ ground states are very closely related on the
level of composite-fermion theories, their electronic properties are
different. It can thus be misleading to extend our intuition
concerning the (completely filled) electronic Landau levels to states
interpreted as (completely filled) composite fermion Landau levels.

\subsection{The half--polarized states at filling factors $\tt$ and $\tf$}







\label{pos-ch03-07}

In the previous section we dealt with the spin singlet and polarized ground
states at filling factors $\tt$ and $\tf$ and it was
mentioned that it is the Zeeman splitting (or better, $E_Z/E_C\propto
\sqrt{B}$) which determines which of them is the actual ground
state. It is the singlet state for vanishing Zeeman splitting (low
magnetic fields) or the polarized state if the Zeeman term dominates
(limit $B\to\infty$). All this can be understood within the composite fermion
concept, Fig. \ref{fig-ch03-01and02}a,
where we even obtain the prediction that there is a {\em direct}
transition (crossing) between these two ground states at some critical
value of $E_Z/E_C$ or equivalently,
at some critical magnetic field $B_C$, if we sweep magnetic field and
keep the filling factor constant, cf. also Sect. \ref{pos-ch04-00}.

However, experiments by Kukushkin et al. \cite{kukushkin:05:1999}
indicate that this picture may be
incomplete. They suggest that some exactly half-polarized state
becomes a stable ground state in the vicinity of $B_C$. In this
Section we will describe one candidate for such a half-polarized
state ground state and discuss its properties.

\subsubsection{Ground state energies by exact diagonalization}

\begin{figure}
\begin{tabular}{ccc}
(a) & (b) & (c)\\
\includegraphics[angle=-0,scale=0.4]{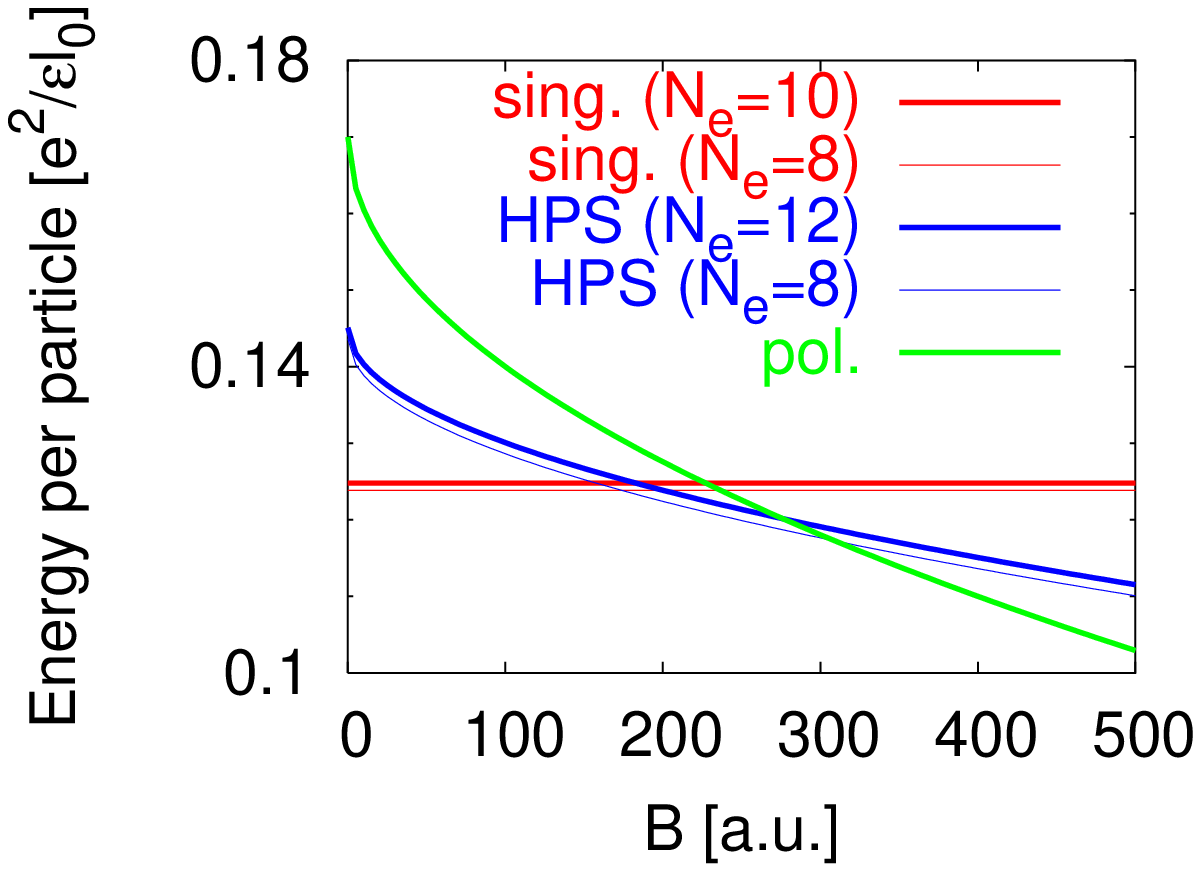} &
\includegraphics[angle=-0,scale=0.4]{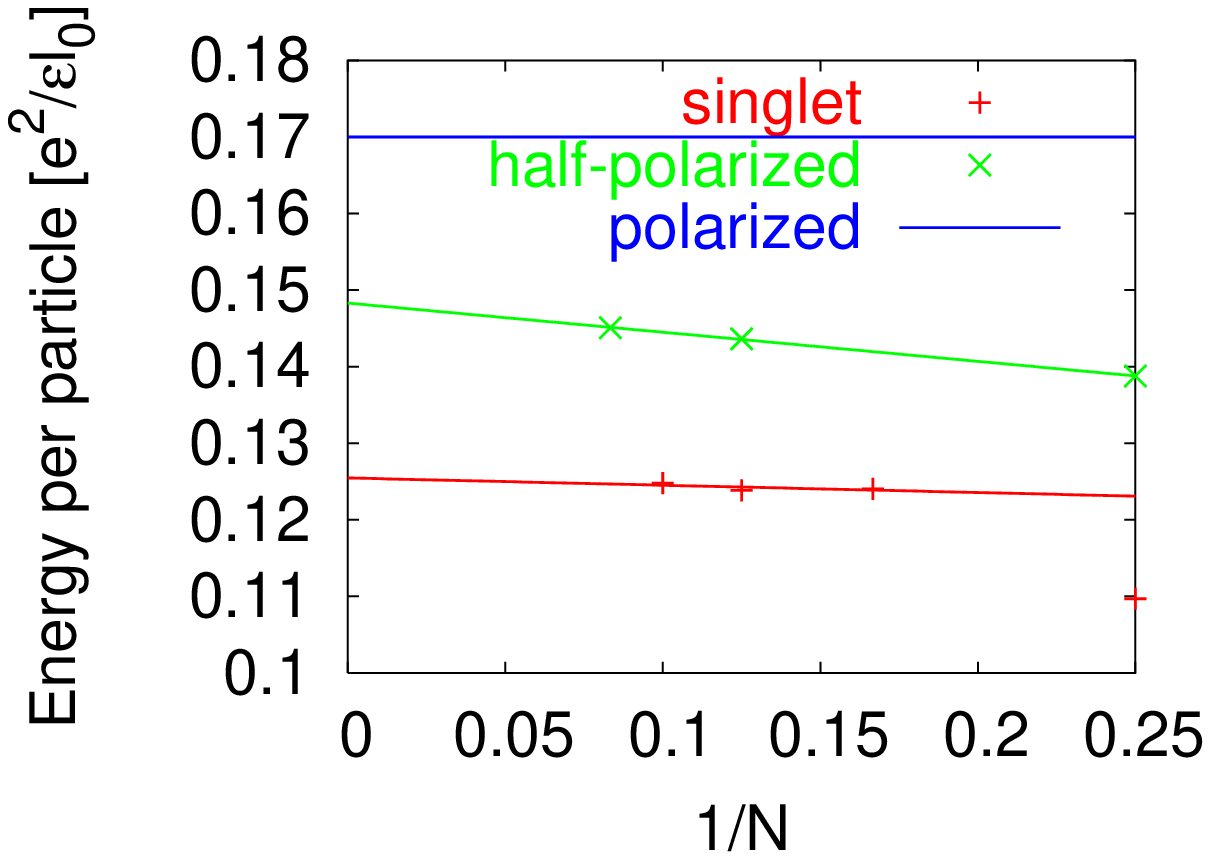}
\raise1cm\hbox{$\Rightarrow$} &
\includegraphics[angle=-0,scale=0.4]{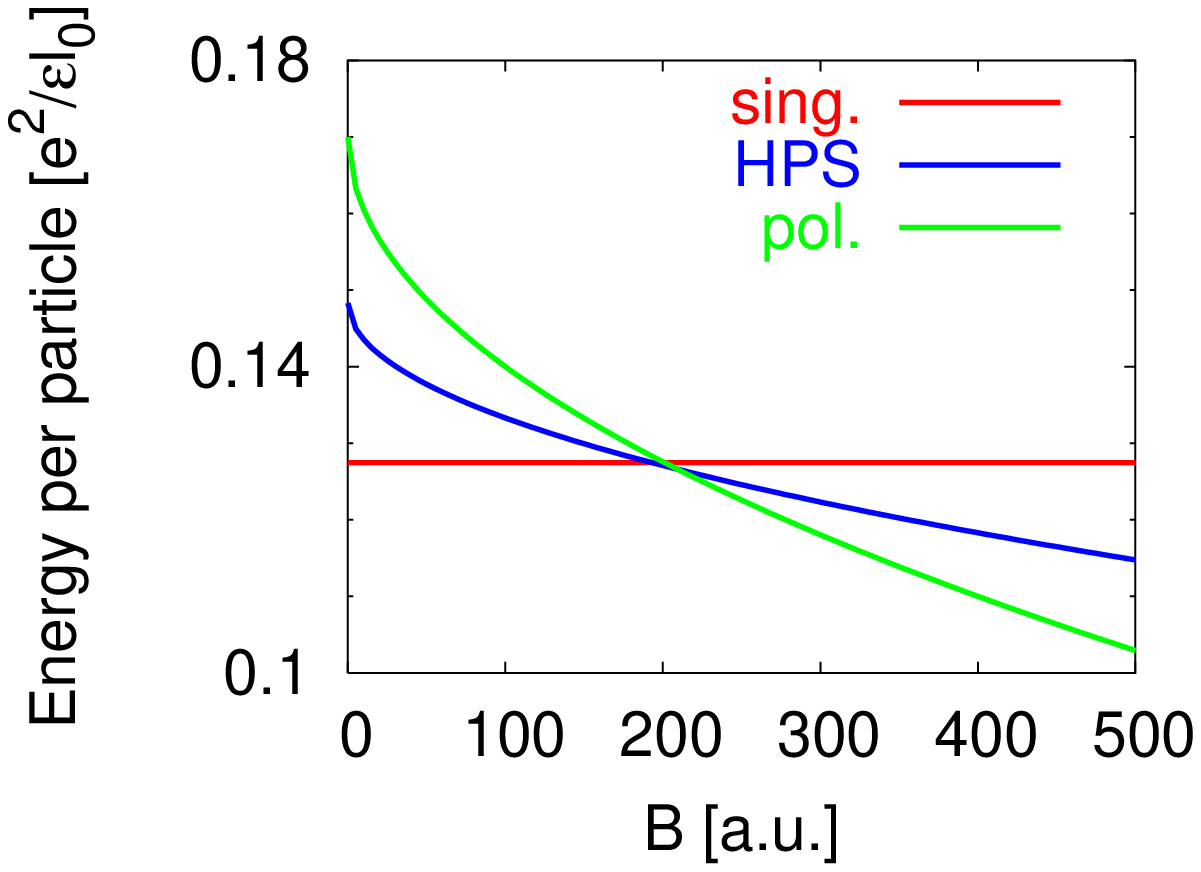} 
\\
\end{tabular}
\caption{Ground state energies at $\nu=\tt$: the half-polarized state
may become the absolute ground state in a narrow interval of magnetic
fields. {\em Left:} Energies of SRI ground states in the subspaces
$S=0$ (singlet), $S=N_e/4$ (half-polarized) and $S=N_e/2$ (fully
polarized) as a function of Zeeman splitting (or magnetic field; note
the energy units $e^2/\ve \ell_0\propto \sqrt{B}$). In all cases,
energies of the two largest systems available to calculations are
shown. {\em Middle:} extrapolation of the GS energies to infinite
systems ($1/N\to 0$). {\em Right:} The energy-versus-magnetic field
diagram for extrapolated ground state energies. This indicates that 
even then the HPS will be a ground
state close to the transition. }\label{fig-ch03-20}
\end{figure}

At first glance, spectra of homogeneous small finite systems with
Coulomb interaction (Sect. \ref{pos-ch04-00}, Fig. \ref{fig-ch04-01}) do not
suggest any intermediate state at the transition. The picture is quite
different when short-range interaction is considered. In
an interval of magnetic fields around $B_C$ 
the GS is a state with total spin equal to $N_e/4$,
Fig. \ref{fig-ch03-20}a, i.e. a half-polarized state (HPS). This
holds for all system sizes accessible to numerical calculation and, by
extrapolating energies to $1/N\to 0$, Fig. \ref{fig-ch03-20}b, it
seems to hold also for infinite systems.

It is probably only through the finiteness of the system that
a half polarized ground state did not appear in Coulomb
interacting systems (Fig. \ref{fig-ch04-01}). The SRI
systems may be less sensitive to this generical
drawback of exact diagonalization models. On the other hand, SRI
models predict wrong values of $B_C$ (see Subsec. \ref{pos-ch04-04})
and thus the scheme presented in Fig. \ref{fig-ch03-20}a must be
checked in systems with Coulomb interaction.

\begin{figure}
\begin{center}
\begin{tabular}{cc}
(a) & (b) \\
\includegraphics[angle=0,scale=0.5]{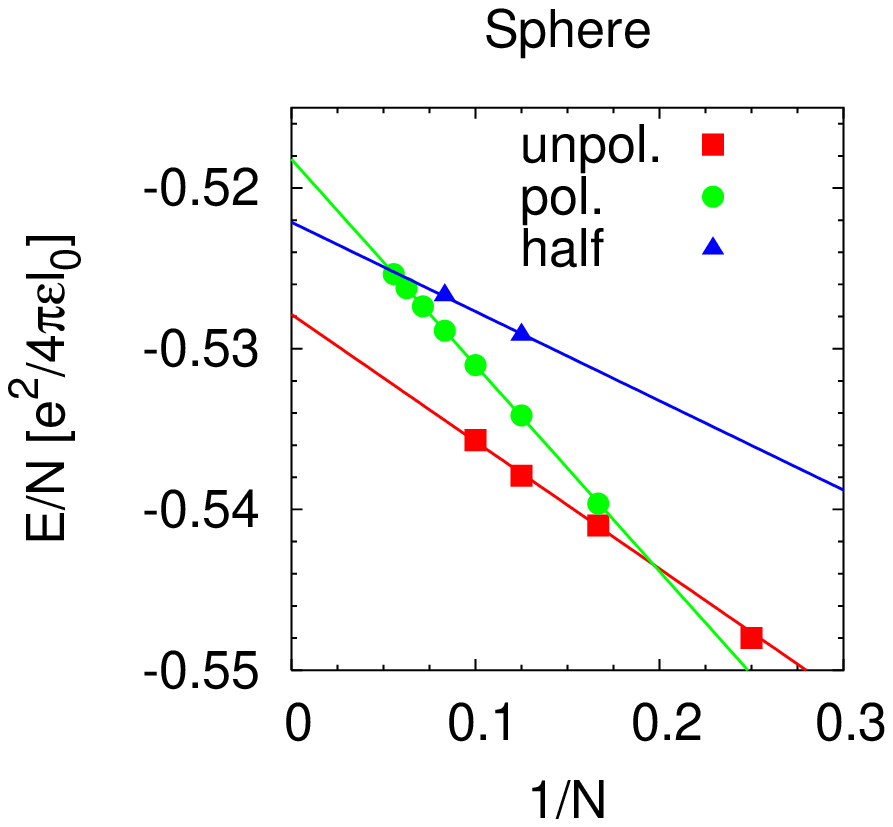} &
\includegraphics[angle=-0,scale=0.5]{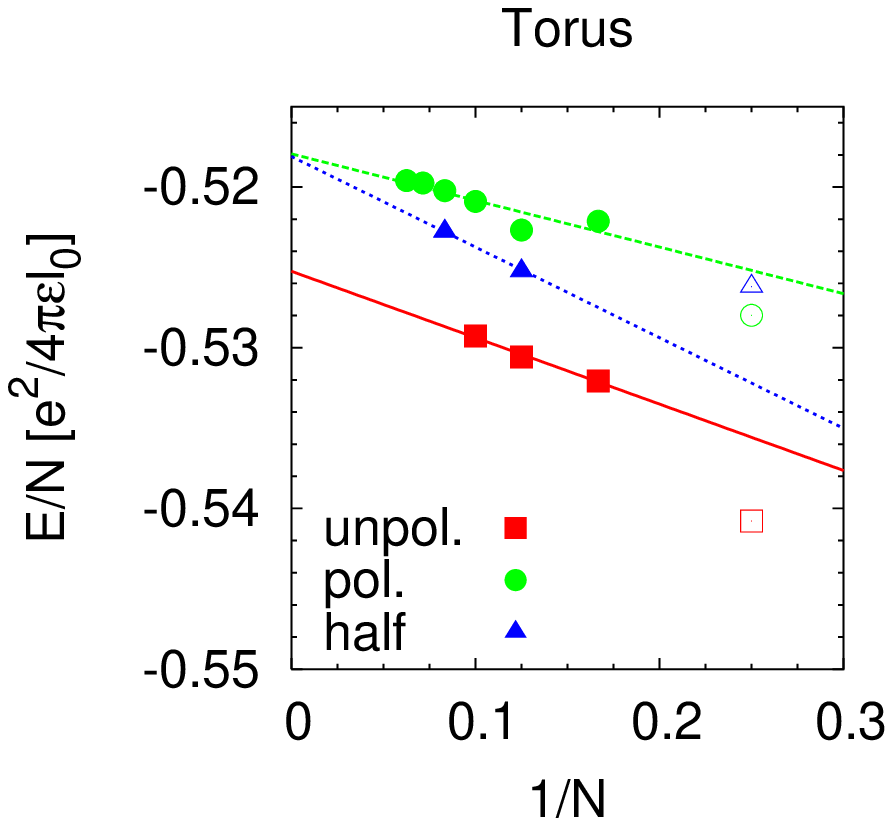} \\
\end{tabular}
\end{center}
\caption{ Extrapolation $1/N\to 0$ of the GS energies for Coulomb
interacting systems on a sphere and on a torus at $\nu=\tt$.} 
\label{fig-ch03-21}
\end{figure}

Considering Coulomb-interacting systems, the scheme suggested in
Fig. \ref{fig-ch03-20}a is supported by extrapolations of GS
energies performed by Niemel\"a, Pietil\"ainen and Chakraborty
\cite{niemela:xx:2000} in spherical geometry, Fig. \ref{fig-ch03-21}a,
and it is not supported by analogous calculations on a torus presented here,
Fig. \ref{fig-ch03-21}b. We would like to stress that the
extrapolation of the energy of the HPS is based only on two points,
the third point ($N_e=4$) in Fig. \ref{fig-ch03-21}b, is not very
reliable, Subsec. \ref{pos-ch03-13}.
Therefore the question of whether the HPS becomes the
absolute GS or not remains basically open until exact diagonalizations
of larger systems become possible. 

Nonetheless let us assume in this Section that a half-polarized state
can indeed lower its energy sufficiently so as to become 
the absolute ground state.
We will therefore focus on the $S=N_e/4$ sector of systems at filling
factor $\tt$, and also at $\tf$ in Subsec. \ref{pos-ch03-14}. 
Studies were mostly focused on the SRI
states where it is easier to identify the best candidate for the
half-polarized ground state.  Its Coulomb-interacting counterpart is
discussed later, in Subsec. \ref{pos-ch03-03}.

By convention a half-polarized state with 12 (8) electrons will
consist of 9 (6) electrons with spin up (majority spin) and 3 (2)
electrons with spin down (minority spin).

\subsubsection{Identifying the HPS in systems of different sizes}

Provided some particular physical half-polarized state \gsI
is the ground state in an infinite system, we may ask what its
realizations in finite systems of different sizes are. Vice versa:
given the half-polarized states calculated in a system of $N_e=12$
(4, 8\dots) electrons, which state corresponds to \gsI? In
this way we can think of states which 'correspond to each other' in
systems of different sizes. The trouble is, of course, that we do not
know \gsI. 

Regarding the computational capacity available, we could study
$\nu=\tt$ systems  with 4, 8 and 12 particles,
the next larger system, $N_e=16$, would require diagonalization in spaces of
dimension many hundred million. It seems likely that the 
analogues to \gsI are the GSs in $N_e=12$ and
$N_e=8$ systems (GS$_{12}$, GS$_{8}$) 
and that it is a low lying excited state (\ttfont{st03}) 
in the smallest
system, $N_e=4$. In the following, reasons for this are proposed.

(i) GS$_{12}$ and GS$_{8}$ belong to the same symmetry class defined by
the 'crystallographic $\krv$' (\ref{eq-ch02-45}). They have both
$\krvd=(\pi,\pi)$, i.e. they lie in the
'corner of the Brillouin zone' (Fig. \ref{fig-ch02-12},
Subsec. \ref{pos-ch02-06}). This is also closely related to the fact
that both GS$_{12}$ and GS$_{8}$ are non-degenerate.

\begin{figure}
\begin{center}
{\unitlength=2.25mm
\includegraphics[scale=0.45,angle=-90]{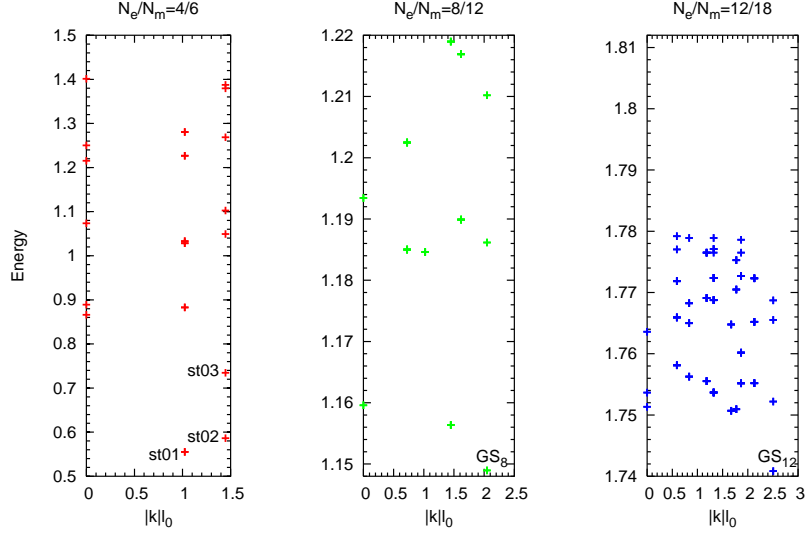}
}
\end{center}
\caption{Low lying energy levels in the $S=N_e/4$ (half-polarized)
sector of $\tt$ systems with ({\em left to right}) $N_e=4$, $8$ and
$12$ particles. The states are sorted according to $|\krv|$ (Subsec.
\ref{pos-ch02-06}). The GS of an infinite system is likely to have
$\krvd=(\pi,\pi)$ which is a point of very high symmetry in the
$\krv$-space. This symmetry class is however distinct from the
one of the singlet and polarized incompressible ground states.}
\label{fig-ch03-22}
\end{figure}

(ii) The states GS$_{12}$ and GS$_{8}$ are well
separated from excitations within the $S=N_e/4$ sector 
and the energy of the lowest excitation is
similar, $0.01\cunit$, in systems of different size,
Fig. \ref{fig-ch03-22}. 

(iii) Though not completely identical, the inner structure of
GS$_{12}$ and GS$_{8}$ is very similar as seen by the
correlation functions,
Fig. \ref{fig-ch03-23}. 

(iv) The GS of the $N_e=4$ system has a lower symmetry than the
formerly described states. Looking for a state of inner structure
(correlation functions) similar to the one of GS$_{12}$ and GS$_{8}$
within the sector $\krvd=(\pi,\pi)$, we find remarkable similarities
with the second excited state ('\ttfont{st03}', marked in
Fig. \ref{fig-ch03-22}), Subsec. \ref{pos-ch03-13}. However, we should
bear in mind that for $N_e=4$ there is only a single electron with
reversed spin in other words the system is indeed extremely
small. A consequence is for example that $g_{\dn\dn}(r)\equiv 0$.
Relevance of such states with respect to infinite systems is
thus doubtful.

\subsubsection{Inner structure of the half-polarized states}
\label{pos-ch03-13}

\begin{figure} 
\begin{center}
\includegraphics[scale=0.55,angle=-90]{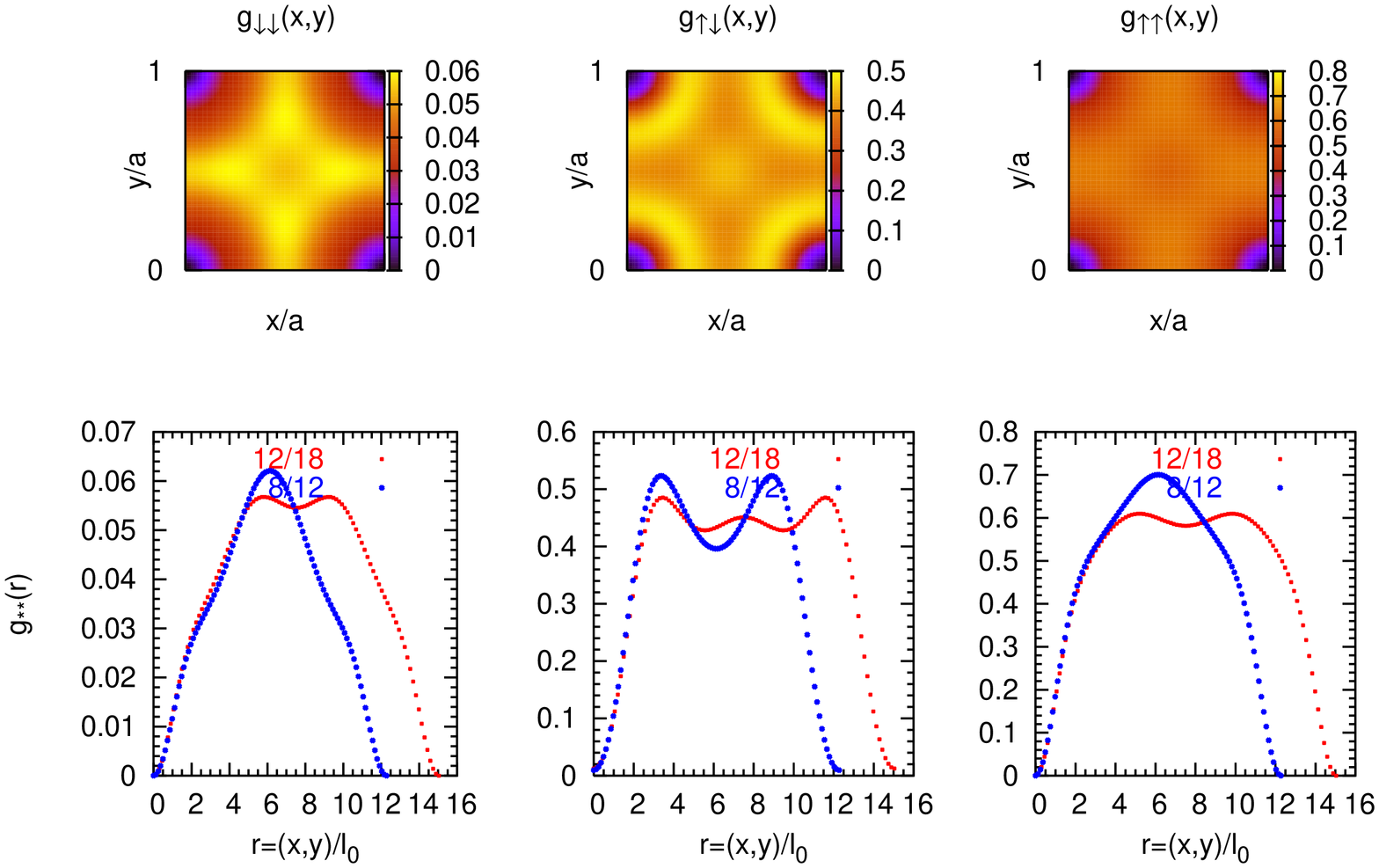}
\end{center}
\caption{The likely analogues of the half-polarized ($S=N_e/4$) \gsI
(ground state in an infinite system) on a torus with $N_e=8$ and
$N_e=12$ particles and filling factor $\tt$. {\em Left to right:}
density-density correlation for $\up\up$ (majority spins), $\up\dn$
and $\dn\dn$ (reversed spins), {\em upper row:} in the whole primitive
cell; {\em lower row:} sections along the diagonal. Note the isotropy
of the state, i.e. visual manifestation of its high
symmetry.}\label{fig-ch03-23}
\end{figure}

Focus of this part will be the correlation functions of the states
GS$_{12}$ and GS$_{8}$ and a brief comment will be made on $N_e=4$
states. As mentioned above and as the kind reader may verify in
Fig. \ref{fig-ch03-23}, GS$_{12}$ and GS$_{8}$ look indeed similar. 

GS$_{12}$ and GS$_{8}$ match in all three spin-resolved correlation
functions, $g_{\up\up}(\vek r)$, $g_{\up\dn}(\vek r)$,
$g_{\dn\dn}(\vek r)$, Fig. \ref{fig-ch03-23}. The match is 
especially good (quantitative) on short distances, $r\lesssim
3\ell_0$. This suggests that states GS$_{12}$ and GS$_{8}$ are not
bound to some particular system size and we can thus hope that if we
could make the system larger, they would eventually develop into the \gsI.

Differences between correlation functions of GS$_{12}$ and GS$_{8}$ at
longer distances $r$ are understandable, given the normalization
(\ref{eq-ch03-14}). The $N_e=12$ system is 'larger' than the
$N_e=8$ one, yet the integral $\int\d r g(\vek r)$ must be the same.
Perhaps the most apparent difference between various correlation
functions is whether they have a maximum or a minimum 'in the middle'
($6\ell_0$ or $8\ell_0$ in Fig. \ref{fig-ch03-23}). In ideal case, a
strong maximum occurs when $g(r)$ is monotoneous in an infinite system 
while a minimum, or a weak maximum following a foregoing
minimum, means that $g(r)$ has some structure, one or more maxima for
finite $r_i$. In reality, however, the former behaviour occurs also
when $r_i$ is larger than the finite system size. A manifestation of
this is seen in $g_{\up\up}(r)$, Fig. \ref{fig-ch03-23}. The flat maximum
at $r_i\approx 5\ell_0$, followed by a minimum,
observed for the $N_m=18$ torus does not occur
for the smaller system ($N_m=12$). Looking only at the 
smaller system we could have wrongly concluded that the correlation
function is almost structureless.

Some further points are worth of notice.

(i) $g_{\up\dn}(r)$ is suppressed nearly to zero
at $r=0$ in spite of the missing Pauli principle, only on account of
the repulsive interaction. It displays strong maxima around
$r\approx 3.4\ell_0$.

\begin{figure}
\begin{tabular}{ccc}
(a) & (b) & (c) \\
\includegraphics[scale=0.5,angle=-0]{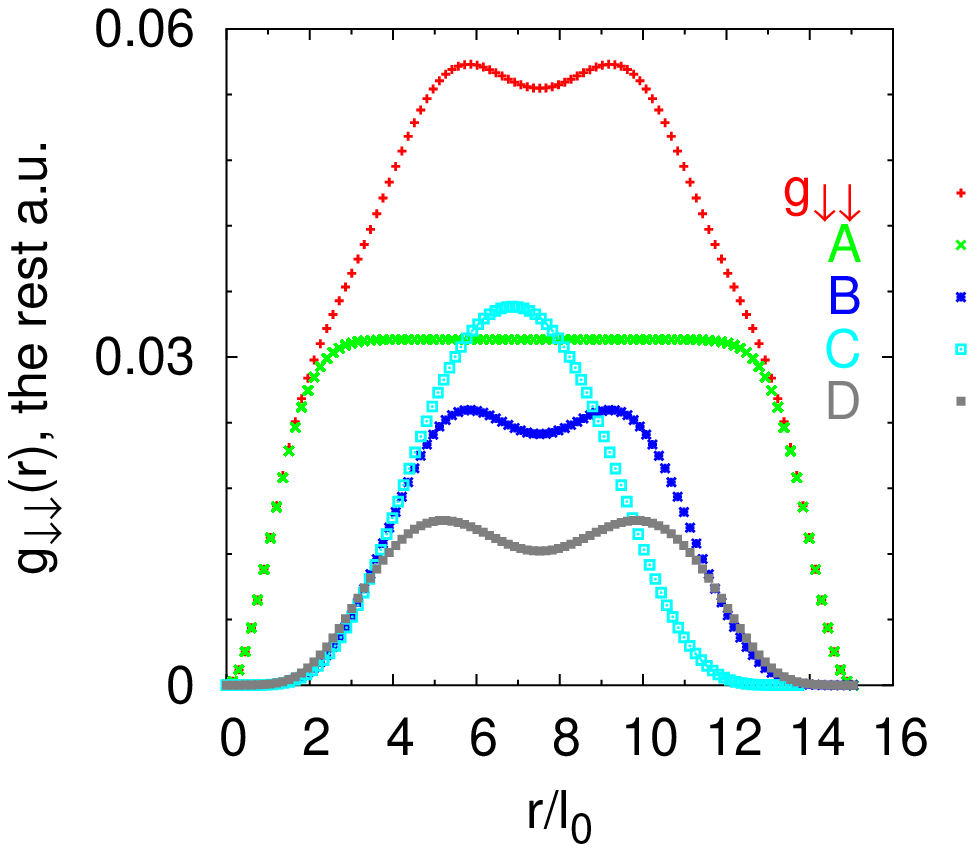} &
\includegraphics[scale=0.5,angle=-0]{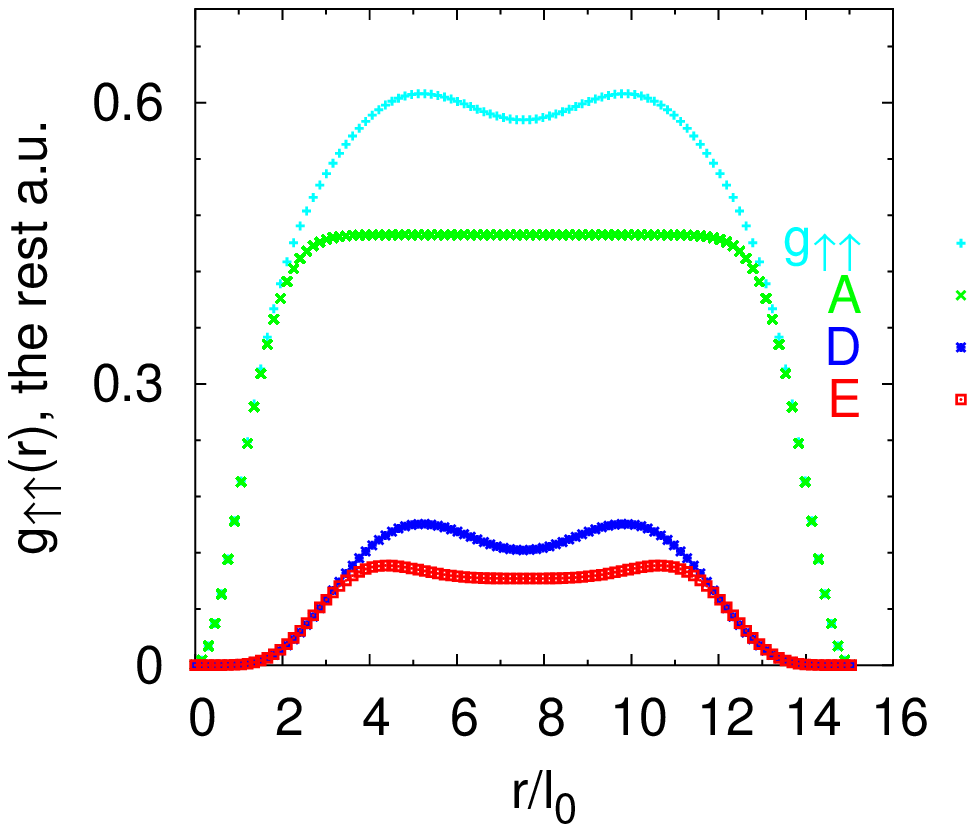} &
\includegraphics[scale=0.5,angle=-0]{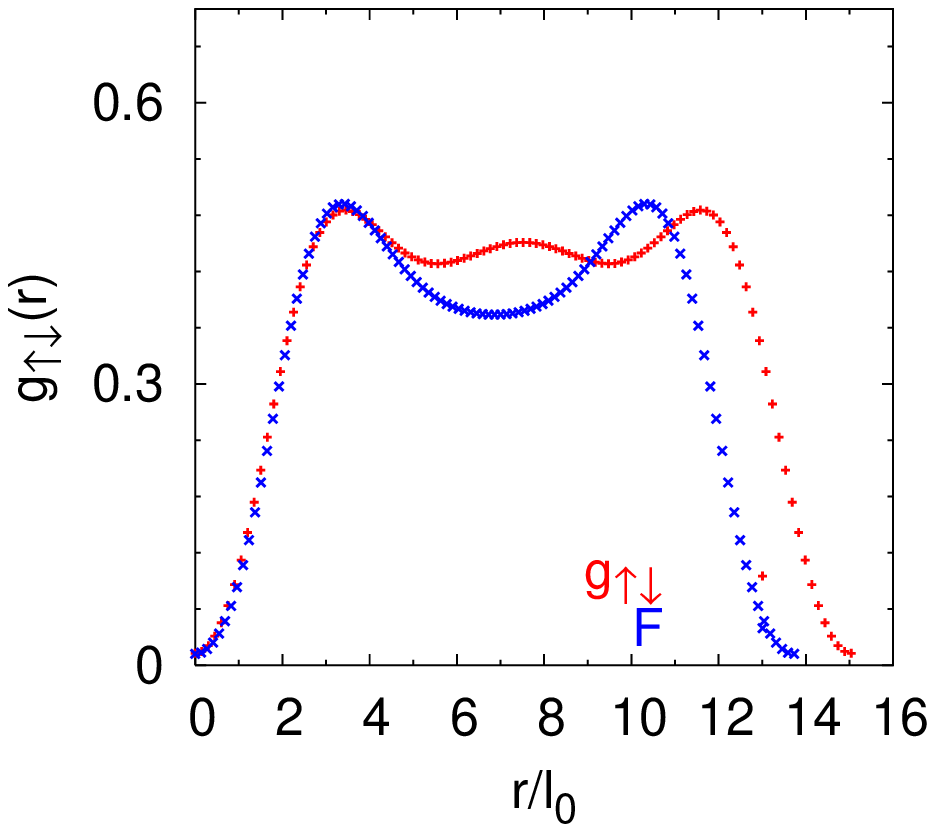} 
\end{tabular}
\caption{The half-polarized state (GS in the $S=N_e/4$ sector) for
$N_e=12$ particles, analysis of the correlation functions (details in text).
{\em Left to right}: (a) $g_{\dn\dn}$ (minority spin), (b)
$g_{\up\up}$ (majority spin) and (c) $g_{\up\dn}$. Legend for
particular curves (in all plots). A:
lowest LL correlation hole, $g_{\nu=1}(r)$ scaled to fit the
shoulder. B and D: $g_{\dn\dn}(r)$ and $g_{\up\up}(r)$ without
shoulder. C and F:
$g_{\dn\dn}(r)$ and $g_{\up\dn}(r)$ 
of the singlet state ($N_e=10$, $\nu=\tt$), $g_{\dn\dn}(r)$ without
shoulder. E: $g(r)$ of the $\nu=\ot$ Laughlin state. 'Without
shoulder' means, that the curve A was multiplied by a suitable
constant to fit the shoulder and subtracted. 
}\label{fig-ch03-25}
\end{figure}

(ii) Even though by far not identical, $g_{\up\up}(r)$ and
$g_{\dn\dn}(r)$ are similar to each other. The clear shoulder around
$r\approx 2\ell_0$ seems to stem from the 'exchange hole' (of the
LLL) $g_{\nu=1}(r)=1-\exp(-r^2/2\ell_0^2)$, see
(\ref{eq-ch03-01}). After subtracting a suitably scaled function
$g_{\nu=1}(r)$ the shoulder completely disappears and the
remaining parts of both $g_{\up\up}(r)$ and
$g_{\dn\dn}(r)$ are $\propto r^6$ close to $r=0$,
Fig. \ref{fig-ch03-25} and discussion below.

(iii) Up to a high precision the 
sum of $g_{\up\up}(r)$, $g_{\dn\dn}(r)$
and $g_{\up\dn}(r)$ (with appropriate scaling, see
Fig. \ref{fig-ch03-08} for explanation) is identical with
$g_{\nu=1}(r)$, however with $\ell_0$ replaced by
$\sqrt{2}\ell_0$. Not shown here.

Let us now turn to the smallest system where $S=N_e/4$ states may
occur (at $\nu=\tt$), i.e. $N_e=4$. Figure \ref{fig-ch03-23} shows
correlation functions of the lowest two states in the sector of
$\krvd=(\pi,\pi)$. Out of these, the
second state (i.e. \ttfont{st03}) seems to be analogous to $S=N_e/4$ GS's
in the two larger systems ($N_e=8$, $12$): $g_{\up\up}(r)$ is again a
sum of the 'correlation hole' and a function $\propto r^6$,
$g_{\up\dn}(r)$ shows a peaked structure with maximum around
$2.8\ell_0$ (both of these features are missing for the lower state
\ttfont{st02}). However, as mentioned above, the $N_e=4$ system is too small
for a reliable study of $S=N_e/4$ states ($g_{\dn\dn}(r)\equiv 0$).

Back to the GS$_{12}$ (called HPS here), it is very interesting to study the
'$\propto r^6$ part' (P6P) of the like-spins correlation functions,
$g_{\up\up}(r)$, $g_{\dn\dn}(r)$. What we mean by 'P6P' is the rest
after we subtract the 'lowest LL correlation hole', i.e. the
$g_{\nu=1}(r)$ part causing the shoulder in $g_{\sigma\sigma}(r)$
around $r\approx 2\ell_0$, curve A in Fig. \ref{fig-ch03-25}a,b.

One of many facts we can extract from Fig. \ref{fig-ch03-25} is that 
P6P$/\up\up$ [refering to $g_{\up\up}(r)$] and P6P$/\dn\dn$ are similar but not
identical. For example, they both exhibit a peaked structure but the
first maxima do not coincide, they occur at $5.0\ell_0$ and
$5.8\ell_0$ for P6P$/\up\up$ and P6P$/\dn\dn$, respectively
(curves B and D in Fig. \ref{fig-ch03-25}a). 

Let us compare the P6P$/\dn\dn$ of the HPS with 
P6P/$\dn\dn$ of the singlet incompressible $\tt$ GS, curves B and C
in Fig. \ref{fig-ch03-25}a).
Match of these two is very good up to
$r\approx 4\ell_0$, the absence of the peak at $5.8\ell_0$ in the
singlet state could be
due to smallness of the system where the singlet state was determined
($N_e=10$). It might appear in the next larger system, $N_e=12$,
cf. similar situation in Fig. \ref{fig-ch03-23}. 

On the other hand, P6P/$\up\up$ of the HPS seems to resemble the
singlet state less than P6P/$\dn\dn$ of the HPS. The form of
P6P/$\up\up$ seems to be not very different from the one of
the correlation function of the Laughlin $\ot$ state,
Fig. \ref{fig-ch03-04+05}b, whose first maximum occurs however
already at $r=4.4\ell_0$ (curves D and E in
Fig. \ref{fig-ch03-25}b). In any case, P6P$/{\up\up}$ of the HPS
matches better $g_{\nu=\ot}(r)$, i.e. the Laughlin state, than P6P of
the $\nu=\tt$ singlet state. Here we mean especially behaviour on
ranges $\lesssim 3\ell_0$.

Last but not least, the correlations between unlike spins are also
very similar in the singlet state and in the HPS,
Fig. \ref{fig-ch03-25}c, in particular positions of the maxima differ
by as little as $0.1\ell_0$ (both are around $r\approx
3.4\ell_0$).

\subsubsection{Discussion}

Findings presented above suggest that the $\nu=\tt$  
half-polarized ground state in short-range interacting systems 
is a gapped state in which the singlet and polarized incompressible
states coexist. Below, some key points regarding the HPS are summarized.

\subsubsubsection{Symmetry and energy}

Both in eight- and twelve-electron systems, the ground state has
$\krvd=(\pi,\pi)$. This is one of two points of the highest symmetry
in the $\krv$-space, another one is $\krv=(0,0)$,
Fig. \ref{fig-ch02-12}.  In particular, the 'highest symmetry' means
that this $\krv$-point is not related to any other point by a
symmetry operation in the $\krv$-space corresponding to relative
translations, Sec. \ref{pos-ch02-06}.  This in turn implies that
states with $\krvd=(\pi,\pi)$ or $(0,0)$ -- and only such states --
are non-degenerate, except for center-of-mass and incidental
degeneracies.  Together with the relatively large lowest excitation
energy $\Delta(N_e=8,12)$  from both GS$_{12}$ and
GS$_8$ (10\% of the gap of
the Laughlin state, Fig. \ref{fig-ch03-22})
, this suggests that the ground state
is gapped. Also the relation $\Delta(N_e=8)<\Delta(N_e=12)$ speaks in
favour of this hypothesis. If the gap were 
to vanish in an infinite system, we would expect the lowest excitation
energy to decrease with system size. Naturally, we must be
careful, since we can compare systems of only two different sizes and
the function $\Delta(N_e)$ may be non-monotonous. 
On the other hand, $\Delta(N_e=12)\approx 0.01\cunit$ 
is much larger than a typical level separation between excited
states, Fig. \ref{fig-ch03-22} and
for a mere finite size effect, this gap seems too large.

In spite of the similarities to the singlet and polarized
incompressible ground states, $\krvd$ clearly distinguishes HPS from
these two states, since they have both $\krvd=(0,0)$. Also in spherical
geometry, where $|\krvd|\propto L$ (end of Subsec. \ref{pos-ch02-06}), 
these incompressible states have $L=0$ while the HPS has
$L=S$, where $S$ is the total spin \cite{niemela:xx:2000}. Thus, even
though we showed that the HPS could be gapped, it is of different
nature than the singlet and polarized ground states. Meaning of this
different symmetry is however not clear.

It would be interesting to study this state in a system with hexagonal
elementary cell \cite{haldane:11:1985} which was unfortunatelly out of
the scope of this work. This geometry
is nearer to an isotropic 2D system than a torus (it has a six-fold
rather than a four-fold rotational symmetry) while it is still
compatible with plane waves (in CDWs). Most importantly, there is only
one point of the highest symmetry in this geometry and a
straightforward question is whether or not the HPS will maintain its
high symmetry.

\subsubsubsection{Inner structure again}

Features of the HPS described by points (i-iii)
in Subsec. \ref{pos-ch03-13} are
actually strikingly similar to those of the incompressible singlet
state at $\nu=\tt$. Investigation of the $g_{\up\up}(r)$ after the
'shoulder' was subtracted (P6P/$\up\up$) suggests again some relation
to the Laughlin state which is the particle-hole conjugate to the
polarized incompressible state at $\nu=\tt$. 

Especially manifest is the hint at pairing between unlike spins, 
the maximum around $3.4\ell_0$ in $g_{\up\dn}(r)$. On the other hand,
the shoulder in correlation functions of like spins seems to be rather a
manifestation of filling factor $>\frac{1}{2}$, since it occurs also
for other states at filling $\nu=\tt$ (than just for the singlet,
polarized and half-polarized GS) and it does not occur at
filling $\nu=\tf<\frac{1}{2}$,
Subsec. \ref{pos-ch03-14}. It suggests that some $\nu=\tt$ states
with less than full polarization
can be interpreted in terms of holes rather than electrons even though
particle-hole symmetry applies only for fully polarized states,
Subsec. \ref{pos-ch03-15}.

In the following Sections we will continue investigating the half-polarized
states at filling factor $\tt$ by other methods and continue
discussing the hypothesis of coexisting singlet and polarized
states. First, however, we look at two different minor issues.

\subsubsection{Half-polarized states at filling $\nu=\tf$}
\label{pos-ch03-14}

\begin{SCfigure} 
\hskip1.5cm\parbox{0.6\textwidth}{%
{\unitlength=1.7mm
\includegraphics[scale=0.34,angle=-90]{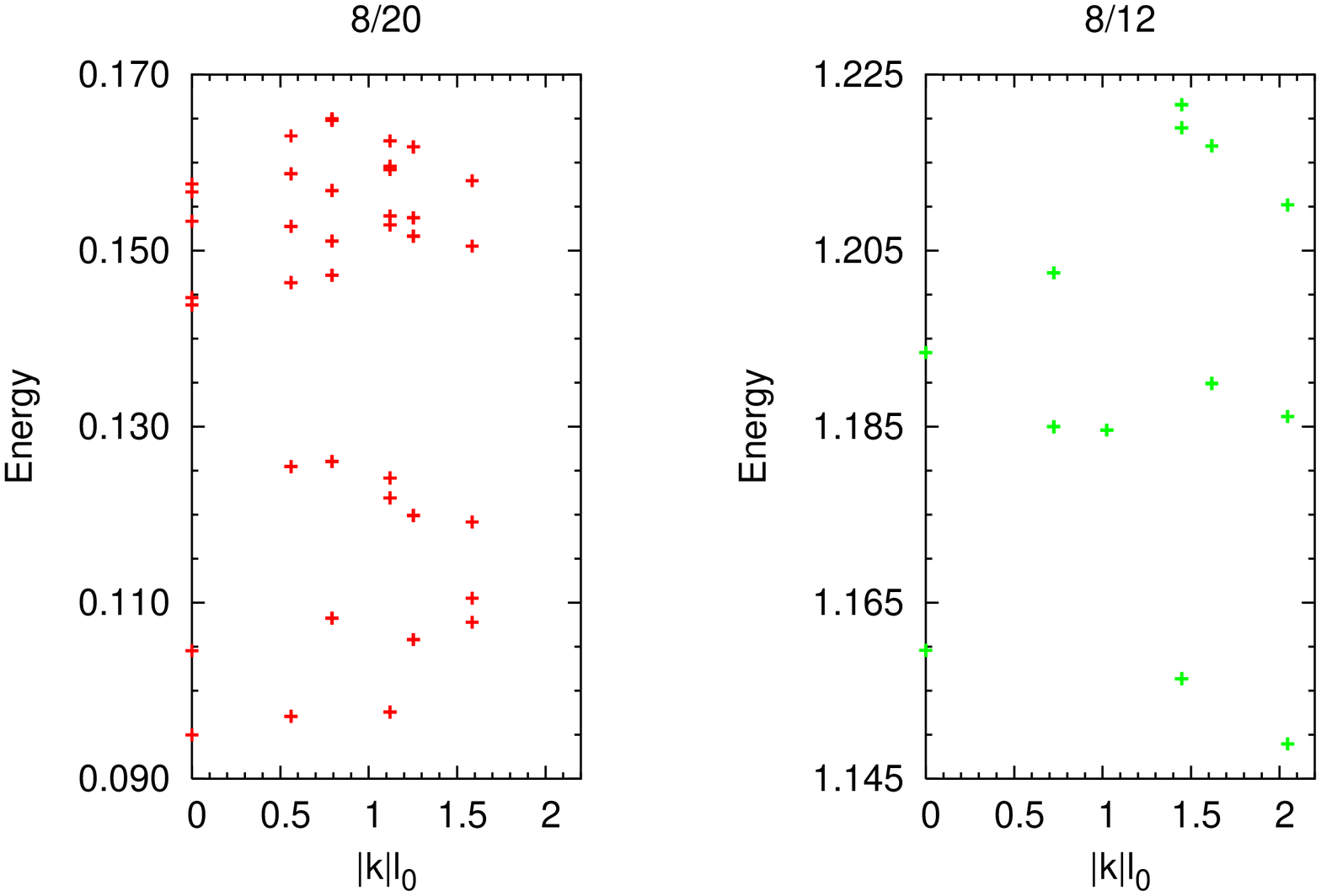}
\put(-33.4,-27.8){{\tiny$\nwarrow$}}
\put(-32,-29){{\tiny$\spadesuit$}}
}}
\caption{Low lying half-polarized (i.e. $S=N_e/4$) states of a $\tf$
system, $N_e=8$. The lowest state with the same symmetry as
the HPS of $\nu=\tt$ is marked by $\spadesuit$. }
\label{fig-ch03-27}
\end{SCfigure}

At filling $\tf$, the situation is much less transparent than at
filling $\tt$. First, only systems with four and eight
particles are accessible to exact diagonalization, the twelve particle
system implies matrix dimensions in the order of hundreds of
millions. Second, the spectrum of the eight particle system in the
$S=N_e/4$ sector is quite different from that of a $\tt$ system,
Fig. \ref{fig-ch03-27}: 

(i) the ground state lies at a
different point in the $\krv$-space, $(0,0)$, than the $\tt$-HPS
having $\krvd=(\pi,\pi)$. 

(ii) The excitation energy from this GS
is very small, less than a third of that one of the $\tt$ HPS.

(iii) The symmetry of the low excited states is lower than for
$N_e=8$, $\tt$ system.

\begin{figure} 
\begin{center}
\begin{tabular}{cc}
(a) & (b) \hbox to 0.1mm{$\spadesuit$} \\
\includegraphics[scale=0.28,angle=-90]{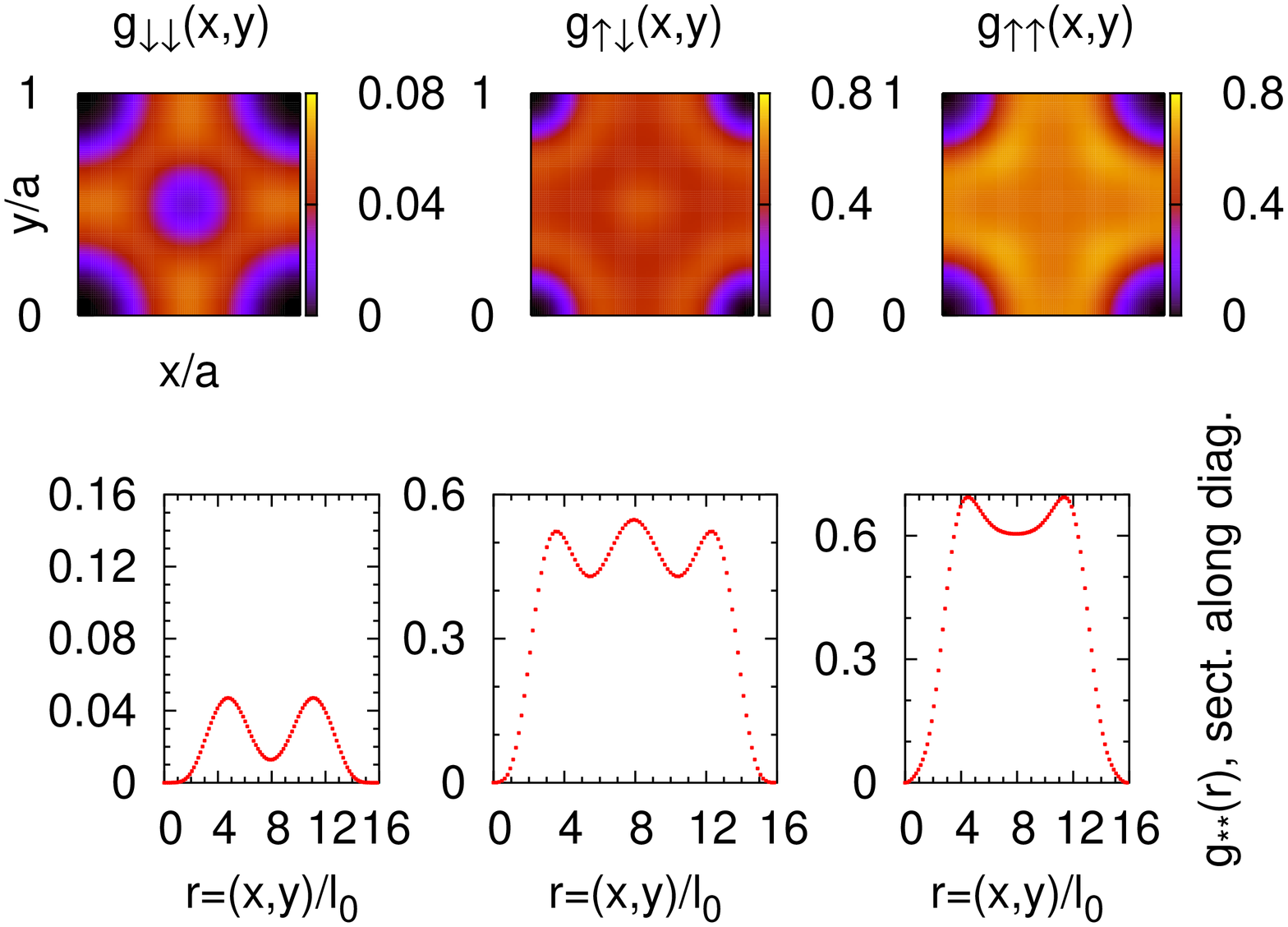} &
\includegraphics[scale=0.28,angle=-90]{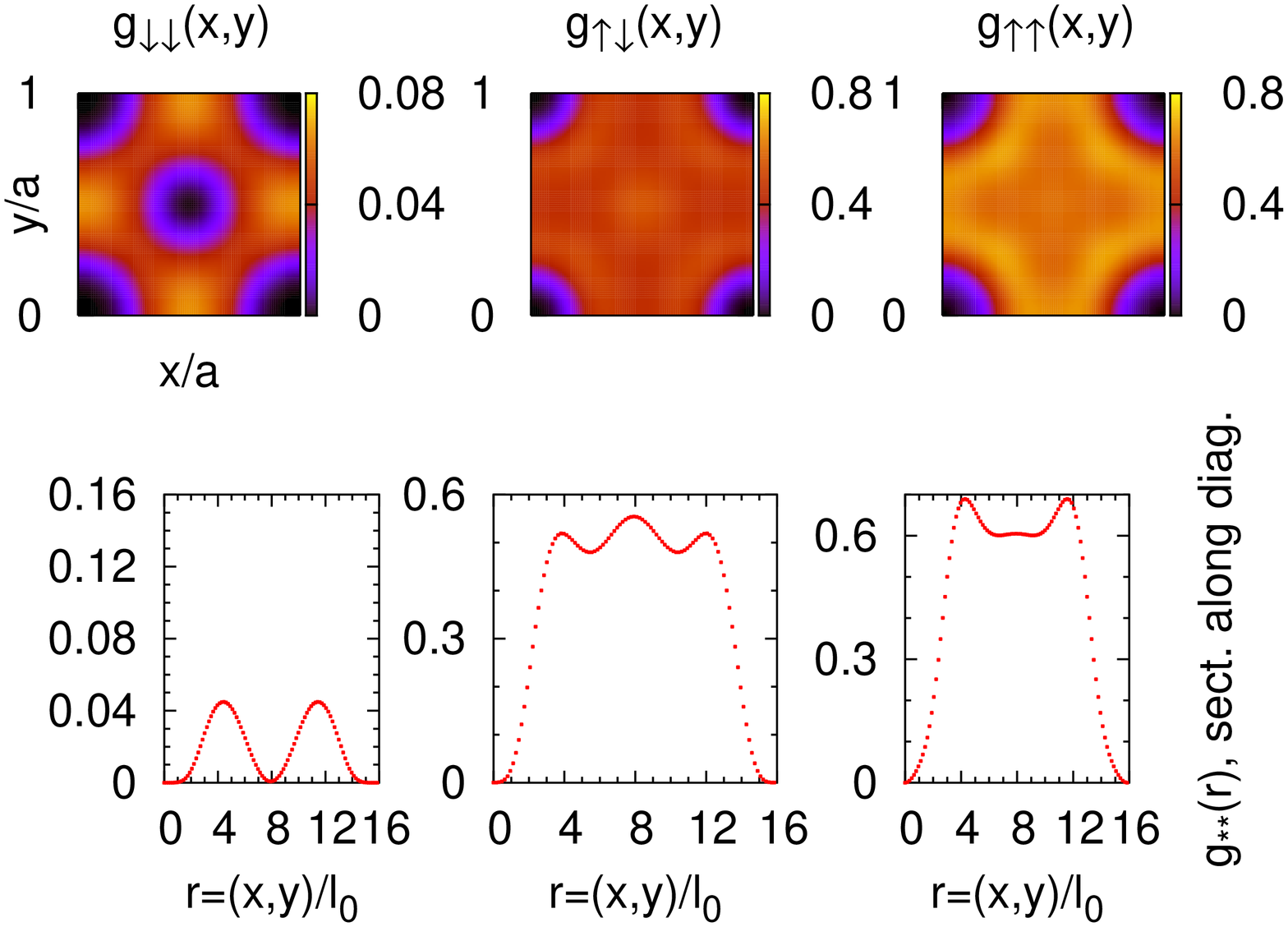} 
\end{tabular}
\end{center}
\caption{Half-polarized states at filling $\tf$ and their inner
structure (density-density correlation functions); eight electron
system. 
(a) the GS in the $S=N_e/4$ sector is non-degenerate but
it has a different symmetry, i.e. $(0,0)$, than the HPS of $\nu=\tt$. (b)
the lowest half-polarized state (at $\tf$) with the same symmetry,
i.e. $(\frac{\pi}{2},\frac{\pi}{2})$ as
the HPS of $\nu=\tt$ (marked by $\spadesuit$ in
Fig. \ref{fig-ch03-27}). 
}\label{fig-ch03-29}
\end{figure}

Regarding the possibility that within the 8 electron calculations
it is not the {\em lowest} energy half-polarized state at $\nu=\tf$ 
to be the counterpart of the HPS at
$\nu=\tt$, there are two $\tf$ states displayed in
Fig. \ref{fig-ch03-29}: (a) the one with the lowest energy in $S=N_e/4$
sector and (b) the lowest state with the same symmetry as the $\tt$ HPS,
i.e. $\wt{\vek k}=(\pi,\pi)$.

Similarly, as for the $\nu=\tt$ states, the $\tf$ HPS bear features of
the polarized and singlet ground states. Let us regard the state in
Fig. \ref{fig-ch03-29}a:

(i) Near $r=0$ the functions $g_{\dn\dn}$
(minority spin), $g_{\up\dn}$ and $g_{\up\up}$ (majority spin) are
$\propto r^6$, $r^4$ and $r^2$, respectively. In this respect, $g_{\dn\dn}$
and $g_{\up\dn}$ resemble the singlet state and $g_{\up\up}$ resembles
the polarized state. 

(ii) Up to the first maximum, $g_{\up\up}$ of the HPS is the same
as in the polarized state, but shifted by about $0.2\ell_0$
outwards. Positions of the first maxima mismatch slightly more (by
$0.4\ell_0$). The strong maximum in the centre of the cell is not present
in the HPS. 

(iii) $g_{\up\dn}$ of the HPS and the singlet GS match very well
even beyond the first maximum. Positions of the maxima are
identical, $r\approx 3.5\ell_0$. On contrary to the previous point,
there is another maximum in the centre of the cell in the HPS state
and nothing in the singlet state, indicating that the similarity
between the singlet and the HPS has certain limits.

(iv) $g_{\dn\dn}$ of the HPS and the singlet GS also match very
well up to $r\approx 4\ell_0$. 
Then there is a deep minimum in the HPS which is absent
in the singlet GS.

Turning to the state (b), we might say that it is less alike to the
singlet state. The minimum in $g_{\dn\dn}$ is much deeper than for
state (a), the first maximum in $g_{\up\dn}$ does not match the maximum
seen in the singlet state. On the other hand, $g_{\up\up}$ seems to be
more similar to the polarized state.

Lowest excitations in the high symmetry sectors show even less
similarities to the singlet and polarized GSs, especially $g_{\dn\dn}$
is quite dissimilar beyond the $r\approx 0$ range
and maxima  in $g_{\up\dn}$ match less well.

In conclusion, if there is a counterpart to the $\tt$ HPS at filling
$\tf$ at all, we may expect it to be the state (a) (the absolute GS), even
though hints for this are not very convincing. Again, this suggests
that differences between filling factors $\tt$ and $\tf$ are not only
of quantitative nature (gap energies, for instance) but may be
as substential as existence or non-existence of some particular ground
state, which is {\em nota bene} completely unexpected on the level of
non-interacting CF picture.

\subsubsection{Short-range versus Coulomb interaction}
\label{pos-ch03-03}

Let us conclude with observations regarding the Coulomb- and
short-range-interacting (SRI) systems in the sector of half-polarized
states.

\begin{figure} 
  \begin{center}
  \subfigure[Spectra. Coulomb and short-range interacting systems
  do not look quite similar, but the ground state is always at the
  maximum $|k|$.]{%
\hbox{\includegraphics[scale=0.3,angle=-90]{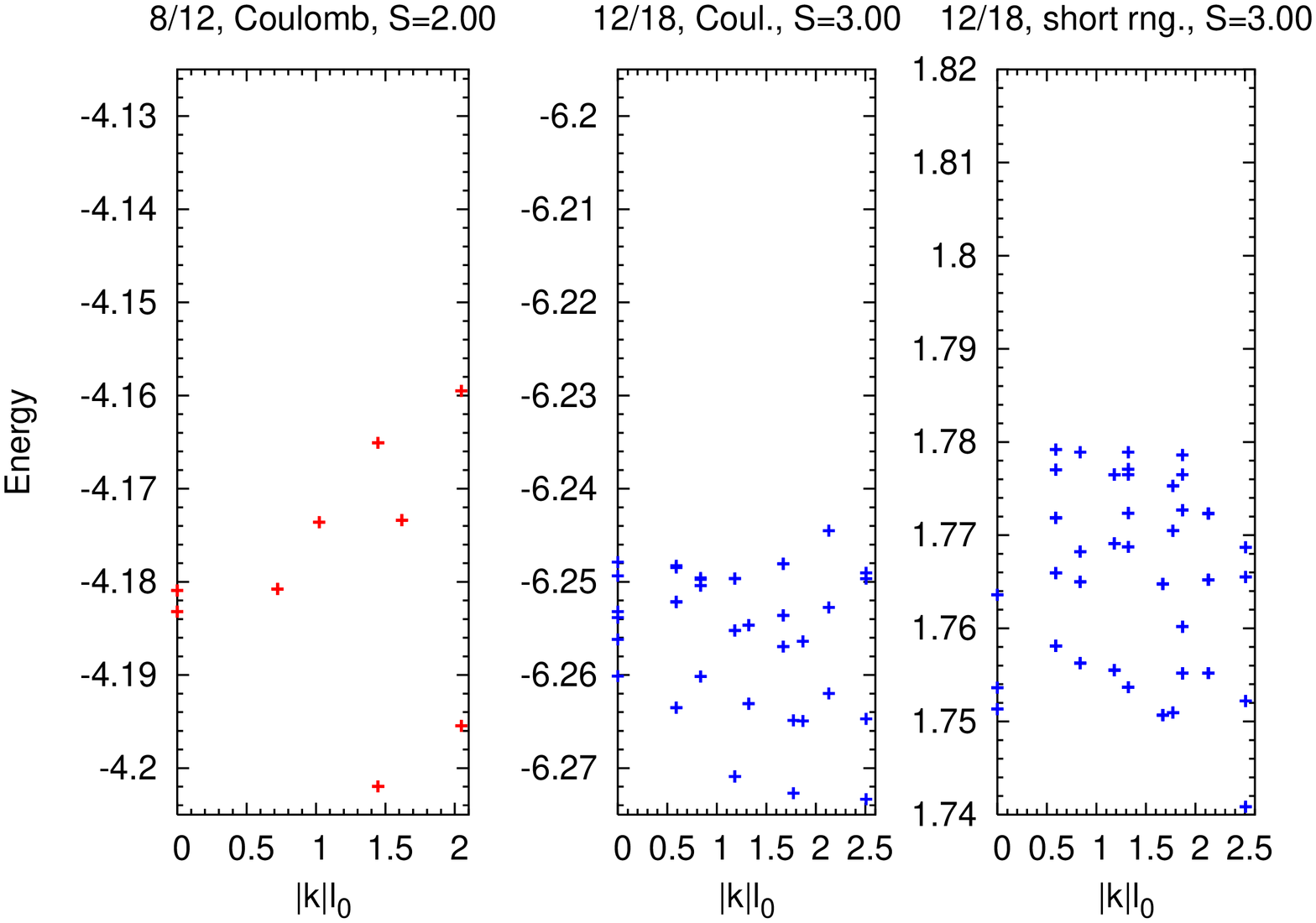}%
\unitlength=1mm\put(-51,-44){$\diamondsuit$}}\quad
    \label{fig-ch03-41a}}
  \subfigure[The ground state]{%
\hskip-.5cm\includegraphics[scale=0.35,angle=-90]{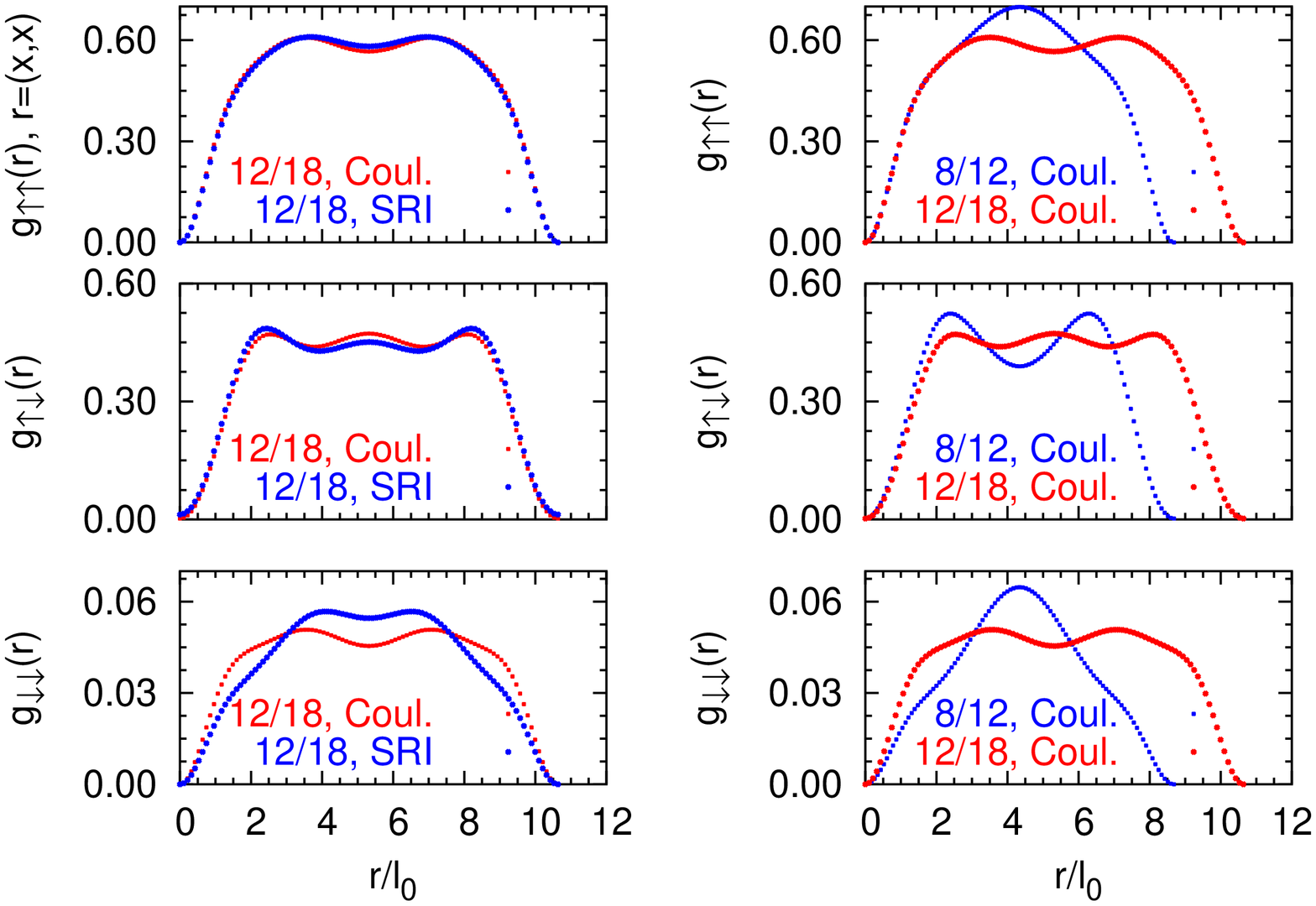}\hskip-2.0cm
    \label{fig-ch03-41b}}
  \caption[Coulomb HPS]{Half-polarized states in Coulomb interacting
  systems. }
  \end{center}
\label{fig-ch03-41}
\end{figure}

The spectra do not look very similar,
Fig. \ref{fig-ch03-41}a. However, the absolute ground states have in
both cases the same symmetry, they lie in the same point of the
$\vek{k}$ space.

The Coulomb and SRI ground states in the largest system
available, $N_e=12$, have very similar structure. The correlation
functions $g_{\up\up}$ and $g_{\up\dn}$ match nicely while
$g_{\dn\dn}$ show some differences between the CI and SRI states. In
spite of this, the overlap between the two states is as large as
$95\%$. This allows for the following conclusions

    (a) The two states 'correspond to each other'. 
    (b) The short-range part of the interaction seems to be essential
    for this state (very similar as for the Laughlin state).
    (c) Deviations in $g_{\dn\dn}$ (minority spin) might come from
    the fact that spin-down electrons are very far separated from
    each other (they have an effective filling of only
    $\nu=\frac{1}{6}$). Thus the long-range part of the interaction
    substantially influences their motion.

In $N_e=8$ systems, the most likely analogue to the $N_e=12$
    ground state is the state $\diamondsuit$, Fig. \ref{fig-ch03-41}a.
    This is the lowest 
    8-electron state with the same symmetry (value of $\krvd$) as the
    $N_e=12$ ground state. Correlation functions of the two states
    ($8$- and $12$-electron ones) match reasonably,
    Fig. \ref{fig-ch03-41b}. Compare also with differences between
    $N_e=8$ and $N_e=12$ short-range ground states,
    Fig. \ref{fig-ch03-23}.

Among the excited states the level order is often modified,
    comparing the $N_e=12$ Coulomb and short-range systems. When
    trying to assign CI to corresponding SRI states, calculating
    overlap between two states seems to be a more reliable tool than
    comparing correlation functions.

In summary, in spite of differences in the excitation spectrum, the
half-polarized ground states of Coulomb and short-range systems seem
to correspond to each other. Differences in the excited states and in the
correlations between the minority spin electrons indicate
that the definition of the short-range interaction should be
improved when we study the half-polarized states. Since the minority
spin electrons are relatively far from each other, non-zero values of
higher pseudopotentials ($V_m$, $m>1$, Subsect. \ref{pos-ch02-11}) 
should probably be considered.

\subsection{In search of the inner structure of states: response to
  delta impurities}





\label{pos-ch03-18}

Now that some candidates for the half-polarized ground state at
filling $\tt$ have been introduced we wish to look at them more
closely and learn more about their inner structure. The ultimate goal
of such efforts can be to propose trial wavefunctions just as the Laughlin
wavefunction at filling $\nu=\ot$. Even though this has not been 
accomplished, the results presented below shed some
light on relations between the half-polarized state and
the singlet and polarized incompressible states.

\begin{figure}
\vspace*{-.3cm}

\subfigure[Density response: incompressible- or com\-pressible-like.
Note that density integrated
over the hatched area remains unchanged for the incompressible system
when the inhomogeneity is switched off.]{\label{fig-ch03-42b}
\hskip.5cm\input{figs/ch03/ch03-fig58.xfig.pstex_t}\hskip.5cm
}
\subfigure[$\delta$-line impurity of the type $\delta_{\sigma
\up}-\delta_{\sigma\dn}$.
]{%
\label{fig-ch03-42a}
\raise.5cm\hbox{\includegraphics[scale=0.28]{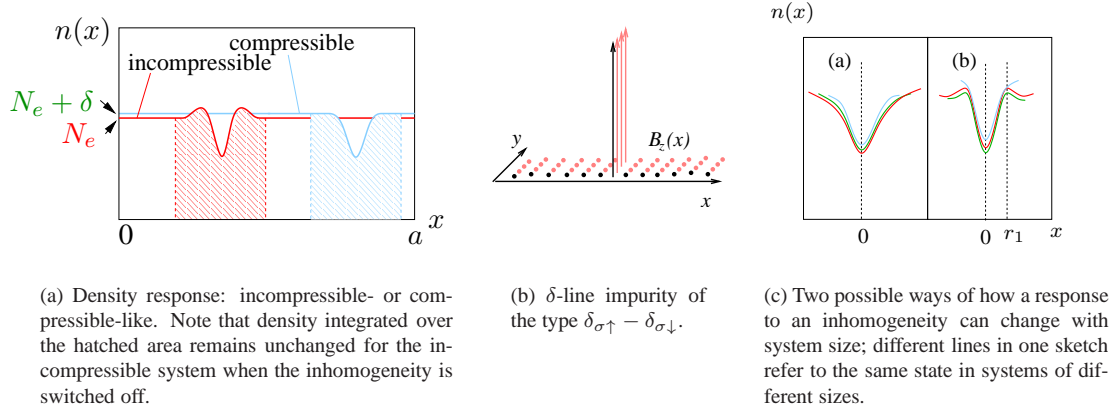}}
}
\subfigure[Two possible ways of how a response to an inhomogeneity can
change with system size; different lines in one sketch refer to the
same state in systems of different sizes.]{\label{fig-ch03-42c}
\hskip.5cm\input{figs/ch03/ch03-fig59.xfig.pstex_t}\hskip1cm
}
\caption{$\delta$-line inhomogeneity and sketches of possible
resulting effects.}\label{fig-ch03-42}
\end{figure}

As a probing tool, the homogeneous states are subjected to a
$\delta$-line impurity and response in density and polarization is
observed. In first quantization,
\begin{equation}\label{eq-ch03-06}
   H_{impurity} = \sum_{i=1}^{N_e} W(\vek r_i)\,,\qquad 
   W(x,y) = \delta(x-x_0)
\end{equation}
This inhomogeneity profile (Fig. \ref{fig-ch03-42}) was chosen since
it is compatible with the torus symmetry. For studies of
point-like impurities, spherical geometry is more suitable since it
preserves the rotational symmetry, cf. references in
Subsec. \ref{pos-ch03-16}. The $\delta$-line form is particularly apt
to unveil a tendency of the state to build plane charge or spin density
waves. We should keep in mind, that due to the restriction to the
lowest Landau level, even a $\delta$-like potential has an effective
cross section of $\ell_0$ \cite{rezayi:11:1985}.

As we are dealing with spinful electrons, inhomogeneities can be
principially of four distinct types:
\begin{eqnarray} \nonumber
  H_{EI} = W(r)\cdot (\delta_{\sigma\up} + \delta_{\sigma\dn})\,, &&
  H_{MI,\up} = W(r)\cdot \delta_{\sigma\up}\,, \\
  H_{MI} = W(r)\cdot (\delta_{\sigma\up} - \delta_{\sigma\dn})\,, &&
  H_{MI,\dn} = W(r)\cdot \delta_{\sigma\dn}\,, \label{eq-ch03-07}
\end{eqnarray}
where the function $W(r)$ describes the spatial form of the impurity,
Fig. \ref{fig-ch03-42} shows the form of $W(r)$ chosen in the present
study. It is
important to note that these impurities fail to conserve $S^2$ but
they do conserve $S_z$. Also, owing to the form of $W(\vek r)=W(x)$,
they conserve $\krv_y$ and thus also $J$ (\ref{eq-ch02-54}) and
they spoil only the $\krv_x$-symmetry. This is very convenient from
the computational point of view as matrix sizes remain tractable. From the
physical point of view, this inhomogeneity is a soft tool which does not
completely destroy the high symmetry of the studied states. For
example, it allows us to stay in the $S_z=N_e/4$ sector when we study
the half-polarized states.

The first type ($H_{EI}$, electric impurity) is an
ordinary non-magnetic impurity or external electric
potential. The magnetic impurity ($H_{MI}$) 
favours particles with correct spin ($\dn$, if $W(r)>0$) and costs
energy for particles with wrong spin ($\up$ in this case). The last
two types describe an impurity which is seen only by one group of spins. 
In case that a system consists of two separated subsystems, one of spin up
particles and another of spin down particles, these impurities allow
to test only one of them without directly disturbing the other one. 

Note that some inhomogeneity types in (\ref{eq-ch03-07}) may be
redundant, depending on the state we apply them to. For instance, the
effect of $H_{MI,\up}$ and $H_{MI,\dn}$ must be the same up to a sign
for all states in the  $S_z=0$ sector.

Before we turn to the exact diagonalization results, let us briefly
think about what types of responses can be expected. 
Most importantly, consider the difference between
compressible and incompressible states. As a classical {\em compressible}
system imagine a playground of fixed size filled with a gas of negatively
charged footballs of density $n(r)$, $\langle n(r)\rangle = N$. A
negative impurity at $r=0$ will repel the gas causing $n(0)<N$,
Fig. \ref{fig-ch03-42}a right. Beyond
some distance $r_h$, the density will reach a constant level again and
this level will be slightly higher than the original density, 
$N+\delta = n(\vek r)>N$, $|\vek r|>r_h$, so that the
constraint $\langle n(\vek r)\rangle = N$ remains preserved. Some charge
has been depleted away from the impurity, thereby compressing slightly
the gas in the rest of the system. If the depleted charge equals the charge
of the impurity, the charge distribution (charge of the footballs plus
charge of the impurity) in the system will remain
constant in spite of non-constant $n(\vek r)$ and the gas particles far
away from $r=0$ will not 'see' the impurity anymore. This is the case
of ideal {\em screening}.

A classical {\em incompressible} liquid, for example again charged footballs,
will not react at all, because it cannot change its
density. Even though particles of the liquid feel repulsion from
$r=0$, the density will remain constant $n(r)=N$. We can also
encounter a different behaviour, Fig. \ref{fig-ch03-42}a (left). 
Though the density decreases directly at $r=0$, an oscillatory
structure develops in $n(r)$, so that the integral density in the
region $|\vek r|<r_h$ remains as it was without the impurity. The density
then also remains at its original value $N$ beyond $r_h$. This is a
non-ideal incompressible behaviour: at very short distances, the density
can vary slightly, but averaged over distances of $r_h$ (or larger), the
density remains constant. Also, since no net charge was depleted from
the region $|\vek r|<r_h$, the impurity is {\em completely unscreened}
on distances larger than $r_h$.

Compressible-like response as shown in Fig. \ref{fig-ch03-42}a can be
combined with quantum interferences (Friedel oscillations) and it is
also possible to think of some overscreening effect which would lead
to an oscillatory $n(r)$. This means
the sole fact that $n(r)$ exhibits oscillations does not necessarily
have to imply incompressibility. A more reliable criterion is that the
integral density over $|\vek r|<r_h$ remains the same with and without
impurity. This procedure is delicate in finite systems where $r_h$ can
be comparable to the system size. 

The last Figure, \ref{fig-ch03-42}c, shows two possible ways of
how responses change with system size. The right panel
suggests that the state is not fixed to a particular size of the
finite system and especially we could expect oscillations with period
$r_1$ also in an infinite system. On the contrary, the left panel
shows a state with no intrinsic length scale and e.g. the width of the
peak is related to the (finite) size of the particular system.

Now, let us proceed to fractional quantum Hall states.

\subsubsection{Electric (nonmagnetic) impurity}
\label{pos-ch03-16}

The effect of electric impurities on incompressible ground states has
been under investigation since the early times of the fractional
quantum Hall effect. The main reason is that some disorder
is needed for the integer
quantum Hall effect to be observable, but on the other hand, 
too strong disorder will destroy the effect \cite{yoshioka:2002}. 
For the fractional quantum Hall effect, two of the basic
questions were, (i) how strong impurity potentials may be so that they
do not destroy the gap and (ii) how does it change the ground
state. Basic studies with the Laughlin state were performed as early
as in 1985 \cite{rezayi:11:1985,zhang:11:1985,girvin:02:1985}.

Since the exact diagonalization is 
limited to finite, and in fact quite small, systems, it
is very delicate to put forward statements about the infinite 2D electron
gas. Therefore, when we use the word 'incompressible' we
mean rather 'incompressible-like' in terms of
Fig. \ref{fig-ch03-42}. In fact, the main purpose of the following
Subsections is to see how the polarized and singlet state respond to
impurities in a {\em finite system} and later to compare them to the
half-polarized state again in a {\em finite system}. We will focus on
{\em short-range interacting} systems here.

\subsubsubsection{The Laughlin state or the fully polarized $\tt$ state}

The fully polarized $\nu=\tt$ state is a particle-hole conjugate to
the $\nu=\ot$ Laughlin state in a homogeneous system,
Subsect. \ref{pos-ch02-03}.  
In this part we will study the latter state. Strictly taken,
  the particle-hole symmetry is lost when an arbitrary
  impurity is considered since the Hamiltonian is no longer
  translationally invariant. Differences between the $\nu=\ot$ and $\tt$
  polarized states are however small if the impurity is weak. In
  particular, for inhomogeneities considered in this paragraph, it has
  been checked numerically that $n(x)-N_e$ are almost the same for the
  two states. Moreover, the larger $N_e$, the smaller are the differences. 

\begin{figure}
  \begin{center}
    \subfigure[The polarized state ($\nu=\ot$ considered, see
    text) in systems with $N_e=4$-$10$ electrons. Note that the
    response 
    does not decay with increasing the system size.]{
      \includegraphics[scale=0.65,angle=0]{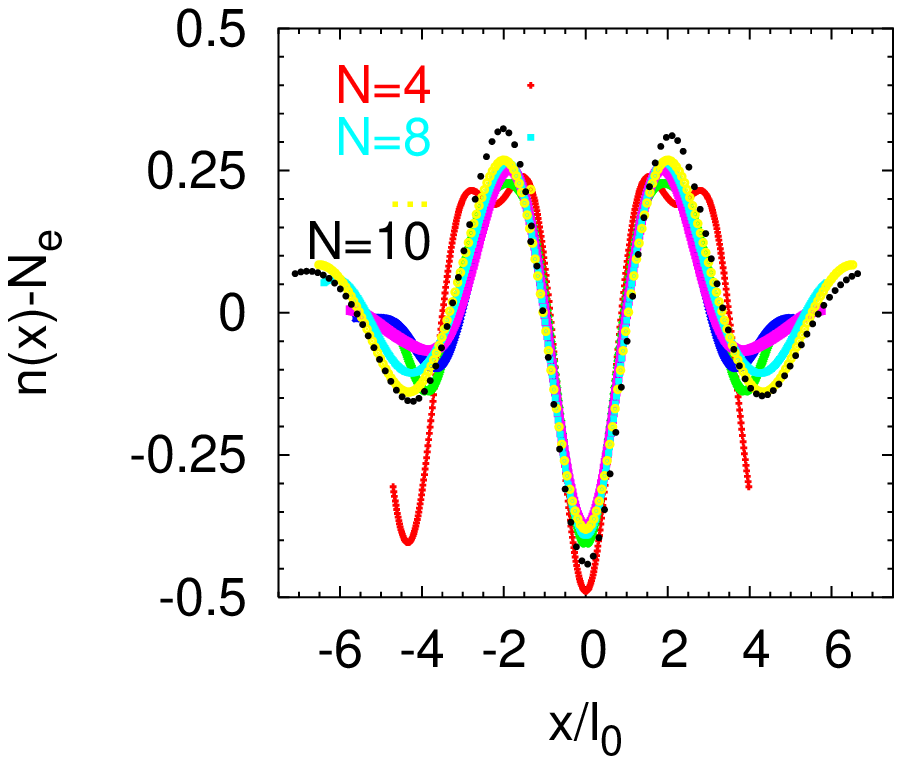}
      \label{fig-ch03-34a}}
    \subfigure[The singlet state. The
    thin line shows the CM oscillations
    in a homogeneous system, thick lines show responses to an impurity
    with CM oscillations subtracted.]{
      \includegraphics[scale=0.65,angle=0]{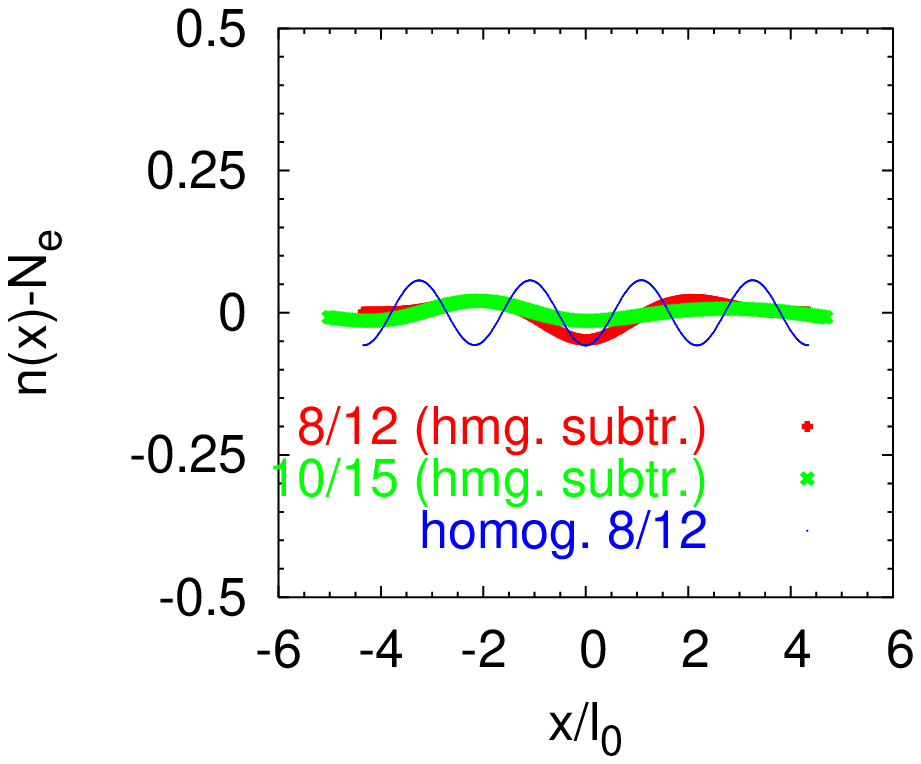}
      \label{fig-ch03-34b}}
  \caption{Polarized and singlet $\tt$ state and (non-magnetic) impurity
  in the form of a $\delta$-line (along $y$). Normalized density along
  $x$ is plotted.}\label{fig-ch03-34}
  \end{center}
\end{figure}

The response of a $\nu=\ot$ system to an impurity of the form
(\ref{eq-ch03-06}), a $\delta$-line along $y$, is shown in
Fig. \ref{fig-ch03-34}. Different curves show the ground state density
$n(x)$ in systems of different sizes ($N_e=4$ to 10
particles). The repulsive impurity is always located 
at $x=0$ and it is weak, its strength
is $\sim 10\%$ of the gap. These results agree very well with the
densities presented by Zhang {\em et al.} \cite{zhang:11:1985}, who
considered a $\delta$ rather than a $\delta$-line impurity, though for
$N_e=4$ systems only. Comparison between rectangular, spherical and
disc geometry showed, in all cases, very similar behaviour
\cite{zhang:11:1985}. Note also that findings in
  Fig. \ref{fig-ch03-34} assume short-range interaction whereas
  \cite{zhang:11:1985,rezayi:11:1985} considered Coulomb interaction.

Results in Fig. \ref{fig-ch03-34}a support the conclusions of Zhang
and Rezayi. The oscillatory response of $n(x)$ is size-independent
and it has a period $r_1\approx 2.5\ell_0$. The response, measured by
$n(0)$, does {\em not} vanish with increasing system size 
but it decays
with distance from the impurity. Comparing $n(x)$ in
Fig. \ref{fig-ch03-34}a to the model cases in Fig. \ref{fig-ch03-42}a,
we may tend to classify the Laughlin state as an incompressible one. 
Incompressibility of the Laughlin state is locally not perfect,
otherwise $n(x)$ would remain constant, at least in infinite systems.
However, if some net charge were accumulated even in a larger region (of
the order $r_h$) around $x=0$, we would expect $n(x)$ at large distances to be
consistently higher than the no-inhomogeneity value $n(x)\equiv
N_e$. Recall the difference $N$ to $N+\delta$
  in Fig. \ref{fig-ch03-42}a. If just a unit charge is depleted from the
  impurity, then $\delta=1/N$. In infinite systems, the difference
  $\delta$ will vanish, but data in Fig. \ref{fig-ch03-34}a come from
  rather small systems $N\le 10$ where $\delta$ is not negligible.
This is not seen in
Fig. \ref{fig-ch03-34}a.

Zhang {\em et al.} suggest that the observed response is a local
charge density wave (\ref{eq-ch02-43}), a strong argument
supporting this idea is given in point
(iv) below. Under this view, it is not surprising that the response to a
$\delta$-line shown in Fig. \ref{fig-ch03-34a} is very similar to the
response to a $\delta$-peak studied by Zhang. Only the envelope
function, not the wavelength depends on the particular form of the
exciting impurity.

We should again add several comments:

(i) oscillations observed in $n(x)$, Fig. \ref{fig-ch03-34}a,
are not related to Friedel oscillations which appear in the Fermi
gas. This is where a sharp Fermi surface exists giving rise to interferences,
just as in correlation functions of a free Fermi gas,
Subsec. \ref{pos-ch03-15}.

(ii) small wiggles on the $N_e=4$ density in Fig. \ref{fig-ch03-34}a
are due to the center-of-mass (CM) part of the wavefunction. Being a
finite size effect, they fall off rapidly with system size as we
indeed see in Fig. \ref{fig-ch03-34}a, Subsec. \ref{pos-ch03-01}.

(iii) the ground state of the 
homogeneous system is triply degenerate in
the CM part, Subsec. \ref{pos-ch03-01}. This degeneracy is lifted by
the inhomogeneity, but energy differences between these three states
remain much smaller than their separation from the lowest excited
states for the inhomogeneity strength considered \cite{zhang:11:1985}. 

The response $n(x)$ of any of the three
states depends slightly on the position of the impurity within the
elementary cell, but this dependence and also differences among the
three states in energy and in $n(x)$ quickly vanish with increasing
system size. In Fig. \ref{fig-ch03-34}a always the impurity giving the
strongest response in $n(x)$ was chosen.

(iv) Period of oscillations: As Rezayi and Haldane
\cite{rezayi:11:1985} note, numerical calculations as in
Fig. \ref{fig-ch03-34}
agree with results of the single mode approximation proposed by Girvin
{\em et al.} \cite{girvin:02:1985}.  The linear
response function $\chi(q)$ (in the $\nu=\ot$ Laughlin state) is dominated by
the magnetoroton collective mode around $q_0\ell_0 \approx 1.4$. Would
it be $\chi(q)=\delta(q-q_0)$, the density response to a point
impurity potential would be $n(r)\propto J_0(q_0r)$. This density
profile looks like damped oscillations with the first node at $r=1.7\ell_0$. 

Regarding a more realistic profile of $\chi(q)$,
this estimate for $n(r)$ is a very good approximation to $n(x)$ in
Fig. \ref{fig-ch03-34}a.

As the purpose of the present work was to study systems with spin, we
will now continue to spin singlet states at $\nu=\tt$.
Some quite new results for the $\nu=\ot$ Laughlin state have been
achieved by~M\"uller~\cite{mueller:2005}.

\subsubsubsection{The singlet state}

The $\tt$ singlet ground state shows basically the same signs of
incompressibility as the polarized state. The period of the 
density oscillations incurred by a $\delta$-line impurity 
is almost the same ($r_1\approx 2\ell_0$), 
and also in terms of classification of Fig. \ref{fig-ch03-42}a, the
singlet state shows an incompressible-like behaviour, cf. discussion
of the polarized state. The striking feature of the singlet state is, 
that the strength of the response is about an order of magnitude less
than in the polarized state. Thus in an $8$-electron system, the
density response is 'hidden' under the center-of-mass oscillations,
Fig. \ref{fig-ch03-34}b. 

This strong difference between the singlet and polarized ground states
is unexpected since 'incompressibility' gaps of both states are
similar.

This hints at unusual stability of the singlet state with respect to
charged inhomogeneities. In terms of perturbation theory, this is
not due to energetic reasons but rather owing to small matrix elements
of $H_{EI}$ between the ground state and excited states. Energy of the
first excited state, however, decreases when impurities are present
and thus, in spite of the quite stable density of the GS, the gap will
eventually collapse. 

Regarding the response in systems of different size, we find a
considerable attenuation when going from eight to ten-electron
systems, Fig. \ref{fig-ch03-34}b. Nevertheless we assume that the
response remains finite even in the thermodynamic limit. To support this
hypothesis, a fact worth of emphasis is that the $N_e=8$ (10) singlet state
occurs in systems with $N_m=12$ (15) flux quanta, i.e. with system area
$A=2\pi\ell_0^2 N_m$ (\ref{eq-ch02-38}).
These are the two smallest systems considered in
Fig. \ref{fig-ch03-34}a. For the these two systems 
we also observe a considerable
attenuation of the $n(x)$ response when going from the $N_e=4$ to $N_e=5$
state, Fig. \ref{fig-ch03-34}a, and 
this reduction in response is  definitely only a finite
size effect. As close as this analogy is, observations presented in
Fig. \ref{fig-ch03-34}b are not conclusive and an investigation of the
singlet state in a larger system ($N_e=12$) would be needed.

Let us just briefly mention, that non-magnetic impurities have no
effect on the polarization of the singlet ground state.

\subsubsection{Magnetic impurity in incompressible $\tt$ states}
\label{pos-ch03-05}

As far as spin polarized states are considered, magnetic impurities
(\ref{eq-ch03-07}) cannot have any other effect than the electric
impurities do. Therefore only the $\tt$ singlet ground state will
be discussed here as the half-polarized states deserve to be
considered separately, Subsec. \ref{pos-ch03-04}.

\begin{figure}
  \begin{center}
    \subfigure[Density.]{
      \includegraphics[scale=0.65,angle=0]{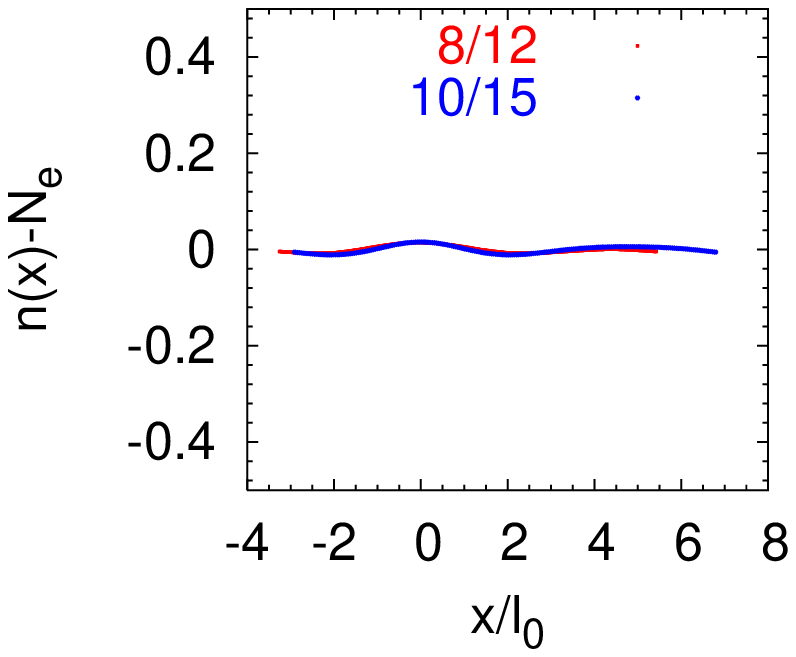}
      \label{fig-ch03-35a}}
    \subfigure[Polarization.]{
      \includegraphics[scale=0.65,angle=0]{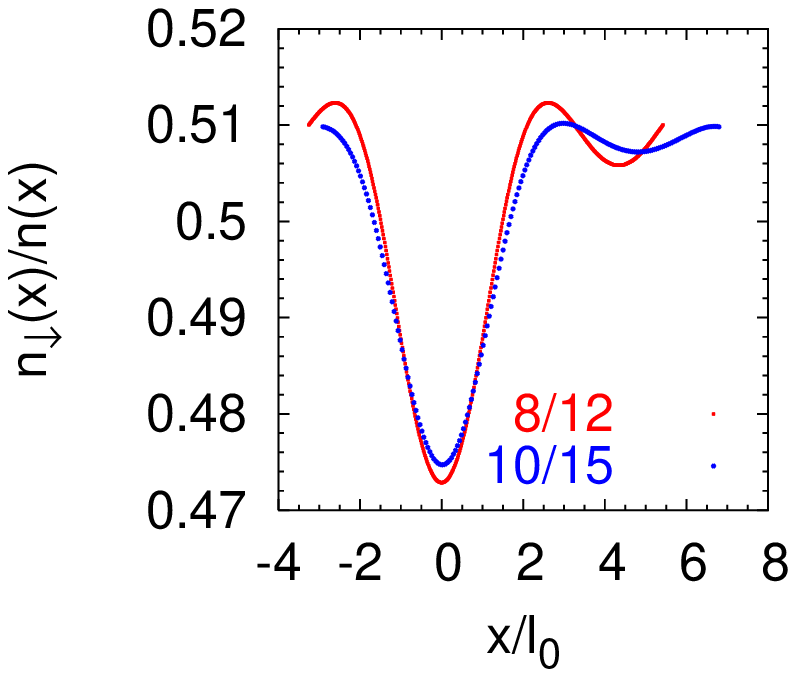}
      \label{fig-ch03-35b}}
  \caption{Singlet $\tt$ ground state for an attractive 
    magnetic impurity $H_{MI,\up}$
    (\ref{eq-ch03-07})
      in the form of a $\delta$-line (along $y$). Normalized density and
      polarization along $x$ are plotted.}\label{fig-ch03-35}
  \end{center}
\end{figure}

Considering the {\em density}, Fig. \ref{fig-ch03-35}a, we find a yet
weaker response than for non-magnetic impurities,
Fig. \ref{fig-ch03-34}b. The response reminds of an incompressible
system, in terms of Fig. \ref{fig-ch03-42}a, and may remain finite
in the thermodynamic limit, cf. discussion of non-magnetic impurities.

{\em Polarization} $n_\dn(x)/n(x)$ behaves quite differently,
Fig. \ref{fig-ch03-35}b: the response
is large and it looks compressible. Again in terms of
  Fig. \ref{fig-ch03-42}a. In particular, note that the polarization
  $n_\dn(x)/n(x)$ in Fig. \ref{fig-ch03-35}b approaches $\approx 0.51$
  as we go 'far away' from the impurity, i.e. a different value than
  the polarization in the homogeneous case, $0.5$..
Electrons with 'correct
spin' ($\up$) accumulate around the impurity, $n_\dn(0)/n(0)$
drops from the homogeneous value ($0.5$) by as much as by $5\%$, 
whereas the average polarization off the impurity slightly increases so
as to keep the overall average value $0.5$ as required by
$S_z=0$. This behaviour differs strongly from the density response,
Fig. \ref{fig-ch03-35}b.

It should also be noted that both density and polarization are here much
less system-size dependent than in the case of non-magnetic impurities.

These are quite remarkable findings. It seems that the singlet state
is {\em locally} much more 'incompressible' than the polarized
state. On the other hand, the singlet state is relatively easily
polarizable which is particularly striking when compared
to the weak response in the density. If we
  assume the density in Fig. \ref{fig-ch03-34}b to be the response of
  two independent liquids, then the polarization in
  Fig. \ref{fig-ch03-35}b should be (i) smaller by a factor of five
  for $N_e=8$ than what is observed 
  and (ii) considerably smaller for $N_e=10$ compared to the $N_e=8$
  case.
This again contradicts the
picture of two uncorrelated $\ot$ Laughlin liquids, one spin up, another spin
down, which we could wrongly infer
from the view of filled composite fermion LLs. Remind, however, 
that it is in fact 
not the claim of CF theories, that particles of $n=0,\up$ and
  $n=0,\dn$ CF LLs are uncorrelated.

\subsubsection{Integer quantum Hall ferromagnets}
\label{pos-ch03-06}

A brief introduction to integer quantum Hall ferromagnets (QHF)
was given in Subsec. \ref{pos-ch02-16}. It is instructive to keep in
mind the scheme of Landau levels, Fig. \ref{fig-ch02-13}.

Here we will focus on the $S_z=0$ sector in prototypes of Ising and
Heisenberg QHFs with neglected LL mixing. These states ($S_z=0$) are
analogues of the half-polarized states at filling $\tt$, the
explanation follows. Disciples of
CF teachings deem the $\nu=\tt$ ground states to have $\nu_{CF}=2$ completely
filled CF LLs, Fig. \ref{fig-ch03-01and02}b. Transitions between the singlet
and polarized GSs occur, when the $(n,\sigma)=(0,\dn)$ CF LL crosses the
$(1,\up)$ CF LL. It is then plausible to neglect the
low lying $(0,\up)$ CF LL and look only at the two crossing CF Landau
levels. The two ferromagnetic Ising states -- the singlet, and
polarized electronic GS at $\nu=\tt$ -- correspond to {\em all} CFs placed in
the $(0,\dn)$, and $(1,\up)$ CF LL, respectively, compare 
to Fig. \ref{fig-ch02-09}b. Hence the half-polarized state ($\nu=\tt$)
corresponds to half-filled $(0,\dn)$ and half-filled $(1,\up)$. 
Disregarding the fully occupied $(0,\up)$ CF LL, i.e. 
counting only particles in the two crossing CF LLs (in total
$N_e$ CFs), the ferromagnetic Ising states are $S_z=\pm N_e/2$ and the
'half-half' state is $S_z=0$.

In this Subsection we study the same situation as the one
occuring at the $\nu=\tt$ ground state transition (within the picture
of crossing CF LLs) but for {\em electronic} Landau levels, i.e. with
electrons instead of composite fermions. We therefore study a
$\nu=1$ system with spin degree of freedom, where spin down (spin up)
electrons lie in the $n=0$ ($n=1$) Landau level,
respectively, and we disregard the fully occupied $(0,\up)$
  level. The physical system we model herewith has thus $\nu=2$. 
Technically, this requires only implementing
  modified values of pseudopotentials, Fig. \ref{fig-ch02-04}.
Without electron-electron interaction, these two Landau levels are set to
equal energy so as to model the LL crossing. Mixing to the fully occupied
$(0,\up)$ LL as well as to all higher LLs is neglected, since all 
these levels are well separated from the two crossing levels.

Heisenberg QHFs are not related to $\nu=\tt$ and we investigate them just for
the sake of comparison between Ising- and some other type of QHF. In the
integer QHE regime, Heisenberg QHF occurs e.g. when $(0,\up)$ and $(0,\dn)$
LLs cross (and $\nu=1$) as it is the case for instance at vanishing Zeeman
splitting. With CFs, this happens at $\nu=\ot$, i.e. $\nu_{CF}=1$,
Fig. \ref{fig-ch03-01and02}b.

We will first briefly discuss homogeneous states in these QHF systems
and then we will turn to their response to magnetic inhomogeneities
($\delta$-lines).

\begin{figure}
\begin{center}
  \subfigure[Spectra. The lowest branch is marked by $*$, the second
    lowest by $\clubsuit$. The state marked in different colour 
    does not belong to the lowest
    branch.]{\label{fig-ch03-45a}
    \includegraphics[scale=0.8,angle=0]{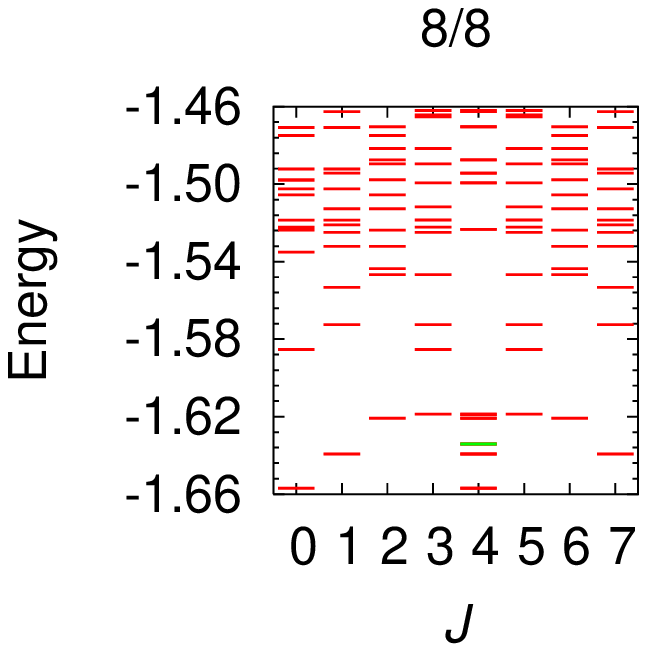}
    \includegraphics[scale=0.8,angle=0]{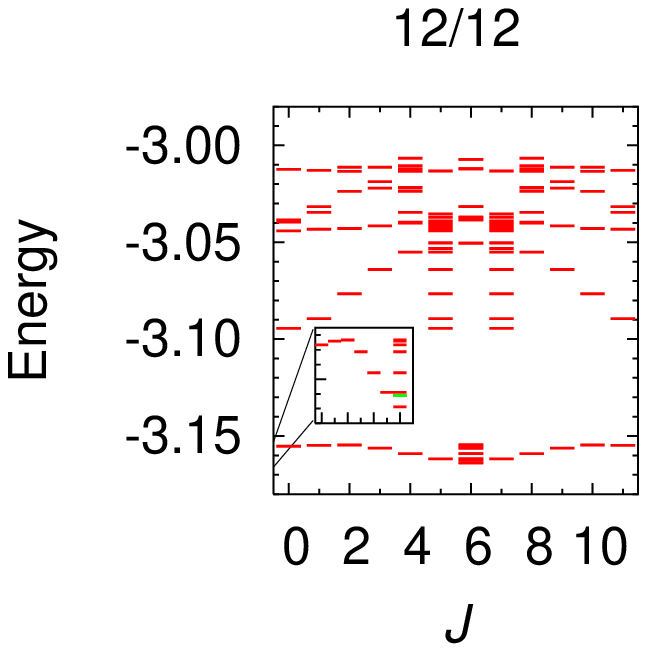}%
\unitlength=1mm
\put(-15,18){$*$}\put(-15,26){$\clubsuit$}}
    \subfigure[{\em Left:} Density in the ground state, 12 particle system;
    {\em right:} change in density in response to a $\delta$-line
    impurity, 
    cf. Fig. \ref{fig-ch03-37}b.]{\label{fig-ch03-45b}
    \includegraphics[scale=0.5,angle=0]{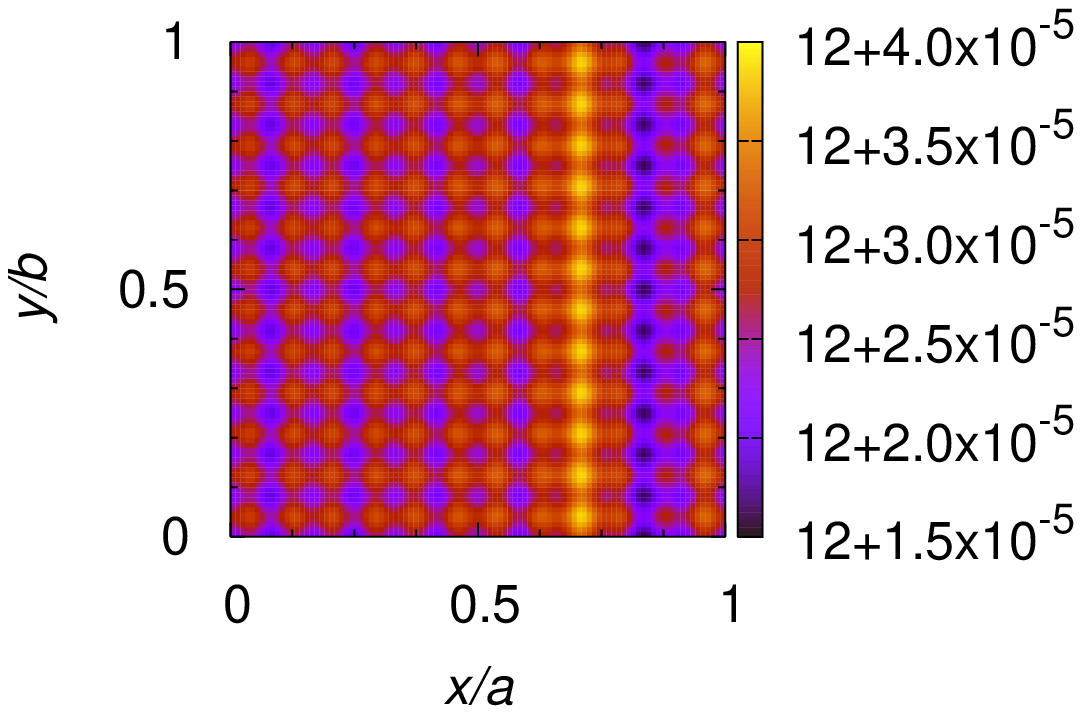}
    \includegraphics[scale=0.7,angle=0]{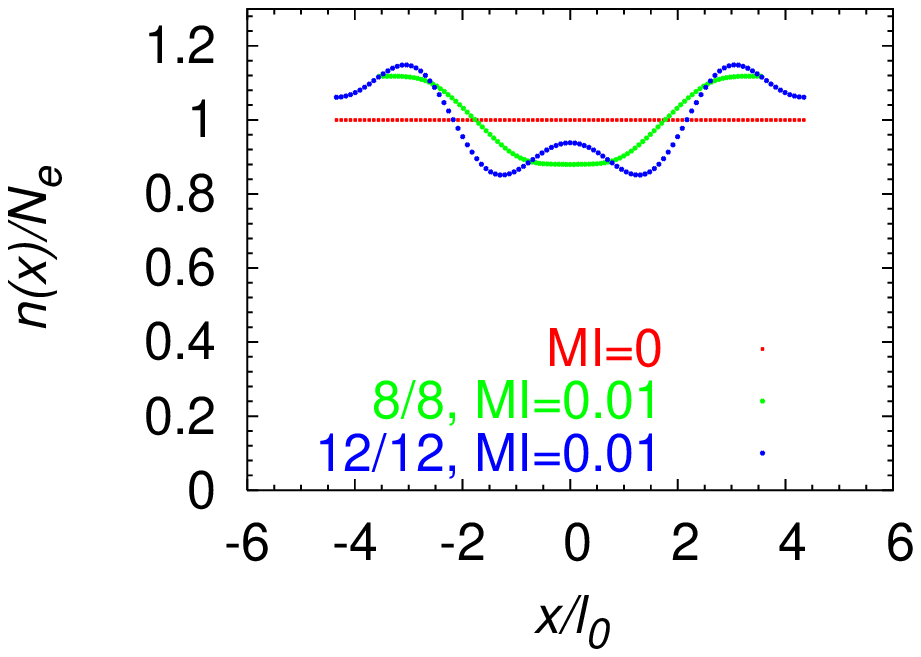}}
    \caption{Half-polarized states ($S_z=0$) of an Ising quantum Hall
    ferromagnet. $\nu=2$}
\label{fig-ch03-45}
\end{center}
\end{figure}

\subsubsubsection{Ising quantum Hall ferromagnet}

There are two degenerate ground states of an Ising ferromagnet: both
with $S=N_e/2$, one $S_z=N_e/2$ and another $S_z=-N_e/2$.  In general,
excited states are no eigenstates to $S^2$ as a consequence of the
omission of the fully occupied $(0,\up)$ level and may only be
classified according to $S_z$. They are all situated well above the
ground states, Fig. \ref{fig-ch02-09}b, and their energy grows with
$N_e/2-|S_z|$. In the following we will only speak about $S_z=0$
states. The whole $S_z=0$ sector is quite high in the complete
spectrum. Unlike for a Heisenberg ferromagnet there is nothing like a
$S=N_e/2,S_z=0$ ground state for an Ising ferromagnet.

Low lying $S_z=0$ states of the considered Ising QHF are apparently arranged
into a flat dispersion branch,
Fig. \ref{fig-ch03-45}a. For a fully occupied Landau
  level, $J$ 'coincides' with $\krv_y$. Precisely, $\krv_y =
  (N_e/2-J)\sqrt{2\pi/N_m}$ for $N_e$ even in the sense of
  (\ref{eq-ch02-45}). Centre-of-mass degeneracy is absent.
The anomalous form of the branch in a $N_e=8$
system, seems to be of finite-size origin, since $N_e=10,12$ and 14
spectra are all similar. States of the lowest
branch have $\krvd$ of the form $(2\pi n/N_e,0)$, $n=0,\pm 1,\ldots, N_e/2$,
or $(0,2\pi n/N_e)$. This is in agreement with the symmetry between
$x$ and $y$ (square elementary cell). It shows that
rotational symmetry is absent in the low energy
sector -- otherwise we would observe also states with
  $\krvd=(k_x,k_y)$, $k_x,k_y\not=0$.
The lowest branch flattens and becomes well separated from
excited states with increasing system size, and the minimum energy remains
at $\krvd=(0,0)$. Also, other branches develop, the second
lowest branch is described by $\krvd=(\pi n/N_e,\pm 2 \pi/N_e)$ (plus
the $x$-$y$ symmetric partner)
and minimum energy at points $(\pi, \pm 2\pi/N_e)$, see the $N_e=12$ spectrum
in Fig. \ref{fig-ch03-45}a.
Apart from these branches an isolated $\krvd=(0,0)$ state is present 
(shown in gray in Fig. \ref{fig-ch03-45}a) and it is hidden within
the branch. We can hypothesise that this 
state becomes the absolute ground state i.e. gets
separated from the lowest branch in sufficiently large systems.

The flat branch is reminiscent of results of Rezayi\footnote{The cited
  work concerns the situation when the lowest and the third Landau levels of
  different subbands cross. Rezayi {\em et al.} had first to show that
  this system is an Ising QHF. See subsection \ref{pos-ch02-16} for
  more details. In the $S_z=0$ sector of his system Rezayi {\em et
  al.} found a multiply (almost) degenerate ground state with $\krvd$ just of
  the sequence  $(2\pi n/N_e,0)$, similar as we see in Fig. \ref{fig-ch03-45}a
  for $N_e=12$.} 
\footlabel{foot-ch03-01}
{\em et al.} \cite{rezayi:05:2003} and could correspond to domain
states, i.e. stripes along $x$ or $y$ with alternating spin polarization, in a
system which does not prefer any particular domain size. The origin of the
highly symmetric isolated (gray in Fig. \ref{fig-ch03-45}a) state is unknown.

As it can be expected, low lying states have homogeneous density
and it is true even for the whole lowest branch,
as an example we show density in the ground state ($\krv=0$), 
Fig.~\ref{fig-ch03-45}b. The anticipated domains would probably be visible
  first in correlation functions.
States in the second lowest branch show
unidirectional charge density waves. 

In summary, in a homogeneous $\nu=2$ Ising QHF we observe

(i) a flat branch of low lying states, which could become
degenerate in infinite systems, Fig. \ref{fig-ch03-45}a. This first branch
probably consists of stripe domains
-- or spin density waves -- of all possible wavelengths $\lambda=a/n$,
$n=0,1,\ldots N_e/2$ just as in the system studied in
\cite{rezayi:05:2003}. Contrary to isotropic states (like Laughlin
liquid), the wave must be parallel to one side of the square
elementary cell.

(ii) second branch with pronounced
dispersion, which could be a charge density wave

(iii) continuum of excited states above the two branches and

(iv) another state, with high symmetry, $\krv=(0,0)$, which lies
among the states of the lowest branch.

\begin{figure}
  \subfigure[Spectra.]{\label{fig-ch03-44a}
\hskip-1cm\includegraphics[scale=0.6,angle=0]{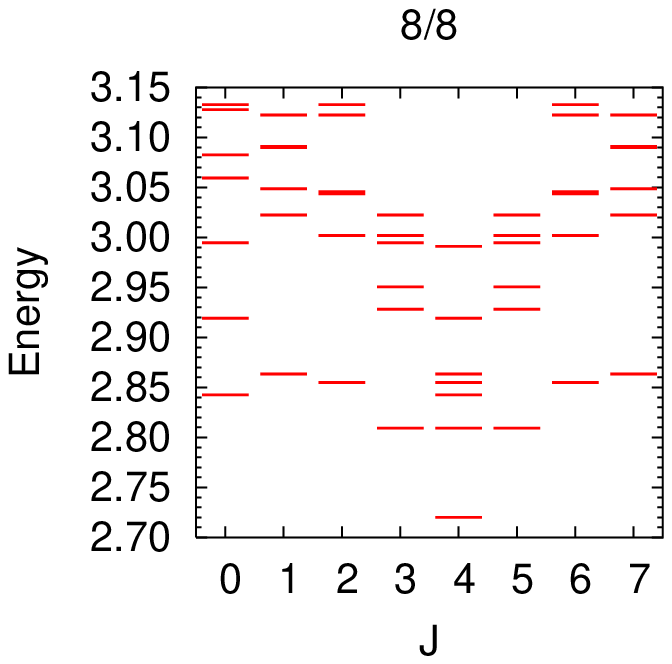}
    \includegraphics[scale=0.6,angle=0]{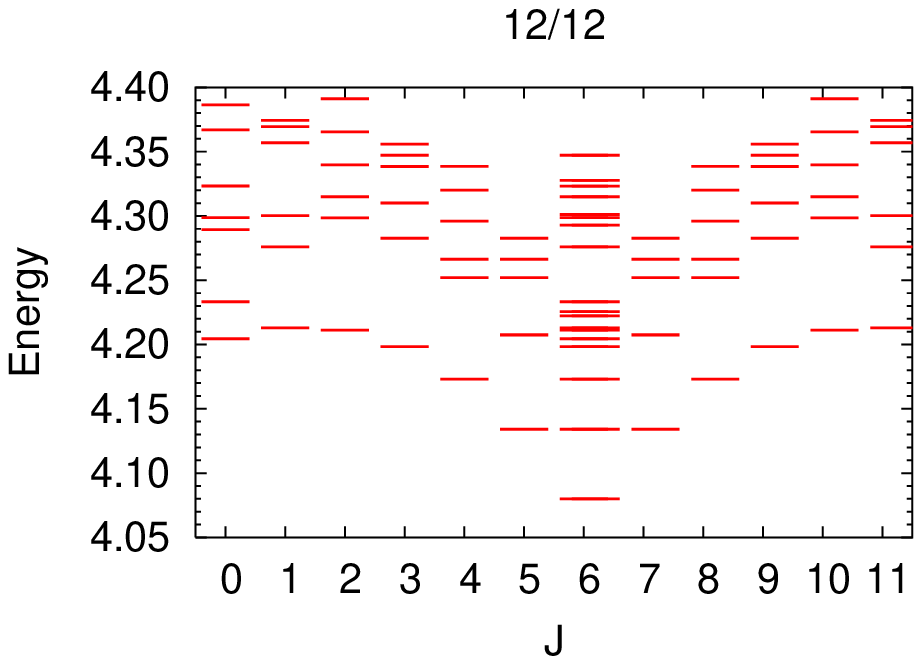}}
  \subfigure[Density, 12 particle system.]{\label{fig-ch03-44b}
    \hskip-1cm\includegraphics[scale=0.75,angle=0]{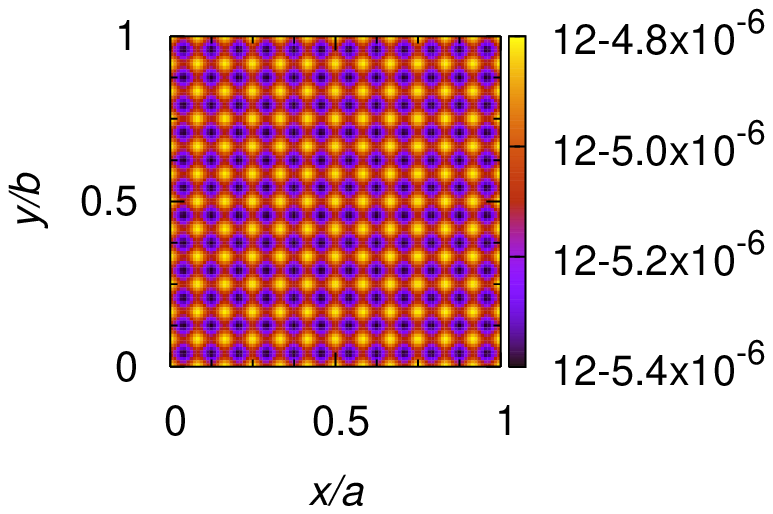}%
\hskip-1cm}
   \caption{Half-polarized states ($S_z=0$) of a Heisenberg quantum
    Hall ferromagnet. $\nu=1$}
\label{fig-ch03-44}
\end{figure}

\subsubsubsection{Heisenberg quantum Hall ferromagnet}

The situation here is quite different from the Ising ferromagnets. The
Hamiltonian (Coulomb interaction projected to the lowest Landau level) 
conserves the total spin $S$ and it even commutes with
$S^+$ and $S^-$ which change $S_z$ while keeping the length of the
total spin. The ground state is fully polarized, $S=N_e/2$, but its
$z$-component of spin is arbitrary, Fig. \ref{fig-ch02-09}b. 

Looking at the sector $S_z=0$, Fig. \ref{fig-ch03-44}b, the lowest
state is thus the ferromagnetic $S=N_e/2$ state. Other low-energy
states form again a branch, $\krvd=(\pm \pi n/N_e, 0)$ and $(0,\pm \pi
n/N_e)$, $n=0,\ldots, N_e/2$ ($x$-$y$ symmetry present, rotational
symmetry absent). Contrary to the Ising QHF, this branch does not seem to
flatten. States in the branch fulfil $S=N_e/2-n$: the ferromagnetic
(ground) state is polarized and going up the branch, the polarization
decreases. In this respect, the excitations of the lowest branch
markedly differ from spin density waves. What we observe in the
Heisenberg QHF are most likely states with $n$ weakly interacting spin
waves which were observed under the same conditions on a sphere by
{W\'ojs} and Quinn \cite{wojs:07:2002}.

\subsubsubsection{Half-polarized QHF states and magnetic impurity}

If a homogeneous state cannot be established and domains formation
is more favourable, then no particular domain size is preferred. This
is the central message of the following paragraph and it applies to
both Ising and Heisenberg QHFs described at the beginning of
Subsection \ref{pos-ch03-06}. 

\begin{figure}
  \subfigure[Heisenberg QHF ($\nu=1$).]{\label{fig-ch03-37a}
    \includegraphics[scale=0.8,angle=0]{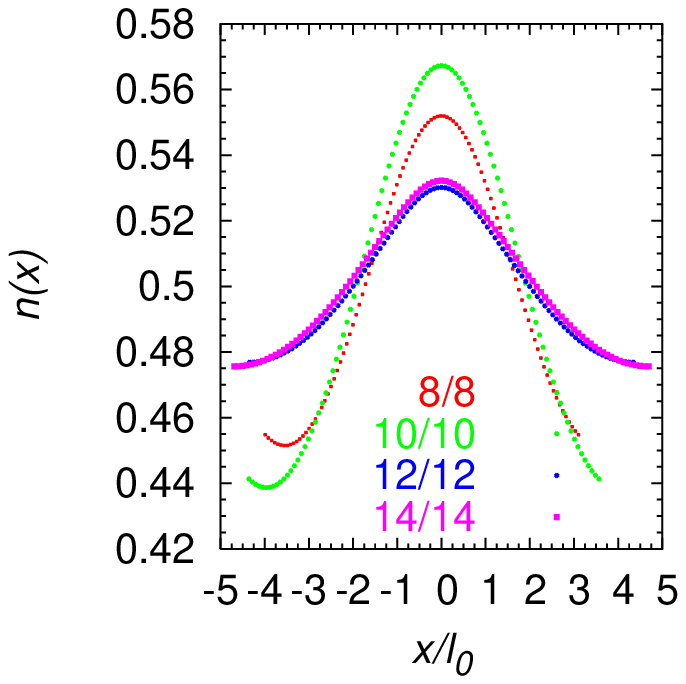}}
  \subfigure[Ising QHF ($\nu=2$, crossing of $n=0,\up$ 
             and $n=1,\up$ levels.]{\label{fig-ch03-37b}
    \includegraphics[scale=0.8,angle=0]{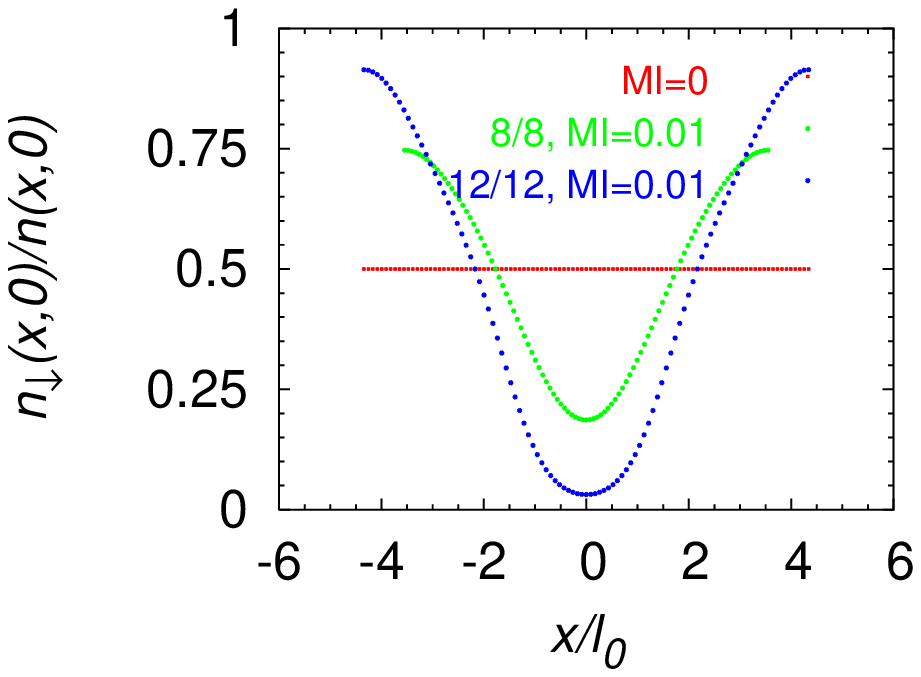}}
\caption{Different quantum Hall ferromagnets (QHF), half-polarized 
  states: polarization 
  response to a magnetic $\delta$-line 
  impurity in systems of 8 and 12 particles.  }\label{fig-ch03-37}
\end{figure}

The two systems were subjected to a $\delta$-line magnetic
inhomogeneity, just as the incompressible singlet ground state in
Subsec. \ref{pos-ch03-05}. However, QHF systems and incompressible
liquid states at $\nu=\ot$ or $\tt$ behave quite differently. 
Looking at a QHF and comparing the response in systems of different sizes,
we observe no intrinsic length scale, Fig. \ref{fig-ch03-37}. 
Rather, the form of the response
reflects the size of the system (cf. the left panel of
Fig. \ref{fig-ch03-42}c). This statement applies both to the Ising,
Fig. \ref{fig-ch03-37}b, and the Heisenberg QHF,
Fig. \ref{fig-ch03-37}a, where we show the polarization of the energetically
lowest state in a system subject to the inhomogeneity.

It is also interesting to study the {\em density} of the disturbed
QHF states. The density of the Heisenberg QHF remains
almost unchanged (it is constant) unlike the density of the Ising QHF state,
Fig. \ref{fig-ch03-45}b right. This is understandable. Whereas in the
Heisenberg QHF spin up
and spin down one-particle states have exactly the same
wavefunction, both spin up and spin down
  states are from the lowest Landau level,
this is not the case for the
Ising QHF. In that case, spin up and spin down states come from
different Landau levels. Thus, even when the magnetic impurity shuffles
the spin up and spin down particles somehow in the Heisenberg QHF, the
density does not change. 

Finally, we comment on densities in the inhomogeneous states in the
Ising QHF.
Results shown in Fig. \ref{fig-ch03-45}b belong to quite small
systems (12 particles at most). In the largest system studied, we
observe a maximum in the density direct at the position of the
impurity ($x=0$) and the maximum approaches the value of density in a
homogeneous system. With some imagination this allows for a
hypothesis that -- if domains are formed in an infinite system --
the density will be inhomogeneous close to the domain boundary while remaining
homogeneous inside a domain. However, we would have to study larger
systems to confirm this speculation.

\subsubsection{The half-polarized states}
\label{pos-ch03-04}

The inner structure of the half-polarized ($S=N_e/4$) ground state at
filling $\tt$ is investigated in this Subsection. We will
argue that this state (assuming short-range interaction) resembles
rather the incompressible singlet and polarized ground states at
$\nu=\tt$ than the Ising quantum Hall ferromagnet in the $S_z=0$ sector as
described in Subsection \ref{pos-ch03-06}.

In this Subsection, by half-polarized ground states we mean  the 8-
and 12-electron $S=N_e/4$ states GS$_{8}$ and GS$_{12}$ as
introduced in Sec. \ref{pos-ch03-07}, cf. correlation functions in
Fig. \ref{fig-ch03-25}. 

\begin{figure}
  \subfigure[Density.]{\label{fig-ch03-47a}
    \includegraphics[scale=0.8,angle=0]{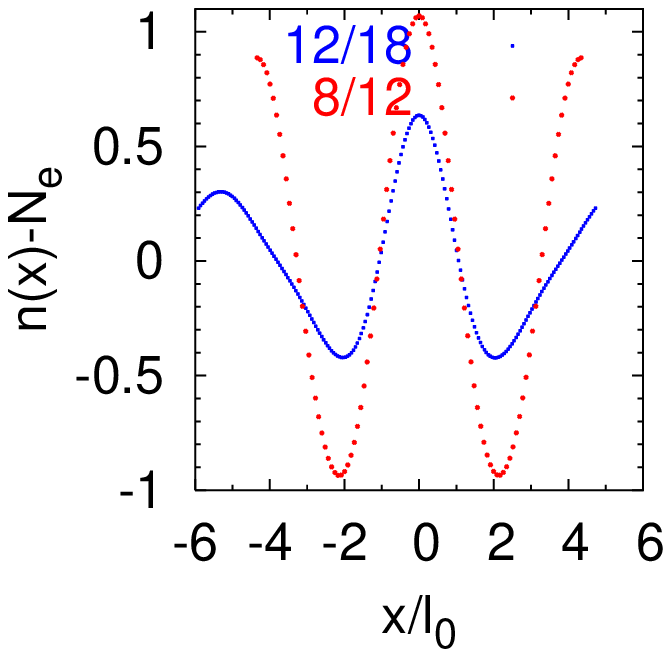}}
    \subfigure[Polarization.]{\label{fig-ch03-47b}
    \includegraphics[scale=0.8,angle=0]{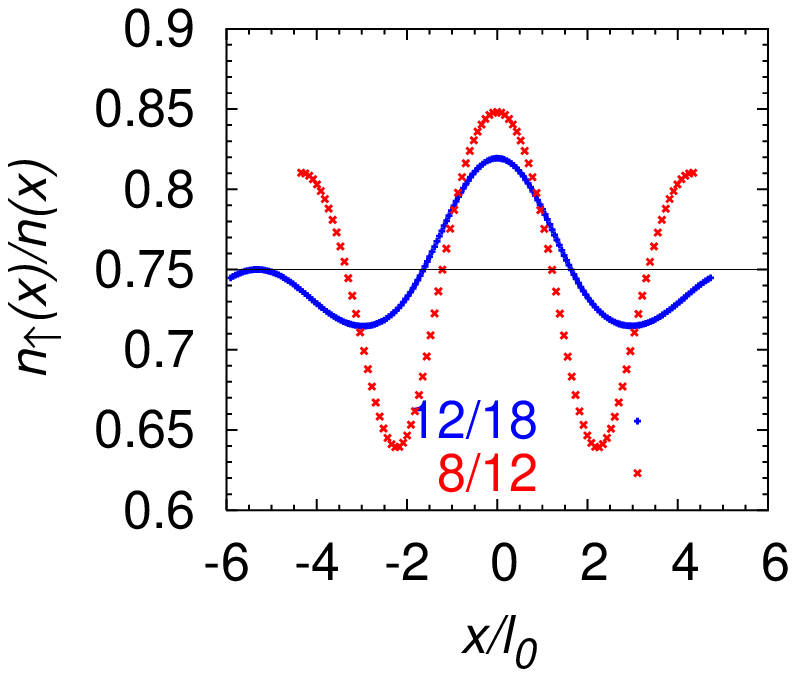}}
\caption{Half-polarized ground state ($S=N_e/4$) 
responding to a $\delta$-line magnetic impurity, $H_{MI,\up}$
(\ref{eq-ch03-07}). }\label{fig-ch03-47}
\end{figure}

The ground state in a homogeneous system has a nearly constant density
(oscillations due to the center-of-mass part wavefunction are less
than $0.1\%$ in the 12-electron system). This changes when a weak 
$\delta$-line magnetic impurity along $y$ is applied. Not only the 
polarization
but also the {\em density} becomes inhomogeneous,
Fig. \ref{fig-ch03-47}. The first
minima of $n(x)$ are at the same position $r_1\approx 2.2\ell_0$ in
the two system sizes considered and decaying oscillations are
likely to follow at larger distances. Comparing the two
system sizes in Fig. \ref{fig-ch03-47}a, we find a much weaker response
in the larger system, but this still does not have to imply a vanishing
response in an infinite system, cf. discussion of the singlet state in
Subsec. \ref{pos-ch03-16}.

Unlike the Ising quantum Hall ferromagnet discussed in
Subsect. \ref{pos-ch03-06}, the half-polarized states seem to have 
an intrinsic length scale in $n(x)$ (of the order of $r_1$),
Fig. \ref{fig-ch03-47}a. 
It is remarkable that this $r_1$ matches
quite well the position of the first maximum in the density of the Laughlin
state ($\ot$) responding to an impurity, Fig. \ref{fig-ch03-34}.

Contrary to the density, the {\em polarization} does not show an intrinsic
length scale as positions of the first minima in $N_e=8$ and $N_e=12$
systems mismatch considerably, Fig. \ref{fig-ch03-47}b. However, the
polarization response here differs from the behaviour of the singlet
state, Fig. \ref{fig-ch03-35}b. Rather, Fig. \ref{fig-ch03-47}b
suggests that $n_\dn(x)/n(x)\to 0.75$ as we go away from the impurity
for the half-polarized states. This  behaviour was classified
as 'incompressible' in Fig. \ref{fig-ch03-42}a.

These observations suggest that the presence of an
impurity will not lead to a splitting of the state into two domains (one
with spin up, second with spin down), which we could expect for Ising
QHF, Fig. \ref{fig-ch03-37}. It seems that an impurity will rather
change the polarization of the system only locally, in an 'incompressible
manner', Fig. \ref{fig-ch03-42}a. The density response has the
same characteristic length scale as the singlet and polarized
$\nu=\tt$ ground states and such a length scale is absent in the
polarization in agreement with behaviour of the singlet state,
Fig. \ref{fig-ch03-35}b.

\begin{figure}
  \subfigure[8 electrons, density.]{\label{fig-ch03-48a}
    \includegraphics[scale=0.8,angle=0]{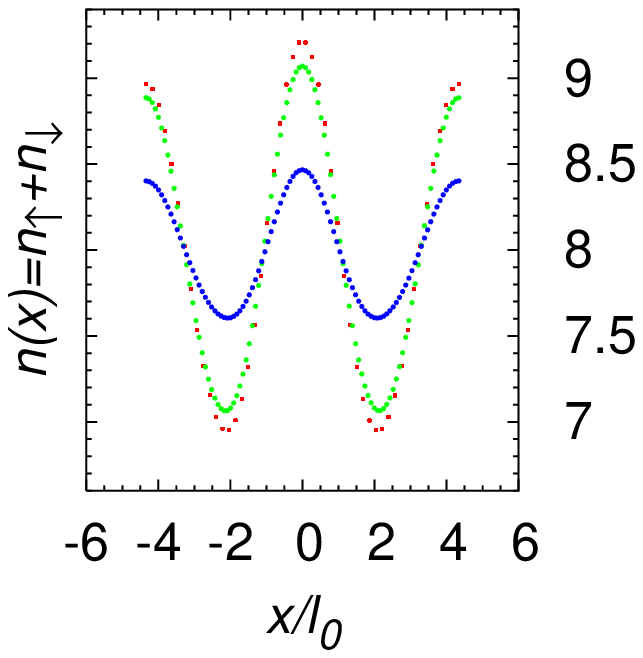}}
  \subfigure[12 electrons, density. ]{\label{fig-ch03-48b}
    \includegraphics[scale=0.8,angle=0]{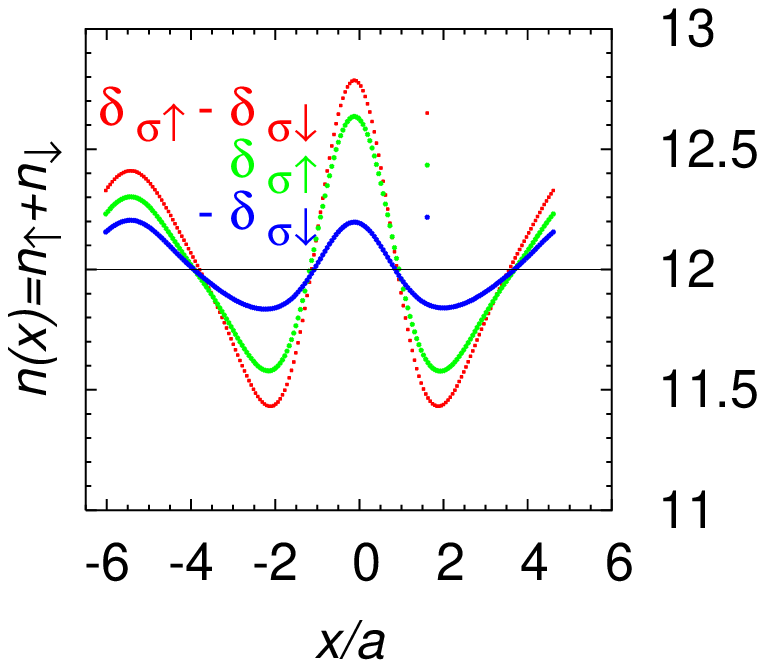}}
\caption{Half-polarized ground state ($S=N_e/4$) 
responding to a $\delta$-line magnetic impurity. Different impurity
types are considered: $H_{MI}$, $H_{MI,\up}$ and $H_{MI,\dn}$ 
(\ref{eq-ch03-07}).  }\label{fig-ch03-48}
\end{figure}

A state with $S_z=N_e/4$ comprises of $\frac{1}{4}N_e$ 
electrons with spin down ('minority spins') and $\frac{3}{4}N_e$
electrons with spin up ('majority spins'). Since the two populations
are not balanced,
we may gain extra information by speaking to them
separately. The simplest concept, assuming non-interacting electrons, 
would be that $H_{MI,\dn}$, $H_{MI,\up}$ and $H_{MI}$ (\ref{eq-ch03-07}) 
give rise to responses in ratio
$\frac{1}{4}:\frac{3}{4}:1$. {\em Very roughly}, this is indeed the
case. Heights of the central peak ($x=0$) for
these three types of inhomogeneities are
indeed approximately in this ratio, both for the density and for the
polarization, Fig. \ref{fig-ch03-48}.
In the following we will discuss investigations with spin-dependent
perturbations in more detail.

\begin{figure}
  \subfigure[8 electrons, repulsive.]{\label{fig-ch03-49a}
    \includegraphics[scale=0.5,angle=0]{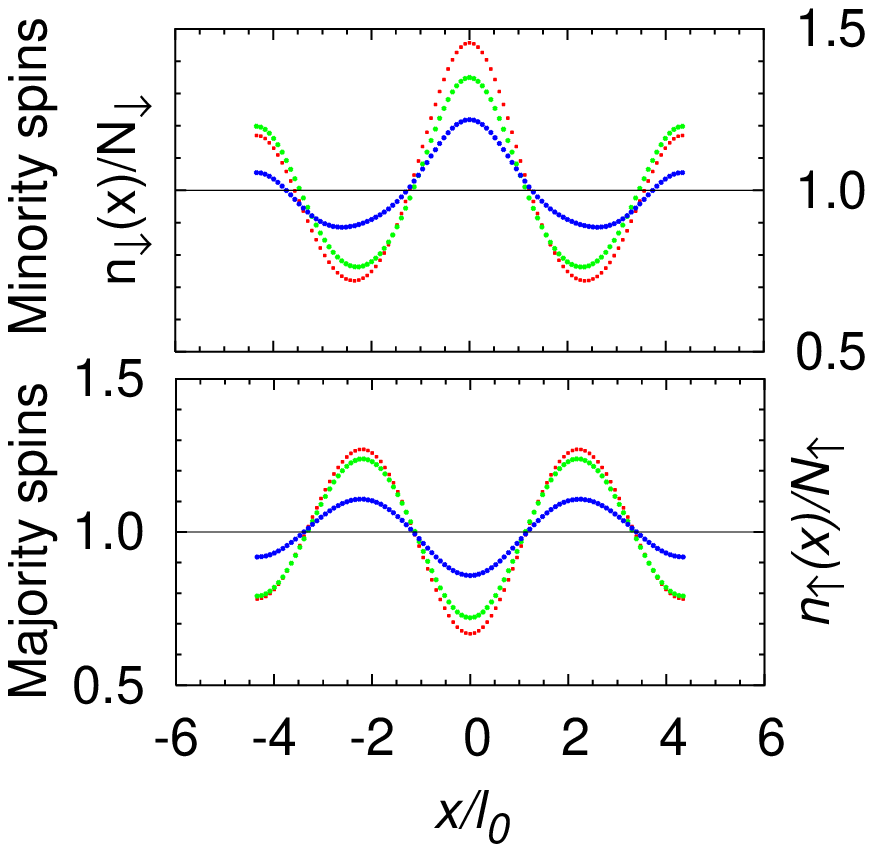}\hskip-1.5cm}
  \subfigure[8 electrons, attractive.]{\label{fig-ch03-49b}
    \includegraphics[scale=0.5,angle=0]{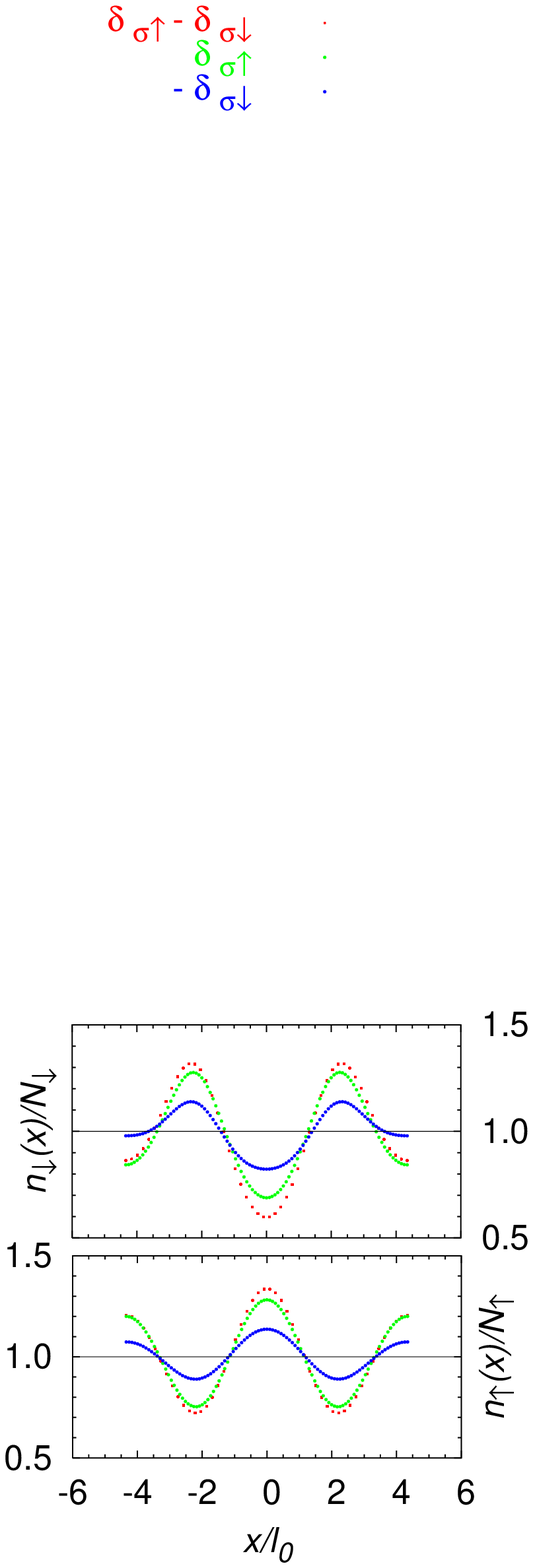}\hskip-1.5cm}
  \subfigure[12 electrons, attractive.]{\label{fig-ch03-49c}
    \includegraphics[scale=0.5,angle=0]{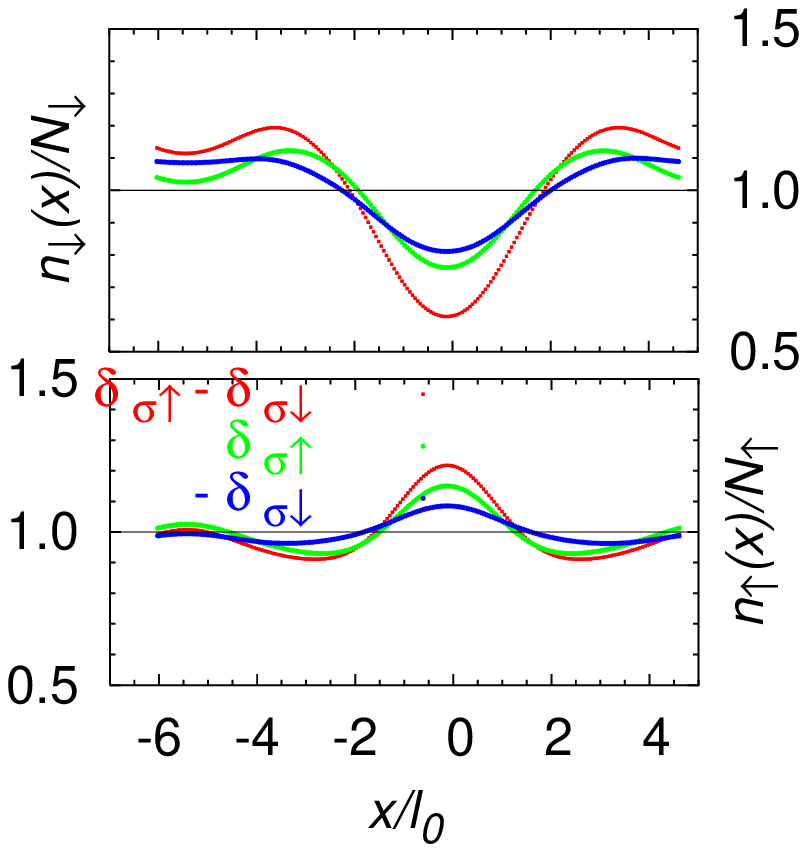}\hskip-1.5cm}
\caption{The same as Fig. \ref{fig-ch03-48}, but the density is decomposed
    into the density of majority spins ($n_{\up}$) and minority
    spins ($n_{\dn}$). By attractive (repulsive) is meant that the
    $\delta$-impurity at $x=0$ is attractive (repulsive) for the
    majority spin.     }\label{fig-ch03-49}
\end{figure}

Let us separate the density of majority and minority spins,
Fig. \ref{fig-ch03-49}. We will argue that the half-polarized
state with $N_e$ electrons consists of two coexisting and weakly
interacting liquids: $N_e/2$ electrons in a fully polarized liquid
(with $S_z^p=N_e/4$) and $N_e/2$ electrons in a $S_z^u=0$
state. Minority spins are thus present only in the $S_z^u=0$ liquid
whereas majority spins occur in both of them. Concentrate on
Fig. \ref{fig-ch03-49}c. 

(i)  Minority spins ($\dn$) react almost equally to
  $H_{MI,\up}$ and $-H_{MI,\dn}$. They reflect only changes in the
  $S_z^u=0$ liquid and there are as many up as down spins in
  it. In fact, the $H_{MI,\up}$ impurity influences also the
    polarized liquid component, but we cannot see it in the density of
    minority spins provided the two liquids do not interact appreciably.
  The combined effect of $H_{MI,\up}-H_{MI,\dn}$ causes a response of about
  the sum of these two.

(ii) Majority spins ($\up$) react differently to $H_{MI,\up}$ and
  $-H_{MI,\dn}$. We should keep in mind that $n_\up$ reflects changes in
  both (polarized and $S_z^u=0$) liquids.  The latter impurity
  inflicts changes only on the $S_z^u=0$ part, whereas the former
  impurity acts on both liquids. If both liquids would have
  the same sensitivity to the considered impurities, we could expect
  responses in ratio $4:3:1$ ($H_{MI}$ to $H_{MI,\up}$ to
  $H_{MI,\dn}$). The fact that responses observed in
  Fig. \ref{fig-ch03-49}c (measured by the height of the central
  maximum) are in ratio $3:2:1$ could be an indication that the
  polarized liquid is less sensitive than the $S_z^u=0$ liquid.   

(iii) Note also, that responses are the same (up to an inversion)
  for attractive and repulsive impurities, Fig. \ref{fig-ch03-49}a and
  \ref{fig-ch03-49}b, provided the impurities are weak.

Studies of the eight electron half-polarized state,
Fig. \ref{fig-ch03-49}b is not in conflict with this interpretation,
responses in densities are quantitatively different though. 
However, we should be cautious in drawing strong conclusions as 
these systems with primitive cell of size
$12$ flux quanta correspond to the smallest system ($N_e=4$) considered in
Fig. \ref{fig-ch03-34} ($\nu=\ot$ state plus an impurity) and in that
case finite size effects are already very strongly pronounced. Thus,
the twelve electron system can be considered as the smallest system
with finite size effects {\em not} playing a major role.

\subsubsubsection{Conclusion}

The hypothesis of the coexistence of the spin singlet and polarized
liquids in the half-polarized states (HPS) seems to be supported. We have
pointed out some similarities between the HPS and the former two
incompressible states. In contrast, response to magnetic impurities
seems to be different for the HPS and the Ising quantum Hall
ferromagnet (in the $S_z=0$ sector) which would be the direct
counterpart of the HPS if composite fermions are substituted by
electrons. 

In general, it is not very surprising that electronic systems ($\nu=2$
Ising QHF) differ strongly from the CF-counterparts. We have already
seen this in correlation functions in
Subsec. \ref{pos-ch03-15}. However, the observed differences seem to be
too deep to allow us to establish a relation between QHF states and
the half-polarized states introduced in Sec. \ref{pos-ch03-07}.

\subsection{Deforming the elementary cell}


\label{pos-ch03-08}

In this Section we discuss another way of how to investigate fractional
quantum Hall states. We will exactly diagonalize $\nu=\ot$ and
$\nu=\tt$ systems  in elongated rectangular elementary cells. The dimensions 
are $a$ by $b$, the aspect ratio is thus $a:b>1$. 
The area of the rectangle is always kept constant, $ab=2\pi\ell_0^2 N_m$
(\ref{eq-ch02-38}), and therefore
\begin{equation}\label{eq-ch03-16}
   ab=2\pi\ell_0^2 N_m\,, \qquad \Rightarrow 
   a = \ell_0\sqrt{2\pi N_m\lambda}\,,\
   b = \ell_0\sqrt{2\pi N_m/\lambda}\,,\
   \lambda = a:b\,.
\end{equation}

What can we expect? In the first approximation, 
we would say (i) nothing happens 
for an isotropic state such as the $\nu=\ot$ Laughlin liquid and (ii)
crystalline or wave-like states will change both in energy and in
density. The reason is, that structures in homogeneous liquid states
(as for example in correlation functions in
Subsec. \ref{pos-ch03-15}) are intrinsic and not incurred by the
finite system size. Consequently, we expect the liquid state to 
change neither their energy
nor their correlation functions, at least not on short distances, if $a:b$ 
is slightly varied.  On the other hand, an integer multiple of the
period of a wave-like or crystalline state must be necessarily equal
to $a$ and/or $b$, hence by varying the aspect ratio we force it to change
its period. In a classical crystal this means compression, or better
deformation, since total 'volume' $ab$ remains constant, and we expect
it to cost energy. 

This investigation of $\nu=\tt$ systems was partly motivated by the
work of Rezayi {\em et al.} \cite{rezayi:05:2003} who investigated one
particular type integer quantum Hall ferromagnet.
Their exact diagonalization on a torus showed an
$N_m$-fold  nearly degenerate ground state and the authors argued
that these states comprised of stripes of alternating spin polarization
(Subsec. \ref{pos-ch03-06}) oriented parallel to one side of
the rectangle, for example $a$. 
As they varied the aspect ratio, the states still remained 
degenerate, and their energy $E(\lambda)$ changed proportional to
$b$. In fact, the degeneracy even improved: the small
energy differences between the $N_m$ states dropped.
This was a strong argument for the stripe order, since then 
$\d E(\lambda)/\d b$ can be interpreted as energy per unit
length of an interface between a spin up and spin down stripe.

\subsubsection{Incompressible ground states}
\label{pos-ch03-09}

As usual, we will start with $\nu=\ot$, being probably the best
understood system. This will also be the only case where we will
discuss Coulomb interacting systems, in the rest we will stay with
short-range interacting systems.

\begin{figure}
  \subfigure[Coulomb interaction ($\ot$).]{\label{fig-ch03-50a}
    \unitlength=1mm
    \includegraphics[scale=.4,angle=0]{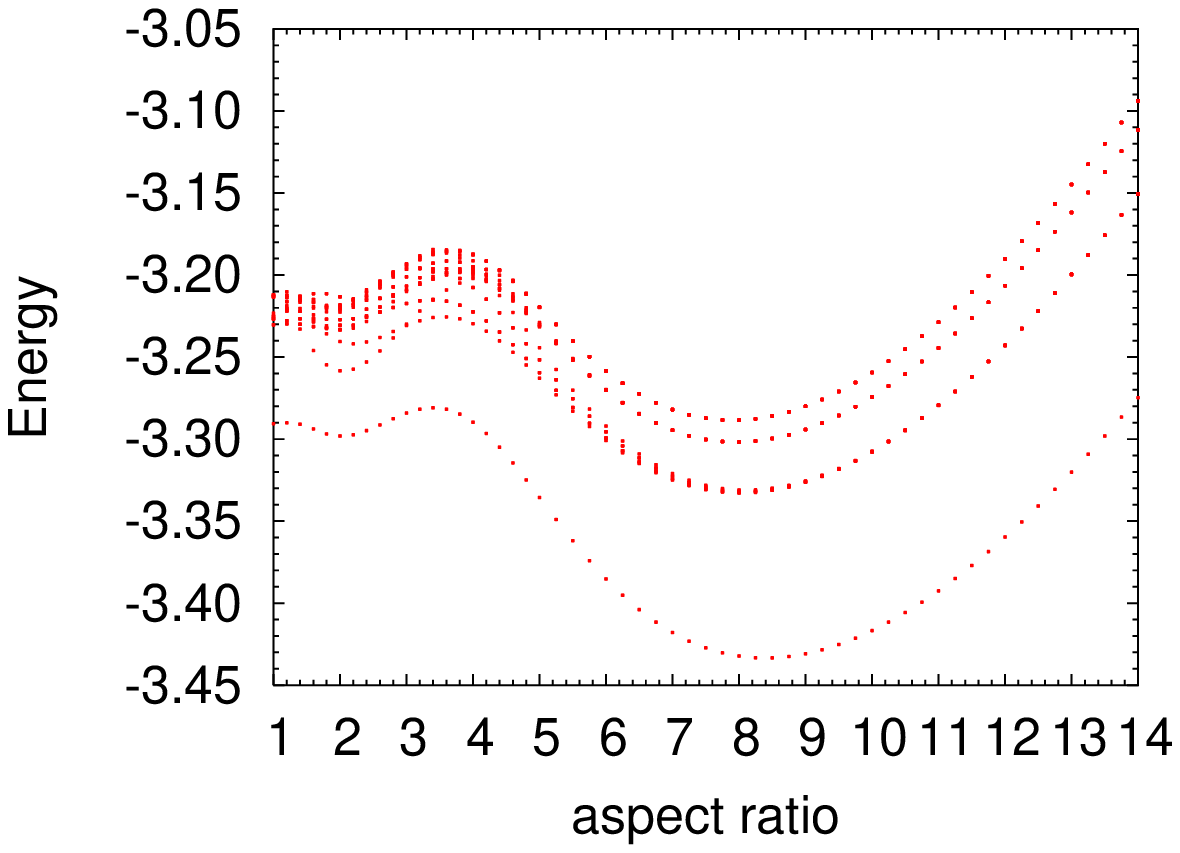}
    \put(-40,16){\small $\nearrow$}\put(-35,17){\small$\nwarrow$}}
  \subfigure[Short-range interaction ($\tt$). Blue crosses mark the
    energy of a full Landau level. See the comment \cite{comm:ch02-05}
    for details.]{\label{fig-ch03-50b}
    \includegraphics[scale=.4,angle=0]{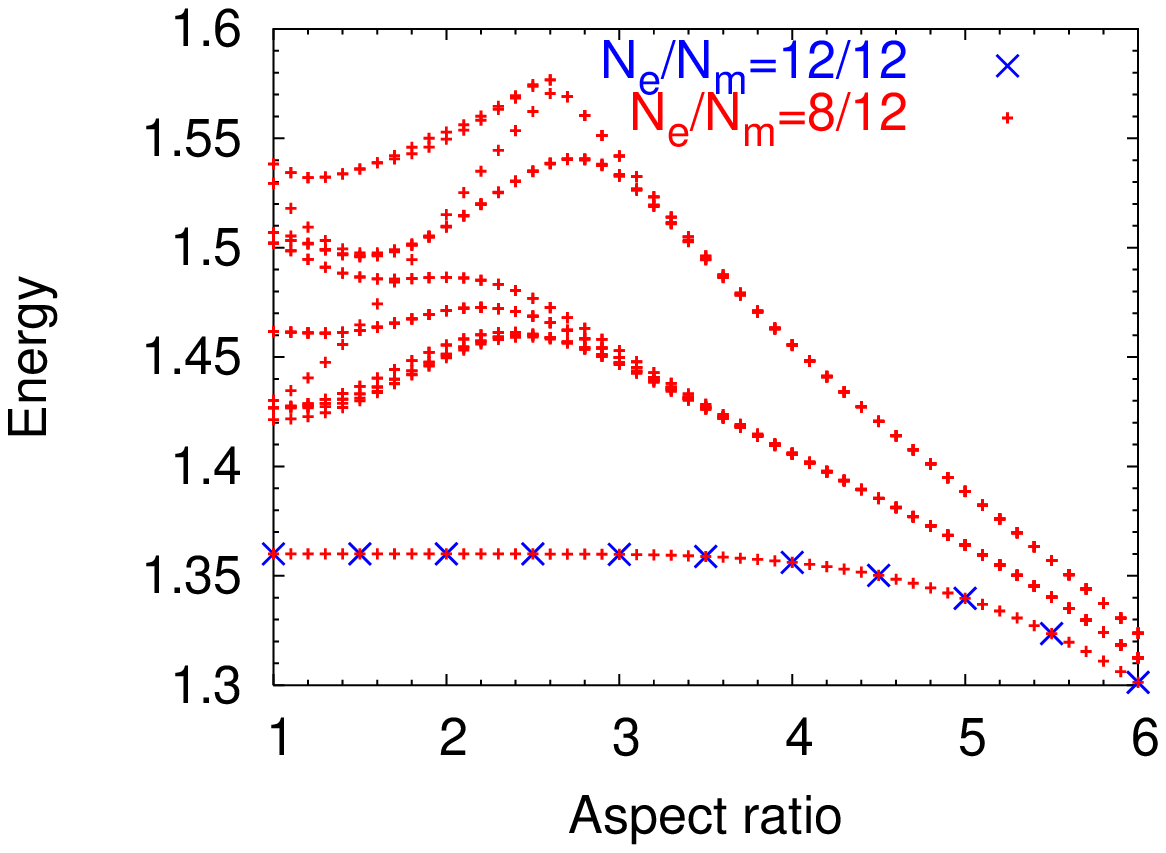}}
  \subfigure[Polarized, singlet and half-polarized states 
  at $\nu=\tt$. Overview.]{\label{fig-ch03-50c}
    \includegraphics[scale=.5,angle=0]{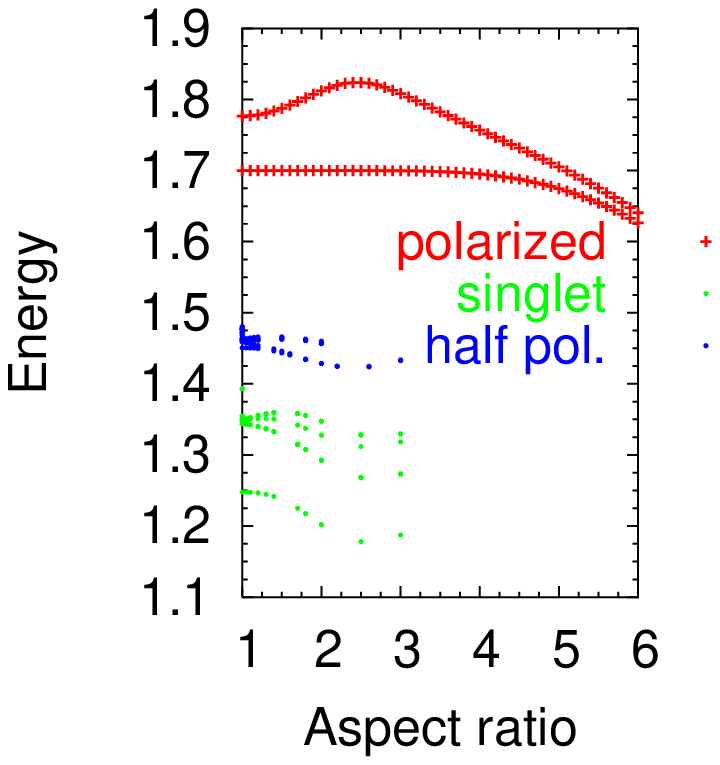}\hskip-0cm}
\caption{Spin polarized eight electrons at filling factor $\ot$ and
    $\tt$. Energy of the lowest states versus aspect ratio of the
    primitive cell.
    An overview of the polarized, singlet and half-polarized
    states in the same scale is presented in the last panel.}
\label{fig-ch03-50}
\end{figure}

\subsubsubsection{Coulomb versus short-range interaction: $\nu=\ot$}

\begin{figure}
  \subfigure[The 'square' crystal state (aspect ratio
    $2$).]{\label{fig-ch03-52a}
    \includegraphics[scale=.5,angle=0]{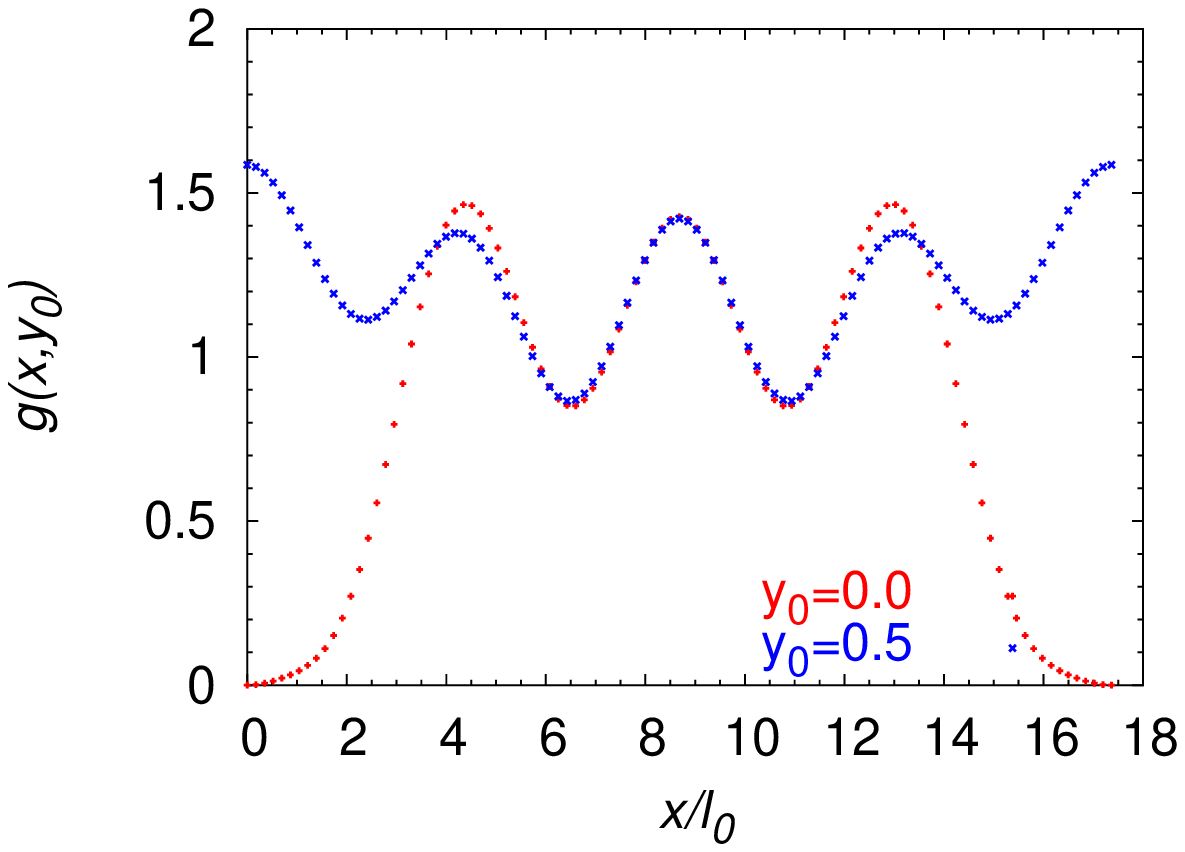}\quad
    \includegraphics[scale=.5,angle=0]{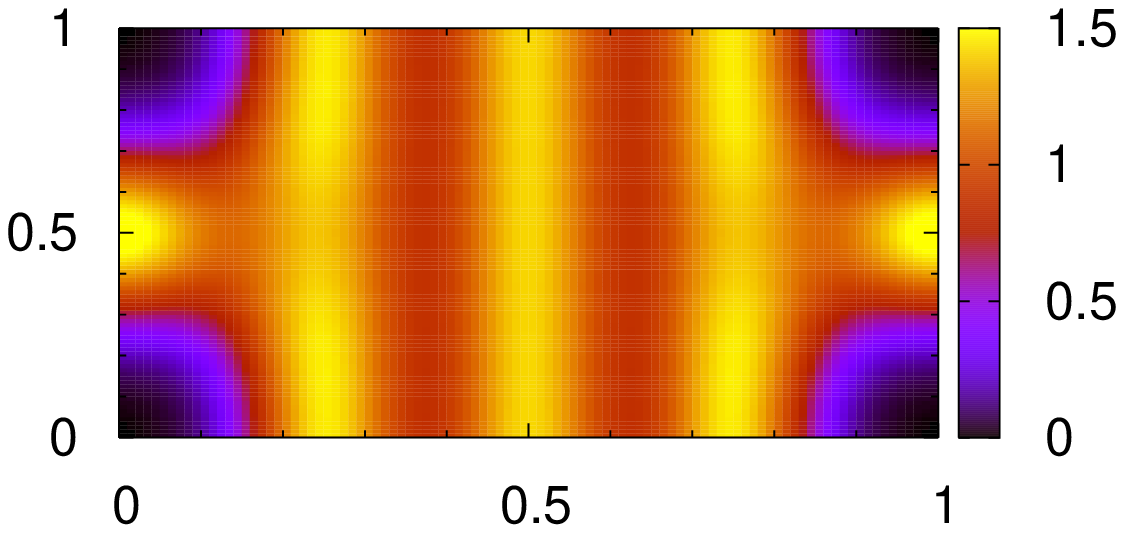}}
    \subfigure[The hexagonal crystal state, aspect ratio
    $4/\sqrt{3}\approx 2.31$. ]{\label{fig-ch03-52b}
    \includegraphics[scale=.5,angle=0]{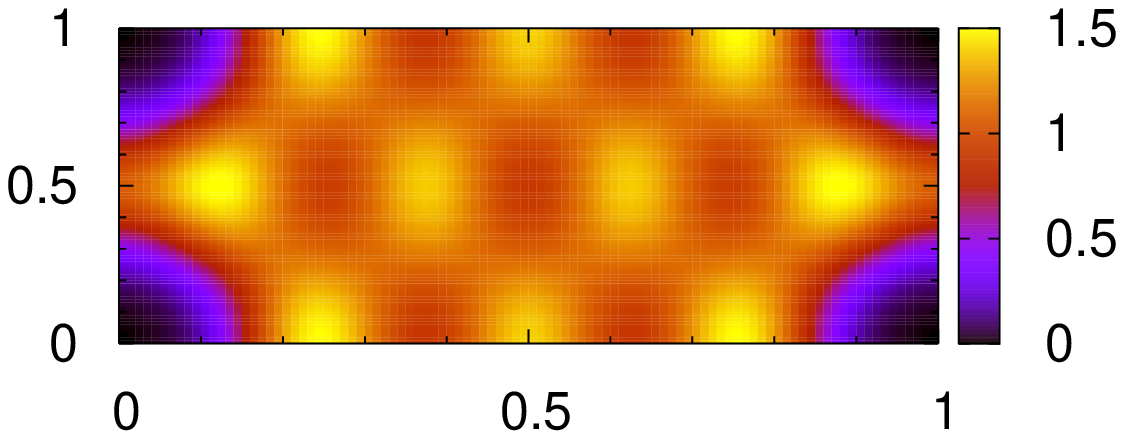}\quad
    \includegraphics[scale=.5,angle=0]{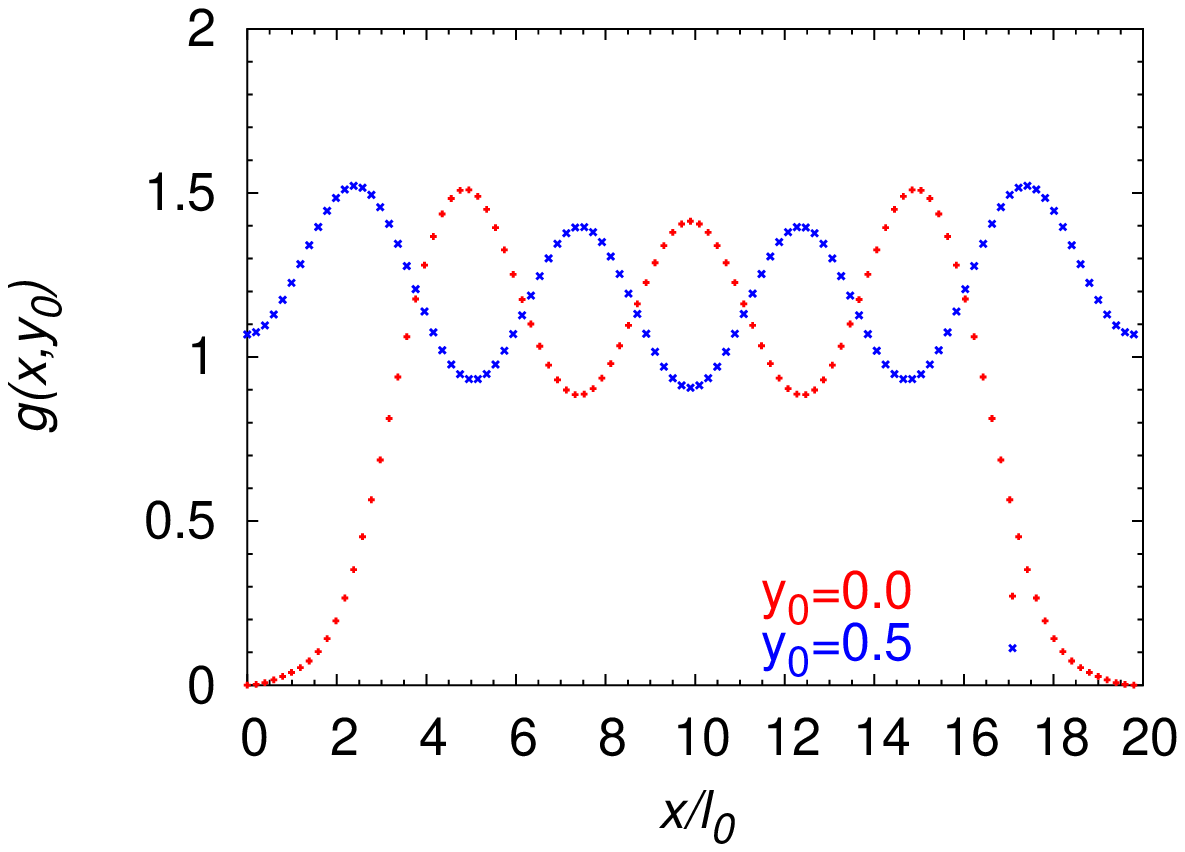}}
\caption{Charge density waves resembling 
    Wigner crystal states 
    are among the lowest excitations in a Coulomb 
    interacting $\ot$ system. Their energy is minimized (as a
    function of aspect ratio) when the elementary cell matches the
    crystal geometry. Correlation functions in eight-electron systems
    are shown, length of $x$- and $y$-sides corresponds to the
    particular aspect ratio.}
\label{fig-ch03-52} 
\end{figure}

The spectrum of a Coulomb-interacting system has a quite rich structure,
Fig. \ref{fig-ch03-50}a. The ground state energy exhibits several minima
as a function of the aspect ratio of the elementary cell. In fact even
more structure seems to appear in larger systems, as far as it could be
inferred from comparing $6$, $8$ and $10$ electron systems. In the
following, we
will show that this structure occurs mainly due to the long-range
part of the Coulomb potential, it should be possible to describe it
mainly by the Hartree part of the total energy or simply that it is due to
formation of charge density waves (CDW) resembling Wigner
crystals. Differences between Wigner crystals and CDWs are
  discussed below. In fact, energy of the states in question, 
  Fig. \ref{fig-ch03-52}, will contain strong exchange
  contributions. Nevertheless, these states are very similar to the
  {\em classical} states which minimize the Coulomb energy. 
In a second step, we will discuss how 
correlations (and energy due to correlations) depend on the aspect ratio,
Fig. \ref{fig-ch03-50}b.

In order to understand the the aspect-ratio dependence of energy 
of the Coulomb-interacting ground state, let us focus on two
low excited states marked by arrows in
Fig. \ref{fig-ch03-50}a. These two states are just the CDWs mentioned
above and they look almost like Wigner crystals: one
hexagonal, another square, as density-density correlation shows,
Fig. \ref{fig-ch03-52}. It is then not surprising that the energy of such
states is minimal, when the aspect ratio matches its geometry. For
eight electrons considered here, this happens for\footnote{$d$ is the
  'lattice constant'. } 
$4d:2d=2$ and $4d:(2\sqrt{3}/2 d)=4\sqrt{3}/3$ for the square and
hexagonal crystal, respectively. Perhaps 
the most apparent difference between a CDW and an
(unpinned) Wigner crystal is that for the latter state 
we expect the correlation function
to drop almost to zero between the 'lattice sites' and this is
not the case here, Fig. \ref{fig-ch03-52}. The reason is that 
at filling factor $\nu=\ot$, the system is too densely populated, or
mean interparticle distance is too small,
$r_{mean}/\ell_0=\sqrt{2\pi/\nu}\approx 4.35$
(\ref{eq-ch02-38}) to allow the electron density (or
correlation function) to vanish between two sites. Remember that an electron
  {\em within the lowest Landau level} cannot be localized more 
  strongly than on a
  length scale of the order of unity (magnetic length $\ell_0$).
Even if we assembled a hexagonal Wigner crystal at $\nu=\ot$, the
wavefunctions at neighbouring sites would strongly overlap and it is
then more favourable for the electrons to retain some features of the
Laughlin correlations. As a result we obtain a CDW (or a 'strongly
correlated crystal' \cite{lam:07:1984}) like the state in
Fig. \ref{fig-ch03-52}b. At lower filling factors, $r_{mean}/\ell_0$
is larger and Wigner crystal states become possible. This can be
interpreted as a quantum phase transition from liquid to solid as the
filling factor is decreased and the extensive studies in this field
suggest the critical value $\nu\approx \frac{1}{7}$, see Sec. 5.7 in
Chakraborty \cite{chakraborty:1995} for a review.

\begin{figure}
  \subfigure[Aspect ratio $1.00$.]{\label{fig-ch03-51a}
    \includegraphics[scale=.4,angle=0]{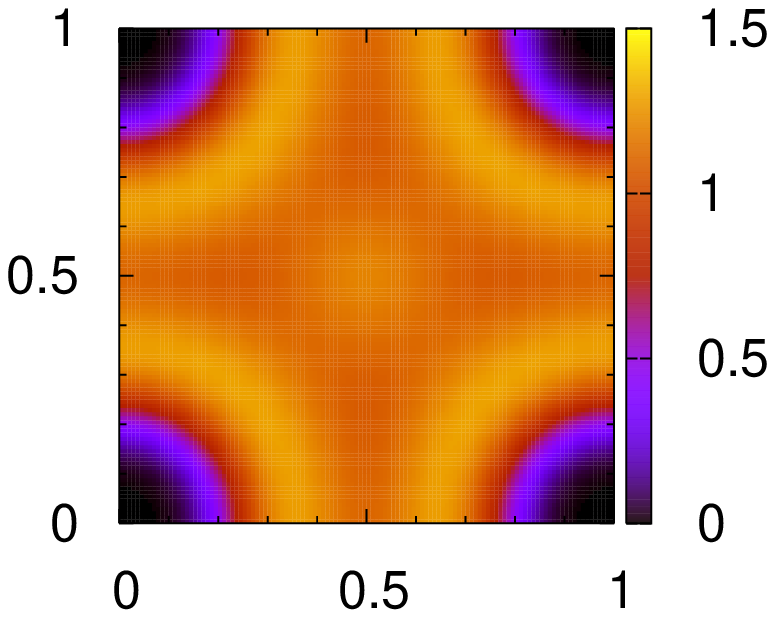}}
  \subfigure[Aspect ratio $2.00$.]{\label{fig-ch03-51b}
    \includegraphics[scale=.4,angle=0]{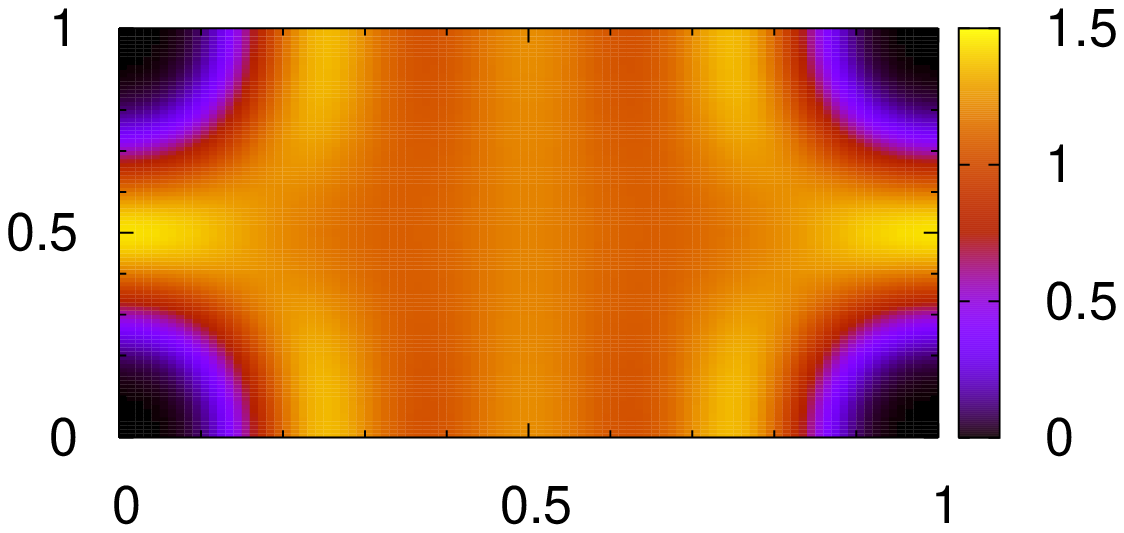}}
  \subfigure[Aspect ratio $2\sqrt{3}\approx 3.46$.]{\label{fig-ch03-51c}
    \includegraphics[scale=.4,angle=0]{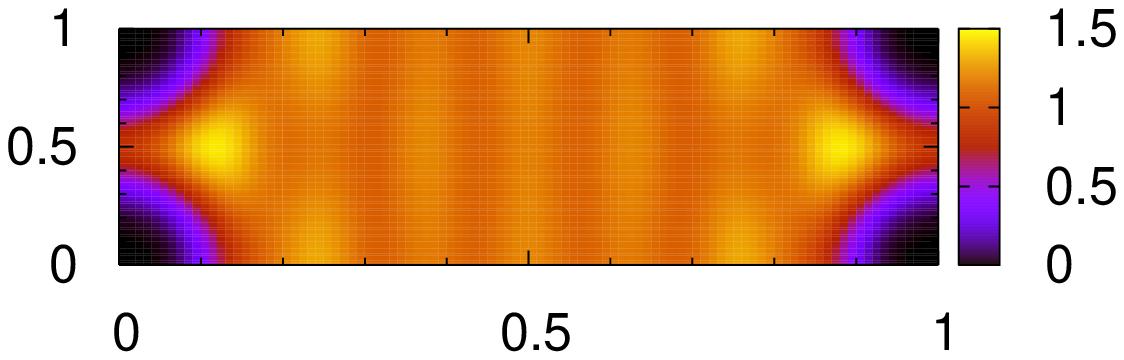}\hskip-1cm}
\caption{Evolution of the Laughlin state with aspect ratio of the
    elementary cell (Coulomb interaction). Correlation functions are shown.}
\label{fig-ch03-51}
\end{figure}

The ground state (GS) energy reflects these geometrical
conditions. This state
also minimizes its energy when the square crystal can easily be
formed, but at short distances it strictly preserves the
liquid-like correlations, Fig. \ref{fig-ch03-51}. It is isotropic
at short distances, in Fig. \ref{fig-ch03-51}b, 
the ring corresponding to the first maximum is
circular and not deformed into an oval for instance,
Fig. \ref{fig-ch03-52}, and also $g(r)\propto r^6$ (not obvious in
Fig. \ref{fig-ch03-51}). It seems plausible that the increase of GS
energy around $a:b\approx 3$, Fig. \ref{fig-ch03-50}a, is due to the loss
of isotropy at shorter distances. The ring of the first maximum in
$g(r)$ disappears, Fig. \ref{fig-ch03-51}c, the feature 
$g(r)\propto r^6$ however remains. It is important
to know, that unlike the energy, the {\em structure}
(correlation functions) of the ground state is quite insensitive to the type of
interaction (Coulomb or short-range).

Let us now proceed to short-range interacting states.  The
crystalline states disappear from the realm of low-energy excitations.  
The ground state energy is completely independent on aspect ratio, it is
zero, Fig. \ref{fig-ch03-50}b. Energies of fully polarized
$\tt$ states displayed therein are equal to those of $\ot$-systems up to a
constant shift. This constant depends on aspect ratio, but the
  dependence is imperceptible up to $a:b\approx 4$ (for 4 electron system).
In fact, the ground state is rigid in the following sense: a state
can have zero energy only if there are three zeroes on the position of
each electron in the wavefunction. The wavefunction is completely
determined by this condition together with the confinement
to the lowest Landau level. It is even surprising, 
that given how strictly determined the ground state is, it once resembles a
liquid (for $a:b\approx 1$) and another time a CDW state (larger aspect 
ratios), Fig. \ref{fig-ch03-51}c.

Assuming fully spin polarized electrons, $\tt$ and $\ot$ systems
(e.g. $8/12$ and $4/12$) are particle-hole conjugated. Thus, spectra
of these systems are identical up to a constant energy shift, which is
just the Coulomb (or short-range interaction) 
energy of a completely filled lowest Landau
level. Note that this energy {\em varies} with aspect ratio (both
   for Coulomb and for short-range interaction). The common statement
   that interaction energy of a full LL is a constant is valid in a broad
   range of aspect ratios, but not everywhere. In
   Fig. \ref{fig-ch03-50}b, this holds up to $a:b<4$,
Subsect. \ref{pos-ch02-03}. This is shown in
Fig. \ref{fig-ch03-50}b. The $\nu=\ot$ Laughlin state has zero
energy for any aspect ratio (not shown), the $\tt$ ground state energy
is then just the Hartree-Fock energy of a completely filled
Landau level. Beyond $a:b\approx 4$, this energy is no longer
constant, indicating that the deformation of the elementary cell
becomes pathologic and beyond this point (at latest), the model 
no longer describes a 2D system but rather an effective 1D system. 

Consider $a/b\gg 1$. Then the $N_e$ electrons are located
on a very thin cylinder \cite{rezayi:12:1994}
of length $\propto \sqrt{a/b}$ (area of the
cylinder is fixed by filling factor, $ab=2\pi N_m$) and 
single electron states resemble 'rings on a pole'. The mean distance
between electrons is then $\propto \sqrt{a/b}/N_e$ and 
Coulomb energy is then proportional to
$(a/b)^{-1/2}$. The increase of the ground state energy for very large
aspect ratios, Fig. \ref{fig-ch03-50}a, is due to the repulsion
between an electron and its own periodic image in the 'short-direction'.

Excited states are sensitive to deformation of the
elementary cell even more. Energy levels group into branches beyond $a/b\approx
2$, Fig. \ref{fig-ch03-50}b. Keeping in mind the transition towards an
effective 1D model, these branches can be attributed to 
0, 1, 2, etc. pairs of 'rings on a pole' sitting at
neighbouring sites. In illustrative terms, there is no longer enough
room for two electrons to be positioned in 'vertical' direction (along
$y$ axis, i.e. the shorter side of the elementary cell) except when they
freeze into a crystal. 

In conclusion, going beyond aspect ratio $\sim 2$ (in a $N_m=12$
system) the system cannot be taken as a faithful model for an isotropic
infinite system.

\subsubsubsection{The singlet state}

\begin{figure}
  \subfigure[Eight electrons.]{\label{fig-ch03-39a}
    \includegraphics[scale=0.5,angle=0]{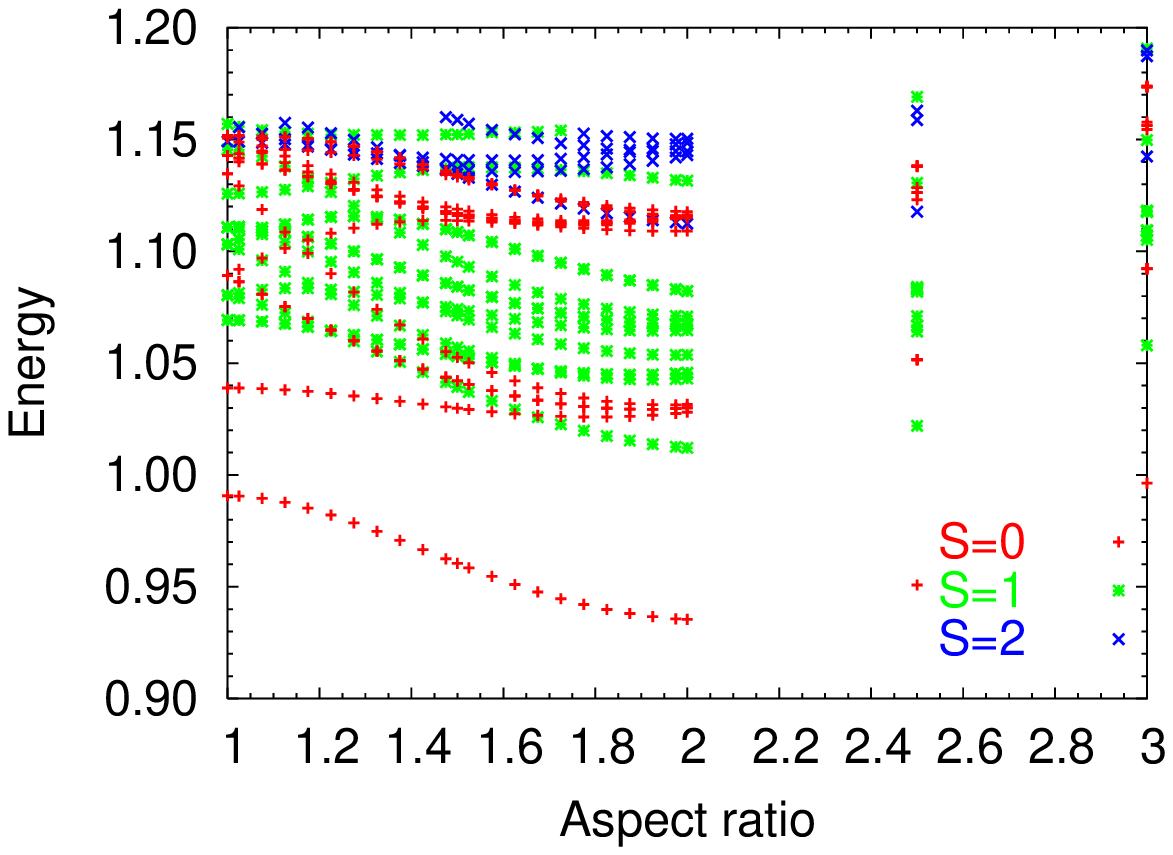}}
  \subfigure[Ten electrons.]{\label{fig-ch03-39b}
    \includegraphics[scale=0.5,angle=0]{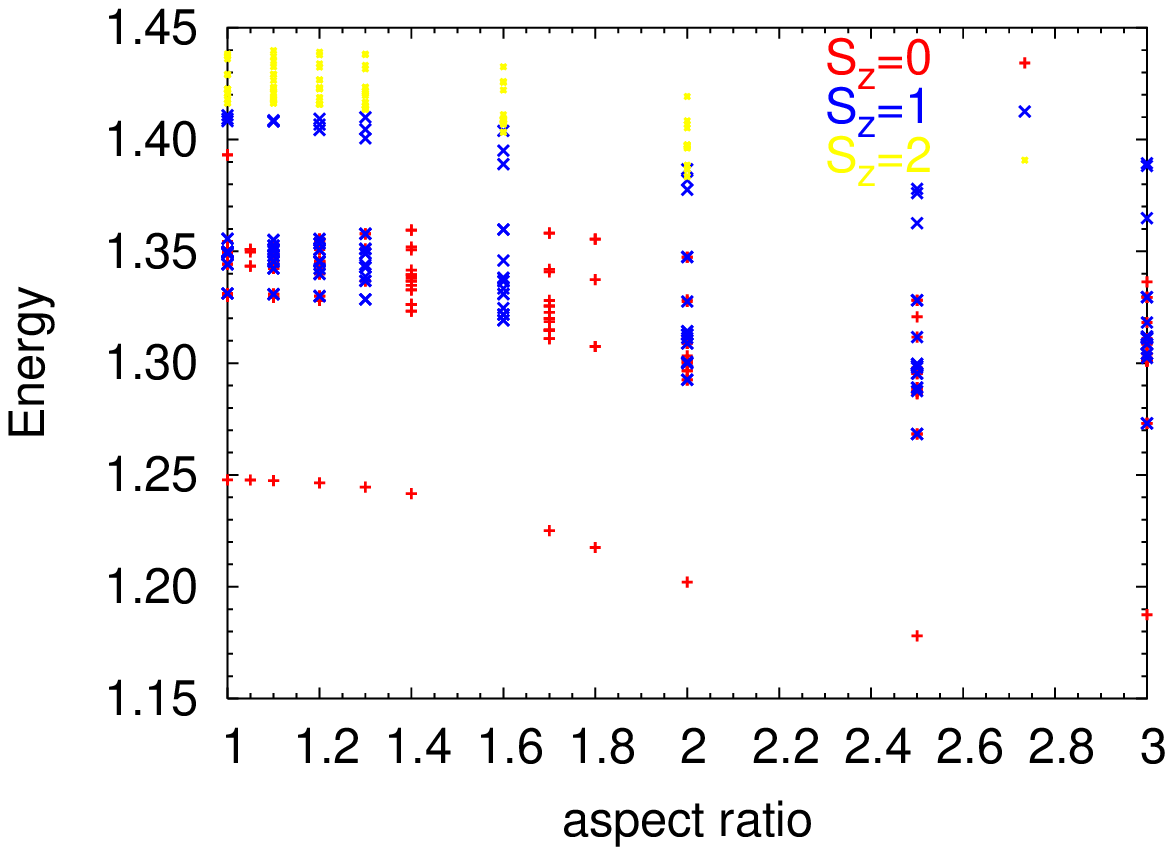}}
\caption{Low lying states at filling $\tt$ under vanishing Zeeman
    splitting versus aspect ratio of the primitive cell. Note that
    energy of the
    singlet ground state remains about constant for aspect ratios
    $\lesssim 1.4$ in the larger system.}\label{fig-ch03-39}
\end{figure}

The singlet ground state is apparently more sensitive to varying the
aspect ratio. Its energy changes at much smaller deformation than that of the
polarized state, Fig. \ref{fig-ch03-50}c. However, comparison between
systems of different sizes shows that its energy is also constant,
provided that the aspect ratio is not much larger than one and the system is
large enough, Fig. \ref{fig-ch03-39}. This is another hint at isotropy
of the state. A crystalline state responds more strongly to a change of
$a/b$, since this is in principle an attempt to compress the lattice
in one direction while expanding it in the other direction. Recall
just the CDW states in $\nu=\ot$ systems marked by arrows in 
Fig. \ref{fig-ch03-50}a.

\begin{figure}
    \includegraphics[scale=.5,angle=0]{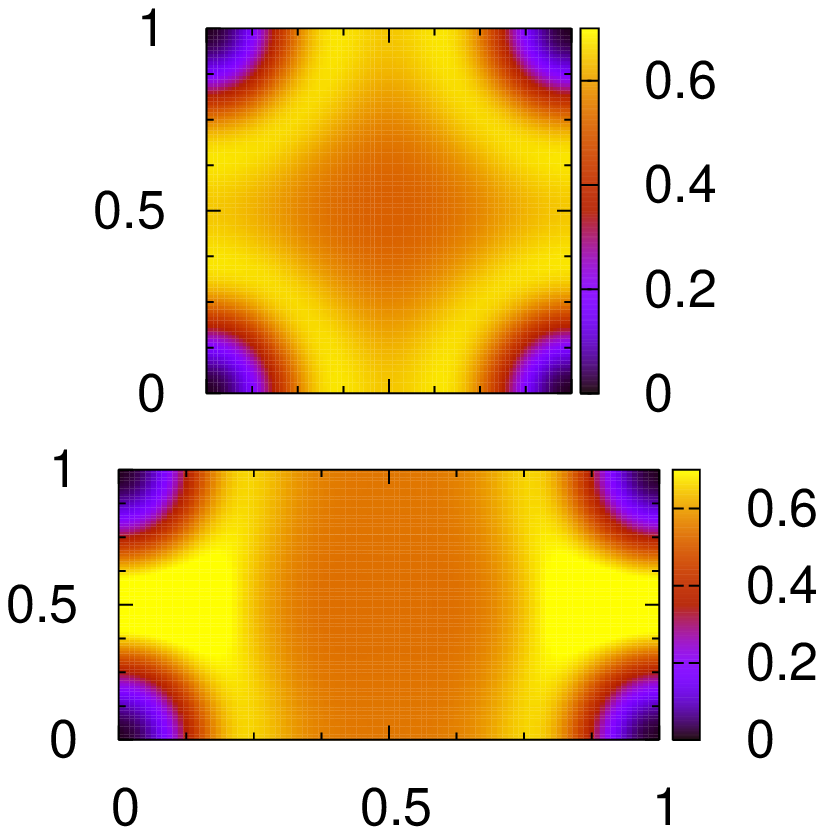}\quad
    \includegraphics[scale=.5,angle=0]{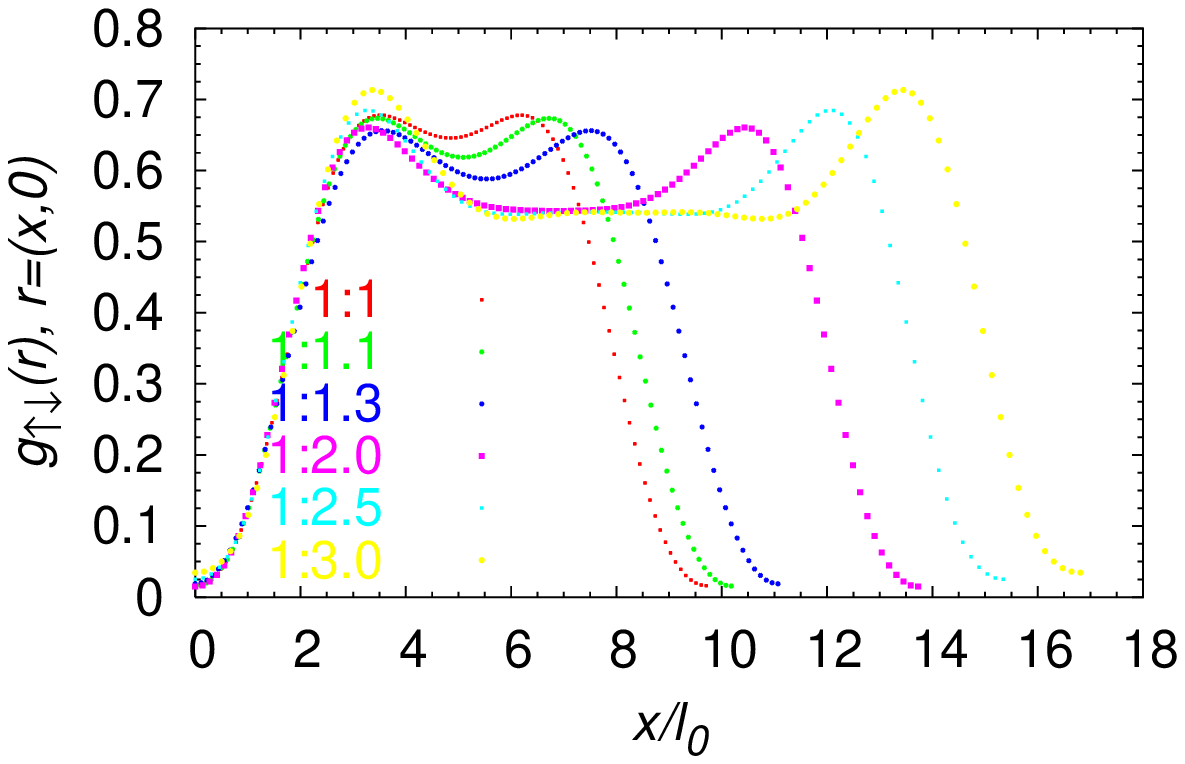}
\caption{The singlet $\tt$ state in elementary cells of different
    aspect ratios: density-density correlation between 
    unlike spins (the two aspect ratios shown in 2D plots are $a:b=1$ and
    $2$).\label{fig-ch03-53}}
\end{figure}

This is in agreement with a direct observation of correlation
functions, Fig.~\ref{fig-ch03-53}. In particular, the ring structure
in $g_{\up\dn}(r)$ (or maximum at $r_0\approx 3.4\ell_0$) remains
preserved even for aspect ratios $a:b\approx 3$,
Fig.~\ref{fig-ch03-53} right. This is similar to how the ring structure of
the first maximum was preserved in the deformed $\nu=\ot$ Laughlin
state, Fig. \ref{fig-ch03-51}b. Also, looking at $g_{\up\up}(x)$ and
$g_{\up\dn}(x)$ in the deformed singlet state, the {\em sum} of these
two seems to remain constant beyond $r_0$ even in deformed systems, in
spite of $g_{\up\dn}(x)$ decreasing beyond $x=r_0$. This was just
the conclusion in $a:b=1$ systems, Fig. \ref{fig-ch03-08}, and it
suggests that the singlet state did not change much even in a quite
strongly deformed system ($a:b\lesssim 3$). 
Moreover, this finding allows us to
use deformed systems to study what happens on slightly larger
distances
than distances accessible in a square cell of a fixed area 
because maximum distance between two electrons in a
  deformed elementary cell,
  $\frac{1}{2}\ell_0\sqrt{2\pi N_m}(\lambda+1/\lambda)$, grows with
  increasing $\lambda$.

Regarding the energy, which seems to react more sensitively to
deformations than the correlation functions, the following speculation
seems plausible. If the singlet state is a liquid of $\up-\dn$ pairs 
of characteristic size $r_0\approx 3.4\ell_0$, Subsec. \ref{pos-ch03-15}, 
it ought to be more sensitive to
aspect ratio variations than the Laughlin state just because such a
pair in the $\nu=\tt$ singlet state is larger than a single electron in
the $\nu=\ot$ Laughlin state.

\subsubsection{Half-polarized states}

The half-polarized states can be expected to suffer severely under
the finite size of the system. A system with eight
electrons contains only two electrons with minority spin.  Contrary to
fully polarized systems (where eight particles is already fair
enough), it is thus the smallest system with $S=N_e/4$ where
many-body effects can be studied.

\begin{figure}
  \subfigure[8 electrons.]{\label{fig-ch03-38a}
    \includegraphics[scale=.5,angle=0]{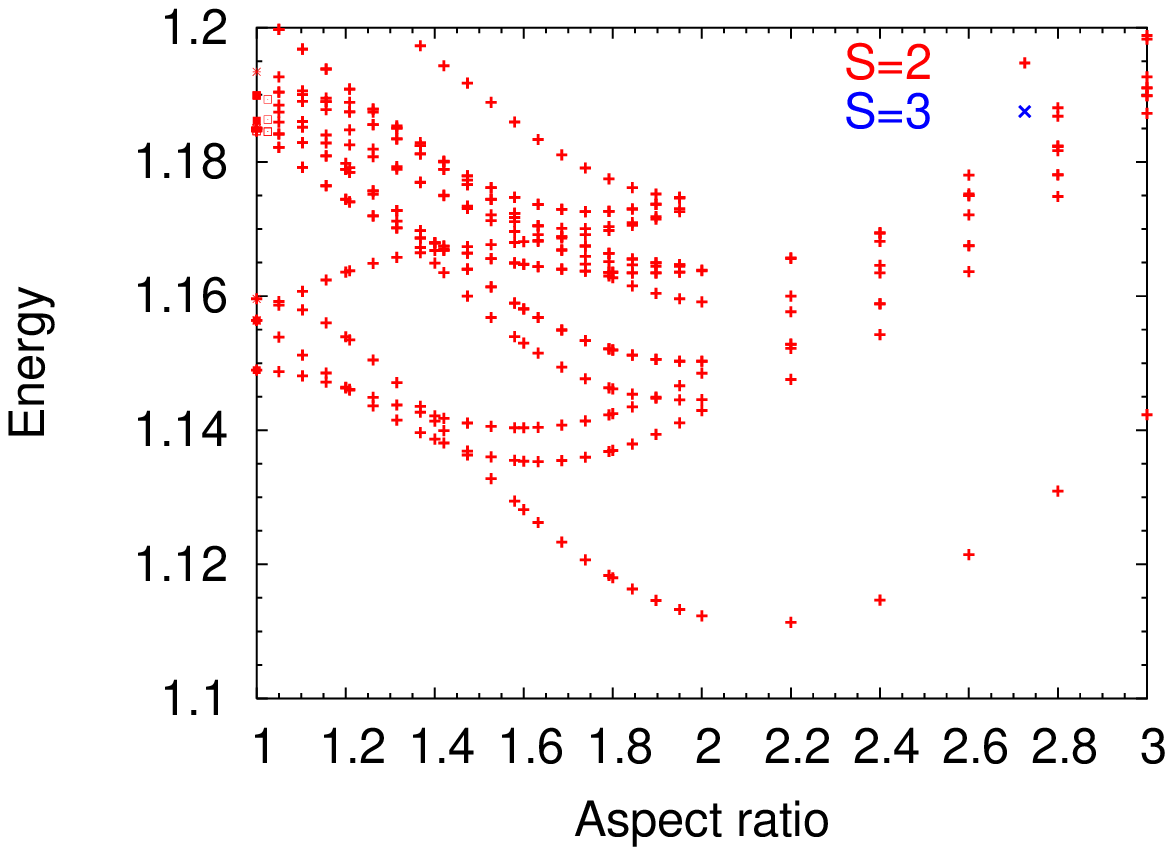}}
  \subfigure[12 electrons. Colours indicate different $J$'s.]
  {\label{fig-ch03-38b}
    \includegraphics[scale=.5,angle=0]{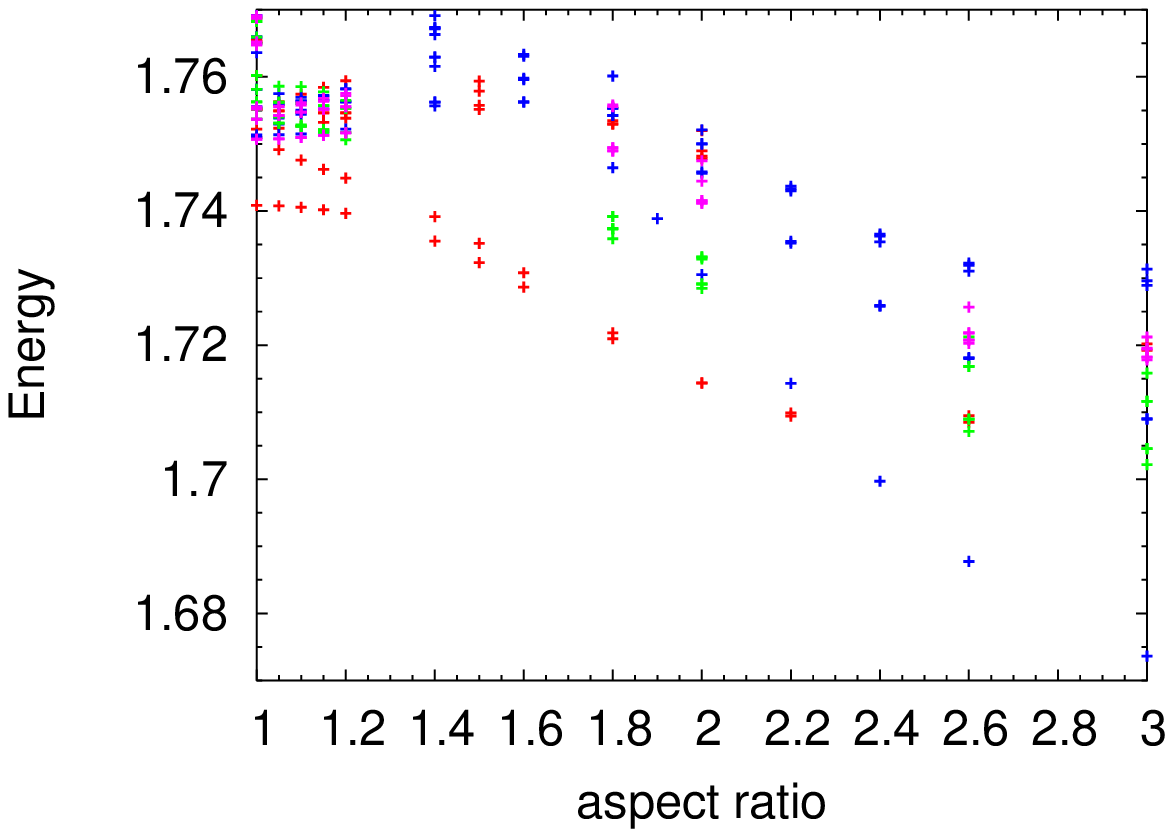}}
\caption{Half-polarized states ($S=N_e/4$) at filling $\tt$ versus
    aspect ratio. States with next larger spin are well above (out of
    scale here).\label{fig-ch03-38}}
\end{figure}

Let us compare how systems of two different sizes respond to varying
aspect ratio. In an eight-electron system, Fig. \ref{fig-ch03-38}a,
there are four low lying states: the ground state at $a/b=1$ with
$\krvd=(\pi,\pi)$, a $(0,0)$ state which
becomes the ground state at $a/b>1.5$ and a pair of degenerate
states, $(0,\pi)$ and $(\pi,0)$ ($\krvd$ is defined in
Subsec. \ref{pos-ch02-06}). The former two
states are isotropic (and lie in high symmetry points of the Brillouin
zone), the other two are spin-density waves in $x$ and
$y$ direction, judging by the correlation functions (not
shown). Moving away from aspect ratio one, degeneracy of the latter
two is lifted -- just as the $90\deg$ rotational symmetry of the
elementary cell is broken -- and the wave along $x$ (the longer side)
becomes energetically more favourable. It is quite conspicious that this
state evolves parallel to the $(\pi,\pi)$ state
for aspect ratios above $\approx 1.4$. For these values of $a:b$,
the inner structure of these two states seems to be very similar, too.

In the low-energy sector, a $(0,0)$ state is absent in a 12-electron
system, Fig. \ref{fig-ch03-38}b. In other respects, however, the situation is
quite similar to the smaller system. There is a
well-separated $(\pi,\pi)$ ground state in a square cell and this
state becomes nearly degenerate with a $(0,\pi)$ state for aspect
ratios $\gtrsim 1.4$. Also, the energy of these two states decreases with
increasing aspect ratio and eventually reaches its minimum. In
contrast to the smaller system, the minimum occurs later,
at $a:b\approx 2.4$ (Fig. \ref{fig-ch03-38}b) compared to $\approx
1.6$ in Fig. \ref{fig-ch03-38}a, but this occurs also for the
incompressible states, e.g. the singlet at $\nu=\tt$ 
(Fig. \ref{fig-ch03-39}a vs Fig. \ref{fig-ch03-39}b). The correlation
functions of these two states, $(0,\pi)$ and $(\pi,\pi)$, are similar to
those of the  $(0,\pi)$ and $(\pi,\pi)$ states in the eight-electron
system (not shown).

\begin{figure}
\subfigure[The lowest state, $\krvd=(\pi,\pi)$.]{\label{fig-ch03-54a}
    \includegraphics[scale=.7,angle=0]{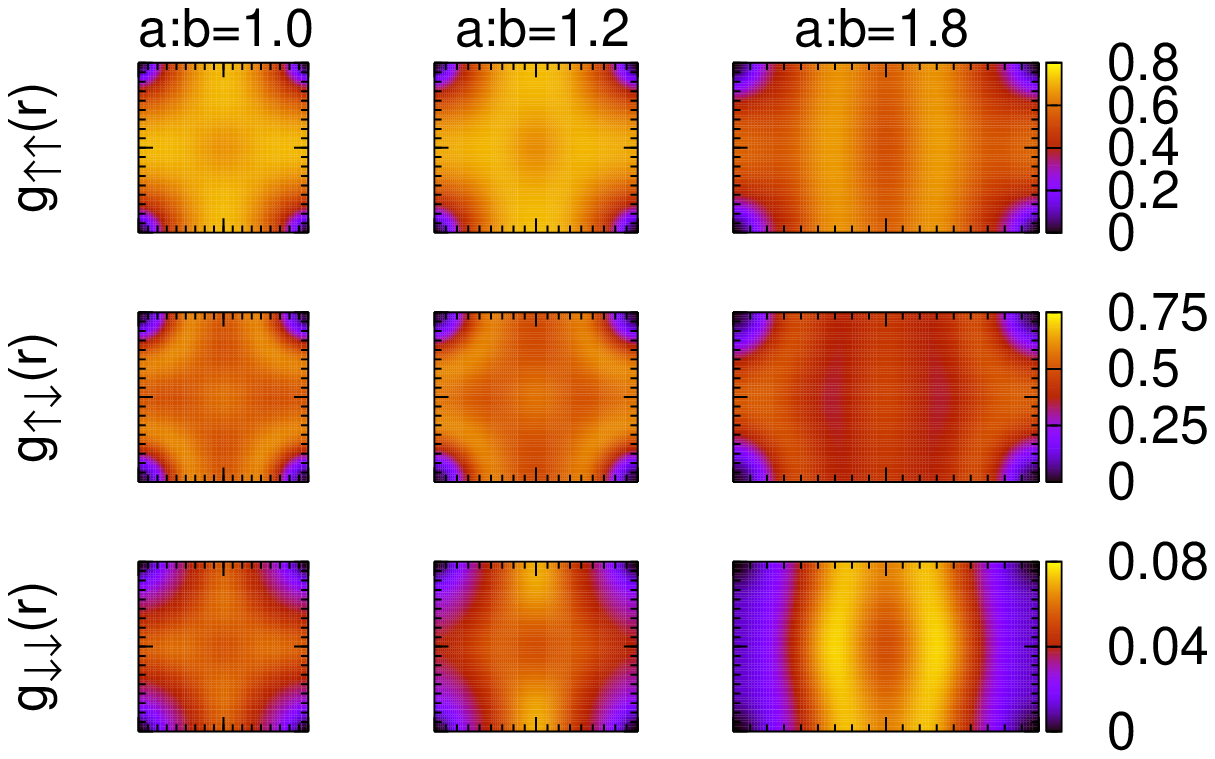}}
\subfigure[The first excited state, $\krvd=(0,\pi)$.]{\label{fig-ch03-54b}
    \includegraphics[scale=.7,angle=0]{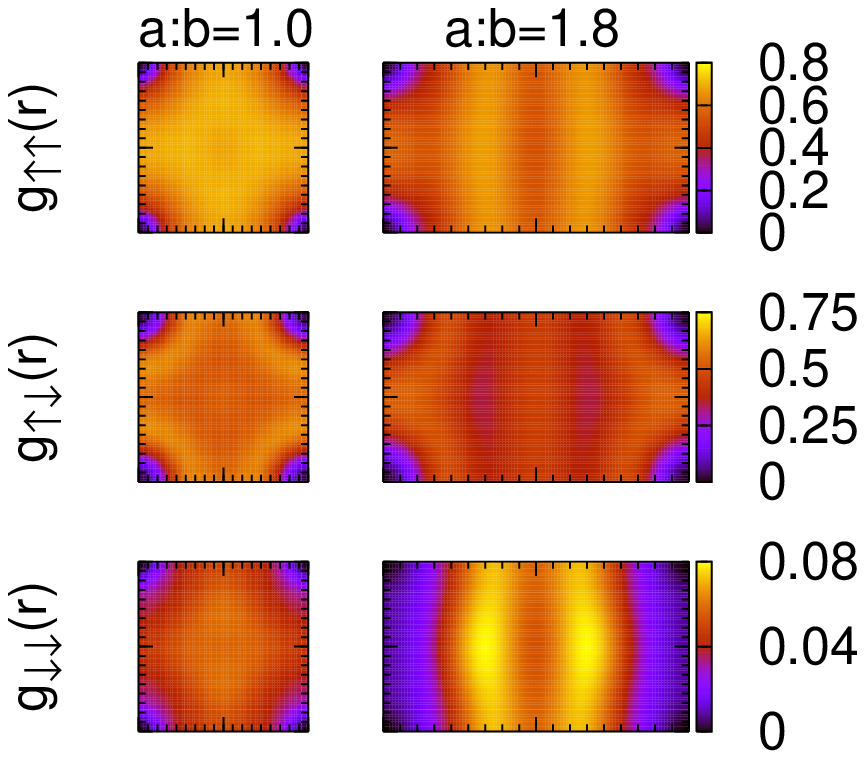}\hskip-3cm}
\subfigure[$g_{\up\up}$, $g_{\up\dn}$ and $g_{\dn\dn}$ of the lowest
    state in a deformed cell.]{\label{fig-ch03-54d}
    \includegraphics[scale=.6,angle=0]{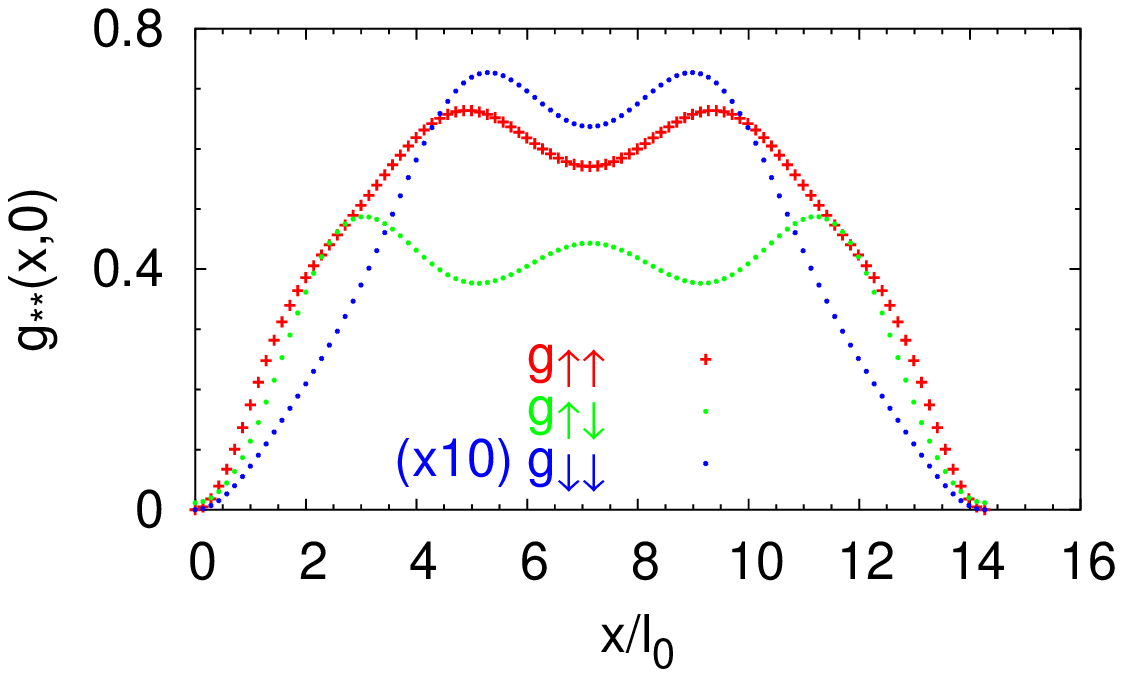}}
\subfigure[$g_{\dn\dn}$, section along $x$ for square and deformed
elementary cell.]{\label{fig-ch03-54c}
    \includegraphics[scale=.6,angle=0]{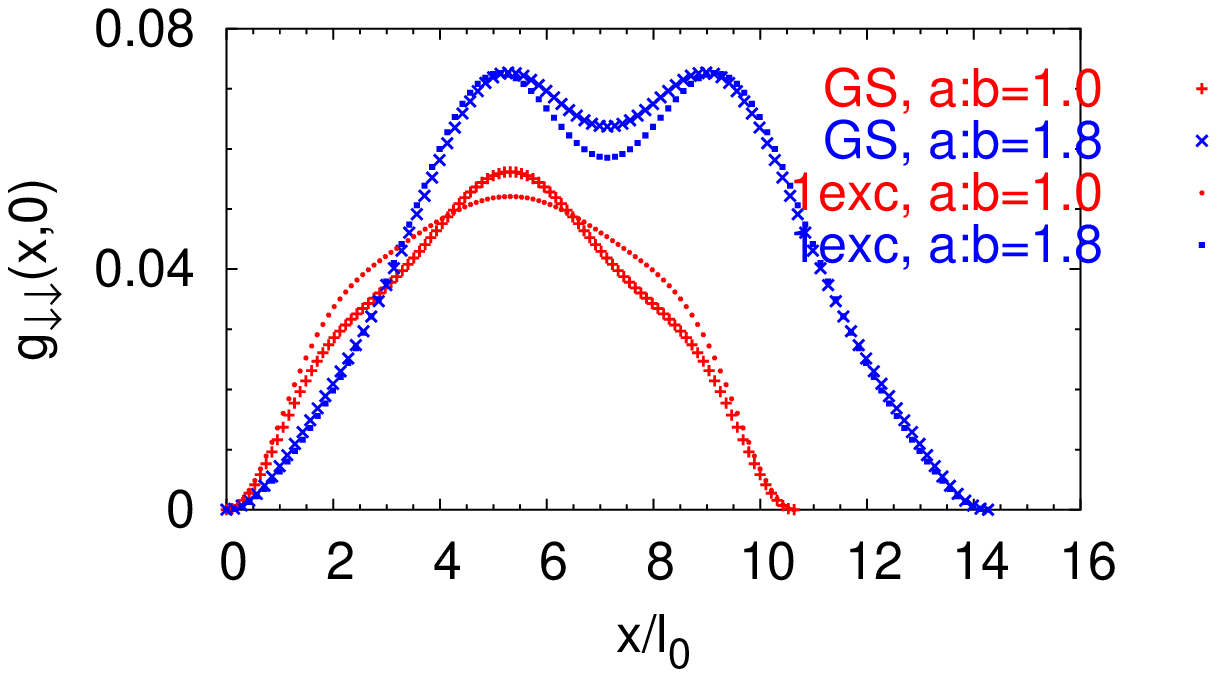}}
\caption{Evolution of two half-polarized states lowest in energy 
with growing aspect
ratio of the elementary cell (12 electrons, $\nu=\tt$, short-range
interaction).  Correlation functions are shown.
\label{fig-ch03-54}}
\end{figure}

Now turn to the correlation functions of the $(0,\pi)$ and $(\pi,\pi)$
states in a 12-electron system, Fig. \ref{fig-ch03-54}. Both states
are quite isotropic, for a square elementary cell, first at
  a very close look, we find a slight $x$ versus $y$ anisotropy in the
  $(0,\pi)$ state.
However, already under slight variation of the aspect ratio,
stripe structures parallel to the shorter side evolve ($a:b=1.2$,
Fig. \ref{fig-ch03-54}a). In this respect, both states look quite
similar, Fig. \ref{fig-ch03-54} (or compare Figs. \ref{fig-ch03-54}a and
\ref{fig-ch03-54}b), and we should stress that the differences 
in the correlation functions between the isotropic (at $a:b=1$) and
the wave-like state ($a:b=1.8$) are very large, both in
isotropy/anisotropy and in the 
short-range behaviour, Fig. \ref{fig-ch03-54}c.
This is in a strong contrast to the behaviour of the
incompressible states, e.g. the Laughlin state which preserves lot of
its original isotropy even at $a:b\approx 2$, Fig.~\ref{fig-ch03-51}.

These observations suggest the following interpretation. The
half-polarized ground state at $\nu=\tt$ is an isotropic state which
however inclines to the formation of a spin-density wave. The wave
has the shortest period allowed by the number of electrons, i.e. it
resembles an antiferromagnetic ordering ($\up\dn\up\dn\dots$ rather
than $\up\up\dn\dn\dots$, for instance) as the correlation functions in the
rightmost column in Fig. \ref{fig-ch03-54}a suggest. Since there are
just three $\dn$-electrons in the system, we expect two stripes (the
third $\dn$-electron is just at the origin) in
$g_{\dn\dn}(x,0)$ in the case of $\up\dn\up\dn\ldots$
ordering. In more detail, see
  Fig. \ref{fig-ch03-54}d: the minima/maxima in
  $g_{\up\up}(x,0)$ match well with the maximum/minima in
  $g_{\up\dn}(x,0)$. In other words, spin up is followed by spin
  down. 
However, the amplitude of oscillations in $g_{\dn\dn}(x,0)$
is moderate, Fig. \ref{fig-ch03-54}c, and hence we should rather classify
the state as a 'spin density wave' than e.g. a state with stripe domains
of alternating spin polarization.

On the basis of the present investigation, it is not clear
whether in a large enough system, this spin wave state is the ground
state, a low-energy excitation or it is degenerate with the isotropic
ground state. Even though the GS at $a:b>1$ (spin wave) has a lower
energy than the isotropic state at $a:b=1$, Fig. \ref{fig-ch03-38}b,
this does not say much about which state would be the ground state in
a larger system. We saw a similar situation for the $\nu=\ot$ Laughlin
state, Fig. \ref{fig-ch03-50}a, or the singlet $\nu=\tt$ state,
Fig. \ref{fig-ch03-39}. The energy of the ground state was not at its
minimum at $a:b=1$, yet the isotropic ($a:b=1$) state
is probably the real ground state in the thermodynamic limit. The question
how to decide which state -- isotropic or anisotropic -- will be
preferred in infinite systems remains open, but comparison between
systems of more different sizes could be very helpful.

\subsubsection{Conclusions}

It has been demonstrated that isotropic states like the fully polarized or
singlet incompressible $\nu=\tt$ ones tend to be insensitive to
slight deformations. The response was observed in the energy of the
state and in its correlation functions, where we saw that especially the
short-range behaviour remains basically unchanged. The insensitivity
improves with increasing the system size (number of particles). We
also  registered some differences between the singlet and polarized
state. In systems of equal size (area) the former state was disturbed
by smaller deformations. This agrees with our previously mentioned
hypothesis (Subsec. \ref{pos-ch03-17}) that the singlet ground state
consists of pairs of electrons with unlike spin. Since the typical
size of such a pair was rather large ($3.4\ell_0$), the singlet state
will suffer under the finite size of the system more than the
polarized state where the 'relevant particles' are still
electrons whose size is about $\ell_0$. Imagine filling a
  container once with ten tennis balls ($\sim$ polarized state) or with five
  footballs ($\sim$ singlet state). Slightly deforming the 
  container will probably
  affect the latter system stronger. 

Investigation of the half-polarized state revealed that while the
state is isotropic in a square cell, it tends to build a
unidirectional spin density wave for aspect ratios not far from
one. In this regime, it also becomes degenerate with {\em one} other
state. Correlation functions of the both states (in deformed
elementary cells) are quite similar to each other. We suggested that
these states have an antiferromagnetic ordering in agreement with both
correlation functions and wavevectors of these two states,
$\krvd=(0,\pi)$ and $(\pi,\pi)$. The question which state (isotropic
or spin wave) is the real ground state in an infinite system remained
unanswered.

\subsection{Summary and comparison to other studies}



\label{pos-ch03-02}

\subsubsection{The incompressible states: the polarized and the singlet ones}

We studied various properties of the fractional quantum Hall states
with spin degree of freedom at
filling factors $\ot$, $\tt$ and $\tf$: correlation functions,
response to magnetic and non-magnetic $\delta$-line impurities or to
deformation of the elementary cell. Briefly summarized:

(i) The results are
in agreement with the concept of incompressibility of these states and
also (in the case of $\nu=\ot$) with some earlier studies,
e.g. \cite{zhang:11:1985}.

(ii) Even though these states can be imagined as composite fermion
systems with integer filling, the analogy to Landau levels completely
filled with {\em electrons} can often be misleading. For instance
electrons of unlike spin are strongly correlated in the $\nu=\tt$
singlet state while they are completely uncorrelated in a
$\nu=2$ singlet state.

(iii) We inferred pairing of spin up and spin down electrons in
the $\nu=\tt$ singlet state. In the spin-unresolved density-density
correlations, this state looks as if the two electrons in each pair
were located exactly at the same position and the pairs then formed a
$\nu=1$ state. This conclusion was not possible for the $\nu=\tf$
singlet state thereby highlighting differences between fillings $\tf$ and
$\tt$ which are very closely related within composite fermion theories.

\subsubsection{Half-polarized states}

We identified a highly symmetric 
half-polarized state at filling factor $\tt$ which
{\em could} become the absolute ground state in a narrow range of Zeeman
energies (or magnetic fields). Such a state is completely unexpected in
mean-field composite fermion theories. Extending earlier studies
with exact diagonalization on a sphere we showed that extrapolating the
energy of this state from finite size exact diagonalizations 
to the thermodynamic limit is problematic and the question whether the
half-polarized state really becomes the absolute ground state remains
open. 

Investigations on this state both for short-range and Coulomb
interacting systems showed strong similarities to the incompressible
singlet and polarized states at $\nu=\tt$. Consequently, we suggested
that the singlet and the polarized state coexist within the
half-polarized state. The state {\em might} be gapped for short-range
interacting electrons but even if yes, it is probably not gapped for
Coulomb interacting systems. These differences in spectra accentuate the fact
that extrapolations to infinite systems should be taken with extreme
caution. It also means, that the definition of the short-range
interaction should be reconsidered. The model may be an oversimplified
if we study the half-polarized states since the mean distance
between two minority spin electrons is rather large, higher pseudopotentials
should also be taken into account.

The half-polarized state forms a
pronounced spin-density wave, or antiferromagnetic order, 
when anisotropy is introduced from
outside (deformation of the elementary cell) but we could not conclude
whether this spin-wave will be more energetically favourable than the
isotropic form in much larger systems.

\subsubsection{Half-polarized states: other studies}

Let us first briefly recall other suggestions which appeared
since Kukushkin {\em et al.} presented their experiment showing
a plateau of the polarization at the value of one half. All
works mentioned below can be applied both to filling factor $\tt$ and
$\tf$ in principle. Unless necessary, we will not distinguish between
these two cases.

Ganpathy Murthy \cite{murthy:01:2000} was attracted by the idea that
correlations favour either the spin singlet or the fully polarized
state. At the point where the two ground states cross (recall Figures
\ref{fig-ch03-20} and \ref{fig-ch03-01and02}a), electrons could prefer to
form a translationally non-invariant state consisting of regularly
alternating areas of (locally) singlet and (locally) polarized states
arranged into a partially polarized density wave (PPDW). He argues
that this structure ought to have square rather than a hexagonal
symmetry. The energy of the
PPDW state is evaluated within the Hamiltonian theory of composite
fermions
\cite{murthy:10:2003} and it is shown that the PPDW state is stable
(against one-particle excitations) and lower in energy than the
(homogeneous) singlet and polarized states. The period of the density wave
should be $2\sqrt{\pi}\ell^\ast$ (\ref{eq-ch02-08}) which is
$7.93\ell_0$ for filling $\tf$ and $6.14\ell_0$ for $\tt$. Charge
modulation in the wave should be quite weak (in the order of $1\%$).

Apal'kov, Chakraborty, Niemel\"a and Pietil\"ainen
\cite{apalkov:02:2001} object that the energy of the PPDW is too high and claim
that a homogeneous Halperin state in the two crossing CF Landau levels
(see below) should have a lower energy. Without invalidating
  the following results, this estimation
  seems to be however incorrect \cite{murthy:10:2001}. 
As the mentioned Halperin
state cannot account for the half-polarized states, Apal'kov {\em et
al.} suggest another candidate for the half-polarized state, a
non-symmetric excitonic liquid. They consider only the 'active
levels' meaning the two CF Landau levels which cross. These ({\em two})
levels have total filling of {\em one}, i.e., there are only $N_m$
electrons for $N_m$ places in the $\up$ level and $N_m$ places in the
$\dn$ level. By convention, they define a $\up$-particle as an
'electron' and a missing $\dn$-particle as a 'hole'; an
'electron'-'hole' pair is an 'exciton' and a pair of a
$\dn$-particle and a missing $\up$-particle is 'vacuum'. Owing to
the constraint $N_\up+N_\dn=N_m$, one-particle states can be mapped
onto a system consisting solely of 'vacua' and 'excitons'. 
The partial filling factor
$N_\dn/N_m\in [0;1]$ then gives simultaneously the polarization and
the number of
'excitons' (by $N_m$). Note, that 'excitons' are bosons by virtue of
an integer spin.

From this viewpoint, the $\nu=1$ quantum Hall ferromagnet (being
described by the Halperin $(1,1,1)$ state) is a
Bose condensate of excitons. In that case, all the excitons are
non-interacting and have
zero angular momentum $L$. This is most easily seen by the fact that
  $g_{\up\dn}(0)=0$. On an 'electron' ($\up$ particle), there is no $\dn$
  particle, i.e. there {\em is} a 'hole'. In an exciton (hydrogen
  atom), the only wavefunctions with $\psi(r=0)\not= 0$ are those with
  $L=0$. On the other hand, $g_{\up\dn}(0)=0$ follows from the fact that the
  Halperin state has maximum polarization and thus the spatial part of the
  wavefunction must be totally antisymmetric.
Apal'kov {\em et al.} suggest that the half-polarized
state at $\nu=\tt$ or $\tf$ could be a condensate of excitons with
$L=1$ for which they call it nonsymmetric.

To support this idea, they perform exact diagonalizations in a $\nu=1$ system
with several model interactions which are meant to describe the two --
active -- crossing CF Landau levels. These interactions are derived
from the Coulomb potential with suppressed short-range component,
probably (without justification) with the intention to describe interacting
composite fermions. Stability of the half-polarized state is
substantiated by showing that the energy versus polarization curve has a
downward cusp at half-polarization. On the other hand,
$g_{\up\dn}(0)\not=0$ in the half-polarized state indicates that
the 'excitons' do not have $L=0$. The particular value of
  $L=1$ is demonstrated by other means.

Finally, the idea of Eros Mariani \cite{mariani:12:2002} should be
presented. Parallel to the previous two works, the two 'active'
crossing CF Landau levels are considered. An assumption is made that
they both have a partial filling of $1/2$ rendering (after a {\em
second} Chern-Simons transformation) two Fermi seas of 'free'
composite fermions (of second generation). Mariani {\em et al.} show
that interaction of these objects with fluctuations of the gauge field
leads to an attractive effective interaction between particles with
opposite spin and momentum. In analogy to superconductive pairing,
this implies a gapped ground state. An estimation of the gap is
given.

\subsubsection{What are the half-polarized states then?}

Presently, it is not clear which (if any) of the candidates proposed
in the previous subsection describes the half-polarized reality. As
Murthy correctly mentions, the final answer should be given by
exact diagonalization of a large enough system. Unfortunatelly, we
dispose of systems not larger than $12$ particles. Nonetheless let us
compare the candidates with what was presented earlier in this
chapter.

The downward cusp in energy-versus-polarization dependence cannot be assured
by the calculations presented here. However, if the lowest
half-polarized state indeed becomes the absolute ground state at the
transition between the singlet and polarized state (see extrapolations
in Fig. \ref{fig-ch03-21}), the cusp is likely to be
present. In the other case, it will turn into an upward cusp, as the
calculated spectra suggest.

Results presented here indicate, that the half-polarized ground state
($\tt$) has $(\krvd_x,\krvd_y)=(\pi,\pi)$ and that it
shows similarities to the singlet and polarized ground states
(Fig. \ref{fig-ch03-23}). In particular $g_{\up\dn}(0)\approx 0$, which
is in contrast with the model of a nonsymmetric exciton
liquid (cf. the correlation functions in \cite{apalkov:02:2001}). 
Comparison between short-range interaction and Coulomb
half-polarized states (Fig. \ref{fig-ch03-41}) suggest that, similar
to the Laughlin state, the short-range part of the interaction plays the
major role. From this point, the model discussed by Apal'kov {\em et
al.} \cite{apalkov:02:2001} seems to be more appropriate rather for
some other systems.

Positioning of the half-polarized state out of the centre of the
Brillouin zone could be an indication that it is indeed a standing
wave. This is also supported by spectral properties when the
elementary cell is deformed, Fig. \ref{fig-ch03-38}. The two lowest
states becoming degenerate at aspect ratios larger than $1.4$ could be a
charge/spin-density wave (note also the correlation functions,
Fig. \ref{fig-ch03-54}). The fact, that
the energy of the ground state lowers with increasing aspect ratio could
indicate that this state is more stable than an isotropic
one. However, caution is advised here, since the singlet
incompressible ground state does the same, Fig. \ref{fig-ch03-39},
while its isotropic form is the true ground state.

Theory of the 'superconductive' pairing was not addressed so
far. Comparisons on the level of correlation functions, possibly in
$\krv$-space, are in principle possible, but quite complicated because
of the two Chern-Simons transformations involved.

\setcounter{footnote}{0}
\section{Quantum Hall Ferromagnetism at $\nu=\tt$?}
\label{pos-ch04-00}





Like the previous Chapter, this 
Chapter also starts from the fact that there are two distinct
ground states at filling factor $\tt$: a spin-singlet and a fully
polarized one. Their structure was studied in Chapter
\ref{pos-ch03-00} and we also interpreted them in terms
of composite fermions, Fig. \ref{fig-ch03-01and02}a. 
Whichever of these two becomes the absolute
ground state depends on the Zeeman splitting which favours spins
aligned parallel to magnetic field. The singlet state is the lowest
in energy for vanishing Zeeman splitting. However, increasing the Zeeman
splitting, its energy remains unchanged while the energy of the fully spin
polarized state decreases and eventually this other state becomes 
the absolute ground state. This simplest scenario, sweeping the Zeeman
energy while magnetic field is kept constant, is not very usual, albeit
it is experimentally possible \cite{leadley:11:1997}. 
However, even if we simply sweep the magnetic
field (and keep constant filling $\nu=\tt$ which requires a simultaneous
change of the electron density), the Coulomb energy of the singlet state
changes $\propto\sqrt{B}$ and that is slower than the Zeeman energy
of the polarized state in the limit of large $B$.
Therefore, the qualitative discussion above is still
valid. The total energy bilance of the two ground states (in SI units)
is thus
\begin{eqnarray*}
  \mbox{polarized:} && E_p(B) = \frac{e^2}{4\pi\ve\ell_0} E_p^C -
              g \mu_B N_e B = -|C_p|\sqrt{B} - |D_p| B\,,\\
  \mbox{singlet:}   && E_s(B) = \frac{e^2}{4\pi\ve\ell_0} E_s^C =
              -|C_s|\sqrt{B}\,,
\end{eqnarray*}
where $N_e$ is the number of particles and $E_p^C>E_s^C$ are the total 
Coulomb energies in units $e^2/4\pi\ve\ell_0$ 
(as calculated by exact
diagonalization, for example; not per particle). 
Obviously, $E_p(B)<E_s(B)$ for $B$ large enough. What the
critical field $B_c$ is, where both energies are equal, depends
on $(E_p^C-E_s^C)/N_e$. This quantity is accessible only
numerically and depends on $N_e$ although we may hope that it
stays nearly constant for large $N_e$.

\begin{figure}
\begin{center}
  \subfigure[$N_e=6$]{\label{fig-ch04-01b}%
  \includegraphics[scale=0.3]{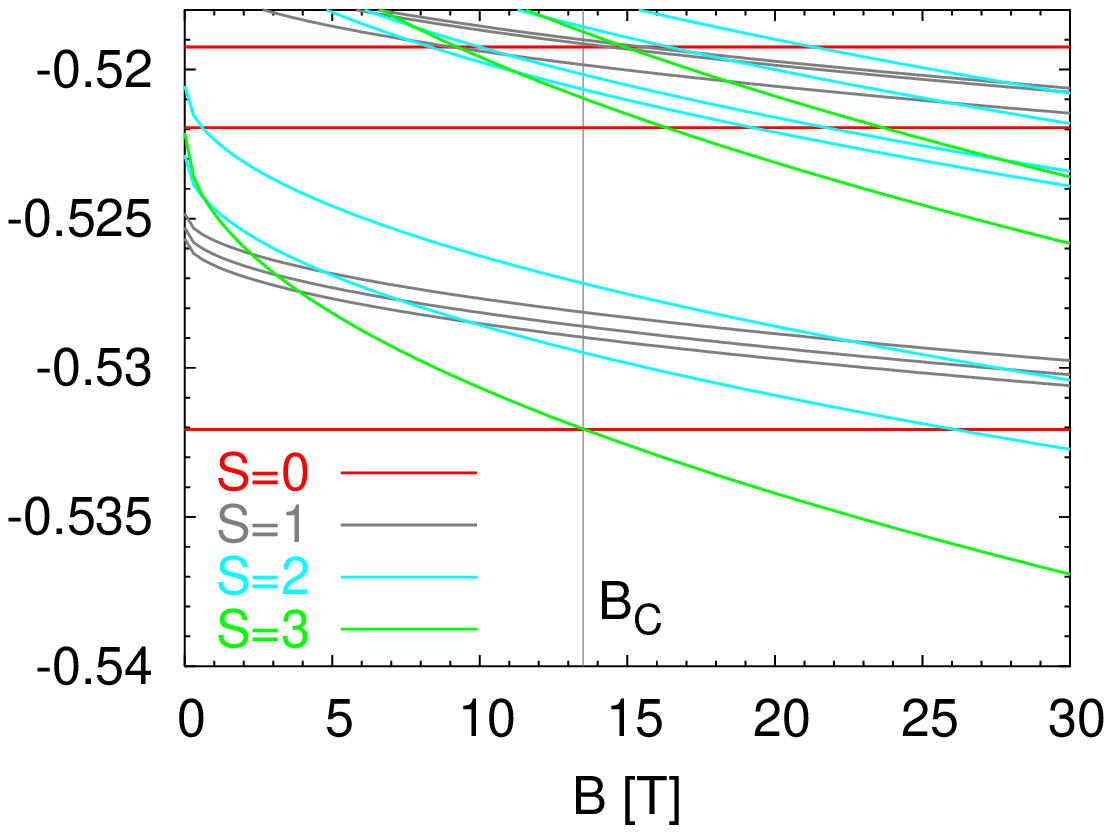}}
  \subfigure[$N_e=8$]{\label{fig-ch04-01c}%
  \includegraphics[scale=0.3]{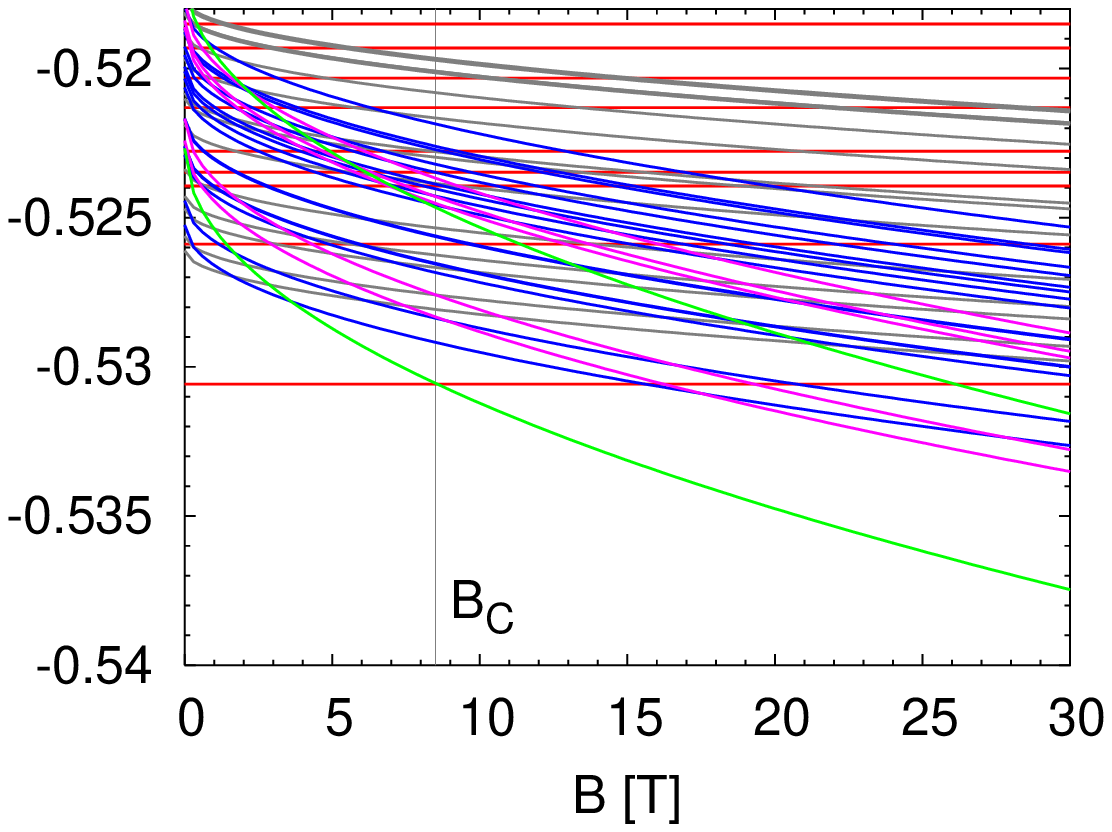}}
  \subfigure[$N_e=10$]{\label{fig-ch04-01d}%
  \includegraphics[scale=0.3]{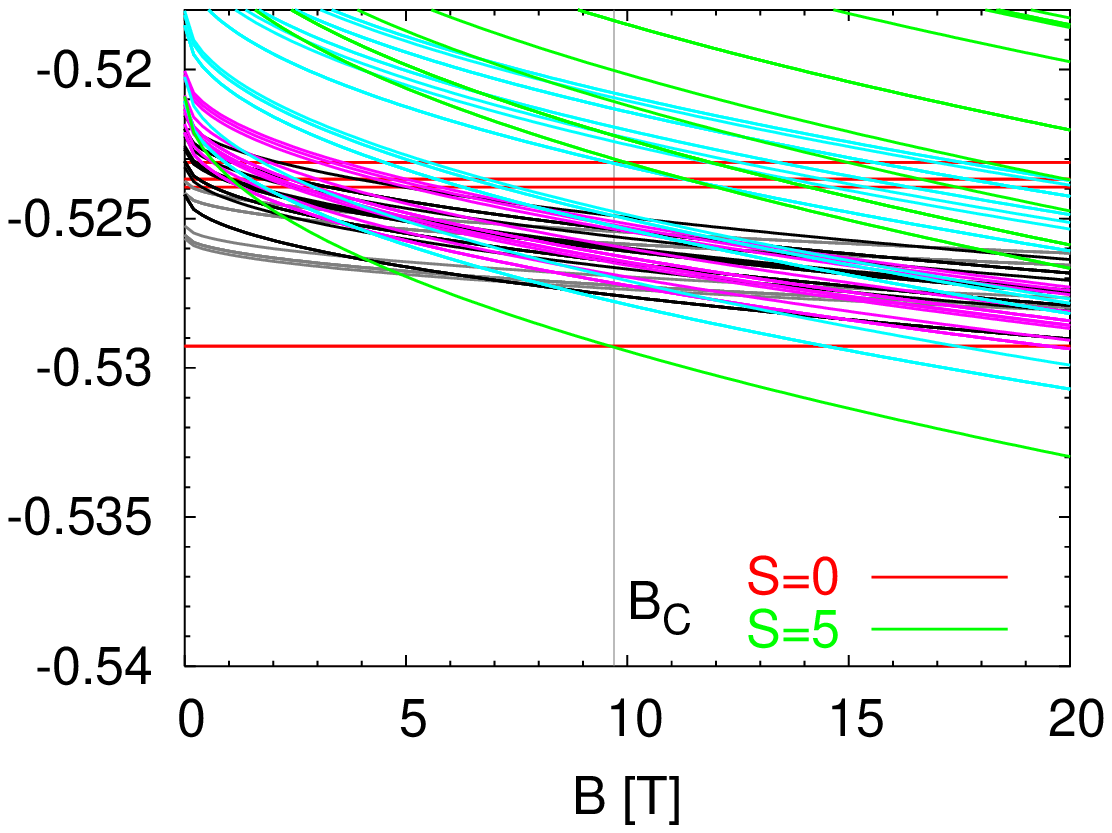}}
\end{center}
\caption{Energies of low lying states at $\nu=\tt$ in a homogeneous
Coulomb-interacting system with Zeeman field: transition from an
incompressible singlet ground state to a fully polarized
incompressible ground state. Different numbers of particles in a
square with periodic boundary conditions are considered, the scenario is
however the same in all cases.}
\label{fig-ch04-01}
\end{figure}

Figure \ref{fig-ch04-01} demonstrates this singlet to polarized
transition for $4,6,8$ and $10$ Coulomb-interacting electrons on a
torus. Note that the energy units in Fig. \ref{fig-ch04-01}, $e^2/(4\pi\ve
\ell_0)\propto \sqrt{B}$, change with magnetic field. In these units,
the potential (Coulomb) energy of all states stays constant (singlet
state) and Zeeman energy scales as $\propto e^2/(4\pi\ve\ell_0)\cdot
\sqrt{B}$.

A close look at Fig. \ref{fig-ch04-01} shows that 
the magnetic field $B_c$, at which the ground state transition
takes place, varies non-monotonically. However, an extrapolation of
energies of the two ground states to $1/N\to 0$ allows for a rough estimate of
$B_c\approx 7\unit{T}$ in an infinite system,
Subsect. \ref{pos-ch03-07}. This is in quite
good agreement with experiments \cite{kronmuller:09:1998}, even though
in some samples $B_c$ as low as $\approx 2\unit{T}$ was observed
\cite{kukushkin:05:1999,smet:01:2002}. 
This could be due to deviations from an ideal 2D system,
Subsect. \ref{pos-ch04-02}.

In the following we want to show that the existence of the spin
structures and the formation of domains are of central interest for
the understanding of the ground state transitions.

The ground state is always either a singlet or fully polarized in a
homogeneous system. The energies of these two states are equal at the
transition. This is similar to an Ising ferromagnet, if we label
the polarized state by pseudospin up and the singlet state by
pseudospin down. In an infinite system at non-zero temperature,
the Ising ferromagnet prefers a state with domains, some with
(pseudo)spin up, others with spin down, to the two homogeneous states. First
because entropy of the former is higher \cite{jungwirth:11:2001}
and second because the total magnetization of 
a domain state is
approximately zero while, locally, most spins are parallel to their
neighbours thereby minimizing the energy of magnetic stray fields
\cite{bertotti:1998}. None of these two mechanisms was included in the studied
model of a $\nu=\tt$ system. Nevertheless, the question was addressed 
how the system
responds if such a domain-inducing mechanism is modeled by a magnetic 
inhomogeneity. Will the ground state split into regions of different
spin polarization? With this question in mind, 
the inhomogeneity should prefer the
singlet ground state in one part and the polarized ground state in
another part of the system.

From the experimental side, there are quite strong hints at
ferromagnetism, mentioned in the introduction. Hysteresis, saturation (in
time) of magnetoresistance, Barkhausen jumps etc. hint at
ferromagnetic states with domain structure near the transition
point. Intention of the present study is to support this interpretation.

\subsection{Attempting to enforce domains by applying a suitable magnetic
  inhomogeneity}


\label{pos-ch04-02}

This and the following sections will be concerned with various
attempts to induce the formation of domains close to the transition point.
At the beginning we must discuss (i) how to enforce domains, what to
add to the Hamiltonian, which form of inhomogeneity and (ii) how
to detect them, which quantities should be observed.

\subsubsection{First attempt: the simplest scenario}

\begin{figure}
  {\unitlength=1mm
  \subfigure[Homogeneous system.]{\label{fig-ch04-02a}
     \includegraphics[scale=0.25]{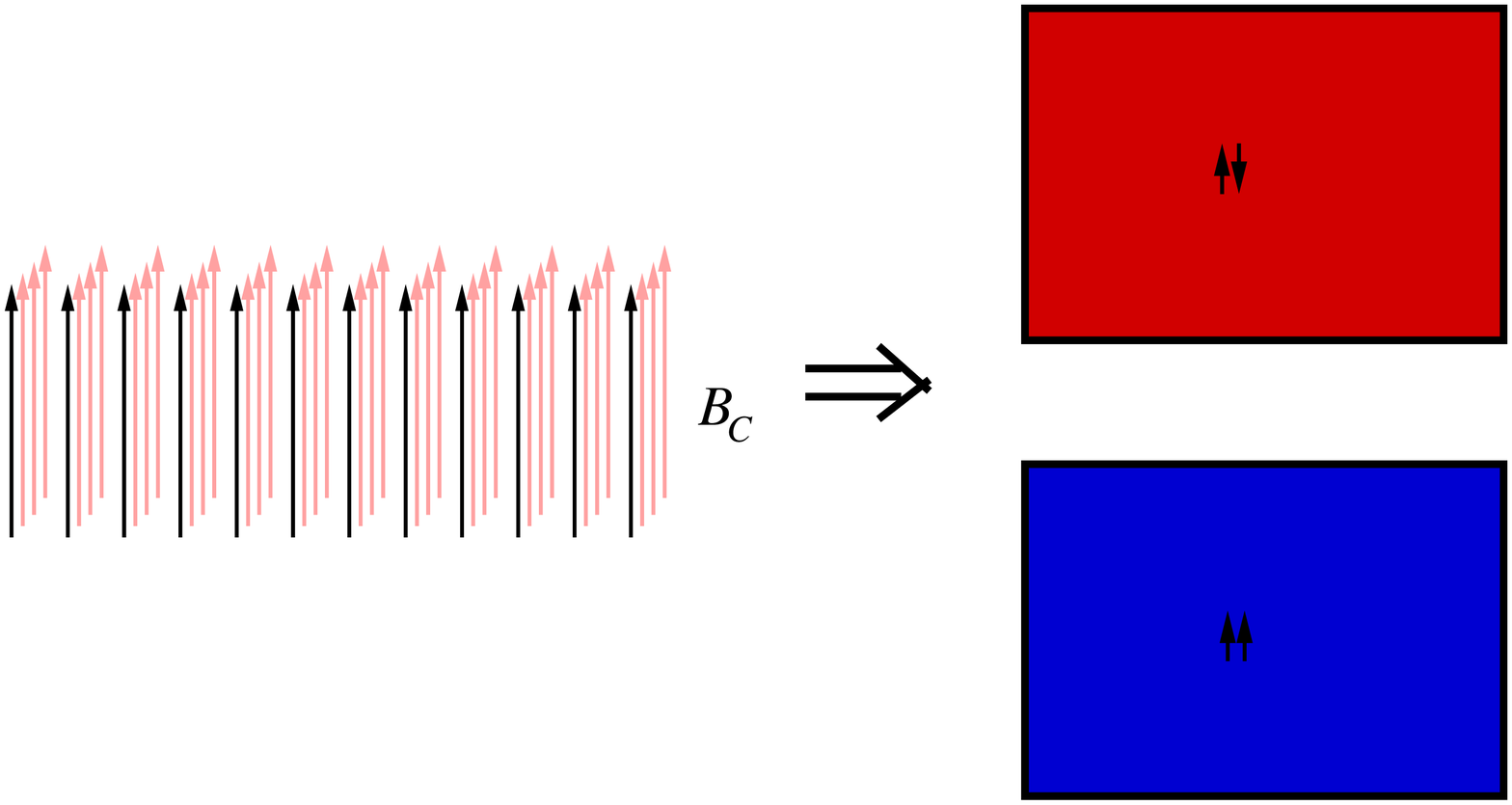}%
\put(-25,17){\small '$E_{Zeeman} = E_{Coulomb}$'}}
  \hskip1cm
  \subfigure[System with 'domain-enforcing' inhomogeneity.]{
     \label{fig-ch04-02b}
     \includegraphics[scale=0.25]{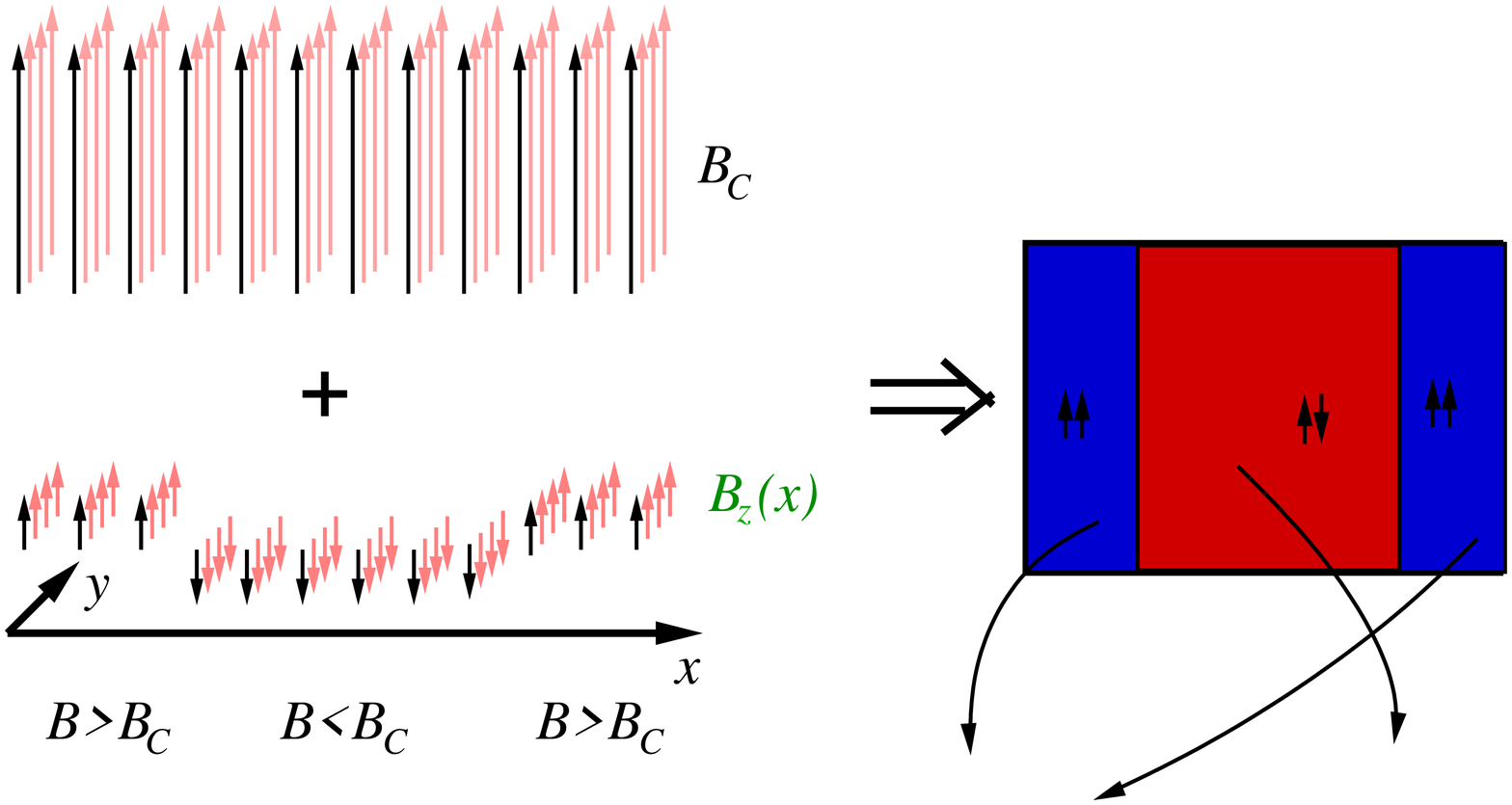}%
\put(-30,-4){\small '$E_{Zeeman} > E_{Coulomb}$'}%
\put(-15,0){\small '$E_{Zeeman} < E_{Coulomb}$'}}
  }
\caption{An idea of how to enforce domains at the crossing of singlet
and polarized ground states of $\nu=\tt$. An average Zeeman field is
chosen so that the both homogeneous states have the same energy.
Modulation of the Zeeman field prefers the singlet state 'in the
middle' and the polarized state 'at the edges' (note however the
periodic boundary conditions). }
\label{fig-ch04-02}
\end{figure}

The simplest scenario is sketched in Fig. \ref{fig-ch04-02}. In the
homogeneous case, the Hamiltonian consists of two terms
\begin{equation}\label{eq-ch04-01}
  H = H_{Coul} + H_{Zeeman} = \frac{e^2}{4\pi\ve} 
  \sum_{i<j} \frac{1}{|\vek{r}_i-\vek{r}_j|} + \sum_j g_0\mu_B B \sigma_z^j\,, 
\end{equation}
the Coulomb interaction and the Zeeman term. If the Coulomb energy is
fixed, the energies of the two incompressible ground states can be shifted
with respect to each other by varying the Zeeman term. If $B$ is fixed
at $B=B_c$ (i.e. the two ground state have the same energy), the Zeeman
energy can be still varied by means of the $g$ factor. Decreasing $g$
slightly, the singlet state will become the absolute ground state,
increasing $g$ the polarized state will prevail.

The idea of a 'domain-enforcing' inhomogeneity is to the turn
the constant $g$ into $g(x_j)=g_0+g_1(x_j)$ in (\ref{eq-ch04-01}) and
$g(x)>g_0$ in one part of the system whereas $g(x)<g_0$ in another.

Or, speaking in terms of Fig. \ref{fig-ch04-01}: we slightly modulate
the magnetic field $B$ and in one part of the system we consider
$B>B_c$ while in another $B<B_c$. By slightly we mean that only the
spin degree of freedom is affected, not the orbital. This is an
approximation.

The full Hamiltonian to consider is thus 
\begin{eqnarray}\label{eq-ch04-02}
  H &=& H_{Coul} + H_{Zeeman} + H_{MI}\,,\qquad \\
  && \nonumber
  H_{MI} = \sum_j g_1(x_j)\mu_B B \sigma_z^j\,,\quad
  \bra{\vp_i}H_{MI}\ket{\vp_j}=\delta_{ij}E_{MI}
  \left\{\begin{array}{l@{\,:\quad}r}
  i=0,1,\ldots,\frac{1}{4}N_m & 1\\[1mm]
  i=\frac{1}{4}N_m+1,\ldots,\frac{3}{4}N_m & -1\\[1mm]
  i=\frac{3}{4}N_m+1,\ldots,N_m & 1\\
  \end{array}\right.
\end{eqnarray}
where $\ket{\vp_j}$ is a one-particle state localized around
$x=(j/N_m)a$, cf. (\ref{eq-ch02-40}). 
This roughly corresponds to $g_1(x)$ having a
'rectangular wave' form ($g_1=1$ for $0<x<\frac{1}{4}a$ and
$\frac{3}{4}a<x<a$ and $g_1=-1$ for $\frac{1}{4}a<x<\frac{3}{4}a$).

The basic results of this model are: 
the ground states slightly change in accord
with the inhomogeneity and nothing peculiar
happens near the transition. As we sweep the magnetic field through $B=B_c$, 
even in the presence of a weak
inhomogeneity, the singlet state evolves 'smoothly and
monotonously' into the polarized state, without any remarkable
intermediate states. 

\begin{figure}
  \subfigure[An overview: spectrum and (inset) polarizations of the singlet,
    transition and polarized ground state (left to right).]{
    \label{fig-ch04-03a}
    \includegraphics[scale=0.5]{figs/ch04/ch04-fig02.epsi}}
    \subfigure[Polarization of the three states in detail
    ($n=10$). Mean polarization subtracted.]{
    \label{fig-ch04-03b}
    \includegraphics[scale=0.35]{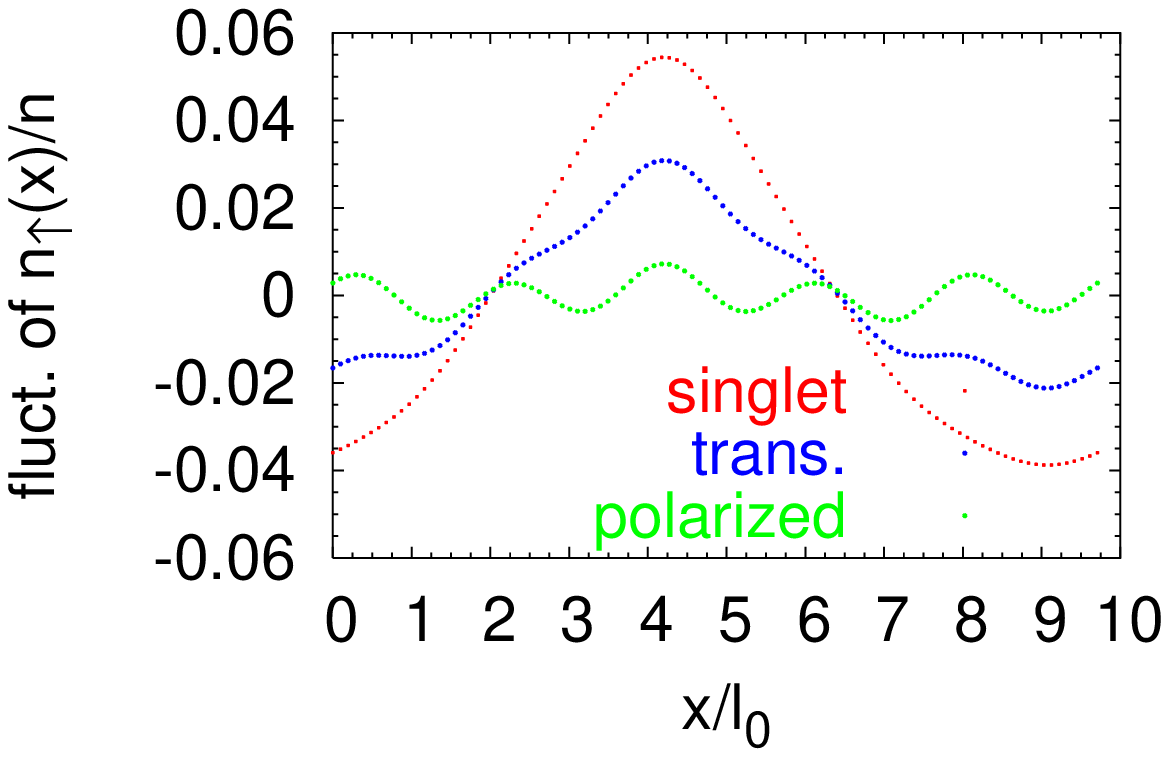}}
    \subfigure[The transition in systems of different
    sizes.]{\label{fig-ch04-03c}
    \includegraphics[scale=0.35]{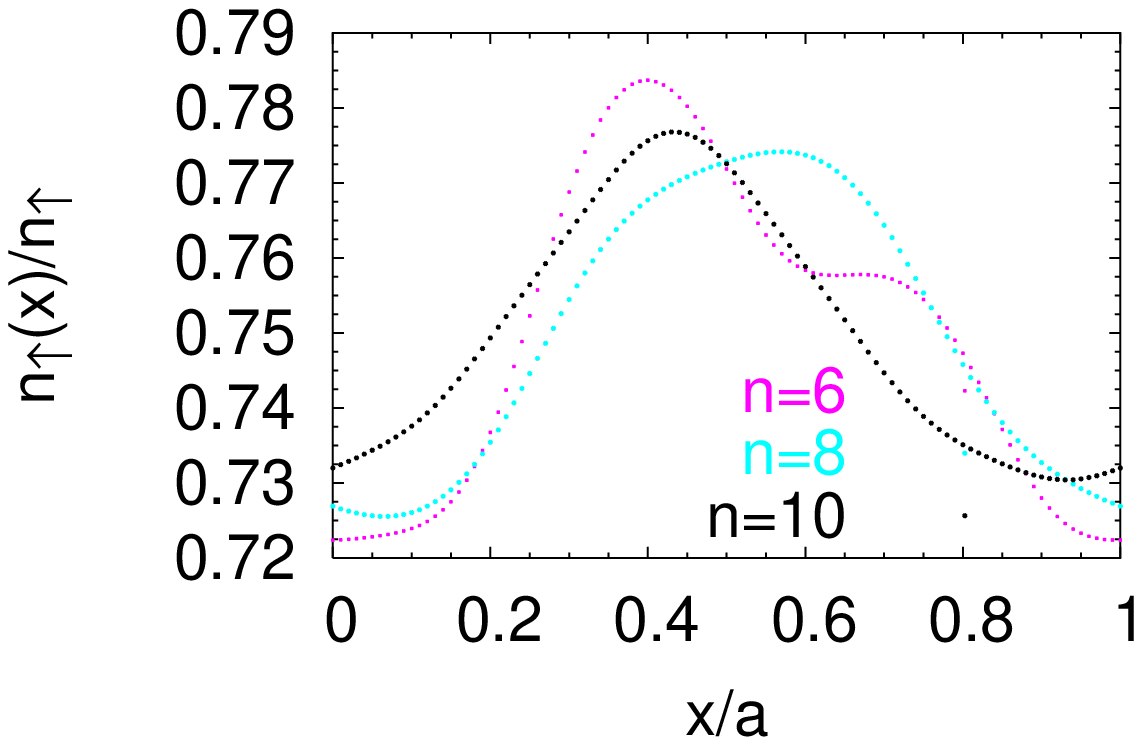}\hskip-1cm}
\caption{Response of a ten-electron system to a weak ($E_{MI}=0.002$) 
magnetic 
inhomogeneity of the form given in (\ref{eq-ch04-02}). No sings of domain
formation observed: the transition state does not respond stronger
than the incompressible states. }
\label{fig-ch04-03}
\end{figure}

Typical results are shown in
Fig. \ref{fig-ch04-03}. A magnetic inhomogeneity
(\ref{eq-ch04-02}) was applied to a ten-electron
Coulomb-interacting system and its strength $E_{MI}$ was chosen to be $\sim
10$\% of the incompressibility gap. Regarding the ground states and
the gap, the spectrum remains virtually unchanged.
Fig. \ref{fig-ch04-05}a shows a comparison of the spectra between homogeneous
and inhomogeneous systems. Looking now at the
singlet and polarized ground states, we find a spatially varying spin
polarization\footnote{Throughout this Chapter, we will refer to
  $p(x)=n_\up(x)/n(x)$ as to polarization. In the literature, another
  definition is more common, $P(x)=[n_\up(x)-n_\dn(x)]/n(x)$, both
  quantities are, however, equivalent: $P(x)=2p(x)-1$.}
$n_\up(x)/n(x)$, Fig. \ref{fig-ch04-03}a. However, the mean values of
the polarization still remains at $0.5$ (1) as it was in the homogeneous
singlet (polarized) state, Fig. \ref{fig-ch04-03}a, leftmost
(rightmost) inset. The polarization of the 
'transition state' has a mean value of $0.75$, i.e. just in
the middle between the polarized and the singlet state. This is
not surprising, since the 'transition state' was taken to be a
symmetric linear combination of the two crossing states (see
Subsect. \ref{pos-ch04-05}). What is more interesting, is the
{\em variation} of the polarization around the mean value,
Fig. \ref{fig-ch04-03}b: in this point, the 'transition state' lies
just between the singlet and polarized states. Contrary to what we
observe in Fig. \ref{fig-ch04-03}a (middle inset), 
formation of domains near the transition
would mean that the polarization of the transition state should
vary between $0.5$ and $1$.

It could be that the system is simply too small for domains to evolve
near the transition. On the other hand, this does not seem to be the
case, since the response to the inhomogeneity does not grow with
increasing system size but rather stays about the same,
Fig. \ref{fig-ch04-03}c.

The particular parameters of the model
presented in Fig. \ref{fig-ch04-03} could have been chosen unluckily
so that domains could not evolve. Let us 
therefore discuss the inhomogeneous $\nu=\tt$ systems more thoroughly.

\subsubsection{Turning crossing into anticrossing: inhomogeneous inplane field}
\label{pos-ch04-05}

At $B=B_C$ there is actually a crossing between the singlet and
polarized ground states, Fig. \ref{fig-ch04-03}a, rendering the
transition jump-like just as in a homogeneous system. For the
transition state (the middle curve in Fig. \ref{fig-ch04-03}b), we
took a fifty-fifty linear combination of these two ground states. One
could say, the transition occurs in an infinitesimally small interval
of magnetic field around $B_c$.

In a realistic system, the transition is unlikely to happen all simultaneously
in the whole system. Two mechanisms causing a
more continuous transition in a finite interval of $B$ are conceivable:

(i) weak inhomogeneities: The spectrum (as a function of $B$)
  looks basically the same as in Fig. \ref{fig-ch04-03}a, {\em but}
  there is an anticrossing at $B\approx B_C$.

(ii) strong inhomogeneities: The energy gap between the pair of
  the crossing ground states and the excited states at $B=B_c$
  (Fig. \ref{fig-ch04-03}a) is reduced
  compared to the incompressibility gaps of the
  singlet and polarized ground states far away from $B_C$, i.e. for
  $B\to 0$ and $B\to\infty$. Under influence of stronger
  inhomogeneities, it could be that some originally excited state (or more
  states) become ground state around $B\approx B_C$ while singlet and
  polarized incompressible states remain lowest in energy only far away
  from $B_C$. If this turns out to be the case, it could be that more
  states (possibly of different $S_z$) can be mixed by the
  inhomogeneity, eventually rendering the ground state compressible.

In this Subsection we will discuss the former possibility, the latter
will be the topic of Subsect. \ref{pos-ch04-03}.

The ground state transition in Fig. \ref{fig-ch04-03}a is a crossing
even in the presence of the magnetic inhomogeneity $H_{MI}$
(\ref{eq-ch04-02}) because the symmetry of $H_{MI}$ is too high and it
does not mix the two crossing ground states. In particular,
$[H_{MI},S^2]\not=0$ but $[H_{MI},S^z]=0$ and the singlet state $\ket{S}$
has $S_z=0$ whereas the polarized state $\ket{P}$ is
$S_z=N_e/2$. Consequently 
$S_z\ket{S}=0$ and $S_z\ket{P}=(N_e/2)\ket{P}$,
  therefore $\bra{P}H_{MI}\ket{S}=(N_e/2)^{-1} \bra{P} S_z
  H_{MI}\ket{S} = (N_e/2)^{-1} \bra{P}H_{MI} S_z\ket{S}=0$.
The inhomogeneity $H_{MI}$ mixes states with different $S$ but only
those with equal $S_z$. The fully polarized ground state ($S=N_e/2$,
  $S_z=N_e/2$) has also an $S_z=0$ counterpart, since the homogeneous
  Hamiltonian commutes with spin lowering operator. This state,
  however, is a highly excited state at $B\approx B_C$, since its
  Zeeman energy is zero.

Which terms added to the Hamiltonian can break this symmetry and
what will then be the response of the ground state?

Weak (inhomogeneous)
inplane magnetic fields will have the desired effect. This scenario is
not unlikely to occur in a realistic system. It merely means, that the
extra fluctuating magnetic field which interacts only with the spins, is not
pointing exactly in the direction of the (strong) external magnetic field
causing the Landau level quantization. The existence of such
symmetry-breaking inhomogeneities is very likely in realistic
systems, although they might be very weak e.g. hyperfine interaction
with nuclear spins.

Let us consider a Hamiltonian with inplane magnetic inhomogeneities
(IMI) of the form
\begin{eqnarray}\label{eq-ch04-03}
  H &=& H_{Coul} + H_{Zeeman} + H_{MI} + H_{IMI}\,,\qquad \\
  &&\nonumber
  H_{IMI} = \sum_j g_0\mu_B B_x(x_j) \sigma_x^j\,,\quad
  \bra{\vp_i}H_{IMI}\ket{\vp_j}=\delta_{ij}E_{IMI}
  \left\{\begin{array}{l@{\,:\quad}r}
  i=\frac{1}{4}N_m & 1\\[1mm]
  i=\frac{3}{4}N_m &-1\\[1mm]
  \mbox{otherwise} & 0\\
  \end{array}\right.
\end{eqnarray}

\begin{figure}
  \subfigure[Spectrum. Solid lines: homogeneous system; points: 
  inhomogeneities switched on.]{
    \label{fig-ch04-04a}
    \includegraphics[scale=0.5]{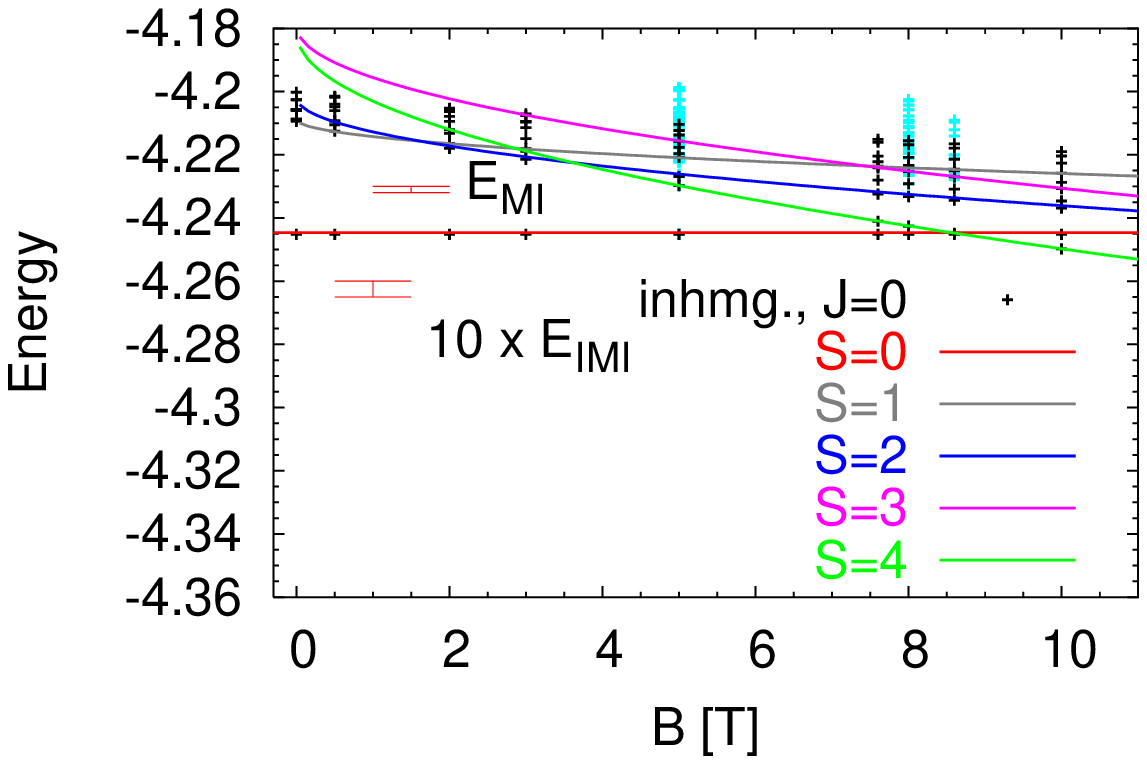}}
    \subfigure[Energy of the lowest two states at the anticrossing as IMI is
  being turned on. {\em Right:} level splitting ]{
    \includegraphics[scale=0.5]{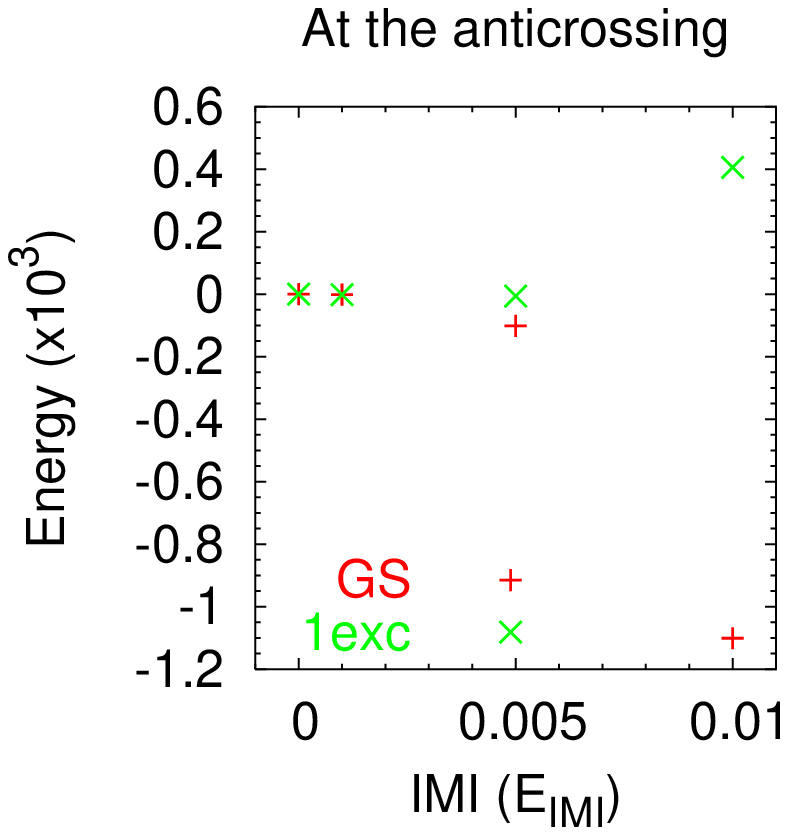}
    \includegraphics[scale=0.5]{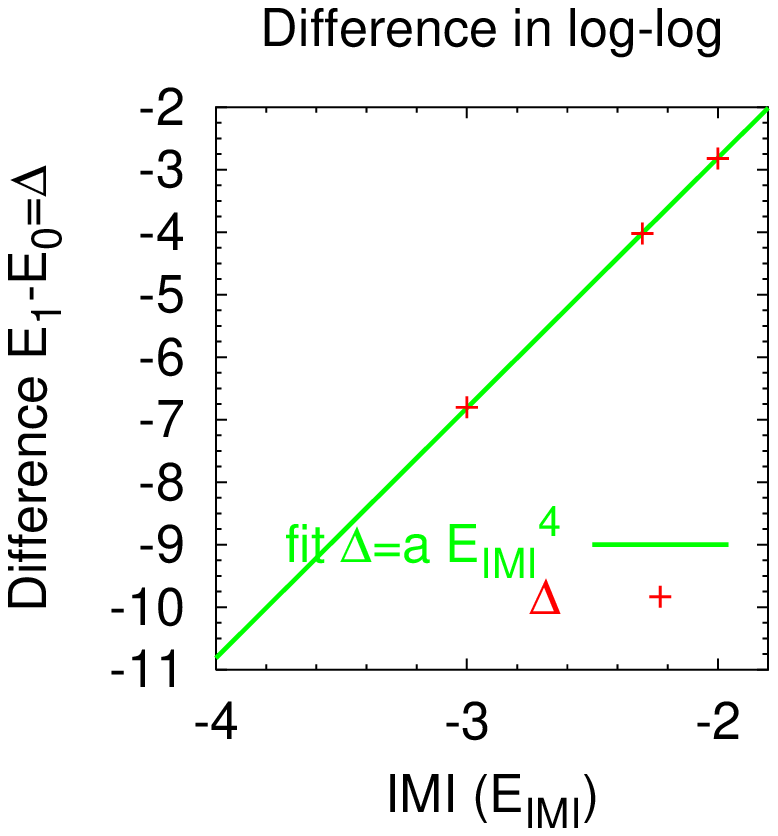}
    \label{fig-ch04-04b}
}
\caption{Response (of an eight-electron system) 
  to a weak inhomogeneity in perpendicular and inplane
  direction (\ref{eq-ch04-03}). Perpendicular component $E_{MI}$
  is the same as in Fig. \ref{fig-ch04-03}. Exponent four comes from
  $N_e/2$, see explanations in the text.}
\label{fig-ch04-04}
\end{figure}

The main claim of this Subsection is that weak $H_{IMI}$ only opens an
anticrossing at the ground state transition. In other words, the
relevant states still basically form 
a two-level system comprising of the (slightly
disturbed) singlet and polarized ground states. The width of the
anticrossing grows with increasing strength of the inplane field
inhomogeneity, $E_{IMI}$. As to the width we refer 
either by the level splitting, 
Fig. \ref{fig-ch04-04}b, or by the range of magnetic field where
$\langle S_z\rangle$ noticeably changes, Fig. \ref{fig-ch04-05}.

This fact is best demonstrated in
Fig. \ref{fig-ch04-04}. Inhomogeneities are weak there (compared to
both incompressibility gaps $E_g^P\approx E_g^S$),
i.e. $E_{MI},E_{IMI}\ll E_g$, and the spectrum remains almost
unchanged, Fig. \ref{fig-ch04-04}a. The energy levels of the
inhomogeneous system (points) are almost equal to the energies in a
system free of impurities (lines). Only directly at $B\approx B_C$
an anticrossing opens and the level separation $\Delta E$ grows with increasing
$E_{IMI}$ ($E_{MI}$ is kept constant),
Fig. \ref{fig-ch04-04}b. Even for a quite strong inplane inhomogeneity
(of the order of $E_g$), the level splitting remains
small ($\ll E_g$). The reason for this is simple: $H_{IMI}$ couples only
states which differ by $\pm 1$ in $S_z$, since it is a one-particle
operator (allowing for only one spin flip at once). Thus, a coupling
of the two ground states occurs for a $N_e=8$ system first
in the fourth order of perturbation theory
($N_e=8$ implies $S_z=4$ for fully polarized system). This
interpretation fully agrees with the finding $\Delta E\propto
(E_{IMI})^4$, Fig. \ref{fig-ch04-04}b. We can therefore expect that,
for an inhomogeneity of constant strength, the level splitting will
vanish exponentially at $N\to\infty$ as long as $E_{IMI}$ is much
smaller than the gap at $B\approx B_C$.

\begin{figure}
  \subfigure[Weak IMI. \hfill\penalty-10000 ($E_{IMI}=0.001$)]{
    \label{fig-ch04-06a}
    \includegraphics[scale=0.4]{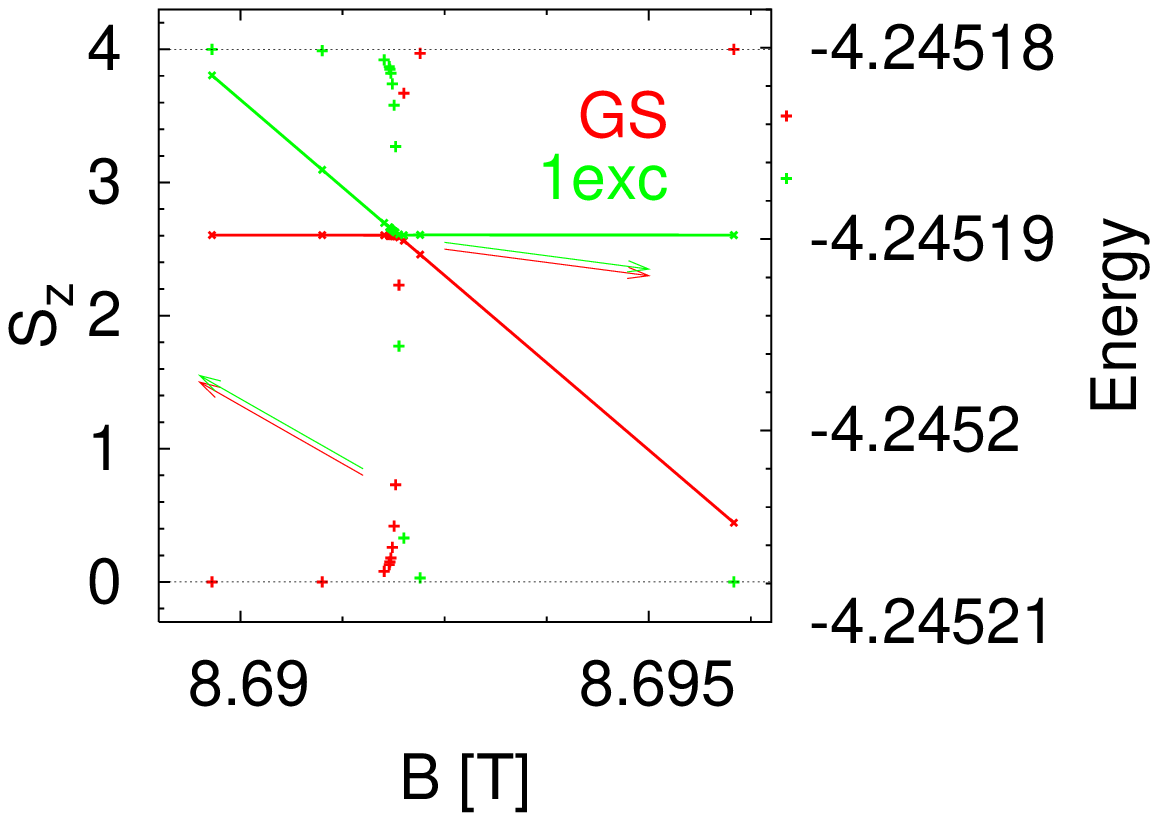}}
    \subfigure[Intermediate IMI. \hfill\penalty-10000 ($E_{IMI}=0.005$)]{
    \label{fig-ch04-06b}
    \includegraphics[scale=0.4]{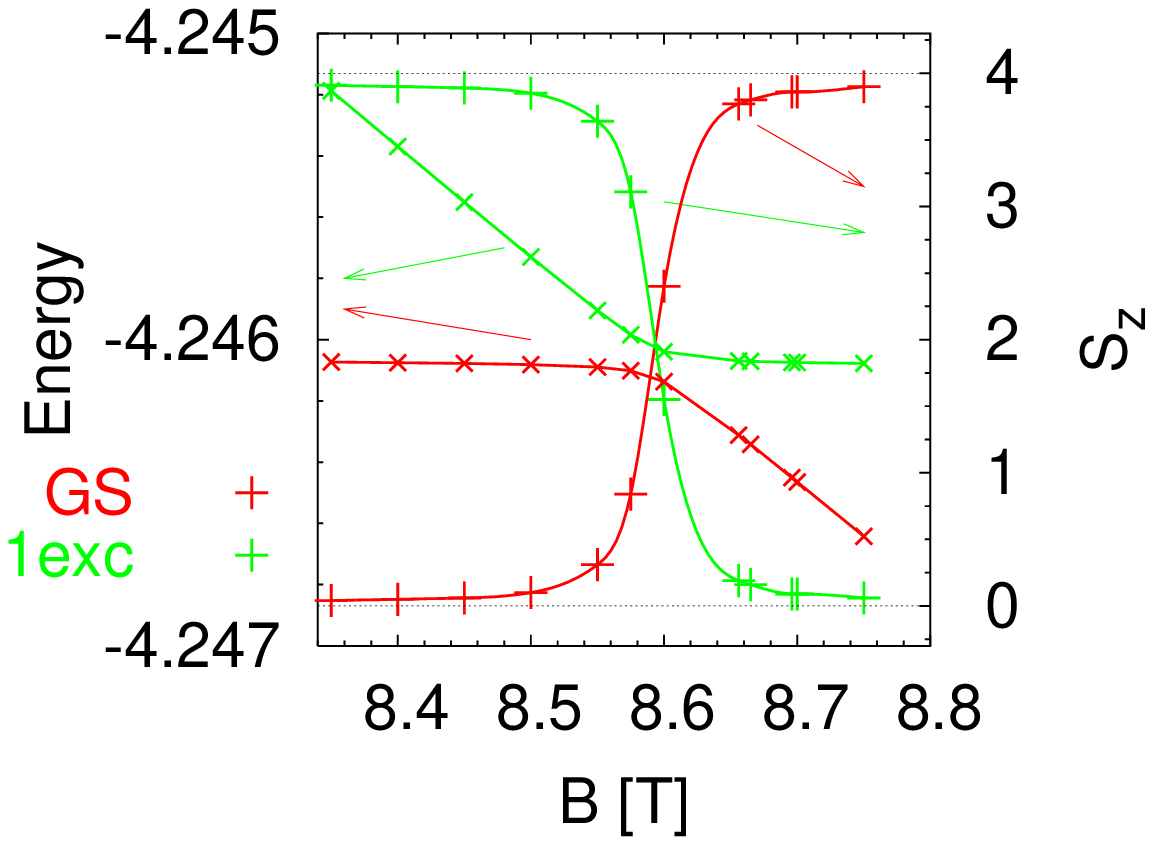}}
    \subfigure[Crossing at different values of IMI.]{
    \label{fig-ch04-06c}
    \includegraphics[scale=0.4]{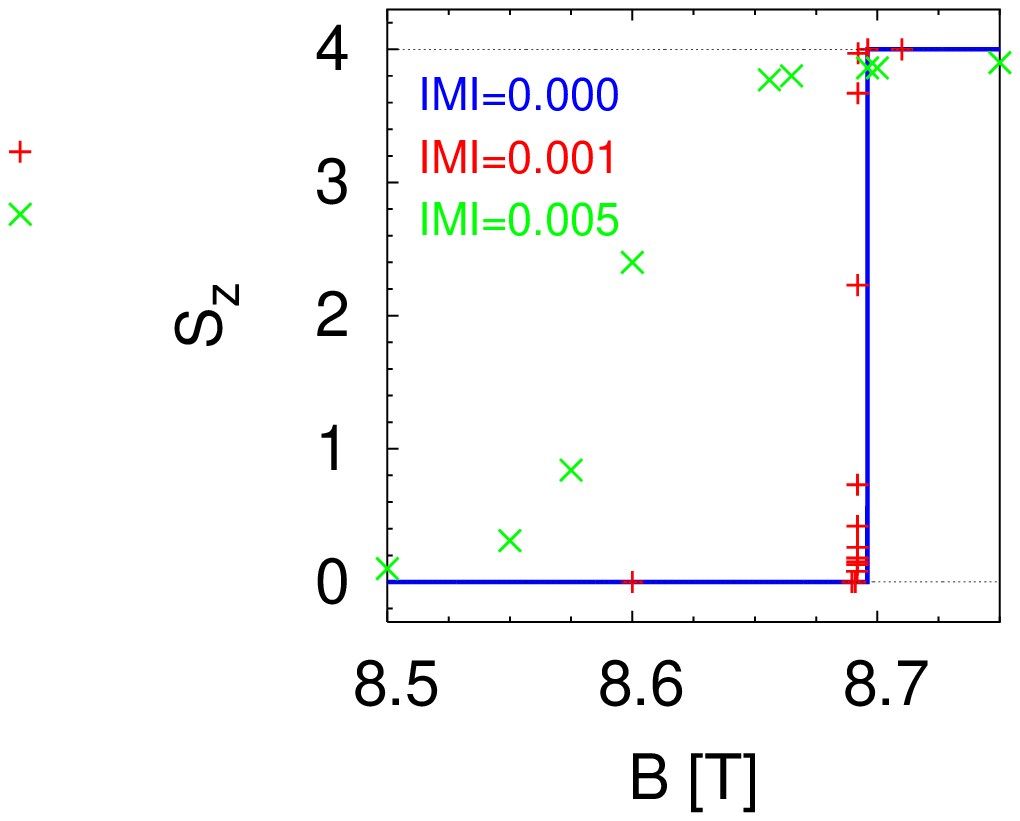}\hskip-1cm}
\caption{Inplane magnetic inhomogeneity (IMI) turns the crossing
    between the singlet ground state and the polarized ground state,
    Fig. \ref{fig-ch04-04}a, into an anticrossing. The cross-over
    between the two ground states can be observed either in the
    spectrum or in $\langle S_z\rangle$ of the ground state as $B$ is
    swept through $B_c$.}
\label{fig-ch04-06}
\end{figure}

Another view at the crossing for $E_{IMI}\not=0$ is presented in
Fig. \ref{fig-ch04-06}. If we focus on the ground state and sweep $B$
through $B_C$, we may observe how $\langle S_z\rangle$ 
(or $\langle S\rangle$) of the ground state
smoothly passes from $0$ to $N_e/2=4$. The transition observed in this
way (i.e. $\langle S_z\rangle\approx 2=N_e/4$) 
coincides with the transition observed in the spectrum (the 'anticrossing
region'), Fig. \ref{fig-ch04-06}a,b. The larger
$E_{IMI}$, the smoother the transition and the broader the range of
$B$ in which the transition occurs, Fig. \ref{fig-ch04-06}c.

So as to conclude: most importantly, an inplane magnetic
inhomogeneity (IMI) transforms the ground state transition into an
anticrossing. This effect should fade away for larger systems 
($N_e\gg 1$). We also remark, that 
the IMI shifts the transition point $B_C$ to lower
fields, Fig. \ref{fig-ch04-06}c, but this effect seems to be rather
small for inhomogeneity strength not exceeding the incompressibility gap.

\subsubsection{Strong inhomogeneities}
\label{pos-ch04-03}

In the following we suppress the inplane inhomogeneities again and let
us study stronger perpendicular inhomogeneities of the form
(\ref{eq-ch04-02}).

\begin{figure}
  \subfigure[Weak MI. ($E_{MI}=0.002$)]{
    \label{fig-ch04-08a}
    \includegraphics[scale=0.47]{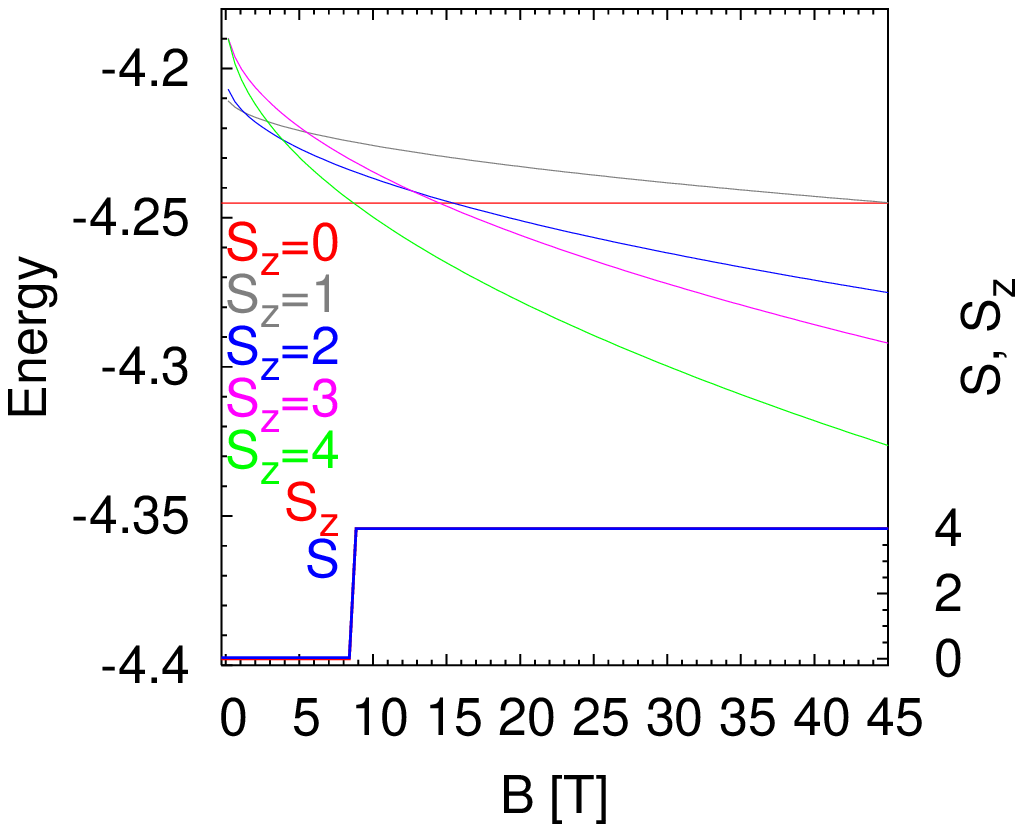}}
    \subfigure[Intermediate MI. ($E_{MI}=0.01$)]{
    \label{fig-ch04-08b}
    \includegraphics[scale=0.47]{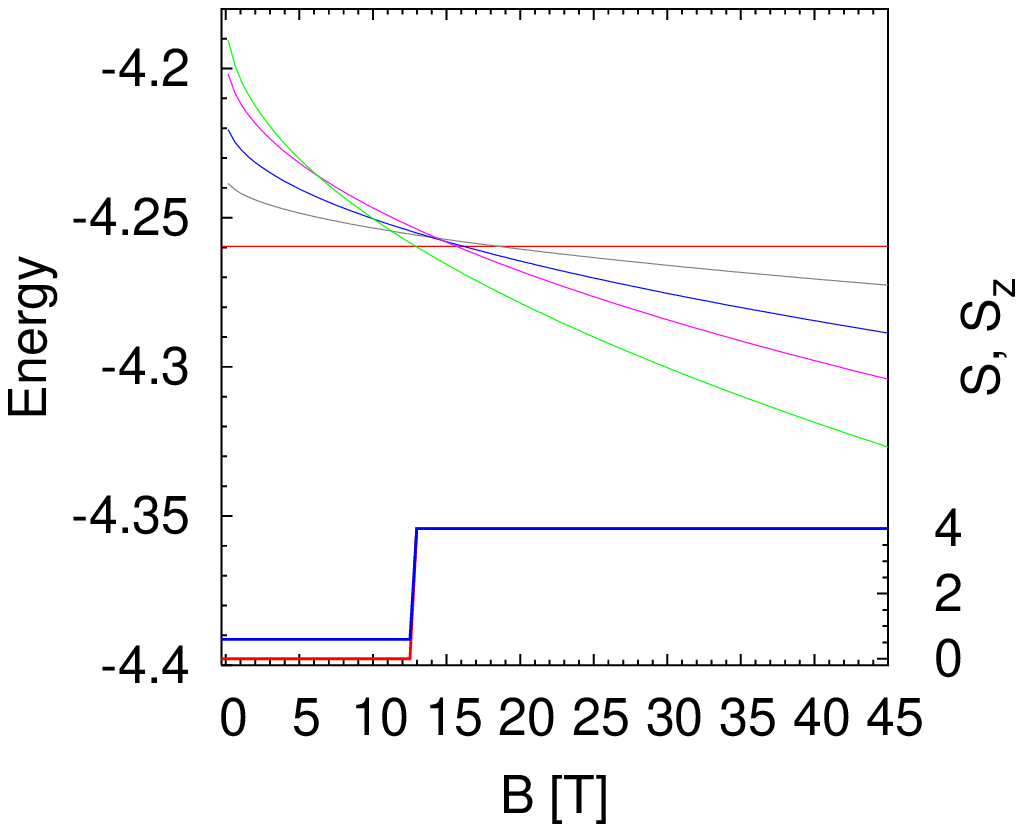}}
    \subfigure[Strong  MI. ($E_{MI}=0.02$)]{
    \label{fig-ch04-08c}
    \includegraphics[scale=0.47]{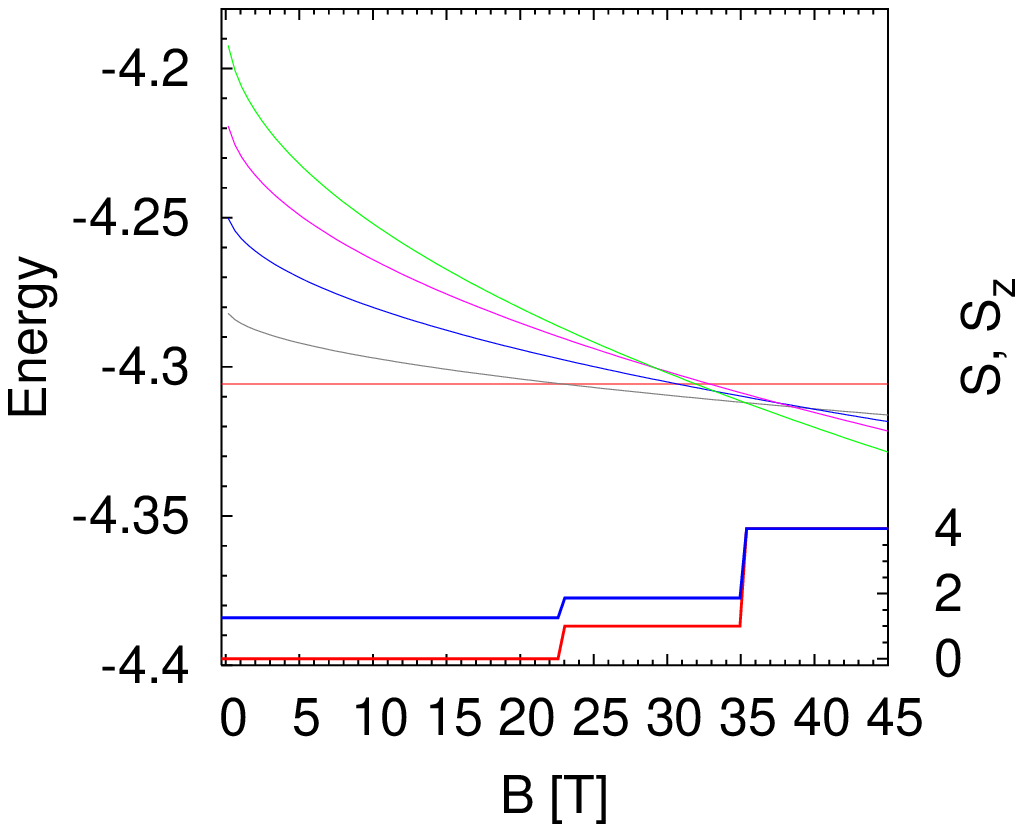}\hskip-1cm}
\caption{Stronger MI's bring another ground state into play
    ($S_z=1$) and the transition from the singlet to the polarized
    ground state becomes more gradual. }
\label{fig-ch04-08}
\end{figure}

If the strength of the 'rectangular wave' impurity becomes comparable to
the gap at $B\approx B_C$, $E_{MI}\approx E_g$, the situation at the
'singlet-to-polarized' transition changes dramatically. The excitation gap
closes and many states of different spin polarizations crowd around
the ground state. Even at zero temperature and in spite of lack of
anticrossings of states with different $S_z$ (i.e. $S_z$ is a good
quantum number again), the transition becomes
more gradual, when measured by $S_z$ of the ground state,
Fig. \ref{fig-ch04-08}c. 

Primarily, this is owing to the $S=1$ state which profits best from
the inhomogeneity. Keeping in mind its value of
$\krv=(1.07,0)\ell_0^{-1}$, this state seems to be a spin density wave
in $x$-direction pinned by the inhomogeneity potential. It is also
important that states with other spins are very near to it.

\begin{figure}
\begin{center}
\parbox{.4\textwidth}{
    \includegraphics[scale=0.5]{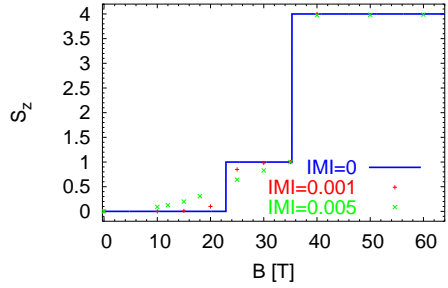}}
\hskip1cm\parbox{.3\textwidth}{
\caption{Strong perpendicular magnetic inhomogeneity, just as in
    Fig. \ref{fig-ch04-08}c, combined with inplane
    inhomogeneity (IMI). $\langle S_z\rangle$ 
    of the ground state. }\label{fig-ch04-07}}
\end{center}
\end{figure}

This $T=0$ transition can be again smoothened by an inplane
inhomogeneity, as shown in Fig. \ref{fig-ch04-07}. Here, the
transition $S_z=0\to 1$ becomes much more gradual than the transition
$S_z=1\to 4$. Reason for this is again that the inplane inhomogeneity
couples directly only states with $\Delta S_z=\pm 1$. Other quantities
than just $S_z$ (e.g. polarization) are
shown in Fig. \ref{fig-ch04-05}.

\begin{figure}
  \subfigure[Polarization $n_\up(x)/n(x)$.]{
    \label{fig-ch04-09a}
    \includegraphics[scale=0.3]{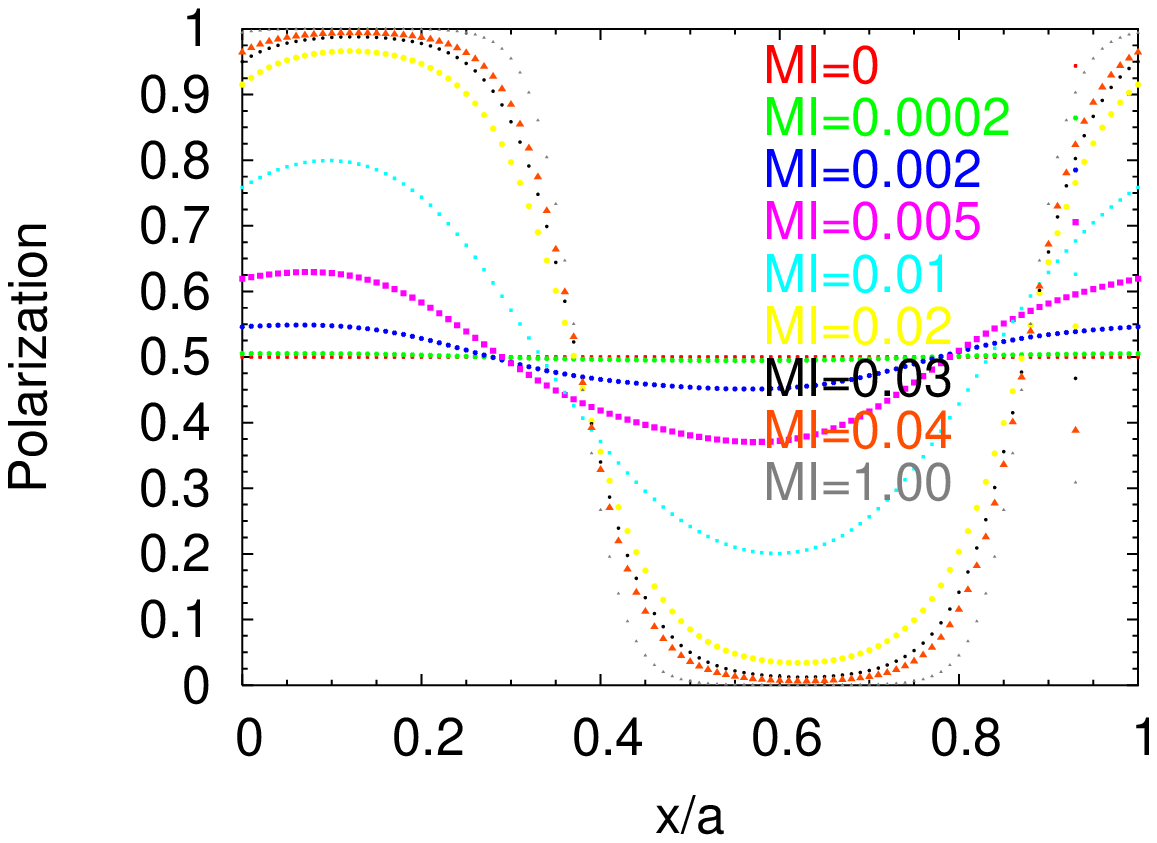}\hskip-2mm}
    \subfigure[Density.]{
    \label{fig-ch04-09b}
    \includegraphics[scale=0.3]{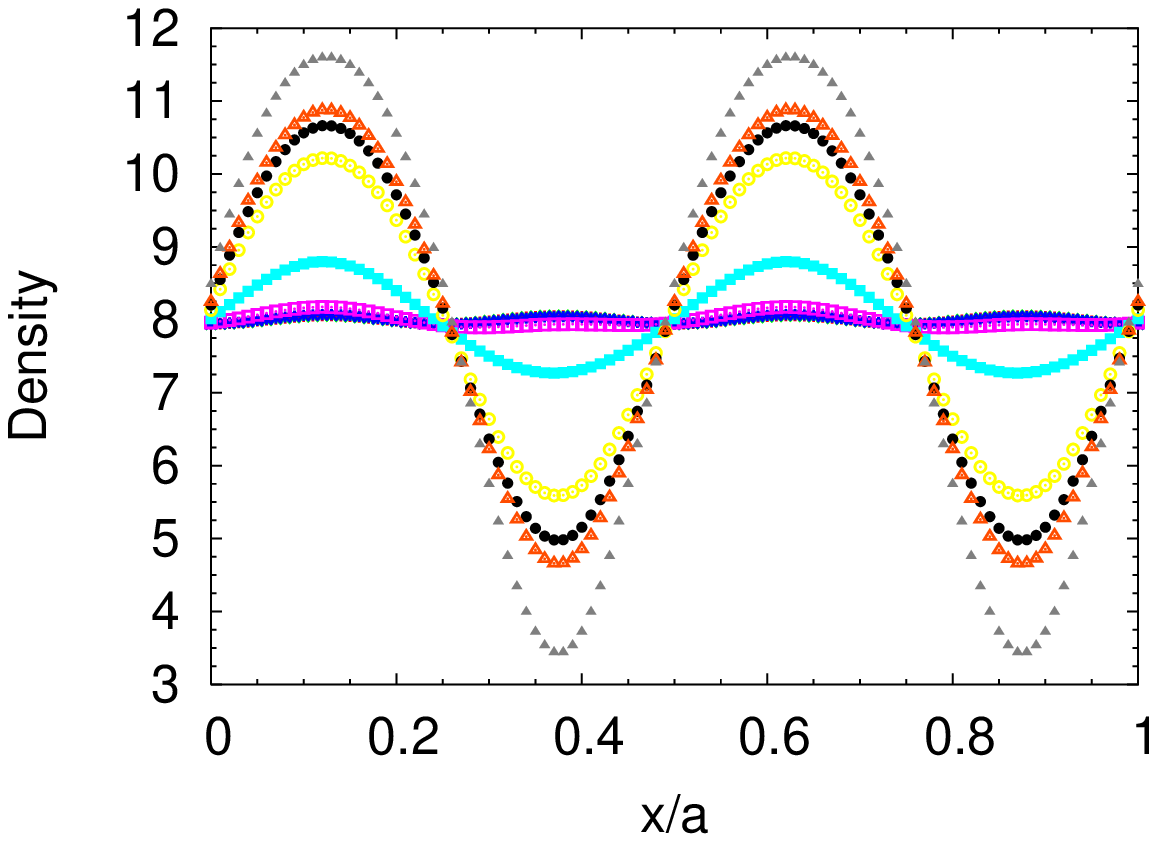}\hskip-2mm}
    \subfigure[Density $n_\dn(x)$.]{
    \label{fig-ch04-09c}
    \includegraphics[scale=0.3]{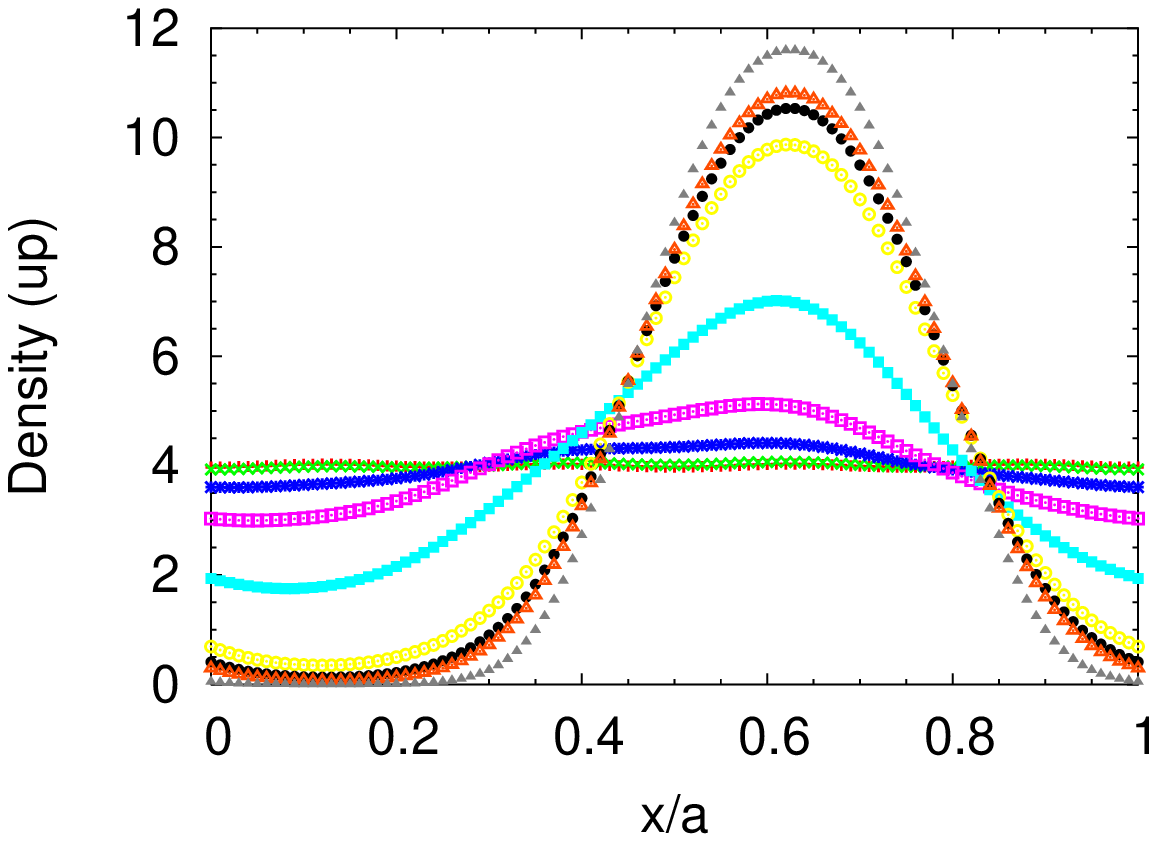}\hskip-2mm}
    \subfigure[Energy and $S$.]{
    \label{fig-ch04-09d}
    \includegraphics[scale=0.3]{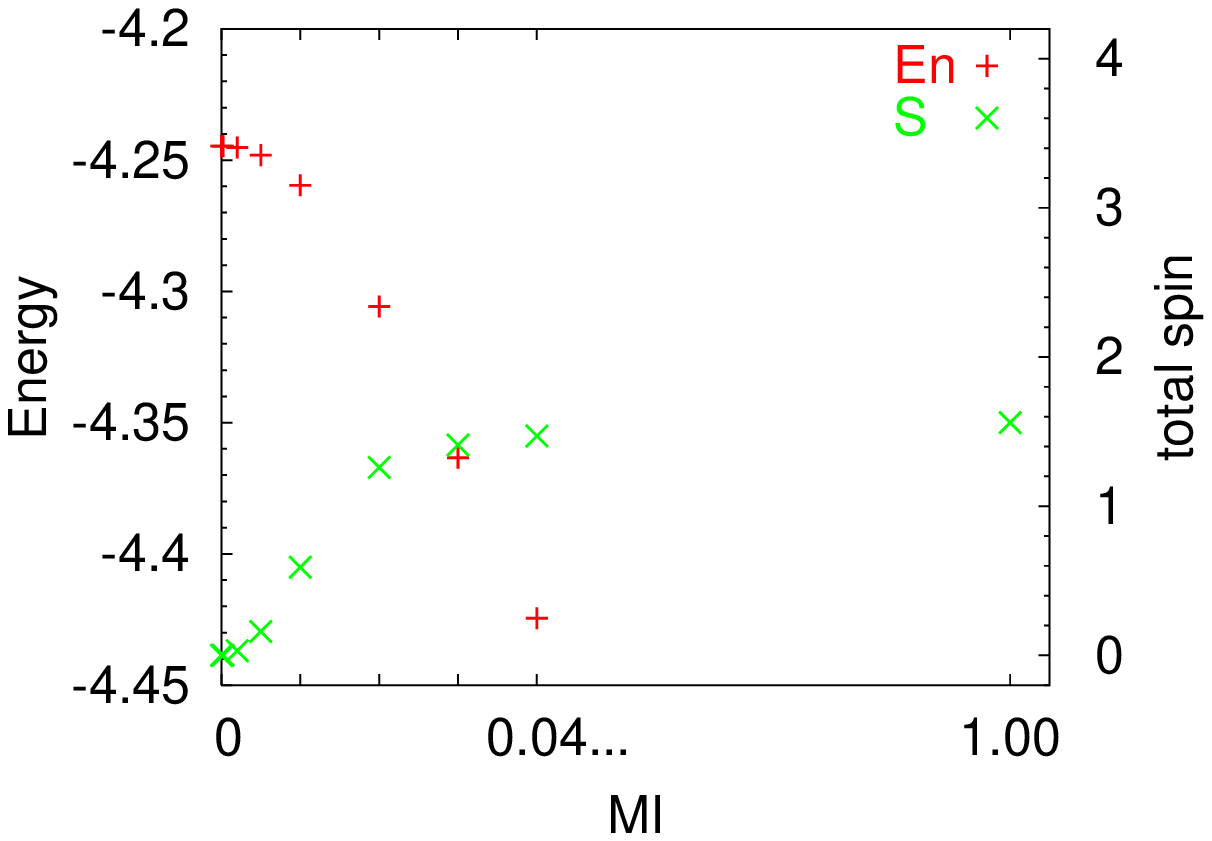}\hskip-12mm}
\caption{Destruction of the singlet state by very strong 
magnetic inhomogeneities: the system splits into two domains, one with
spin up, another with spin down and the electrons avoid the 'interface
region' (minima in the density).}
\label{fig-ch04-09}
\end{figure}

A strong magnetic inhomogeneity has also another quite pronounced
feature: the singlet-polarized ground state transition $B_C$ shifts to
higher magnetic fields, Fig. \ref{fig-ch04-08}. This effect is
considerably stronger than the shift to lower fields in case of the
inplane inhomogeneity (Fig. \ref{fig-ch04-06}). Origin of this shift to
higher $B$ is the decreasing energy of the singlet ground state,
Fig. \ref{fig-ch04-08} or Fig. \ref{fig-ch04-09}d.

Let us look at this issue more closely. Increasing $E_{MI}$, there is
no apparent transition (crossing) in the ground state of the $S_z=0$
sector (not shown). 
The total spin of the ground state increases smoothly from zero
and saturates around $S\approx 1.6$ 
for $E_{MI}\approx 0.02$, Fig. \ref{fig-ch04-09}d. 
Beyond this point, the label 'singlet ground state' becomes
inappropriate. At such values of $E_{MI}$, the polarization achieves the
maximum variation between zero and one, Fig. \ref{fig-ch04-09}a. The
eight electrons, four with spin up, four with spin down, split into
two nearly independent groups: the 
spin up (down) electrons gather in the region
where $g_1(x)$ is positive (negative), see (\ref{eq-ch04-02}). Such a
state where e.g. no spin up electrons occur in the 'wrong region'
(Fig. \ref{fig-ch04-09}c, $E_{MI}=0.02$) is no longer even
remotely related to the homogeneous incompressible state, even though
it has $S_z=0$. Rather, we could interpret it as two $\nu=\ot$ systems
living next to each other: one with spin up, another with spin
down. The strong spatial variation of density in this system,
Fig. \ref{fig-ch04-09}b, indicates that electrons try to avoid the
'interface region'; an alternative point of view is to compare the
'spin-down domain region' (seen in the polarization,
Fig. \ref{fig-ch04-09}a) with the density of spin down electrons,
Fig. \ref{fig-ch04-09}c. However, we must always be aware that we
investigate only a finite system which is too small to observe the 'inside'
of a domain where we expect the density to be constant. In a
sufficiently large system, the maximum in Fig. \ref{fig-ch04-09}c
should spread into a plateau. Therefore, also conclusions  about the
interface region must be interpreted with caution.

\subsubsection{Quantities to observe}

Polarization is the most natural quantity to study when
looking for domains of polarized and singlet states. Nevertheless,
it could be useful to search for other observables as they might bring
some more information on what is happening in the states.

Here, we suggest to study the {\em local} expectation values (or densities) of
otherwise 'global' operators such as $S_z$ or $S^2$. These are defined
by
$$
  S_{x,z}(\vek r) = S_{x,z} \otimes n(\vek r)\,, \qquad
  S^2 (\vek r) = S^2 \otimes n(\vek r)\,,\mbox{ where}\qquad
  n(\vek r) = \sum_{i=1}^{N_e} \delta(\vek r-\vek r_i)\,
$$
and they should be plotted in the form $S_z(\vek r)/n(\vek r)$. Their
meaning is the following: Imagine an $n$-electron state which is an
equal-weight superposition of two states: one localized in the region
$0<x<a/2$ which is fully spin polarized ($S_1=n/2$) and another
localized in $a/2<x<a$ which is a spin singlet. This state is $S_z=n/4$,
yet its $S_z(x)/n(x)$ is equal to $n/2$ or $0$ in the two respective
regions.

\begin{figure}
  \subfigure[Spectrum and $\langle S_z\rangle$, $\langle S\rangle$ of
    the ground state.]{
    \begin{tabular}{c}
    \includegraphics[scale=0.45]{figs/ch04/ch04-fig08a.eps} \\
    \includegraphics[scale=0.45]{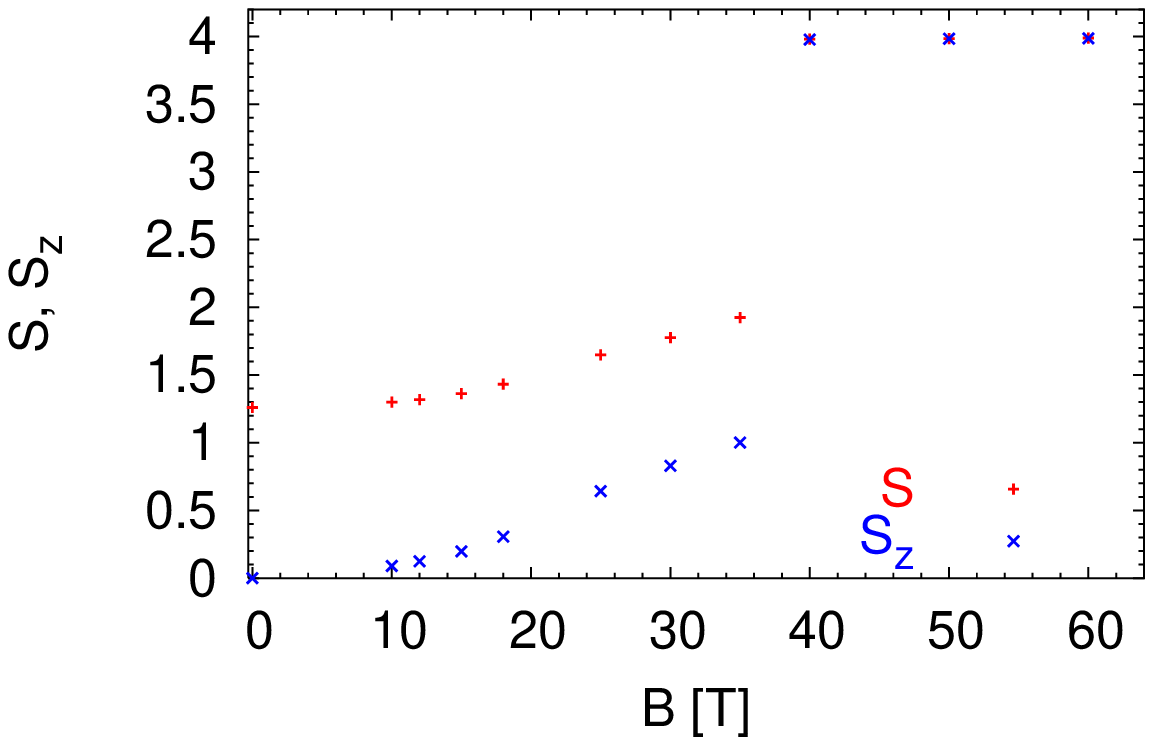}
    \end{tabular}
    \label{fig-ch04-05a}
    }
  \subfigure[The ground state in different quantities.]{
    \begin{tabular}{cc}
    \includegraphics[scale=0.4]{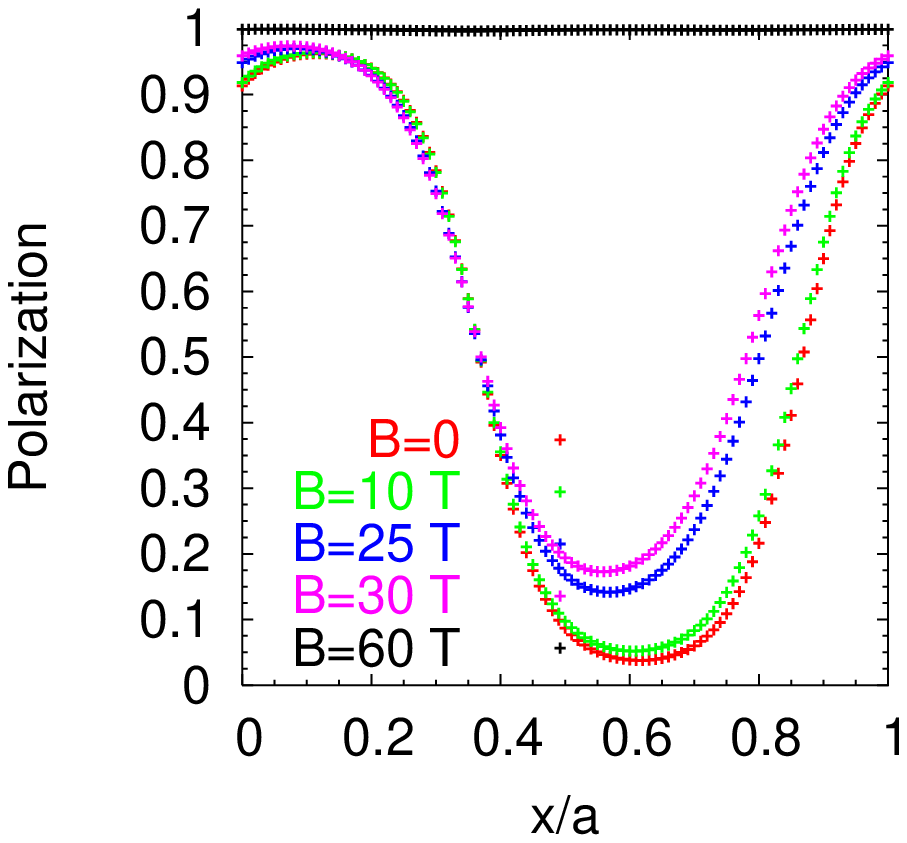} &
    \includegraphics[scale=0.4]{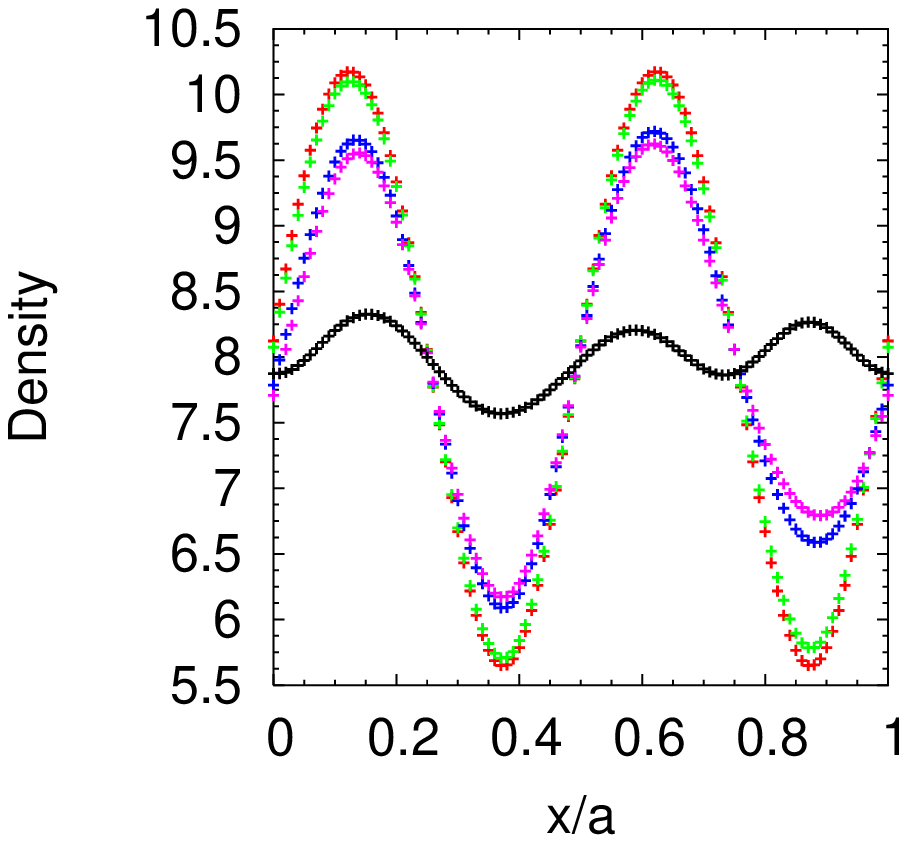} \\
    \includegraphics[scale=0.4]{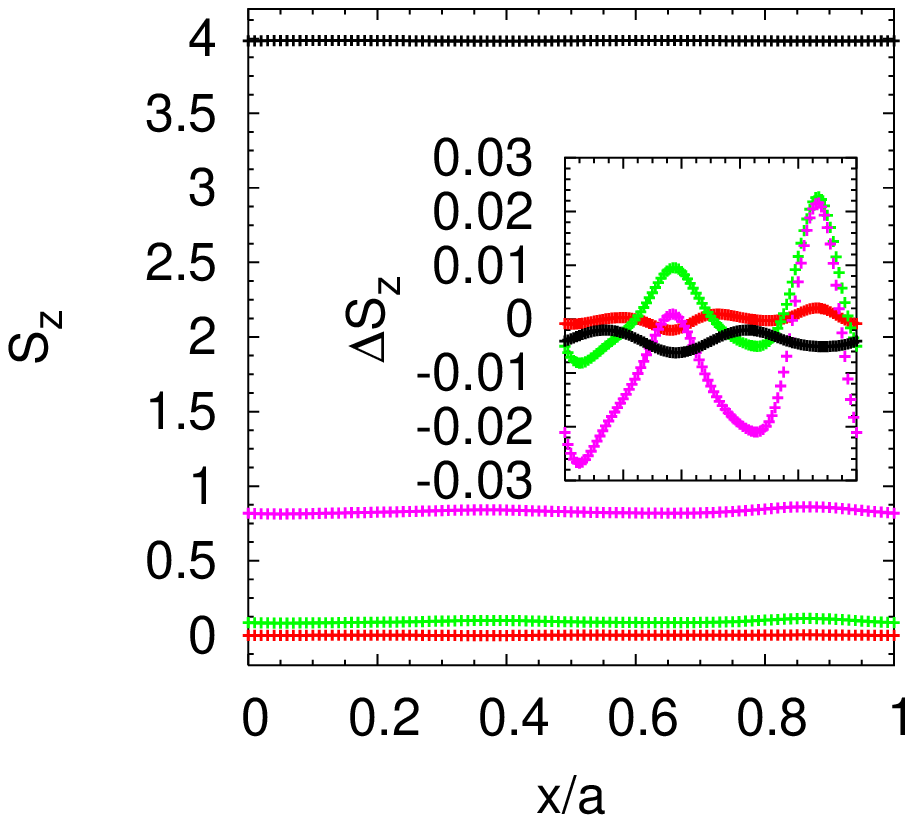} &
    \includegraphics[scale=0.4]{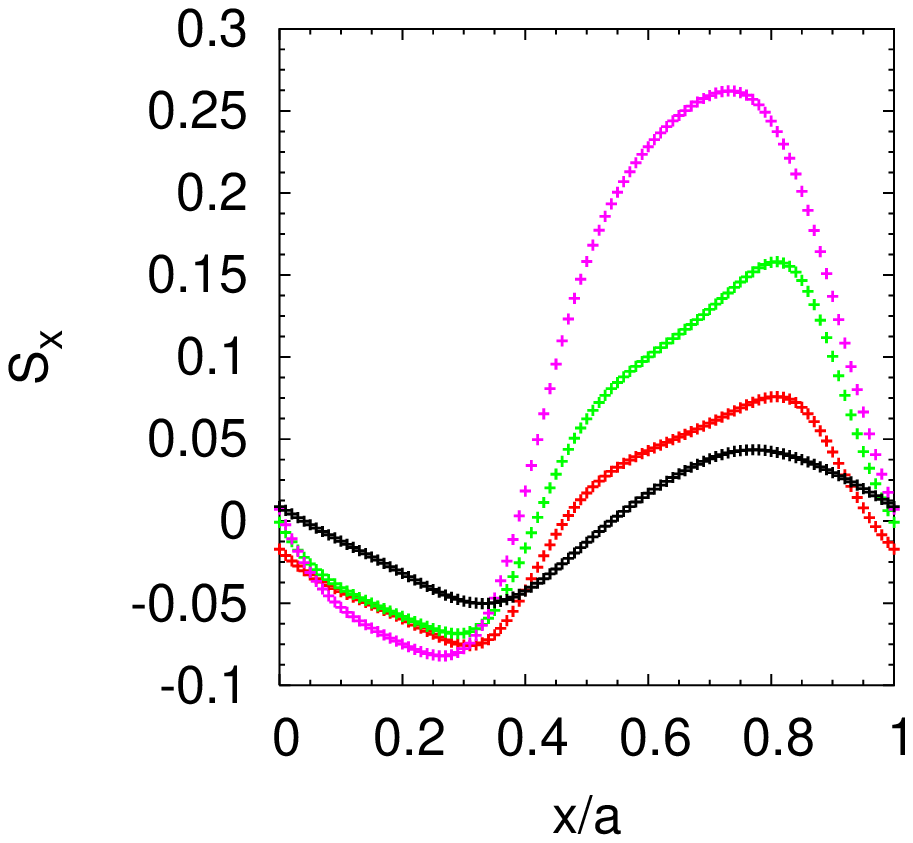} \\
    \end{tabular}
    \label{fig-ch04-05b}
    }
\caption{Response to strong inhomogeneity of the form in
    (\ref{eq-ch04-03}). $E_{MI}=0.02$, $E_{IMI}=0.005$. }
\label{fig-ch04-05}
\end{figure}

What these quantities reveal is demonstrated in the case of a strong
magnetic inhomogeneity, both perpendicular and inplane,
Fig. \ref{fig-ch04-05}. The low-energy part of the spectrum does not
change considerably, Fig. \ref{fig-ch04-05}a, even though the singlet
state is separated almost completely into a spin-up and a spin-down
domain by the inhomogeneity, seen in the polarization,
Fig. \ref{fig-ch04-05}b. Note that again the states near the
transition have smaller variations in the polarization than the
'inhomogeneous singlet-state'.

The lower two plots of Fig. \ref{fig-ch04-05}b show $S_z(x)/n(x)$ and
$S_x(x)/n(x)$. Obviously, $S_z$ stays quite constant with $x$, at
least on the scale ranging from $S_z=0$ (singlet) to $S_z=4$ (fully
polarized). Albeit polarization (or relative density of spin down
electrons) varies strongly, $S_z(x)$ remains nearly 
constant. This indicates that the
state does not really separate into domains of locally different
$S_z$. Observation of the quantity $S^2(x)$ (not shown) 
points in the same direction.

Local expectation values of $S_x$ indicate that states near the
transition are more susceptible to inplane inhomogeneities. At any
$B$, this response is much stronger than for perpendicular
inhomogeneities. The following picture explains this behaviour: 
if we imagine a 'classical' spin
vector pointing in $z$-direction and accept that it fluctuates by
a small angle $\Delta\vp$, then $S_z\propto \cos\Delta\vp\approx 1$
whereas $S_x\propto \sin\Delta\vp\approx \Delta \vp$.

\subsubsection{Different geometries of the inhomogeneity}

Disregarding entropy, it is unlikely that a domain state will be the
ground state in a homogeneous system. If it is an excitation we can
hope to encourage it energetically by including a suitable inhomogeneity
like $H_{MI}$ in (\ref{eq-ch04-02}). However, we do not {\em a priori} know
what 'suitable' means. So far, we divided the system into two
{\em equal} parts by $H_{MI}$.

\begin{figure}
  \subfigure[Polarization.]{
    \label{fig-ch04-11a}
    \includegraphics[scale=0.5]{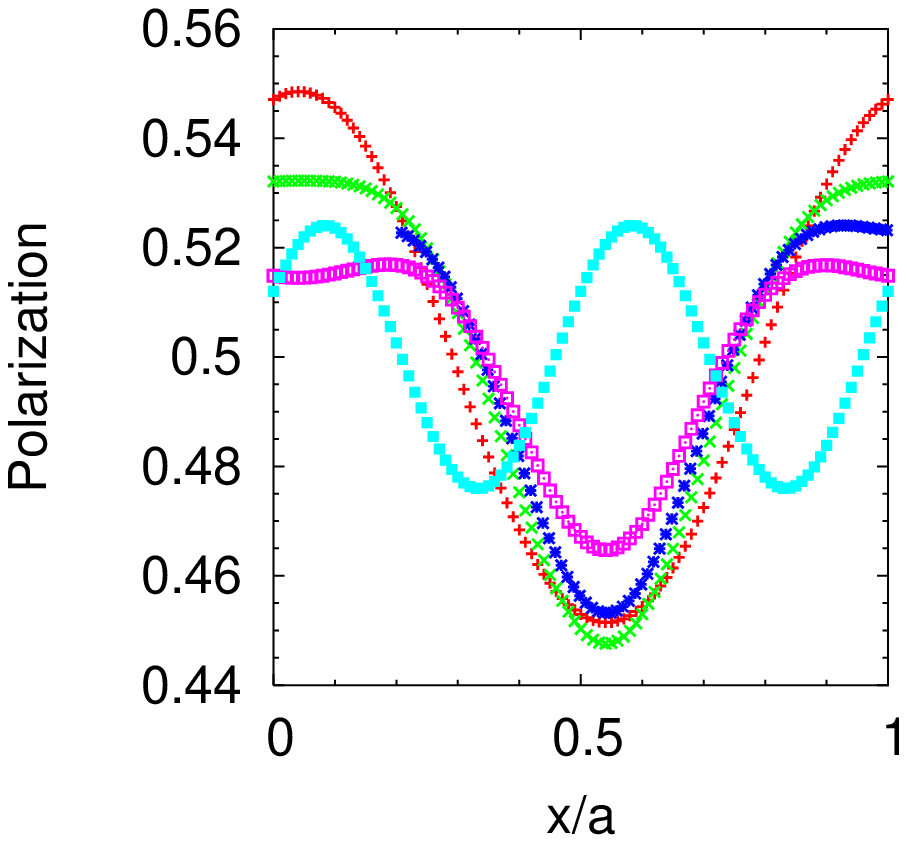}}
    \subfigure[Density.]{
    \includegraphics[scale=0.5]{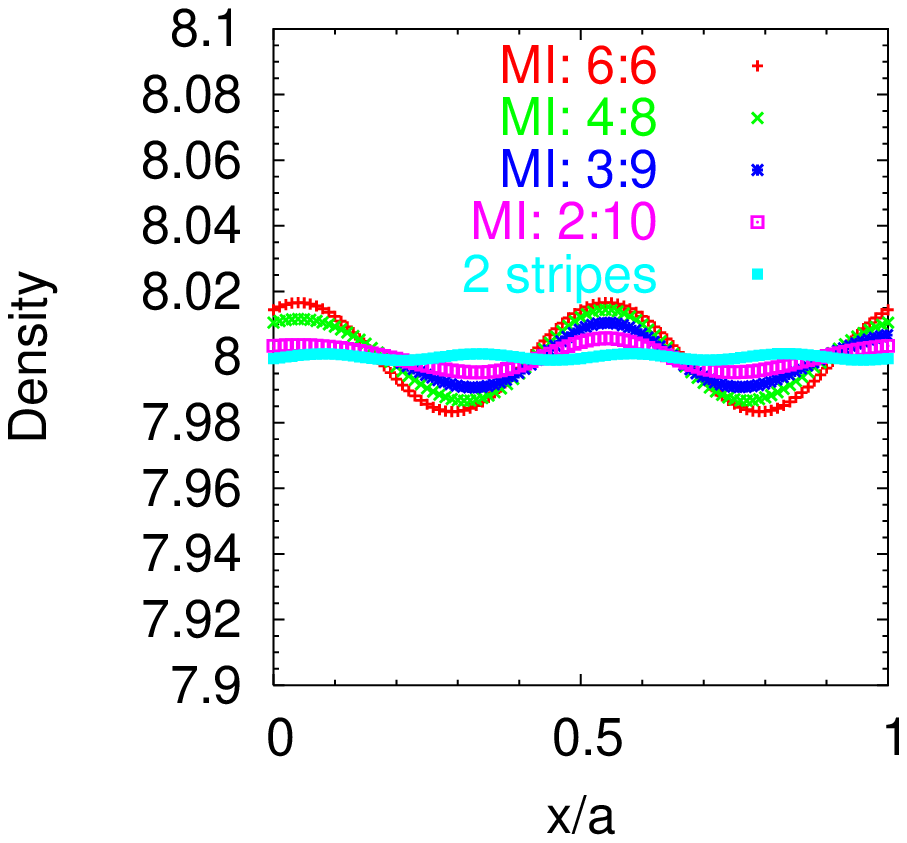}
    \label{fig-ch04-11b}}
  \subfigure[Various inhomogeneities $H_{MI}$ used.]{
  \hbox to 3cm{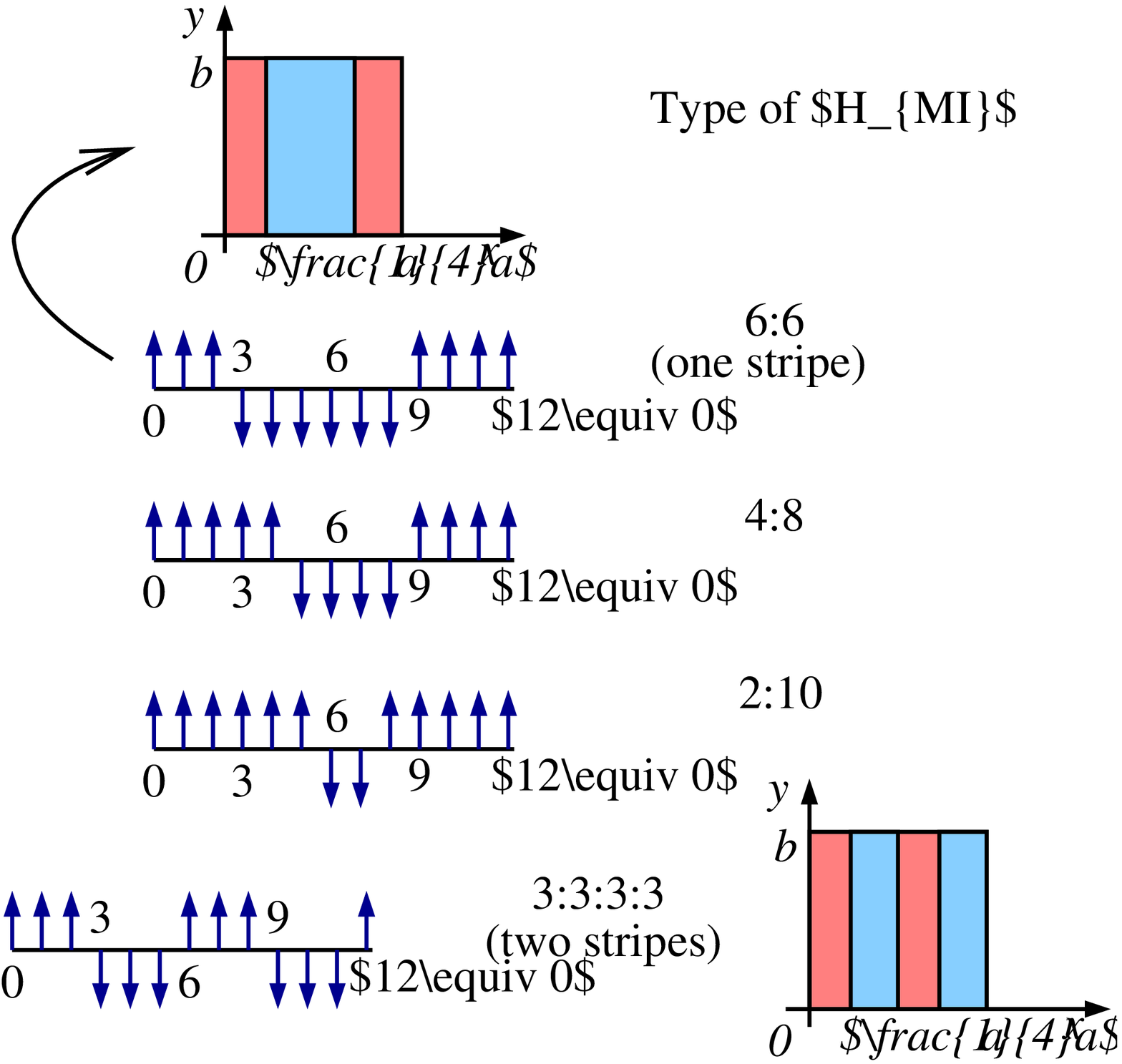\hskip-2cm}
  \label{fig-ch04-11c}}
\caption{Response of the singlet to various weak magnetic
    inhomogeneities. The inhomogeneity is similar to the one in 
    (\ref{eq-ch04-02}) but it divides the system into two areas (stripes) of
    various ratios ($6:6$ through $2:10$) or it consists of two
    stripes with size $3:3:3:3$.}
\label{fig-ch04-11}
\end{figure}

How the singlet state responds to inhomogeneities of different forms
is shown in Fig. \ref{fig-ch04-11}. Different lines correspond to the
'rectangular wave' inhomogeneities 
with different ratios of the 'plus' and 'minus'
parts. All these inhomogeneities are thus a single stripe of various
width (per elementary cell) parallel to $y$ . 

Also, response to $H_{MI}$ consisting of two stripes is shown
(i.e. 'rectangular wave' with half period).

Responses are basically very similar to each other and it seems by
having focused on $H_{MI}$ of the form of (\ref{eq-ch04-02}) we did
not choose a particularly clumsy one. One particularly interesting
information which can be extracted from Figure \ref{fig-ch04-11} is
that the polarization response is always at least an order of magnitude
larger than in the density, $10\%$ against $0.3\%$ in the present
case. This confirms the conclusion of Subsection
\ref{pos-ch03-05}: even though singlet incompressible
states try to maintain constant density, they can be fairly easily
polarized.

\begin{figure}
\subfigure[Energies.]{\label{fig-ch04-12a}
      \includegraphics[scale=.6]{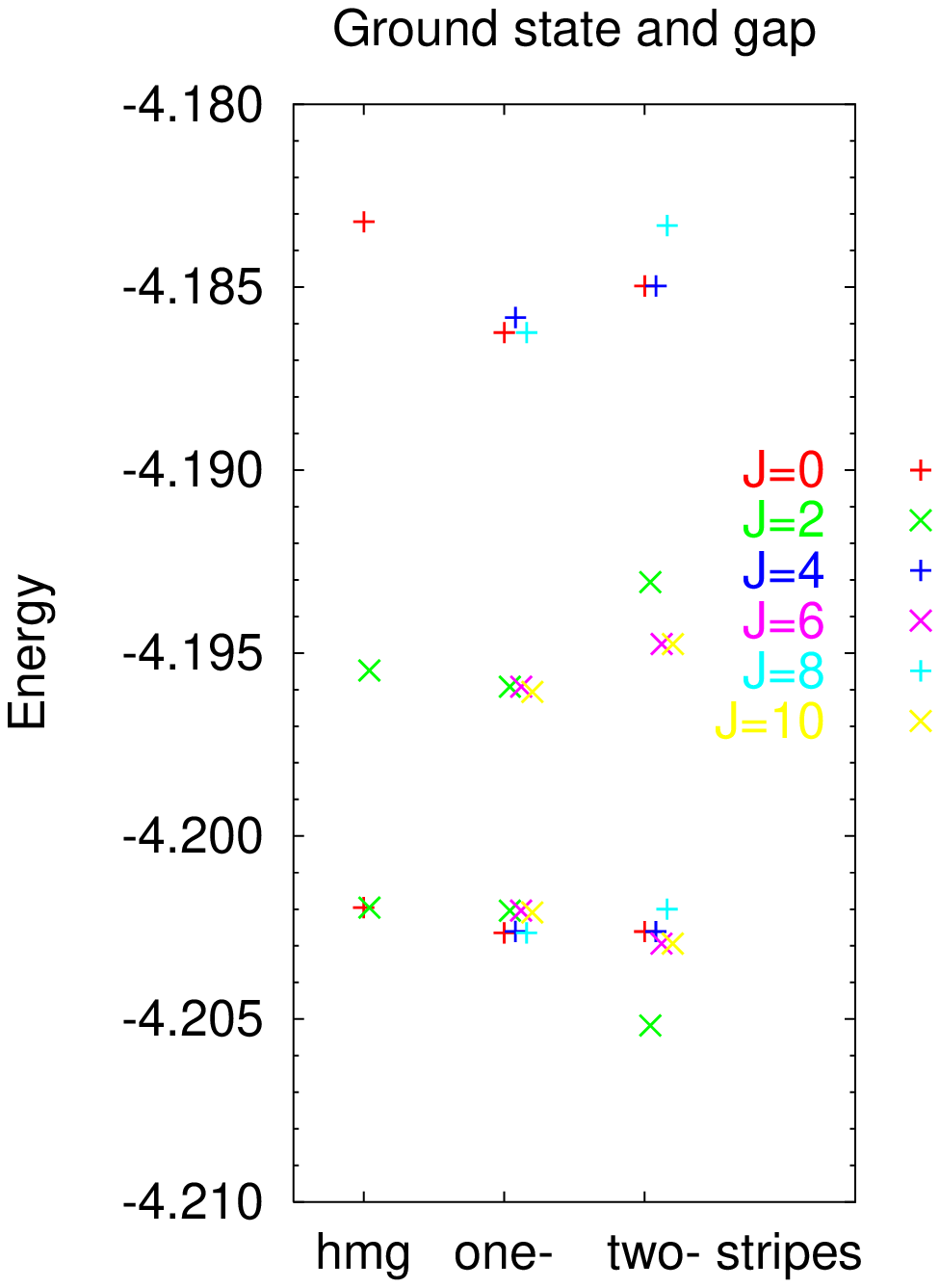}}
\subfigure[Polarizations.]{\label{fig-ch04-12b}
\raise4cm\hbox{\begin{tabular}{cc}
      \includegraphics[scale=0.5]{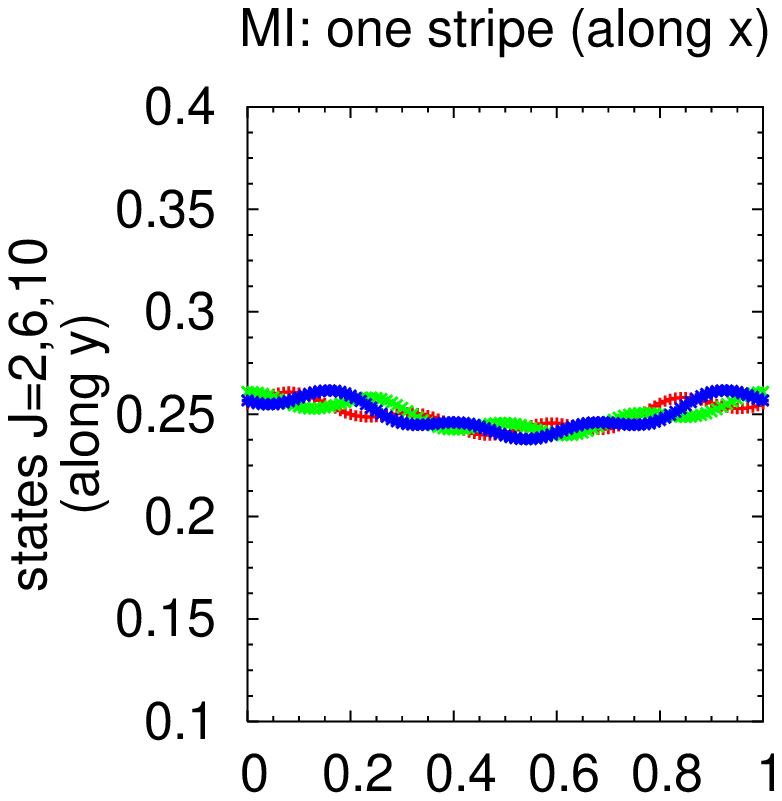} &
      \includegraphics[scale=0.5]{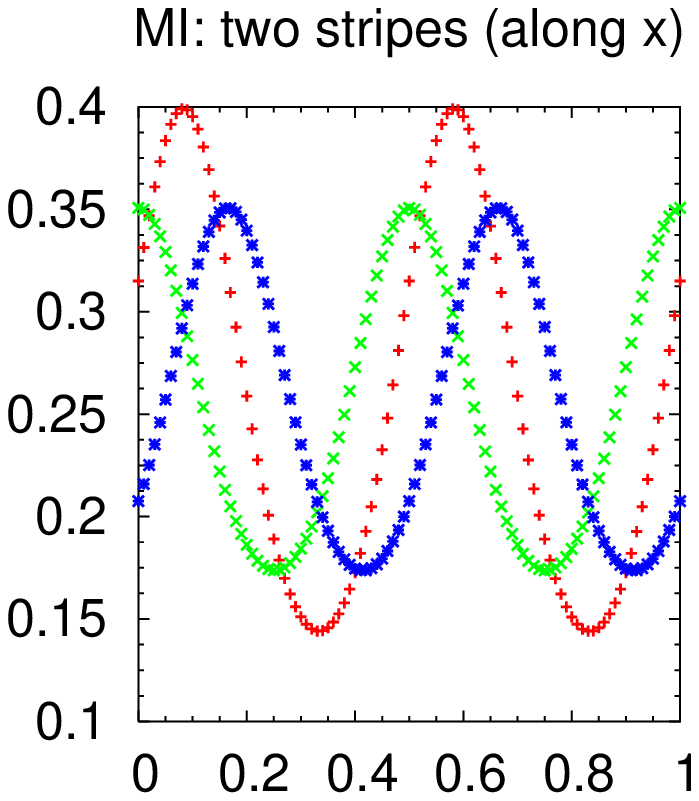} \\
      \includegraphics[scale=0.5]{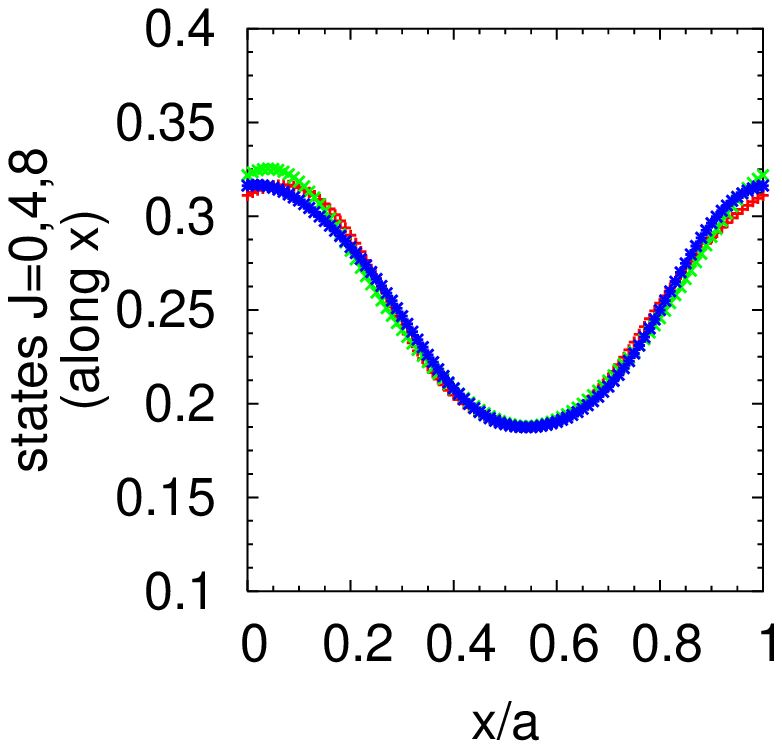} &
      \includegraphics[scale=0.5]{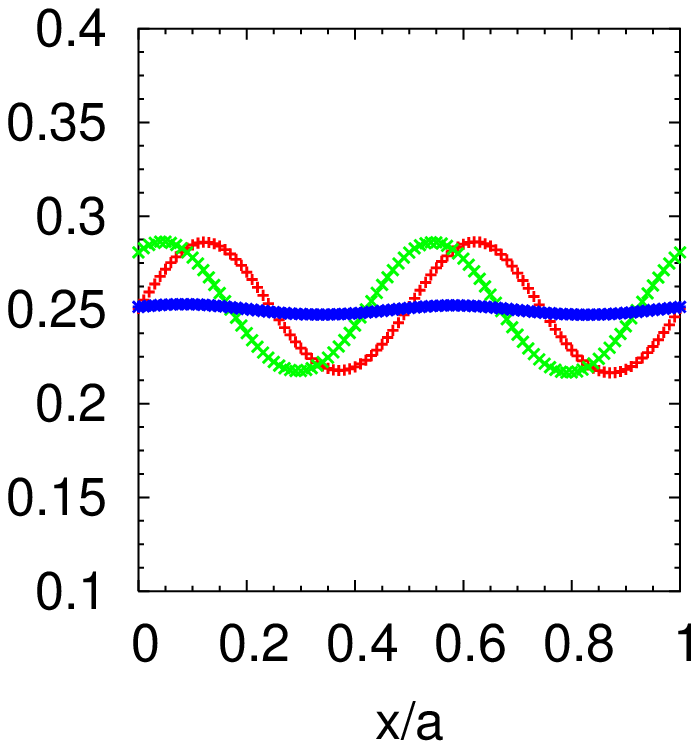} \\
\end{tabular}}
}
\caption{The half-polarized states ($S=2$) and their response to
a magnetic inhomogeneity of the form of one or two stripes.}
\label{fig-ch04-12}
\end{figure}

A good example of the influence of the form of the inhomogeneity are the
half-polarized states, Fig. \ref{fig-ch04-12}. The lowest level in the
$S=2$ sector is six-fold degenerate (factor of three from the
centre-of-mass and factor of two from the relative part). The two
states are mirror images of each other with respect to the diagonal of
the elementary cell. We
split them into two groups $J=2,6,10$ and $J=0,4,8$ (within each group
the states differ only by the center-of-mass part) and subject each
group to one-stripe and two-stripe inhomogeneities,
Fig. \ref{fig-ch04-11}c.

One group ($J=2,6,10$) responds strongly to the two-stripe $H_{MI}$
and is left almost unchanged by the one-stripe $H_{MI}$, upper row in
Fig. \ref{fig-ch04-12}b. Nearly the opposite is true for the other
group. It gives a clear picture of the structure of these states. They
are spin density waves with two periods in one direction and one
period in another direction. This is in full agreement with the
spin-spin correlation functions, not shown here. The
conclusion is also underlined by the markedly lowered energy of the
$J=2$ state when it is addressed by the two-stripe inhomogeneity,
Fig. \ref{fig-ch04-12}a. This is a practical demonstration of one
spin-density-wave state selected by an impurity from a degenerate
manifold.

\subsubsection{Transition at nonzero temperature}

Regardless of how intensely we try to help an eventual domain state to
become the ground state, it may still be, that it is hidden among the
excitations. Therefore we may try to take  the excited states into account
by means of thermal averaging. 

The strong impurity mode ($E_{MI}=0.02$) was chosen for this
study. Three-fold degeneracy in center-of-mass of the incompressible
ground states is lifted. The level splitting is however still smaller than
the incompressibility gap, compare the black and grey points in the upper
plot of Fig. \ref{fig-ch04-05}a around $B=10\unit{T}$.

\begin{figure}
  \begin{center}
  \begin{tabular}{ccc}
  \includegraphics[scale=0.35]{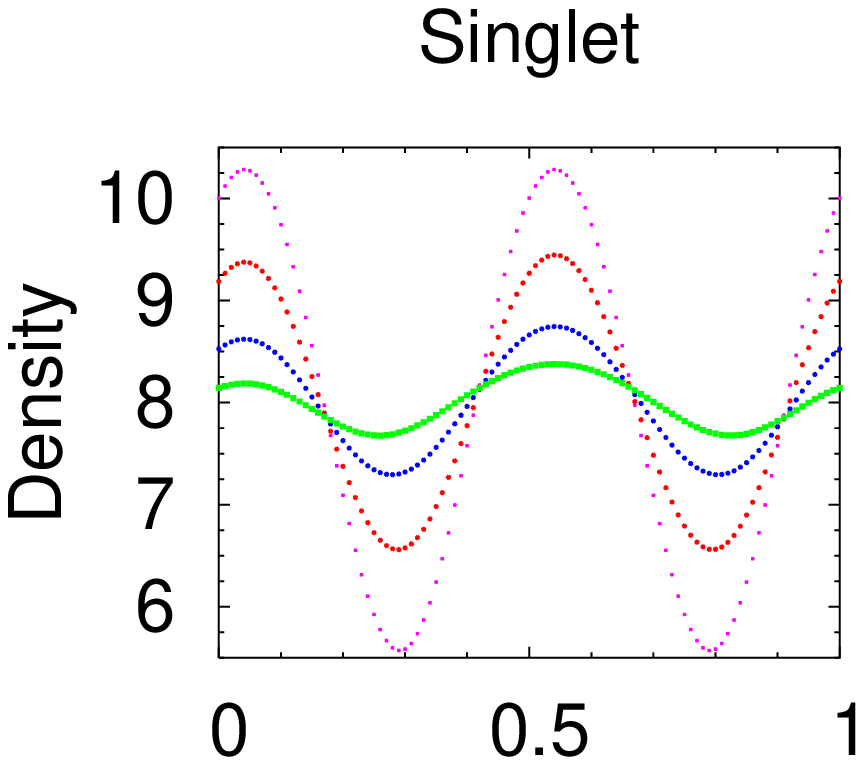} &
  \includegraphics[scale=0.35]{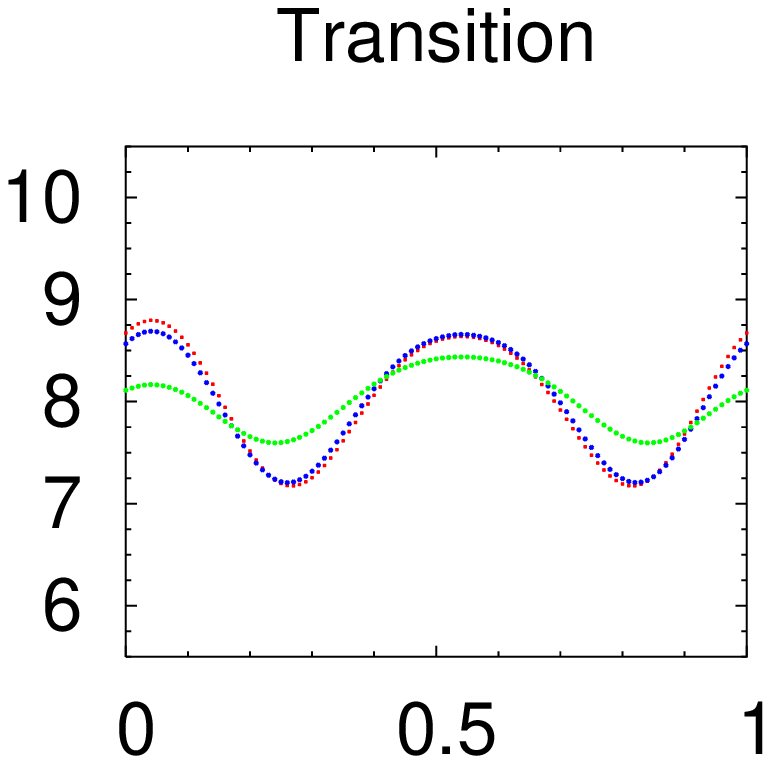} &
  \includegraphics[scale=0.35]{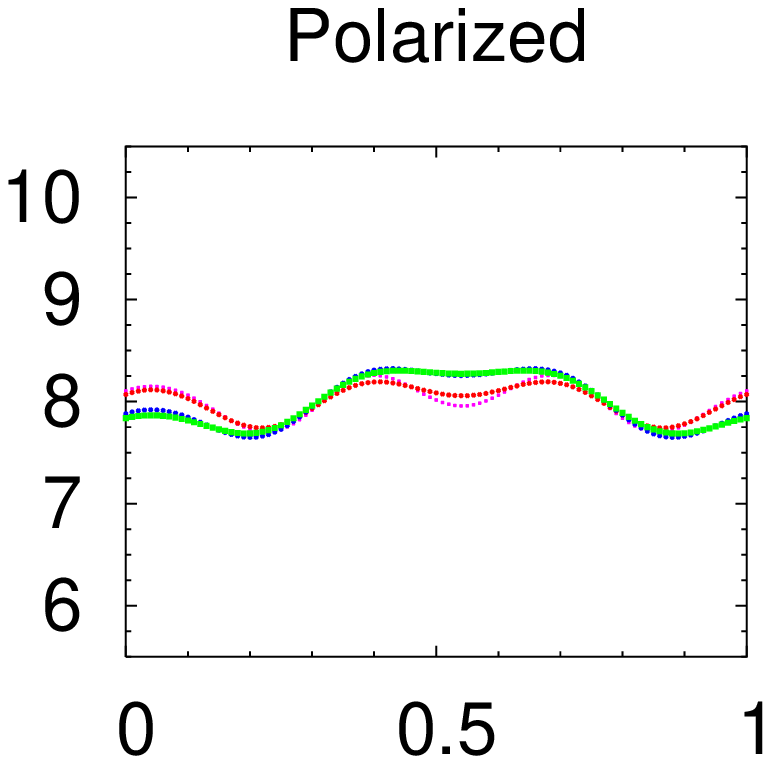} \\
  \includegraphics[scale=0.35]{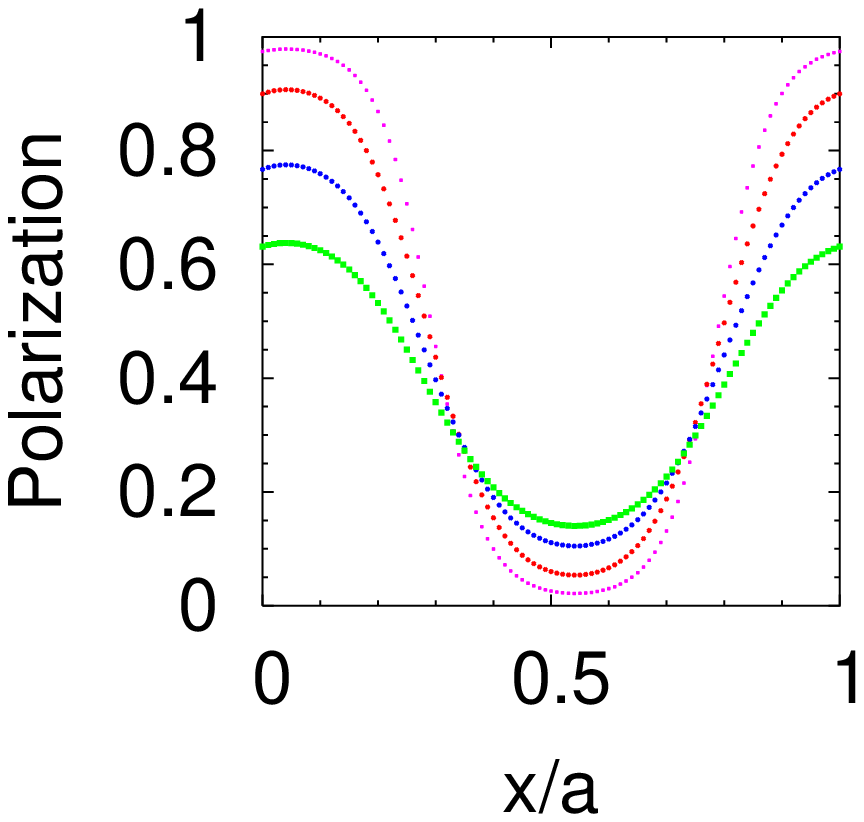} &
  \includegraphics[scale=0.35]{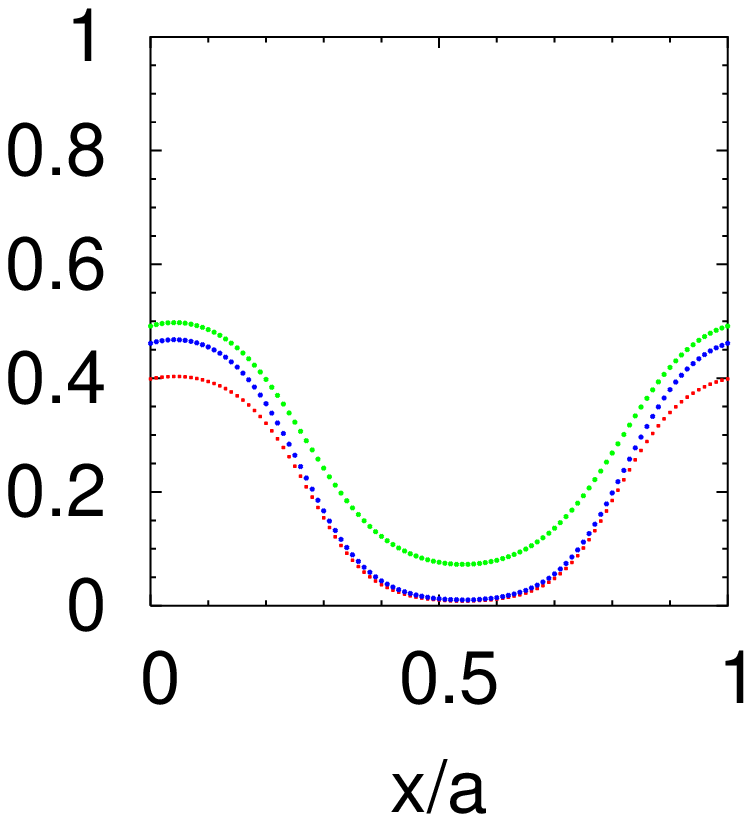} &
  \includegraphics[scale=0.35]{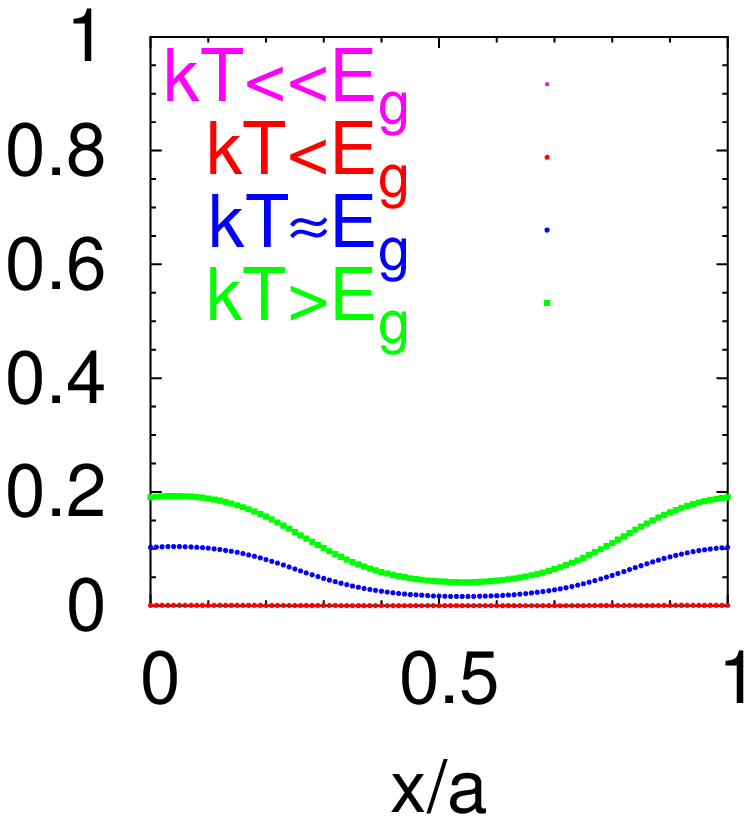} \\
  (a) & (b) & (c) \\
  \end{tabular}
  \end{center}
\caption{Density and polarization at different temperatures: filling
  factor $2/3$ with magnetic inhomogeneity (corresponding to
  Fig. \ref{fig-ch04-08}c). {\em Left to right:} before, at and after
  transition, i.e. $B\to 0$, $B\approx B_c$ and $B\to\infty$.}
\label{fig-ch04-10}
\end{figure}

Various temperatures were considered: $kT\ll E_g$ means than we do not
average even over all states of the originally degenerate
triple. Knowing that $E_g$ means the gap energy at $B\to 0$ or
$B\to\infty$, the other temperatures shown in Fig. \ref{fig-ch04-10} are
self-explaining.  
  
Judging by polarization $n_\dn(x)/n(x)$, the state at the transition approaches
a situation which we could call 'domain'. In the middle ($x/a\approx
0.5$), the polarization drops to zero and only spin up electrons are
present. In the other region ($x/a\approx 0\equiv 1$), polarization is
about $0.5$, meaning that the number of spin up as spin down electrons in
this area is the same. 

We should note though that an inhomogeneity which is strong enough to
produce such nice 'domains' is also strong enough to change the
originally incompressible singlet state completely,
Fig. \ref{fig-ch04-10}a. In other words, the response of the system at the
transition is still weaker than the response of the singlet state. This
manifests that spontaneous build-up of domains is not very likely
within this model. 

In the following Section we will try to suggest slightly different
models which may put us on the trace of states which exhibit
nontrivial behaviour at the transition between the incompressible singlet and
polarized ground states.

\subsection{Systems with short range interaction}


\label{pos-ch04-06}

\begin{figure}
\begin{center}
\subfigure[Coulomb interaction.]{\label{fig-ch04-15a}
  \includegraphics[scale=0.5]{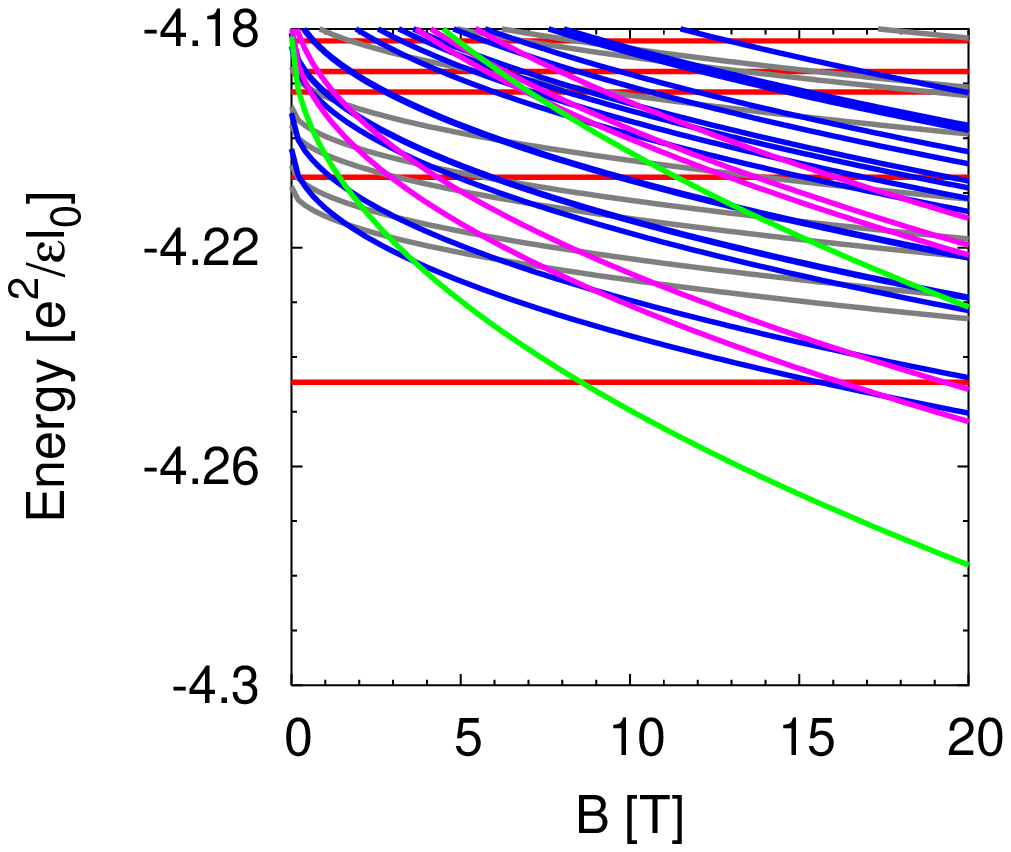}}
\subfigure[Short-range interaction.]{\label{fig-ch04-15b}
  \includegraphics[scale=0.5]{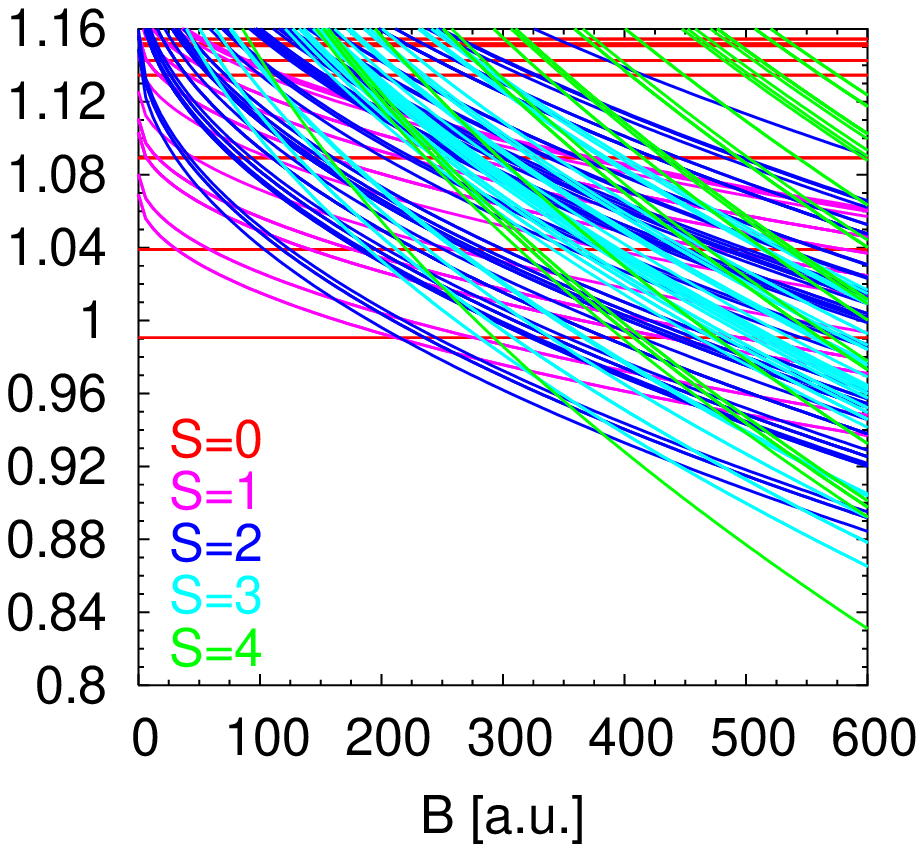}}
\end{center}
\caption{Spectrum of a homogeneous system with Zeeman splitting (8
  electrons, $\nu=\tt$).}
\label{fig-ch04-15}
\end{figure}

As far as the transition between singlet and polarized ground state is
concerned, the most obvious feature of the $\nu=\tt$ Coulomb-interacting
systems is the energy 'gap' which separates the two degenerate ground
states from excited ones even at the very crossing,
Figs. \ref{fig-ch04-15}a, \ref{fig-ch04-01}. In the previous
section we demonstrated that this picture may change when fairly
strong magnetic inhomogeneities are applied, Fig. \ref{fig-ch04-08}. 
We can cause a similar drastic change by replacing
the Coulomb by the short-range interaction, Fig. \ref{fig-ch04-15}b.

\begin{figure}
\subfigure[Homogeneous.]{\label{fig-ch04-13a}\hskip-1cm
  \includegraphics[scale=0.45]{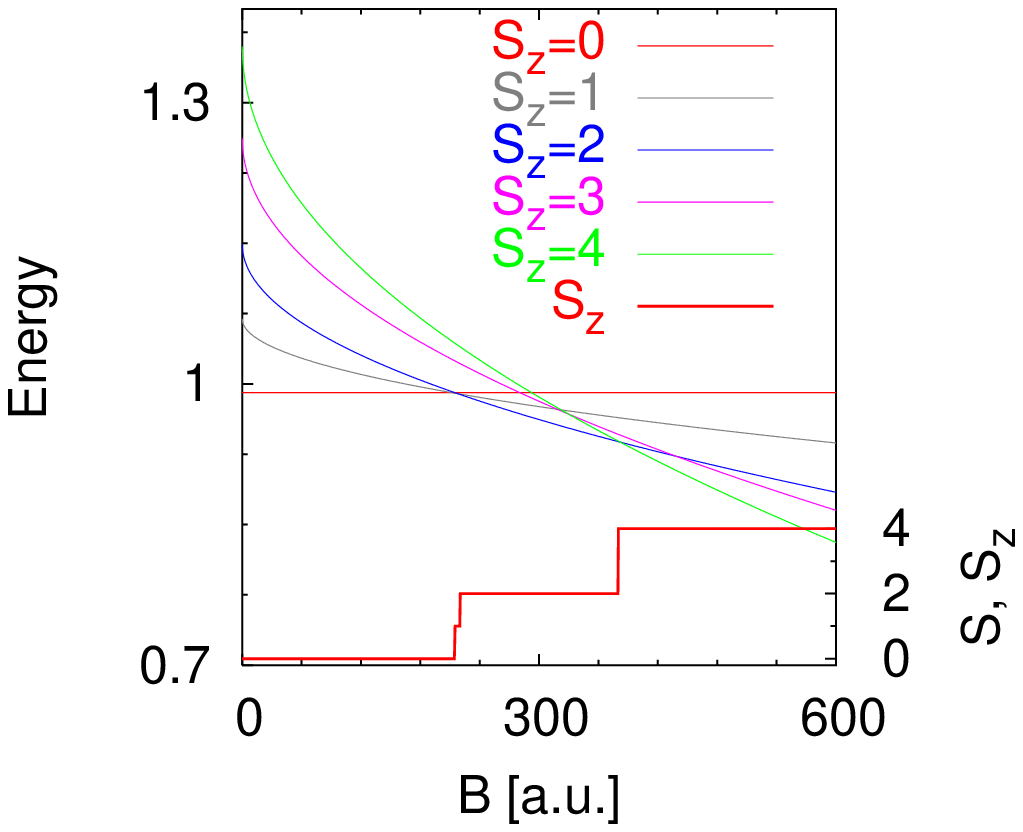}\hskip-1cm}
\subfigure[Intermediate inhomogeneity ($E_{MI}=0.01$).]{\label{fig-ch04-13b}
  \includegraphics[scale=0.45]{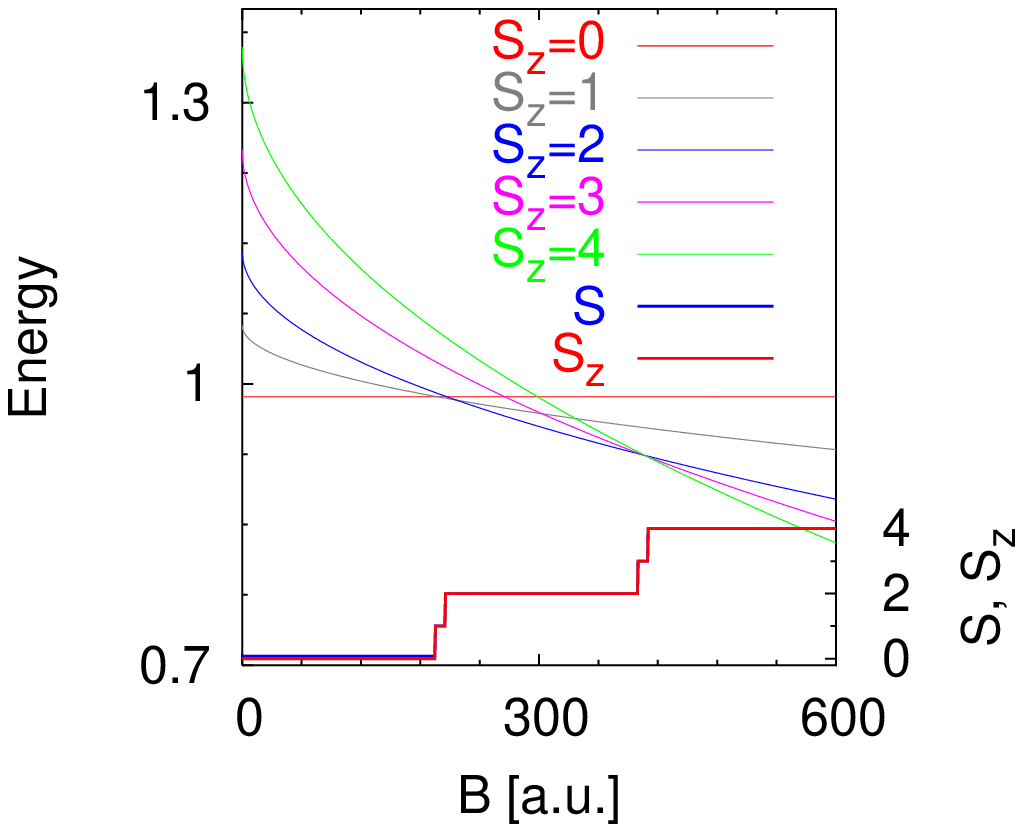}\hskip-1cm}
\subfigure[Strong inhomogeneity ($E_{MI}=0.02$).]{\label{fig-ch04-13c}
  \includegraphics[scale=0.45]{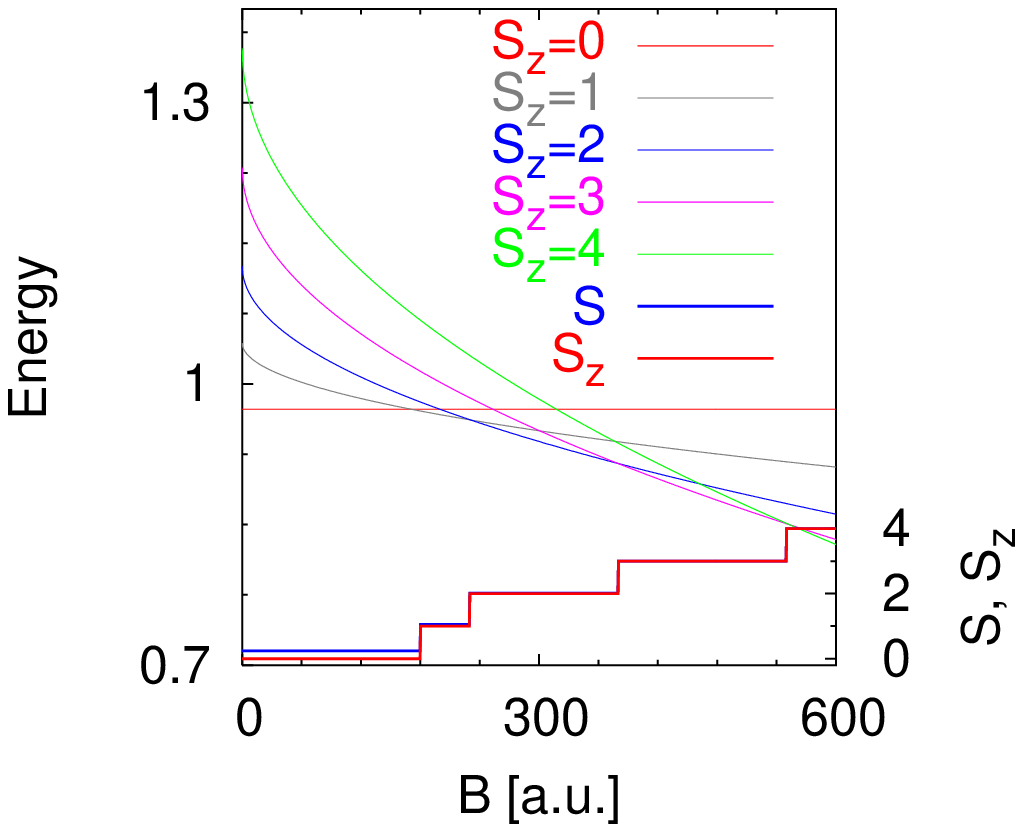}\hskip-1cm}
\subfigure[Spin at finite temperature, cf.
  Fig. \ref{fig-ch04-13}c.] {\label{fig-ch04-13d}
\includegraphics[scale=0.5]{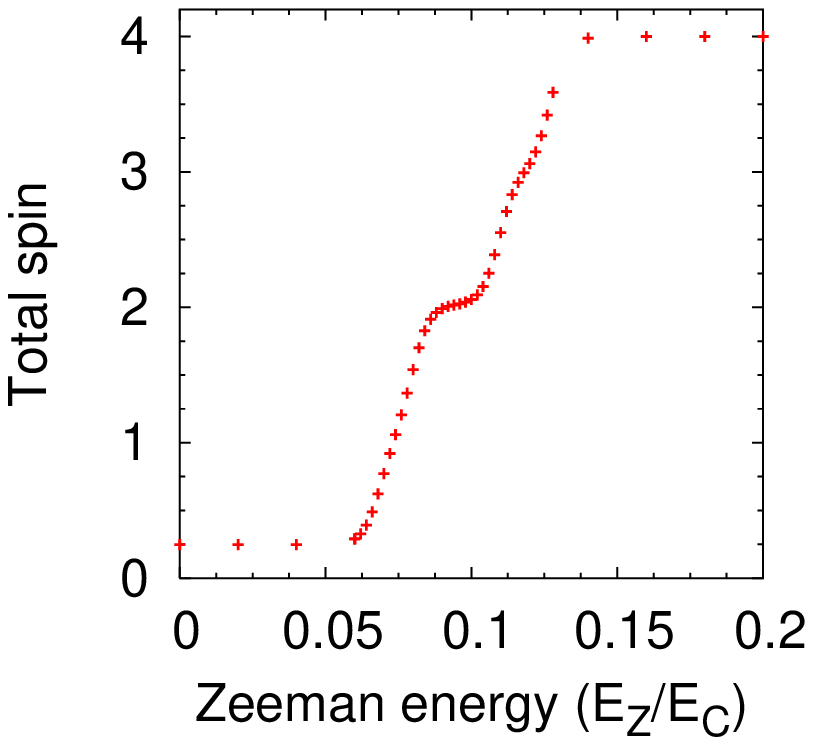}\hskip-1cm}
\caption{Spectrum and the expectation value of the spin 
in the ground state in short-range interacting systems
(eight electrons).}
\label{fig-ch04-13}
\end{figure}

Let us first concentrate on the calculated spectrum of the {\em
homogeneous} system with short-range interaction,
Fig. \ref{fig-ch04-15}b. Again, we observe a gapped ground state with
maximum spin and zero spin in the limit of $B\to\infty$ and $B\to 0$,
respectively. In between, states with different spins become
the absolute ground states. Aforemost, it is the half-polarized state
($S=2$), although states with other spins ($S=1$ and 3) are not very
far. Alternatively, this can be expressed by the $B$-dependence of
the spin of the ground state, Fig. \ref{fig-ch04-13}a.  

The half-polarized states have been extensively discussed in
Sec. \ref{pos-ch03-17} where they were studied as 'zero-temperature
candidates' for the ground state in homogeneous systems.  
However, since inhomogeneities couple the ground state to the excited
states, the properties of the lowest-lying state will not be
determined solely by the those of the ground state.

Spectral properties of the short-range interacting (SRI) system subjected
to a 'perpendicular' magnetic inhomogeneity (\ref{eq-ch04-02}) are
summarized in Fig. \ref{fig-ch04-13}. In a similar fashion as for the
Coulomb-interacting systems, states with other spins become the
absolute ground state in some range of the magnetic field,
Fig.~\ref{fig-ch04-08}. This is also manifested in the
expectation value of spin (or $S_z$) of the system even at nonzero
temperatures. Most significant is still the
'half-polarized' plateau $S_z(B)\approx 2$, Fig. \ref{fig-ch04-13}d.

\begin{figure}
\hskip-1.5cm\subfigure['Low' temperature.]{\label{fig-ch04-14a}
  \includegraphics[scale=0.5]{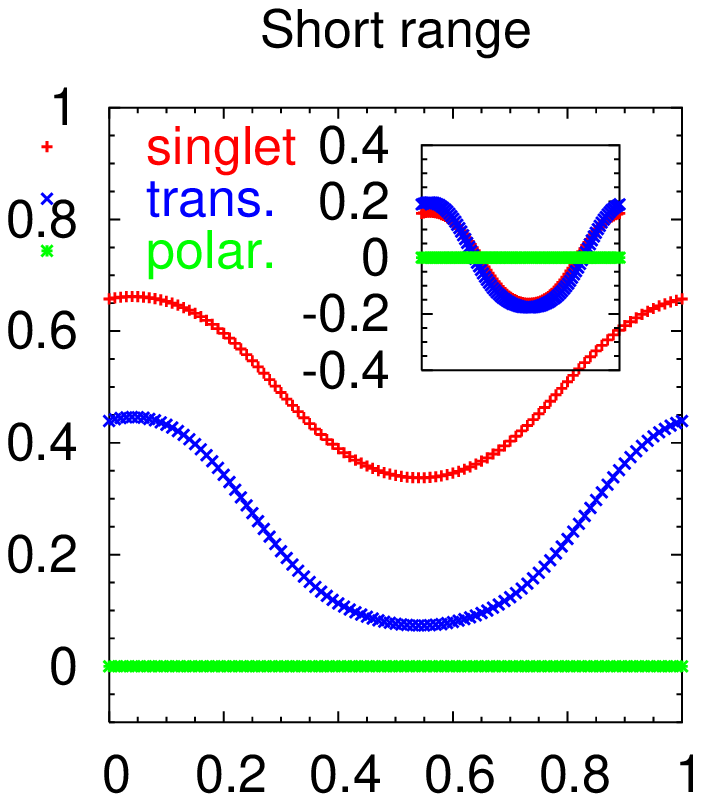}
  \includegraphics[scale=0.5]{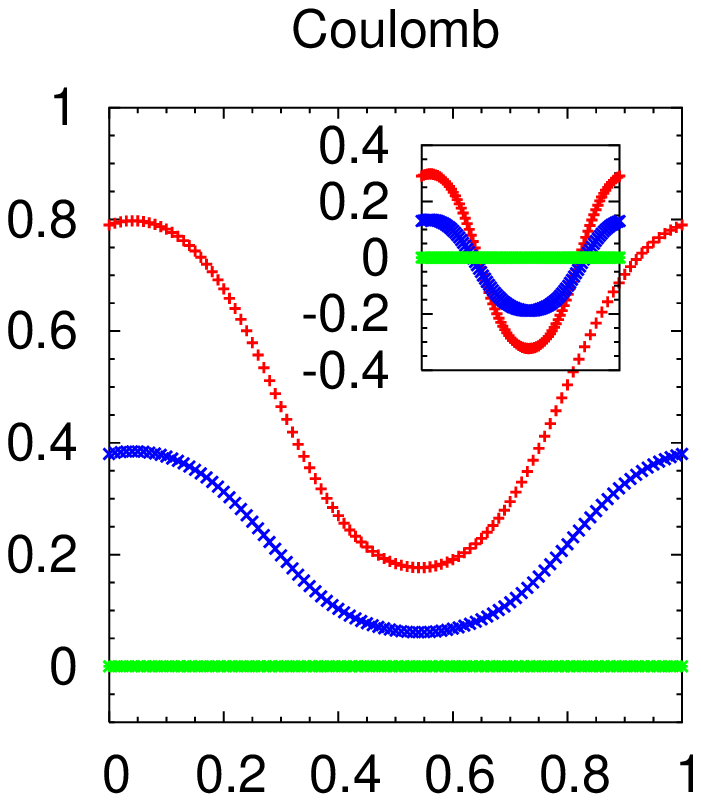}
}
\subfigure[Slightly higher temperature.]{\label{fig-ch04-14b}
  \includegraphics[scale=0.5]{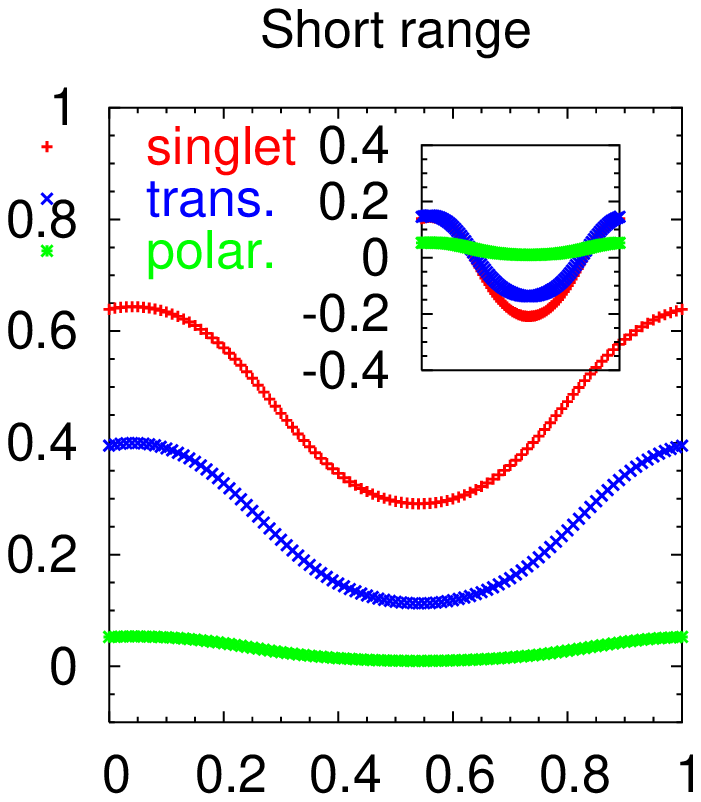}
  \includegraphics[scale=0.5]{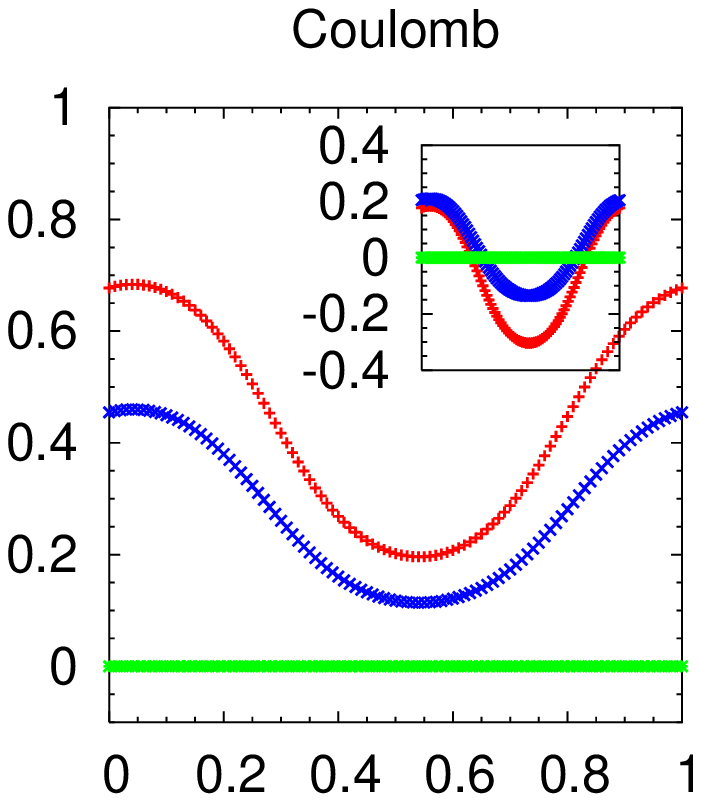}
}
\caption{Short-range interacting system with strong inhomogeneity,
  (\ref{eq-ch04-02}), $E_{MI}=0.02$: polarization in the singlet
  sector, around the transition and in the polarized sector,
  cf. spectrum in Fig. \ref{fig-ch04-13}c. Insets show how
  polarizations fluctuate around anticipated mean values ($0.5$, $0.25$
  and $0$ for the singlet, transition and polarized state).}
\label{fig-ch04-14}
\end{figure}

After this introduction let us look at the inhomogeneous states
themselves. Their properties are highlighted especially in comparison
to the Coulomb interacting (CI) states. When subjected to a 'rectangular
cosine' magnetic impurity, those CI states showed a smooth
monotonous transition from the singlet to the polarized state. The
singlet was most strongly affected by the inhomogeneity, the polarized
state was not affected at all, it was frozen by its
symmetry. The polarized state has all spins up,
  $S_z=N_e/2$. Since magnetic inhomogeneity of the form in
  (\ref{eq-ch04-02}) preserves $S_z$, it does not couple the
  polarized state with any states which contain spin down electrons,
  since such a state must have $S_z<N_e/2$.
The transition state was just in the middle. This is the finding
both at $T=0$, Fig. \ref{fig-ch04-03}b, and at temperature low enough
to average only over the three degenerate states of the
homogeneous system, Fig. \ref{fig-ch04-14}a. 

The SRI systems give a different view. The response to the inhomogeneity is
slightly stronger at the transition than in the singlet state,
Fig. \ref{fig-ch04-14}a, inset.  This is not very surprising
given that there are quite many states near the ground state in the
transition region. At slightly higher temperature where we average
over about 10 states in the singlet and polarized limit, the
distinction between Coulomb and short-range interacting systems
weakens, Fig. \ref{fig-ch04-14}b.

Regarding Figure \ref{fig-ch04-14}, it should be stressed once
again, that the 'transition states' for the Coulomb and the
short-range interaction have completely different character. In the former
case, this state is basically a superposition of the singlet and the
polarized states, whereas it is a half-polarized state
($S_z=N_e/4$) for the SRI.

It seems we are on the track of the domain build-up here. In an ideal
case, we would expect negligibly affected singlet and polarized
states while the polarization of the transition state varies between 0
(polarized domain) and 0.5 (singlet domain). In real systems,
we are still very far from such behaviour as the difference between
polarizations of the singlet and transition state is quite small.
Nevertheless, the direction seems correct, in contrast to the
Coulomb interacting systems. We may therefore conclude:

(i) If nontrivial effects at the transition are expected, there
  must be more states involved than just the singlet and polarized
  ground states;
(ii) it is likely that the half-polarized states play a major
  role;
(iii) at low temperatures inhomogeneous states as in
  Fig. \ref{fig-ch04-14}b can be observed simultaneously with a
  plateau in $S_z(B)$, Fig. \ref{fig-ch04-13}d.

The last point is a consequence of the fact that not only the ground
state but also the lowest excited states have $S_z=2$ in a part of the
transition region, Fig. \ref{fig-ch04-15}b.

\subsubsection{Comments on the form of the short-range interaction}
\label{pos-ch04-04}

For a short-range interaction, the form described in
Subsect. \ref{pos-ch02-11}, Fig. \ref{fig-ch02-04}d was chosen. 
The basic idea there was to keep the pseudopotentials
$V_0$ and $V_1$ at their normal values while setting the others to zero.
$V_0$ and $V_1$ give the
  energy of two interacting particles in the state with angular
  momentum 0 and $1$. These are the states with smallest and second
  smallest interparticle separation possible and only the latter one
  is accessible if the particles have the same spin.

As far as incompressible liquid states are concerned, not much
happens during such 'pseudopotential engineering'. The best measure
for this are directly the density-density correlation functions,
Fig. \ref{fig-ch03-12}. The good match between correlation functions of
Coulomb- and short-range-interacting states 
agrees with the common claim that 
their energy is determined mostly by effects occurring at short
distances. Also excitation energies remain essentially unchanged as
long as 'zero momentum' states are considered (as opposed to charge or
spin density waves). 

What strongly changes is the energy difference between the polarized
and singlet state: it is $0.0632$ for Coulomb and $0.3693$ for
short-range interaction in an eight-electron system, with zero
Zeeman energy. This happens because the average Coulomb potentials felt
by electrons in a singlet state and in a spin polarized state differ.
Roughly, we take the average over set $\{V_0,V_1,0,0,\ldots\}$ in the
former case (all $m$'s allowed) and over $\{V_1,0,0,\ldots\}$ in the
latter case (only odd $m$'s allowed). This is a fundamental
problem. The requirement of equal averages is not compatible with
preserving the ratio of $V_0$ and $V_1$ as in a Coulomb interacting
system. We would have to use higher Haldane potentials to
  achieve this, losing thereby the simplicity of the definition of
  short-range interaction.
Therefore, with short-range interactions, we must be very cautious
whenever we compare absolute energies of states with different spins
(and thus parity properties of the wavefunction). Namely, position of
the singlet-polarized transition depends directly on the energy
difference of the singlet and polarized ground state,
Sec. \ref{pos-ch04-00}. 

This difficulties apply to spectra in this subsection,
Fig. \ref{fig-ch04-13}, \ref{fig-ch04-15}b. Fortunatelly, the
polarizations in Fig.~\ref{fig-ch04-14} do not suffer from this,
provided that
half-polarized states indeed become the absolute ground state
somewhere around the transition.

\subsection{Systems with an oblong elementary cell}


\label{pos-ch04-07}

So far in this Chapter, we have 
only considered square elementary cells $a$ by $a$
so far. If we somehow e.g. by means of a magnetic inhomogeneity,
manage to split such a system into two domains of the same size, this 
will be $a/2$ by $a$, cf. Fig. \ref{fig-ch04-11}c. Consequently, the
spin singlet and spin polarized states which we expect to appear in
these domains would necessarily have to be deformed as in a cell of
aspect ratio $1:2$. In principle, this could even suppress the
formation of such domains or at least shift them to higher excited
states. The energy of any of the two
  incompressible ground states depends on aspect ratio (the stronger
  the smaller the system is), Subsect. \ref{pos-ch03-09}. 
  There is no reason to expect that the energy
  of a domain wall between two such states is constant.
In the following Section we will investigate
systems in a rectangular cell with aspect ratio $2:1$ which have the
possibility of splitting into two square domains. All results in this
Section refer to Coulomb interacting systems.

\subsubsection{Overview of the transition: which states play a part}

\begin{figure}
\begin{center}
\subfigure[Aspect ratio $1:1$.]{\label{fig-ch04-16a}
  \includegraphics[scale=0.5]{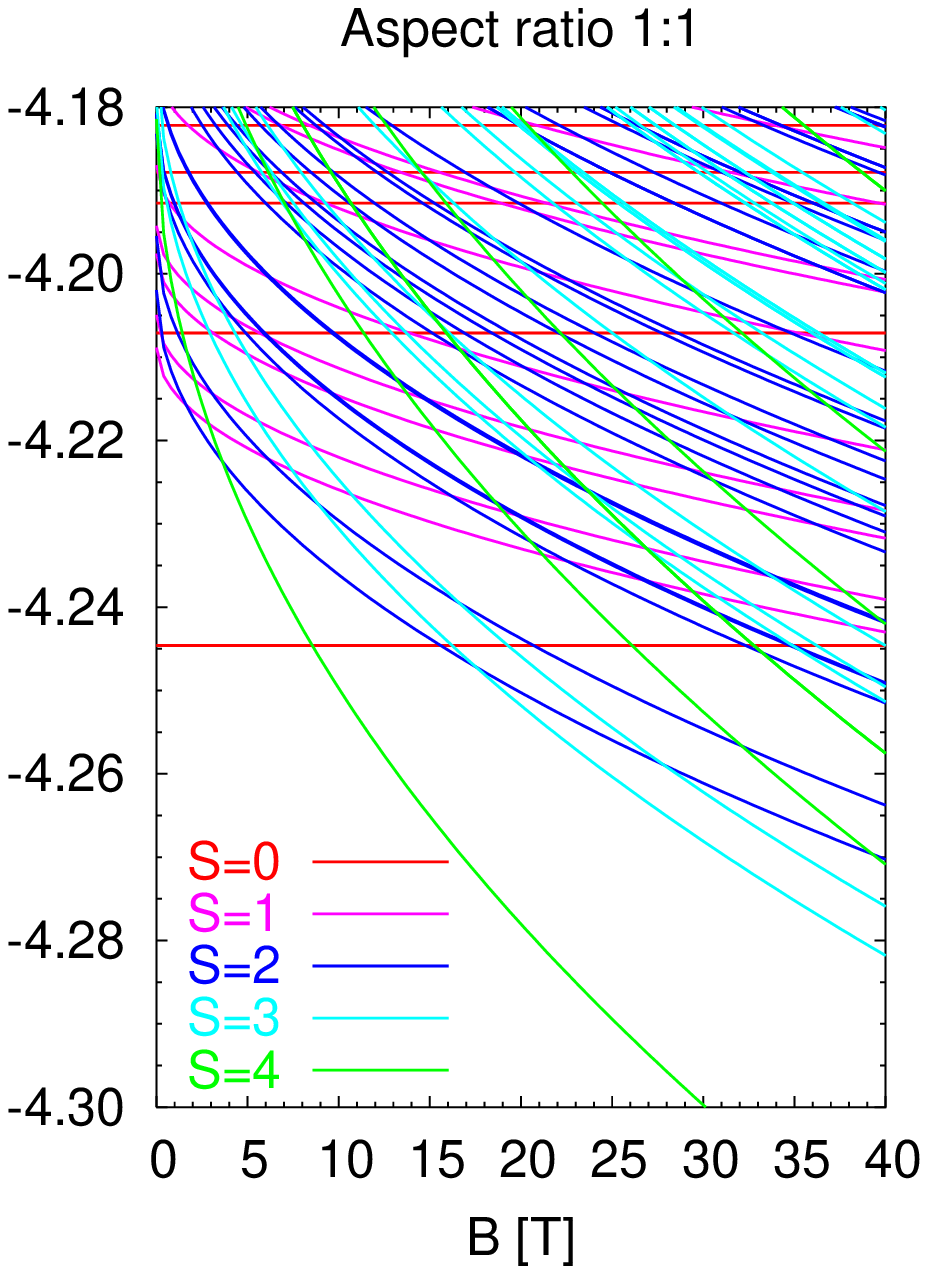}}
\subfigure[Aspect ratio $2:1$.]{\label{fig-ch04-16b}
  \includegraphics[scale=0.5]{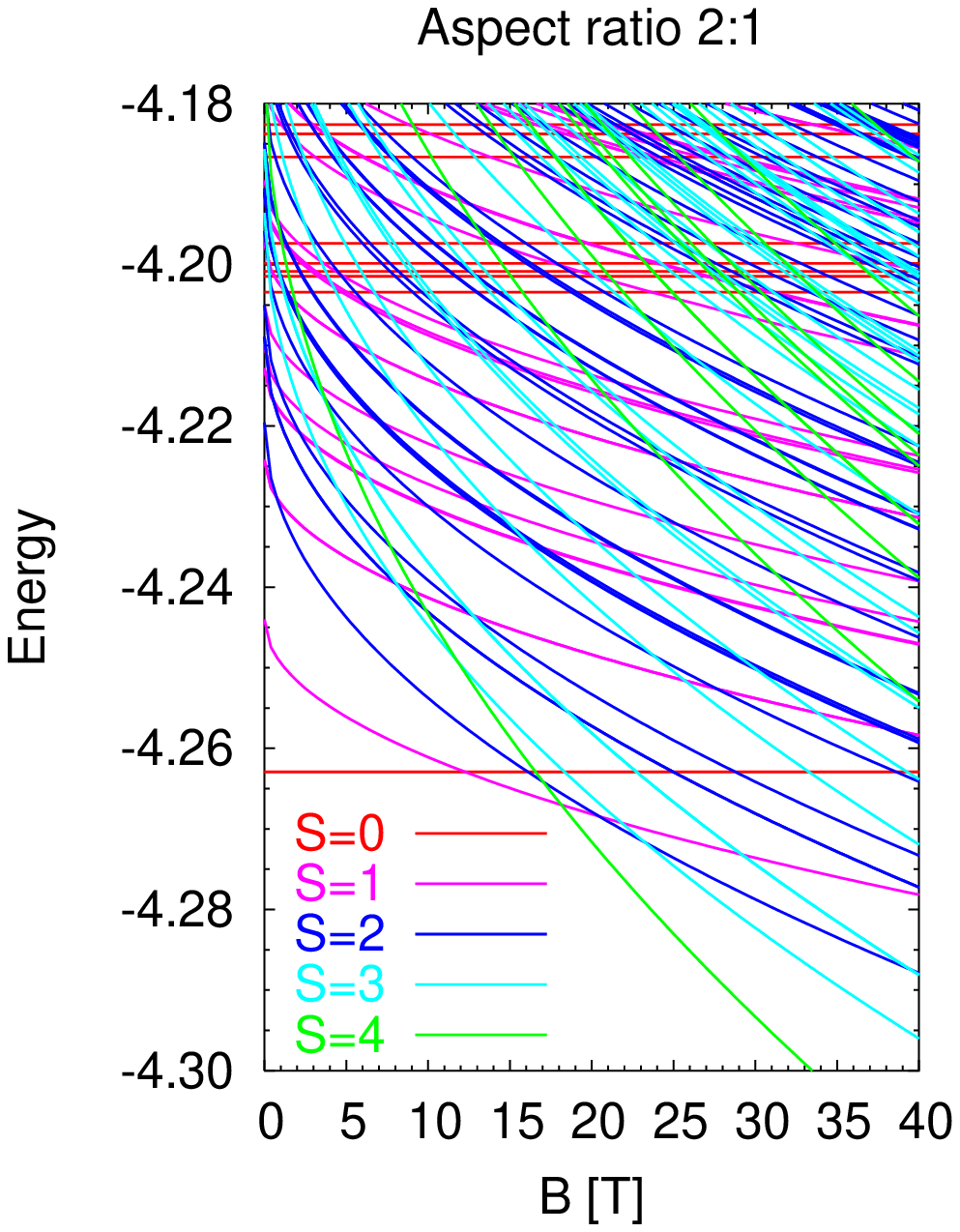}}
\end{center}
\caption{Spectra of homogeneous eight electron systems (with Zeeman
splitting) for square and oblong elementary cell.}
\label{fig-ch04-16}
\end{figure}

Going from square elementary cell to aspect ratio $1:2$, the overall
view of the transition changes. The crossing between singlet and
polarized incompressible states is no longer well separated from
excited states, Fig. \ref{fig-ch04-16}. 

\begin{figure}
\subfigure[Homogeneous.]{\label{fig-ch04-17a}
  \hskip-.5cm\includegraphics[scale=0.5]{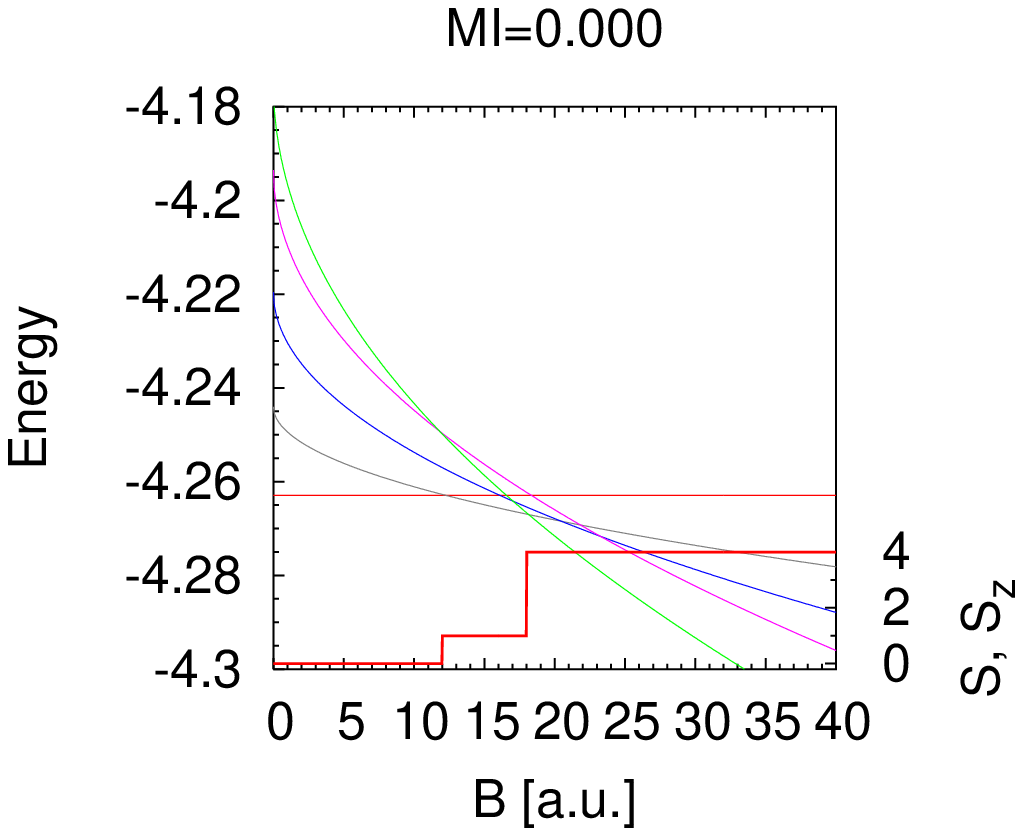}}
\subfigure[$E_{MI}=0.002$ (weak).]{\label{fig-ch04-17b}
  \hskip-.5cm\includegraphics[scale=0.5]{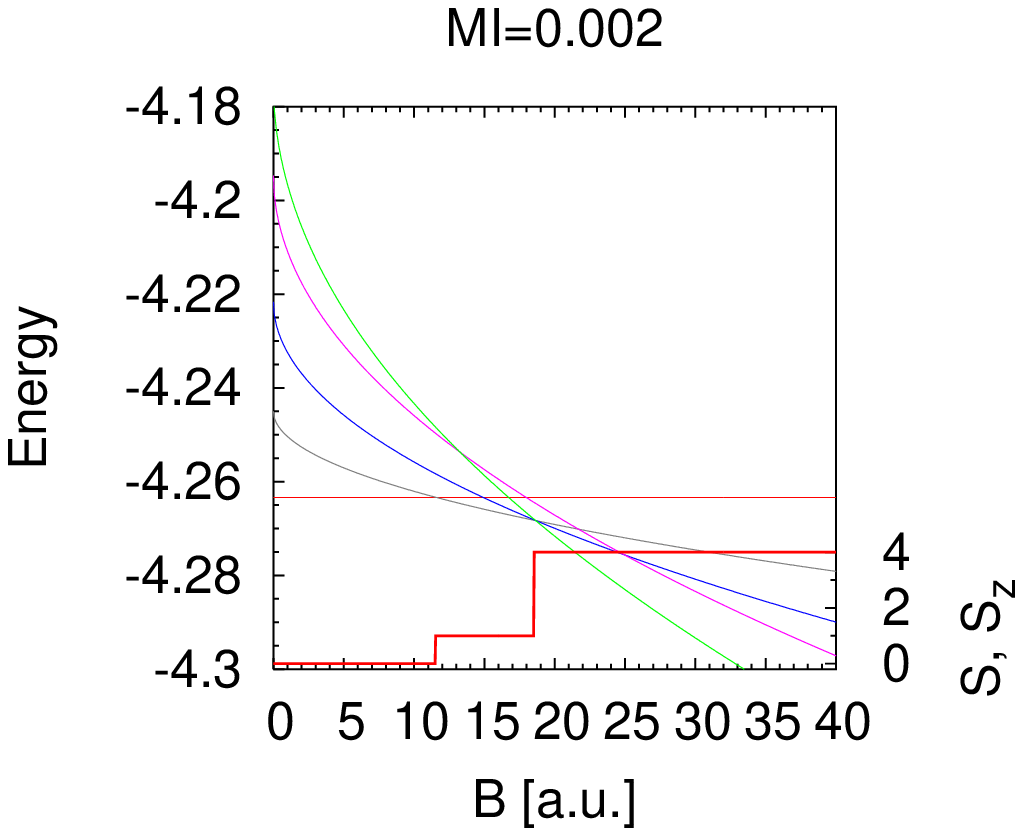}}
\subfigure[$E_{MI}=0.004$ (intermediate).]{\label{fig-ch04-17c}
  \hskip-.5cm\includegraphics[scale=0.5]{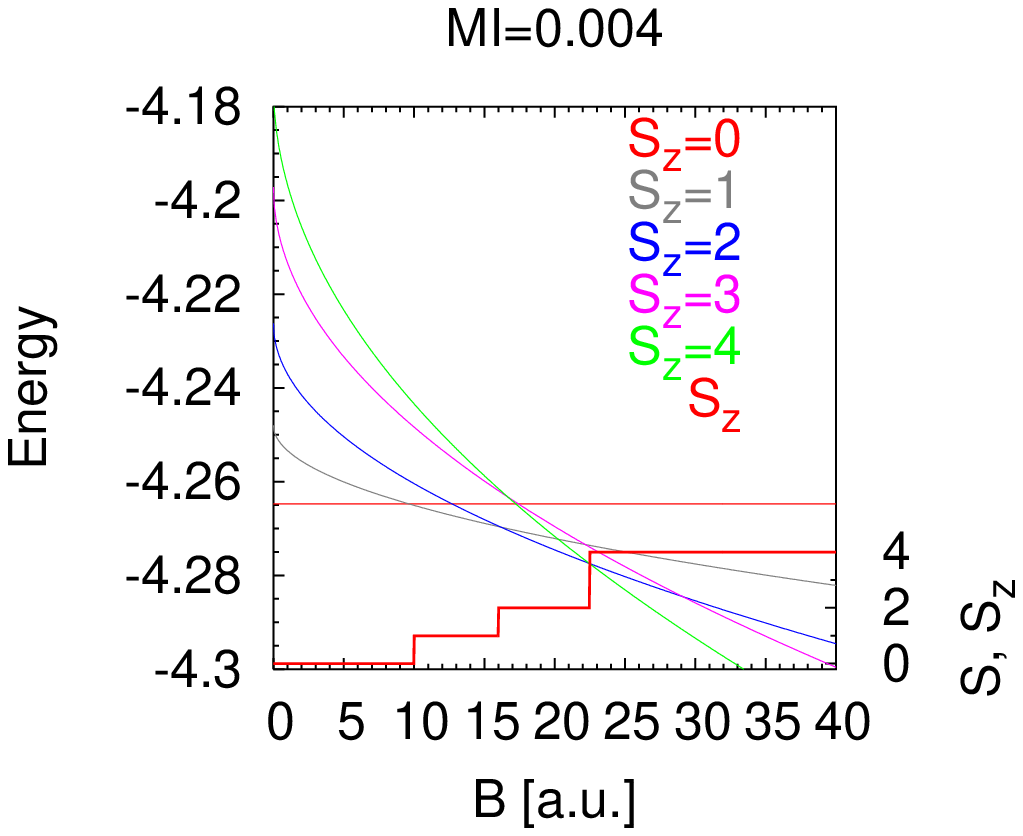}}
\caption{Spectra and $S_z$ of the ground state in a system with
oblique rectangular elementary cell (aspect ratio $2:1$, eight
electrons). Magnetic inhomogeneities (\ref{eq-ch04-02}) of
different strengths are considered.}
\label{fig-ch04-17}
\end{figure}

Similarly, as for short-range interaction, states with different spin
appear near the transition: most prominently $S=1$ and $S=2$.
Again, Figs. \ref{fig-ch04-08}, \ref{fig-ch04-13}, these
states are promoted by the  
perpendicular magnetic inhomogeneities in the form of a 'rectangular
wave', Fig. \ref{fig-ch04-17}.
A consequence of this is a gradual change of the spin in the
ground state as we sweep magnetic field (or simply increase Zeeman
energy). Here, we should point out the difference between the present case
and the Coulomb interacting system in a square elementary cell,
Fig. \ref{fig-ch04-08}. For an oblique elementary cell, (i) the $S=1$
state becomes the absolute ground state near the transition even in
homogeneous systems. (ii) A much weaker inhomogeneity is needed to
make the $S=2$ state the absolute ground state in some range of the
magnetic field. Fig. \ref{fig-ch04-17}c shows that
  $E_{MI}=0.004$ is sufficient for this to happen in a $2:1$ system,
  while $E_{MI}=0.02$ is not strong enough for a square elementary
  cell, Fig. \ref{fig-ch04-08}c.

By changing the elementary cell geometry we support possible domain
states, but it is adequate to ask how much the incompressible singlet
and polarized states are affected by this procedure. The inner structure
of these states under elementary cell variations was addressed in
Subsect. \ref{pos-ch03-09} and we saw indications that the states are
liquid like (and very similar to the original states from square
elementary cell) even at aspect ratio $1:2$. However, overlaps
between the square-cell and deformed states are noticeably below
unity and hence their behaviour is not representative if we are
interested in infinite homogeneous systems. Recall, that the
square-cell polarized state is extremely close to the Laughlin state
(overlaps $\approx 99\%$).

\begin{table}
\begin{center}
\begin{tabular}{l|p{3cm}p{3cm}p{3cm}}
 & asp. $1:1$, hmg. & asp. $2:1$, hmg. & asp. $2:1$, $E_{MI}=0.004$ \\ \hline
$S_z=0$ GS  
 & \leavevmode\kern2.5cm\hbox{0.713}& \leavevmode\kern2.5cm\hbox{0.976} \\
$S_z=N_e/2$ GS  
 & \leavevmode\kern2.5cm\hbox{0.750}& \leavevmode\kern2.5cm\hbox{0.9996} \\
\end{tabular}
\end{center}
\caption{Incompressible ground states (polarized and singlet) in an
eight-electron system. Overlaps between states in a square elementary
cell, oblong elementary cell and oblong elementary cell with
intermediate magnetic inhomogeneity.}
\label{tab-ch04-01}
\end{table}

\subsubsection{States at the transition}

\begin{figure}
\subfigure[Polarization.]{\label{fig-ch04-18a}
  \includegraphics[scale=0.5]{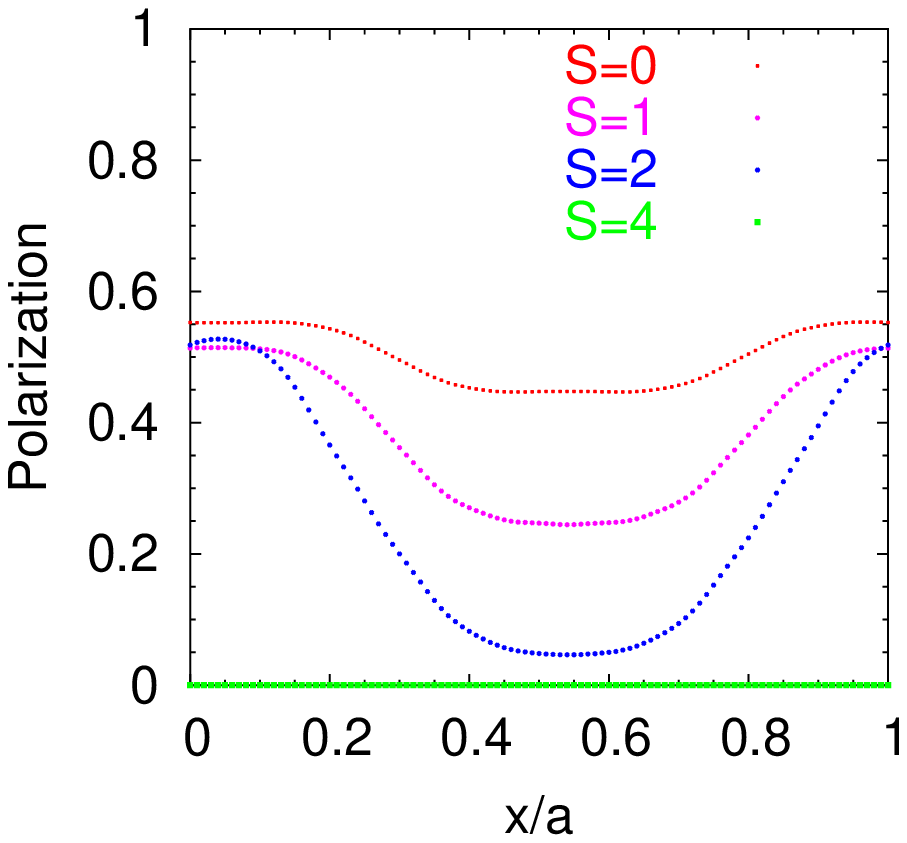}}
\subfigure[Densities of electrons with spin up (or down).]{\label{fig-ch04-18b}
  \includegraphics[scale=0.5]{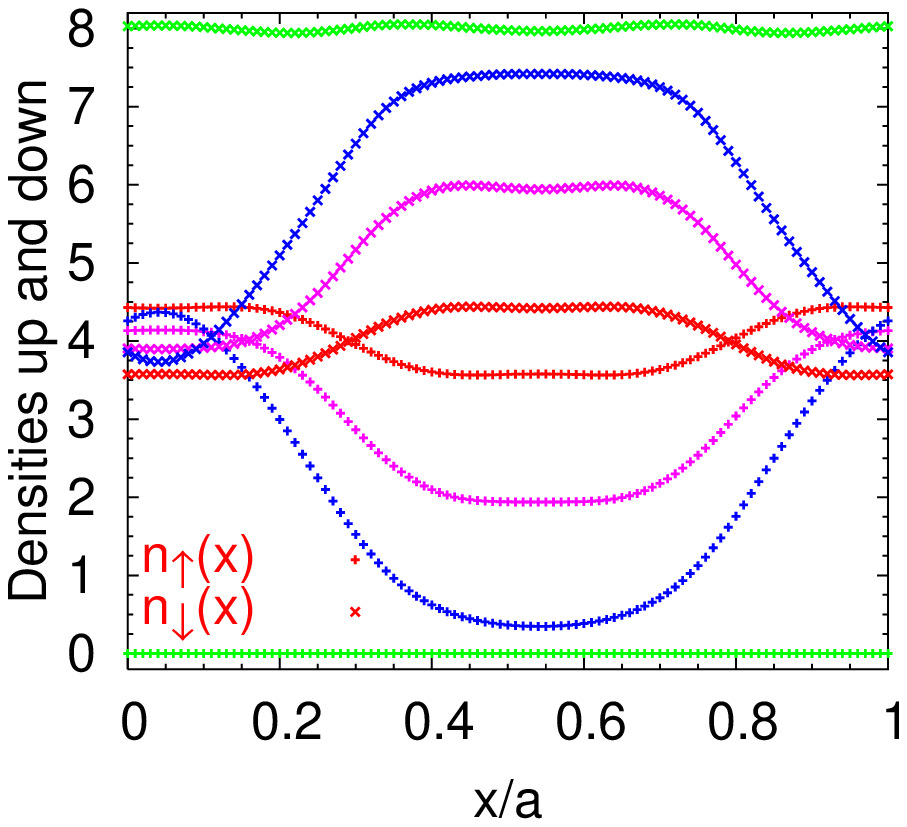}}
\subfigure[Total density.]{\label{fig-ch04-18c}
  \includegraphics[scale=0.5]{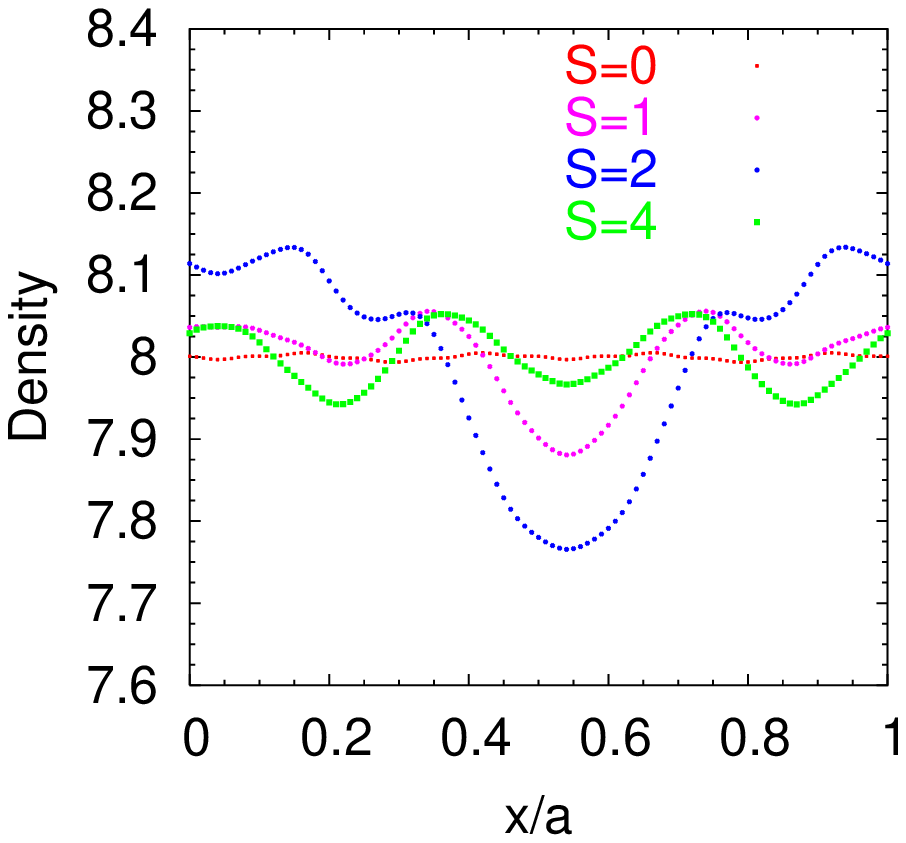}\hskip-1cm}
\caption{'Domains' imprinted by a magnetic inhomogeneity of type
'rectangular wave' into a system with oblong rectangular elementary
cell (aspect ratio $2:1$). The strength of the inhomogeneity is about 20
\% of the gap in the limit $B\to 0$ (in particular
$E_{MI}=0.004$). Plotted quantities are averaged over the three states
which were degenerate in the homogeneous system (in the center-of-mass part).}
\label{fig-ch04-18}
\end{figure}

The following paragraph deals with he central result of the 
investigations on systems with
aspect ratio $2:1$. The low-energy states near the 
transition ($S=1$ and $S=2$ in
Fig. \ref{fig-ch04-17}) respond very strongly to a 'rectangular
wave' magnetic inhomogeneity, Fig. \ref{fig-ch04-18} (the middle 
two lines). Already for intermediate strength of the inhomogeneity
like $15\%$ of the singlet incompressibility gap in a square cell,
polarization varies between $\approx 0.5$ and $\approx 0.05$,
Fig. \ref{fig-ch04-18}a (values of
$0.5$ and $0$ would mean a state with $S_z=0$ and $S_z=N_e/2$,
respectively). Equivalently,
Fig. \ref{fig-ch04-18}b shows that (a) the density of spin down electrons drops
below $25\%$ of its average value in the spin polarized region and (b)
spin up and spin down densities are balanced up to $10\%$ variations
in the 'spin singlet region'. At the same time, variations of the
total density remain small (less than $5\%$), but there is a clear
deficit of electrons in the 'polarized region',
Fig. \ref{fig-ch04-18}c.

In order to check that the inhomogeneity is not too strong ('destructive')
compared to the Coulomb interaction responsible for the formation of
the incompressible ground states (far away from $B_c$), 
we should observe the incompressible $S=0$ and $S=N_e/2$
states, Fig. \ref{fig-ch04-18}. For both of
them, responses are much weaker than for the transition states.

\begin{figure}
\begin{center}
  \includegraphics[scale=0.8]{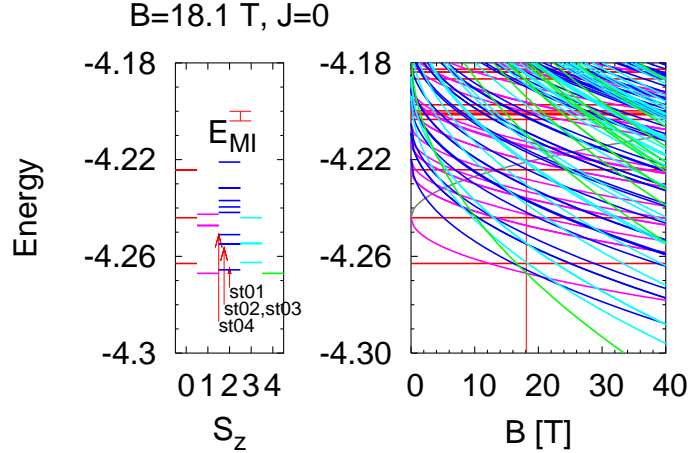}
\end{center}
\caption{Low lying states near the transition (in absence of
inhomogeneity, eight electrons). 
The $J=0$ sector is where (a) both incompressible
ground states (singlet and polarized) occur and (b) the lowest
half-polarized state occurs.}
\label{fig-ch04-22}
\end{figure}

\begin{figure}
\subfigure[The lowest state (\ttfont {st01}).]{\label{fig-ch04-19a}
 \hskip-1cm\includegraphics[scale=0.45]{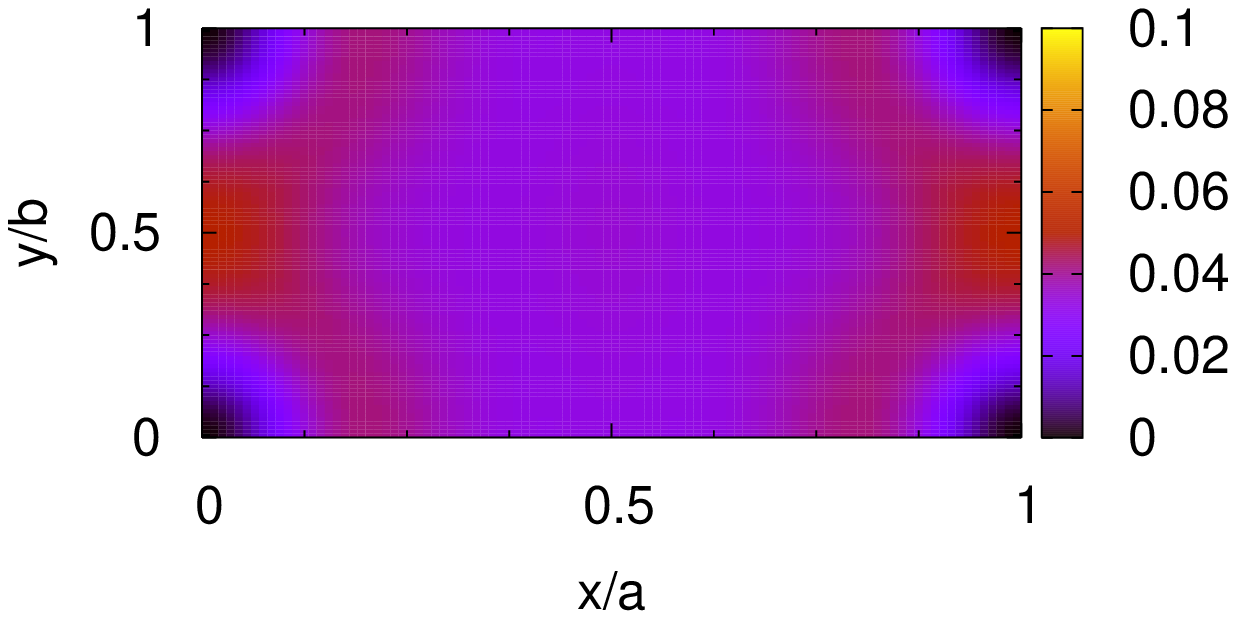}\hskip-.0cm}
\subfigure[An SDW state (\ttfont{st02,st03}).]{\label{fig-ch04-19b}
  \hskip-.2cm\includegraphics[scale=0.45]{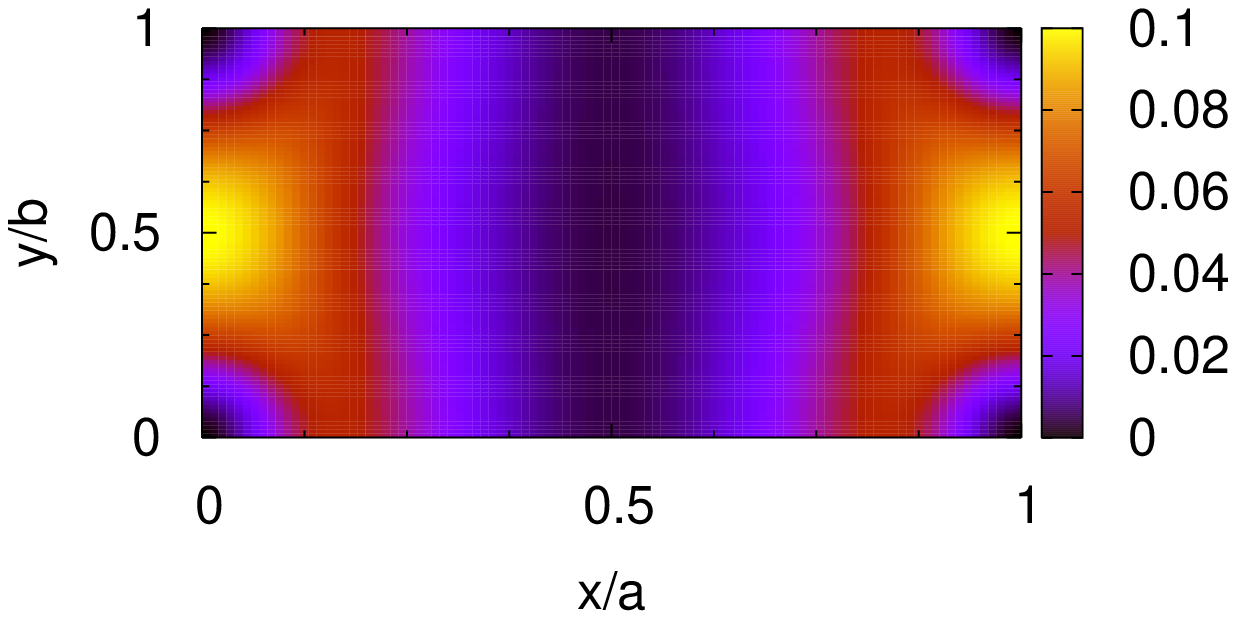}}
\subfigure[Another SDW (\ttfont{st04}).]{\label{fig-ch04-19c}
  \hskip-0cm\includegraphics[scale=0.45]{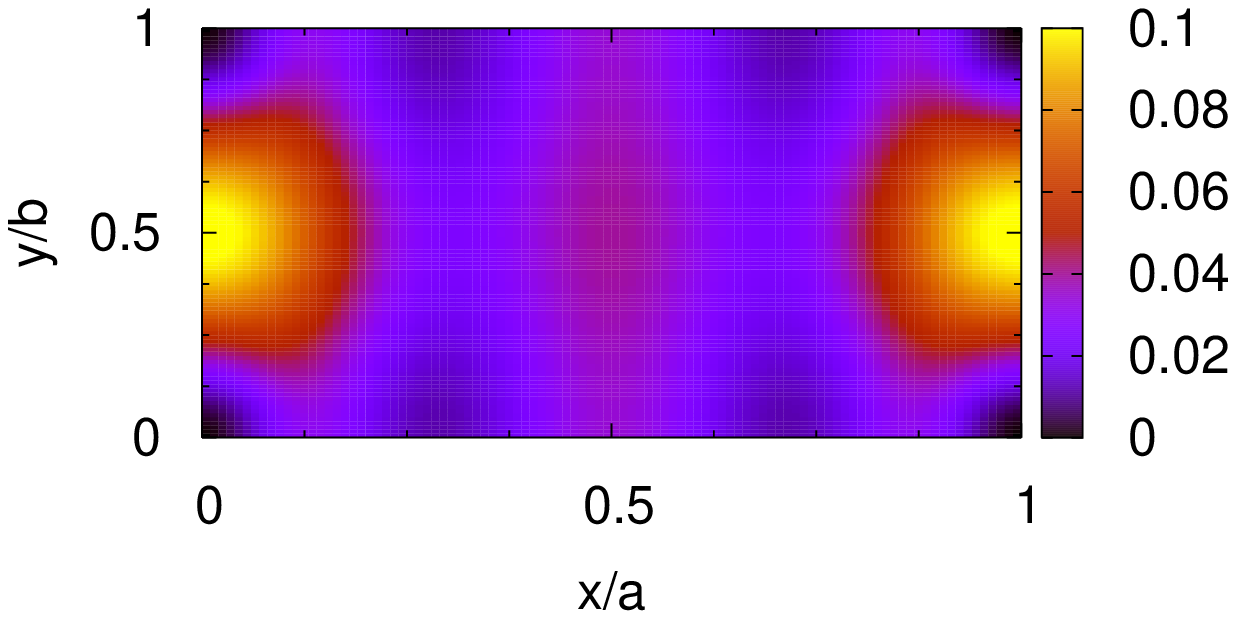}\hskip-2.cm}
\caption{Density-density correlations ($g_{\dn\dn}$, i.e. minority
spin) in the lowest half-polarized states. Coulomb interaction,
a homogeneous system, aspect ratio $2:1$.}
\label{fig-ch04-19}
\end{figure}

Let us now concentrate exclusively to the half-polarized states and try to
analyze their nature. Observe first the homogeneous system near the
transition, Fig. \ref{fig-ch04-22} and focus on the 
half-polarized sector ($S_z=N_e/4=2$) with one particular
value of $J$, Subsect. \ref{pos-ch02-06}. The low lying states show
pronounced spin structures and, moreover, several distinct types of spin
structures appear in the low energy part of the spectrum. This is
heralded by different values of $\krv$ which are $(0,0)$, $(\pm 1,0)$
and $(2,0)$ for the lowest three states
(\ttfont{st01},\ttfont{st02}+\ttfont{st03}, \ttfont{st04}, the middle
pair is degenerate) and the different spin structures 
can be seen best in the density-density
correlation of minority spin, $g_{\dn\dn}(\vek r)$,
Fig. \ref{fig-ch04-19}. Half-polarized states contain six
  spins up and two spins down here, which we choose to call majority
  and minority spins respectively.

The lowest state looks
isotropic as far as the rectangular elementary cell allows, the other
two (\ttfont{st02}+\ttfont{st03} and 
\ttfont{st04}) are different kinds of spin density waves 
in the 'long direction'
($x$). Keeping in mind that these states are energetically close to
each other as compared to incompressibility gaps at $\nu=\tt$ in a square
elementary cell, for example, we can indeed expect strongly modulated
polarization in response to suitable not very strong
inhomogeneities. Polarizations in Fig. \ref{fig-ch04-18} were a
good demonstration of this prediction.

\begin{figure}
\subfigure[Polarization. Impurity types are the same as in 
Fig. \ref{fig-ch04-11}.]{\label{fig-ch04-20a}
\includegraphics[scale=0.5]{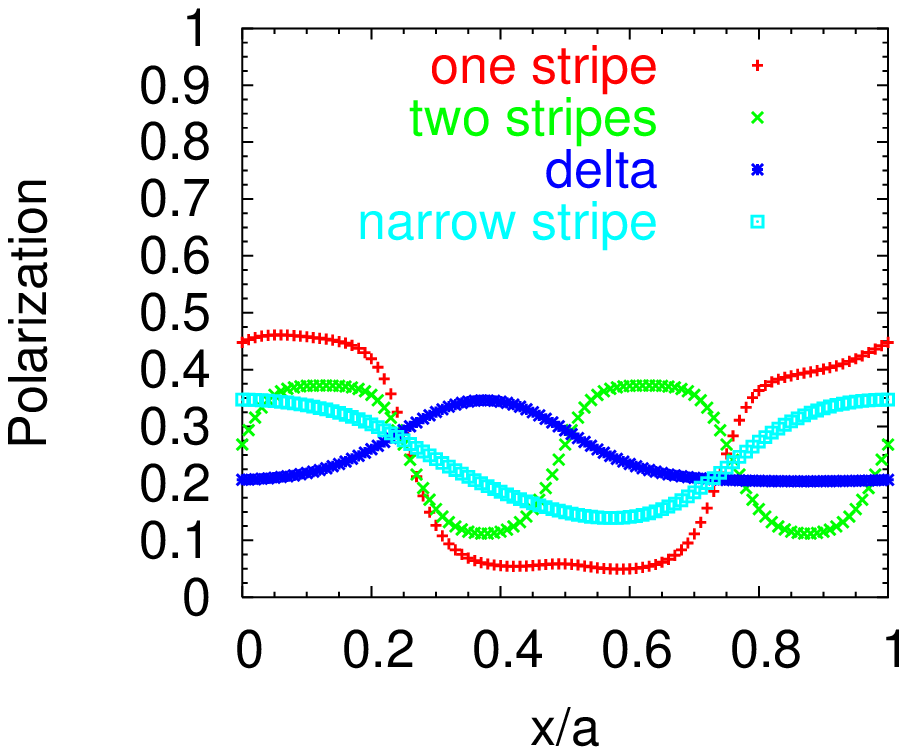}}
\subfigure[Projections of the inhomogeneous ground states to the
lowest four homogeneous states (in Fig. \ref{fig-ch04-19}).]{
\label{fig-ch04-20b} 
\raise2cm\hbox{%
  \begin{tabular}{c|cccc} 
  State    &  hmg. & one stripe & two stripes & $\delta$-peak \\ \hline\hline
  Energy & -4.2656& -4.2713   & -4.2695     & -4.2439 \\ \hline 
\ttfont{st01}& 1  & -0.82     & -0.92       & -0.04   \\
\ttfont{st02}& 0  & -0.51     &	 0.00	    &  0.03   \\
\ttfont{st03}& 0  &  0.21     &	 0.00	    &  0.07   \\
\ttfont{st04}& 0  &  0.03     &	 0.22	    &  0.64   \\ \hline
cummul. $\Sigma (\braket{\cdot}{\cdot})^2$ 
             & 1        &  0.98     &	 0.90	    &  0.41\\
  \end{tabular}}}
\caption{Half-polarized state and different forms of
inhomogeneity. Strength of the inhomogeneity is the same in all cases,
$E_{MI}=0.004$.}
\label{fig-ch04-20}
\end{figure}

This leads us to the question
what types of spin structures can be
imprinted into these states. Are they completely 'soft' or are some
particular structures preferred? An answer
is given by polarizations in response to various types of
inhomogeneities, Fig. \ref{fig-ch04-20}. Briefly summarized: a variety
of spin structures is possible but 'periodic' structures are
preferred. Among the 'periodic' structures, the largest period
available is preferred. This means one stripe or just the 'domains' as in
Fig. \ref{fig-ch04-18}. By periodic we mean  commensurate with the
elementary cell period, for instance a 'rectangular wave'
in contrast to a delta peak since otherwise, any structure is periodic
in our system due to periodic boundary conditions.

Looking only at polarizations, Fig. \ref{fig-ch04-20}a, responses to
all types of inhomogeneities considered here seem to be the same (in
strength) within a factor of two. However, a closer look reveals some
differences between those which are 'periodic' and the
others, Fig. \ref{fig-ch04-20}b. 
The one-stripe and two-stripe inhomogeneities mix mostly
only the lowest four states: (sum of squares of) projections of the
inhomogeneous state to states \ttfont{st01}-\ttfont{st04} give in
these cases $\ge 90\%$. It seems that a 
one-stripe structure, or domain state in
Fig. \ref{fig-ch04-18}, stems from the $\krv=(\pm 1,0)$ states
(\ttfont{st02}+\ttfont{st03}) and the two-stripe structure comes from
the $\krv=(2,0)$ state (\ttfont{st04}). In both cases, however, 
projections to the lowest state (\ttfont{st01}) remain high.

A different situation occurs for 'non-periodic' structures like a delta
peak. Inhomogeneous states are then 'constructed' in the main from
states which were originally energetically higher in a homogeneous
system. Such states (e.g. with a delta peak in the polarization) 
only have a strong
projection to the $\krv=(2,0)$ state (\ttfont{st04}), but still more
than $50\%$ of weight comes from higher states, Fig. \ref{fig-ch04-20}b.

This scheme, 'periodic-welcome, others-less welcome', is confirmed
also in terms of energy. While the 'periodic' states (one- and
two-stripes) profit energetically from the inhomogeneity, the
delta-peak state is shifted to higher energy, Fig. \ref{fig-ch04-20}b.

Finally, the following conclusion about the $\nu=\tt$ system near the
transition seems to be possible. The softening against
magnetic inhomogeneities
of different forms, as observed in Fig. \ref{fig-ch04-18}a, stems 
not only from the spectral properties of the system (small level
spacing, Fig. \ref{fig-ch04-22}) but also from the fact that more
different (inner) spin structures occur among the low lying
states. States belonging to a single value of $S$ (e.g. $S=N_e/4$) are
capable of generating a response as shown in Fig. \ref{fig-ch04-18}a.

\subsubsection{What is inside the domains?}
\label{pos-ch04-08}

We will now only consider the 'one-stripe' inhomogeneity, in sense of
Fig. \ref{fig-ch04-11}, which has lead us to the states with
polarization varying almost between zero and one half,
Fig. \ref{fig-ch04-18}. In other words, we could distinguish between two
domains of about equal areas in that state: one, with only
spin up electrons and another with as many spin up as spin
down electrons, whereby the total density is spatially nearly
constant. Now we are interested in the inner structure of these
domains. One of the aims of this thesis was to find side-by-side domains
comprising of the incompressible singlet and incompressible polarized
states. Unfortunately, the results presented in this Subsection cannot
conclusively answer whether the states discussed in the
previous Subsection are of this type or not. Also, it would be
surprising if they could in the view of the small systems (eight electrons) we
study. One of the reasons why studies of finite systems on a torus
  or on a sphere were so successful was that these models contain no
  edges. On contrary, there are 'edges' in the state with 'domains':  
  the domain walls.
Nevertheless, these results provide at least some basis for comparing
the inside of the domains to the incompressible states and, in
particular, highlight some differences between these two.

As a probing tool we chose density-density correlation functions. As
we are dealing with inhomogeneous states, we must use 
$g(\vek r,\vek{r}_0)\propto \langle
\delta(\vek r_1-\vek r) \delta(\vek r_2 -\vek{r}_0)\rangle$ rather than
$g(\vek r)\propto \langle\delta(\vek r_1-\vek r_2 -\vek r)\rangle$.
The former quantity is the conditional probability to find an electron at
$\vek r$ given there is an electron at $\vek{r}_0$ and we will
separately address the cases when  both electrons have spin up or
when they both have spin down. By convention, majority electrons are
spin up (expected to be present in both domains) and minority
electrons are spin down (absent in the fully polarized
domain).

\begin{figure}
\begin{tabular}{ll}
(a) & (b) \\[-8mm]
\includegraphics[scale=0.5]{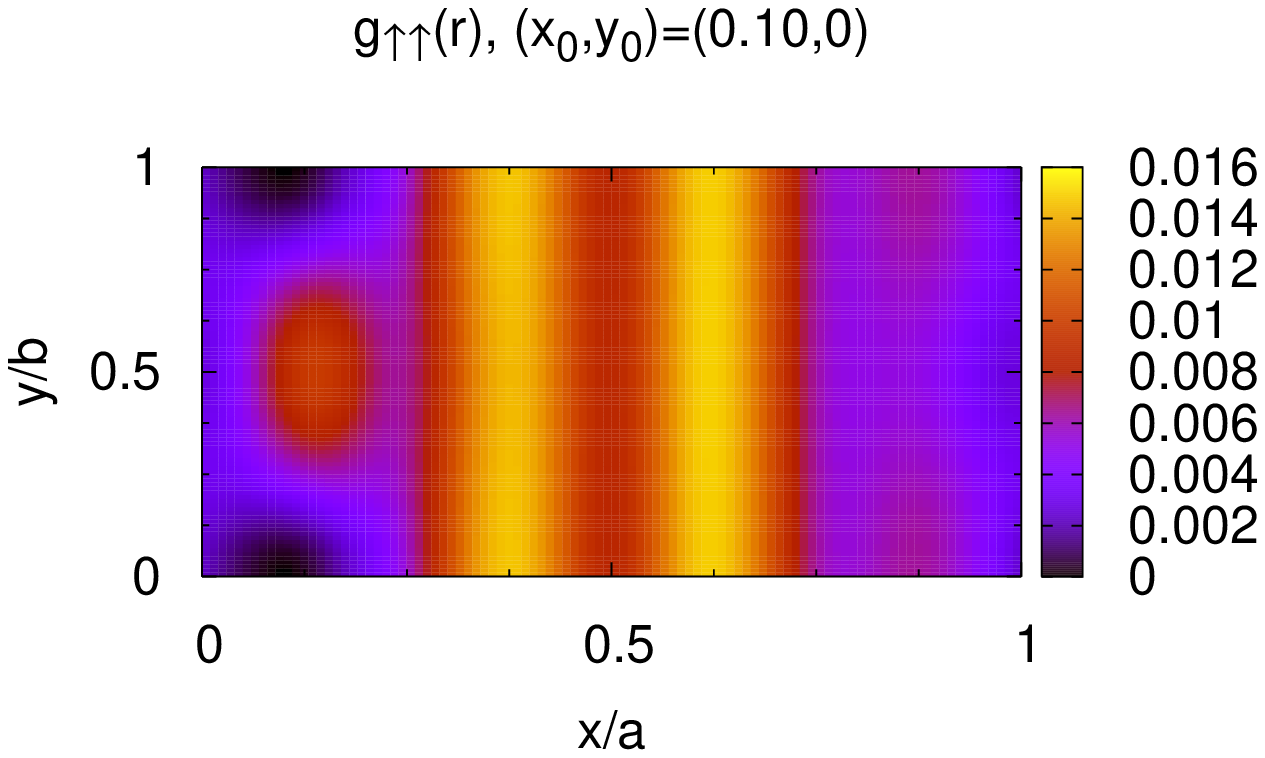}
&
\includegraphics[scale=0.5]{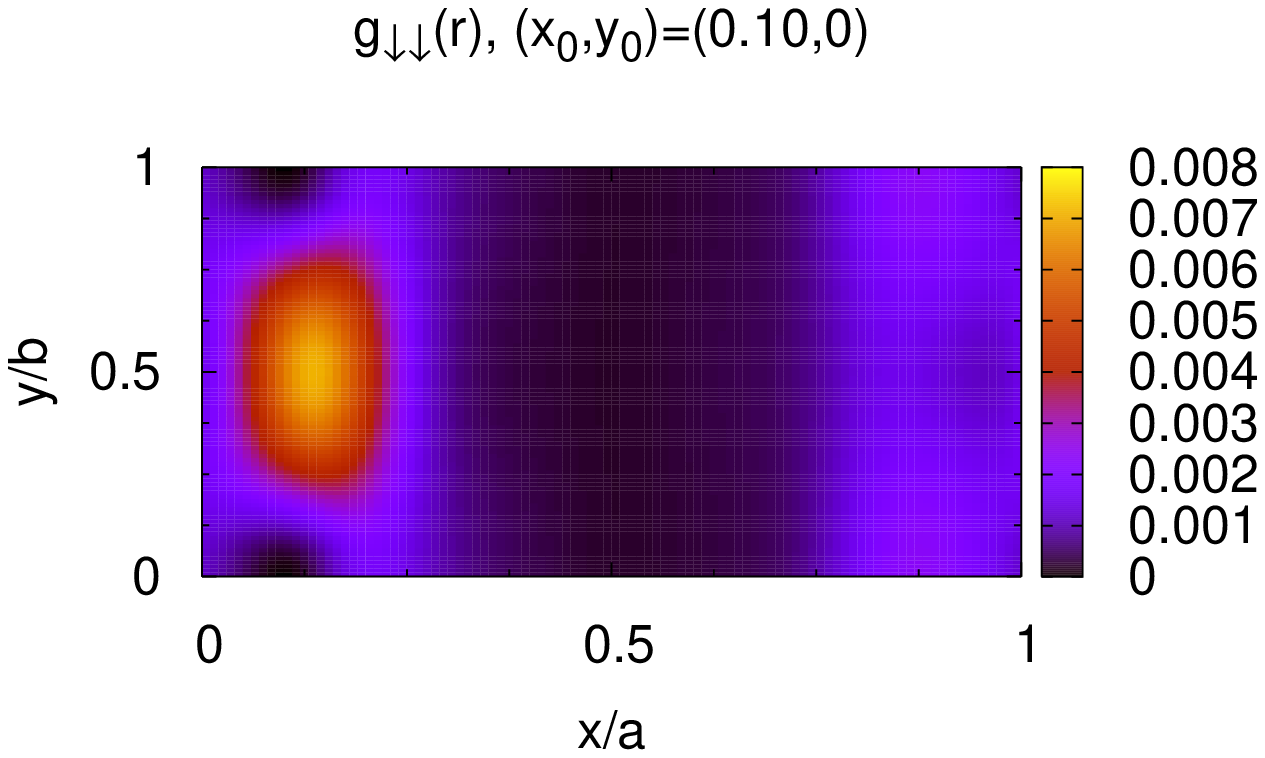}
\\
(c) & (d) \\[-8mm]
\includegraphics[scale=0.5]{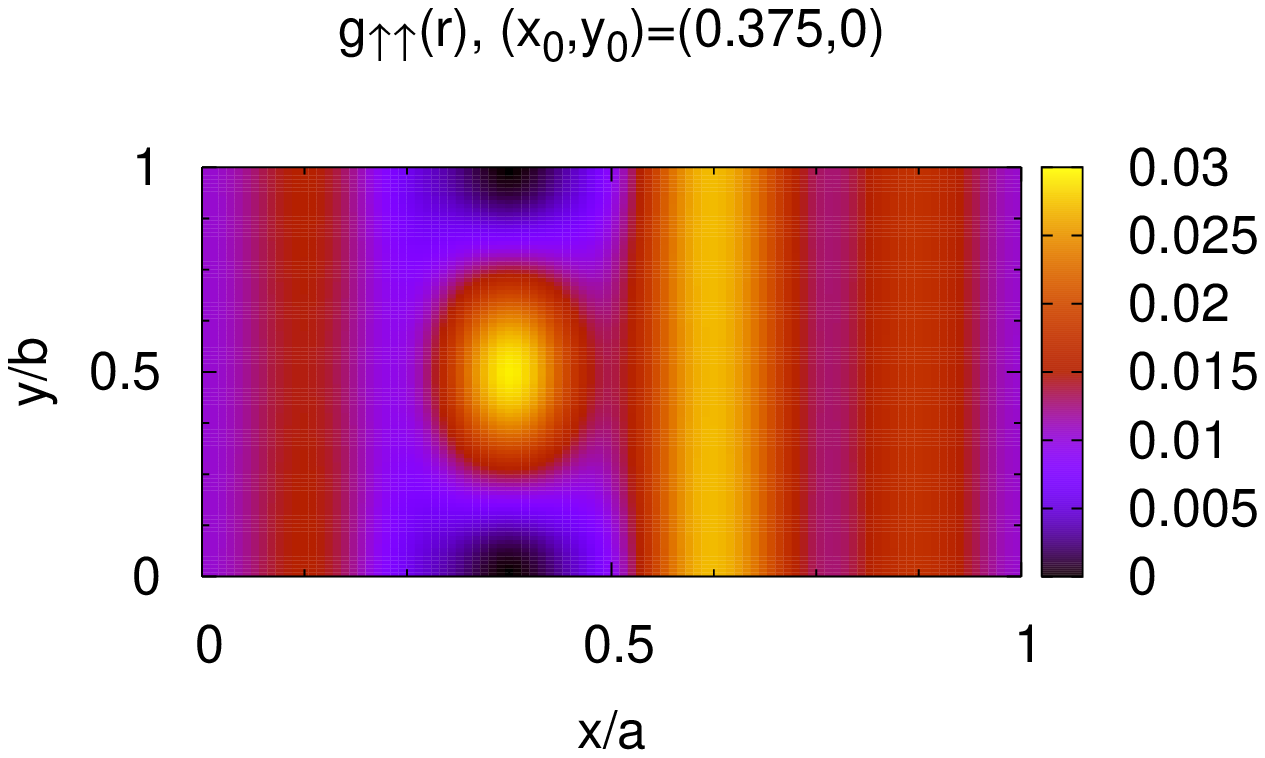}
&
\includegraphics[scale=0.5]{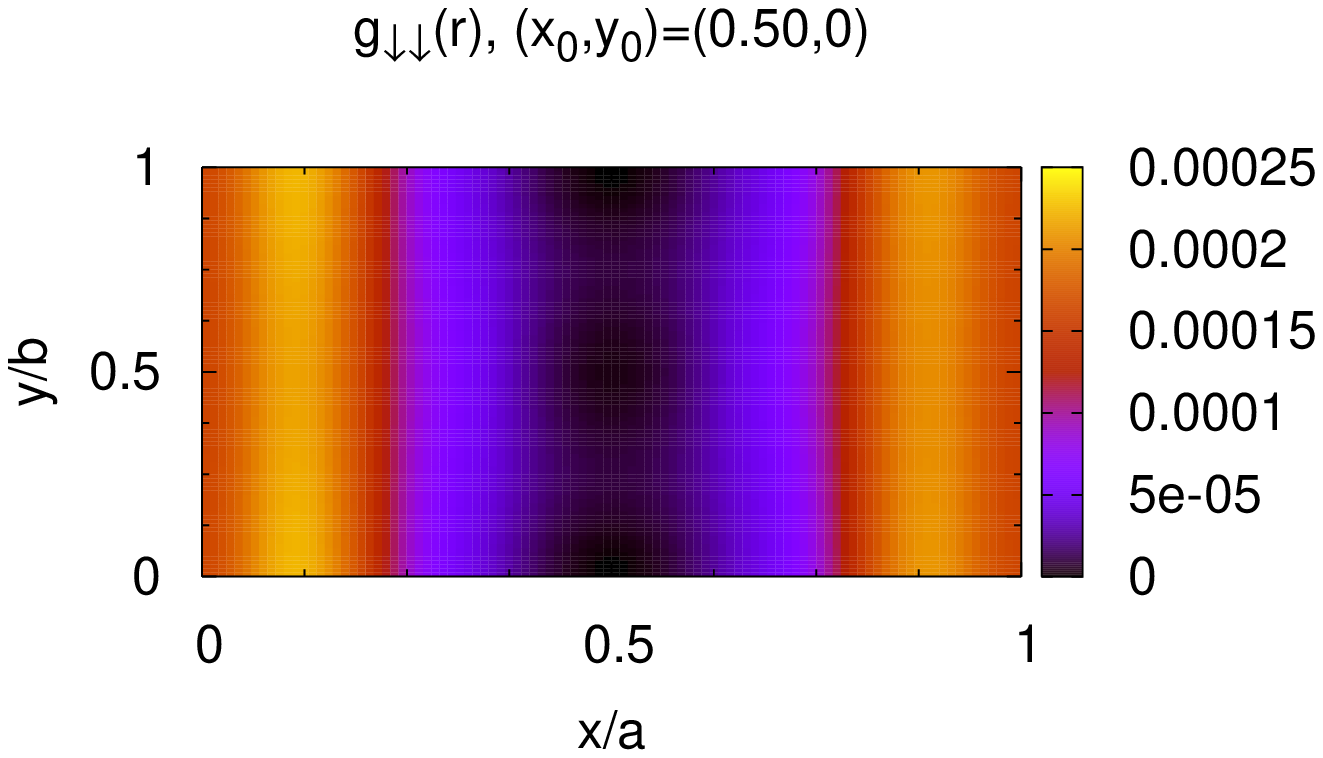}
\\
(e) & (f) \\[-5mm]
\includegraphics[scale=0.5]{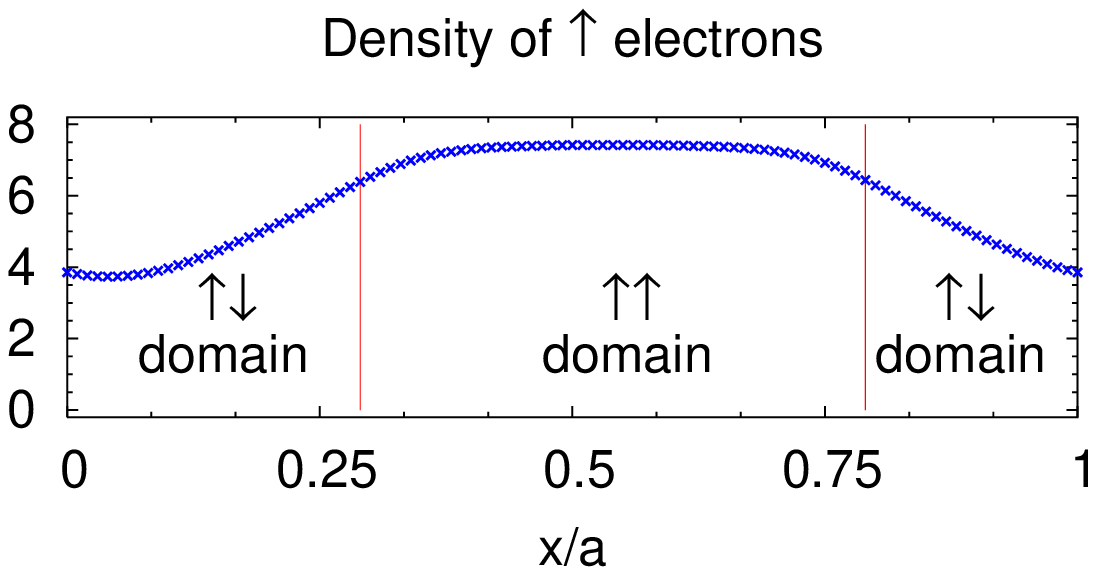}
&
\includegraphics[scale=0.5]{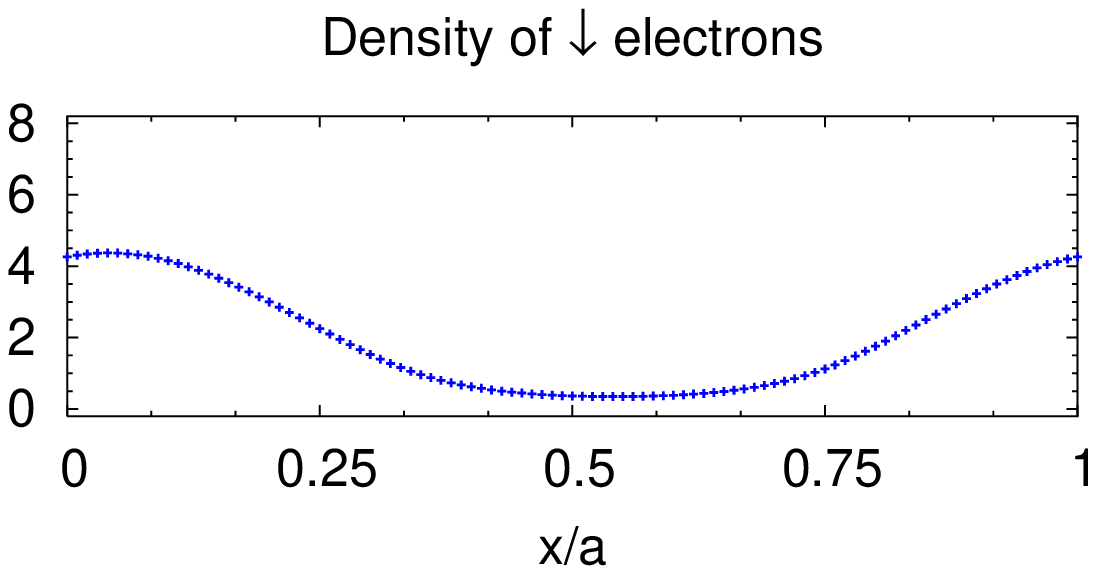}
\\
\end{tabular}
\caption{Half-polarized state with intermediate magnetic inhomogeneity
('rectangular wave', $E_{MI}=0.004$), non-averaged density-density
correlation functions.}
\label{fig-ch04-21}
\end{figure}

Roughly, we can say that the eight electrons are organized in four
vertical stripes: two in the polarized and two in the unpolarized
domain.  For instance, if we catch a majority spin electron in the
left stripe in the polarized domain, we will see another quite sharply
localized (majority spin) electron in the same stripe and two
delocalized electrons in the other stripe of the polarized domain,
Fig. \ref{fig-ch04-21}c. In the unpolarized domain, we will see the
two majority electrons distributed nearly equally among the two
stripes.

Similarly, if we pin a majority spin electron in one stripe of the unpolarized
domain, Fig. \ref{fig-ch04-21}a, we find another (majority spin)
electron in the same stripe. Four electrons in the polarized domain
are distributed homogeneously to the two stripes. We will see almost
the same picture with minority spin electrons, if we catch a {\em minority}
spin electron at the same place. Naturally, we will
see almost nothing in the polarized domain, Fig. \ref{fig-ch04-21}b.

{\em Summary.} In eight electron systems, the domain states comprise
of four vertical stripes (i.e. parallel to the short side of
elementary cell), each occupied by two electrons. In the
polarized domain, each stripe contains two electrons separated by
$b/2$, and the two stripes can 'freely slide' besides each
other. Stripes of the unpolarized domain are preferentially
occupied by electrons of the same spin and both spins (majority and
minority) seem to be equivalent: schematically
$\bra{\up\up}_L\bra{\dn\dn}_R+\bra{\dn\dn}_L\bra{\up\up}_R$. The
domains seem to be rather independent. For instance, 
regardless of where, within the unpolarized domain,
we pin the majority spin electron, the density of electrons seen in
the polarized domain does not change much.

It should be emphasised that although the stripe structure is well
pronounced in conditional probabilities, the density varies only
weakly, Fig. \ref{fig-ch04-18}c. But, even so,
it contains indications of the four stripes. This structure suggests
that the interior of any of the domains is rather anisotropic and this
is quite distinct from the liquid states at $\nu=\tt$ (polarized and singlet,
Fig. \ref{fig-ch03-07}) where at least the first
maximum in $g(\vek r)$ occurs for all $\vek r$ with $|\vek r|=r_1$
(Subsect. \ref{pos-ch03-15}) and not only in the $x$- or $y$-direction.
The study of finite size effects comparing the averaged and
non-averaged correlation functions suggests that for liquid states, the
anisotropy of non-averaged correlation functions should be much
smaller than what we observe in Fig. \ref{fig-ch04-21}. On the other
hand, the results shown in Fig. \ref{fig-ch04-21} refer to a state
which is inhomogeneous and strongly influenced by
the aspect ratio being far from unity. A more thorough study of the
non-averaged correlation functions in systems of various aspect
ratios and comparison to systems of different sizes is therefore
necessary to allow more definite conclusions.

\subsubsection{Comment on homogeneous half-polarized states} 

It should be mentioned that the half-polarized state we study here
is not the same as the half-polarized states 
discussed in Sect. \ref{pos-ch03-07}. 

The lowest state here, in an elongated elementary cell, has
$\krv=(0,0)$, whereas the half-polarized ground state in a square cell
has $\krv=(2,2)$, Sect. \ref{pos-ch03-07}. These are the two
inequivalent points of the highest symmetry.

In the present system (aspect ratio $2:1$) all low lying
half-polarized states belong to the $\krv_y=0$ sector. States with
other values of $\krv_y$ lie well
above the four discussed states \ttfont{st01}-\ttfont{st04}, the
lowest of these other states has an energy of $-4.240$, cf. the spectrum
on the left in Fig. \ref{fig-ch04-22}. Being interested in the states low
in energy we may therefore stay restricted to the sector 
$\krv_y=0$. This is a pleasant fact since there is no need to consider
inhomogeneities of very low symmetry (implying handling larger Hilbert
spaces). 

Also note that a spatially fixed inhomogeneity couples the relative
and center-of-mass (CM) coordinates. At filling $\nu=\tt$ there are three
possible CM states on a torus, which however remain mutually decoupled
owing to the high symmetry of the inhomogeneity. The chosen sector $J=0$
corresponds to a combination of $\krv_y=0$ sector and the one CM
state, which leads to the lowest energy. Differences to other CM
states are, however, not too large.

%

\subsection{Summary of studies on the inhomogeneous systems}

\label{pos-ch04-09}

Perhaps the most important conclusion of this Chapter is that the two
incompressible ground states at $\nu=\tt$, the polarized and the
singlet one, alone are not enough to create a state with 'domains',
i.e. regions of polarization
zero, corresponding to $S_z=0$, 
existing side-by-side with regions of polarization one. We have
demonstrated this in Subsection \ref{pos-ch04-02}. When
a 'domain-inducing' magnetic inhomogeneity is applied, the singlet
ground state is more strongly affected than states near the
transition. This claim remained true for different types of
magnetic inhomogeneities, for different quantities used to detect the
domains (apart from the polarization, also for $S_z(x)$, $S_x(x)$, etc.)
and also for non-zero temperature. 

Different conclusions apply when more than just the polarized and the
singlet ground states are present in the low-energy sector. We have
demonstrated, that near the transition, the gap could actually close
in several different situations. In the present study, this happens
for very strong magnetic inhomogeneities (Subsect. \ref{pos-ch04-03}),
for systems with an elongated elementary cell
(Sect. \ref{pos-ch04-07}) and for short-range
interacting systems (Sect. \ref{pos-ch04-06}). The states which closed
the gap always belong to an intermediate value of spin, most
prominent are those with $S=1$ and $S=2$ and since we considered only
eight-electron systems, the latter value of spin corresponds to the
half-polarized sector $S=N_e/4$. These states are considerably softer
against magnetic inhomogeneities 
than the incompressible singlet and polarized states. On one hand, this fact
follows from a small level spacing in the low energy sector when the
gap closes. However, the magnetic inhomogeneities were also found to
have large ($\sim 0.1$) matrix elements between most of the low lying
states. 

The states with the strongest tendency to form domains i.e. the
'softest' states, were found in systems with Coulomb interaction and
an elongated elementary cell. Near the transition, even a moderately
strong magnetic inhomogeneity (weaker than the incompressibility gap)
was enough to make the polarization approach the values
corresponding to the singlet and polarized states inside the domains.
For these domain states, we have investigated
the 'inside' of the domains by means of non-averaged correlation
functions, Subsect. \ref{pos-ch04-08}. We could not yet confirm
that the domains comprise of an incompressible liquid but this
system defintely deserves a more detailed study. Especially in this case,
a comparison with larger systems would be very helpful.

\addtocontents{toc}{\protect\enlargethispage*{1000pt}}
\section{Conclusions}

Fractional quantum Hall systems at filling factors $\nu=\tt$ and $\tf$
have been studied numerically by means of exact diagonalization
techniques on a torus. In both systems, the existence of two different ground
states is well established: one is fully spin polarized, another is
a spin singlet and they are both strongly correlated. All four states
can be visualised as composite fermion systems at integer filling
factor ($\nu_{CF}=2$). A transition
between these two ground states can be induced by changing the Zeeman
energy while keeping the filling factor constant, Chapter \ref{pos-ch04-00}.

At the beginning of Chapter \ref{pos-ch03-00}, we investigated the
polarized and the singlet incompressible ground states in terms of
their density-density correlation functions. First, we highlighted
the fact that -- even if these states were {\em exactly} described by
some composite fermion model -- the inner structures of the ground states
at $\nu=\tt$ and $\tf$ differ strongly from the inner structure of a
state comprising of two fully occupied Landau levels. In other words,
in a composite-fermion state (e.g. $\nu_{CF}=2$), the correlations
between the {\em electrons} are different than in a corresponding
electronic state ($\nu=2$). A more important result is, however, that
the electronic correlations differ strongly also between the $\nu=\tt$
and $\tf$ states themselves. This is surprising, since both filling factors map
to the same filling factor of composite fermions ($\nu_{CF}=2$) and
only the orientation of the effective field is different. Study of the
correlation functions allowed to suggest a new interpretation of the singlet
$\nu=\tt$ ground state. The electrons move along in pairs of opposite
spins and the pairs form a state equivalent to a fully occupied lowest
Landau level. This conclusion does {\em not} apply to the $\nu=\tf$
singlet ground state.

The central focus of the present work was on the low-energy 
states occurring near the transition between the singlet and the
polarized ground states. Some experimental results indicate
that another ground state distinct from the two ground states already mentioned
could exist near the transition \cite{kukushkin:05:1999}. In Sections
\ref{pos-ch04-09} and \ref{pos-ch03-07} we found several arguments in favour
of a half-polarized state ($S=N_e/4$) 
becoming the absolute ground state in a narrow
range of the magnetic field. The systems available to exact
diagonalization were however too small to allow for an unswerving
prediction. Two candidates for the half-polarized ground state were
identified. In Section \ref{pos-ch03-07} we concentrated on the
'isotropic candidate'. A study of its inner structure
(correlation functions) combined with an investigation of the response
to probing magnetic inhomogeneities (Sect. \ref{pos-ch03-18}) 
produced results resembling both 
the singlet and polarized incompressible ground state. A hypothesis
that both these states coexist within the half-polarized state has
been presented.

Calculations with elongated rectangular elementary cells
(Sect. \ref{pos-ch03-08}) suggested another candidate for the
half-polarized ground state: a spin-density wave along the longer
side of the elementary cell. A comparison
between two systems of different size indicated that this state has
the shortest period allowed by the finite size of the considered
system (e.g. one third of the length of the cell for a state which
contains three minority spins). Based on the present calculations it
is not possible to decide which of the two candidates (if any) 
evolves into the ground state of an infinite system.

At $\nu=\tf$, no obvious analogue to the half-polarized state at
$\nu=\tt$ was found.

Employing magnetic inhomogeneities
to enforce domains of different spin polarization near the
transition at $\nu=\tt$ (Chapter \ref{pos-ch04-00}) we found that no
signs of domain formation occur unless the energy gap closes. The
loss of incompressibility could however still be compatible with the
experimental observation of a plateau of polarization one half during
the transition. It is enough if there are many states with $S=N_e/4$
and no (or only few) states with other values of $S$ in the low-energy
sector (Sect. \ref{pos-ch04-06}). 

The 'best' candidates for domain states were found to appear in
systems with an elongated rectangular cell. The fundamental idea here
was that the elementary cell with aspect ratio $2:1$ is divided by the
inhomogeneity into two square parts which could be more convenient for
the formation of isotropic states (the singlet and the polarized
incompressible liquid). Examination of the domain state however showed
that the inside of the domains does not resemble the incompressible
ground states at $\nu=\tt$. Nevertheless, a more detailed study is
necessary here, since systems with aspect ratio far from unity can
suffer more from finite size effects than what was demonstrated in
Sect. \ref{pos-ch03-17}. 

At the very end, I would like to acknowledge Daniela Pfannkuche for
her support during my PhD studies 
and many fruitful discussions, Benjamin Kr\"uger,
Philip von Ende and Ond\v rej \v Cert\'\i k 
for some of the calculations presented here and
Matti Manninen for his hospitality in the period when this
work was being completed. Finally, I am much obliged to a
group of my colleaugues who carefully read the very long
manuscript and helped to substantially 
improve both its contents and language: Frank Hellmuth,
Rob Knapik, Katrin Malessa, Christian M\"uller, Michael Prouza,
Arek W\'ojs and Jan Zemen.

\bibliography{notes,diss,diss-adp}

\begin{thebibliography}{10}

\bibitem{comm:ch02-03}
Going once around an $s$-fold vortex gives phase $2\pi s$. Exchange of two
  particles corresponds to one half of such a loop (for $\psi(r_1,r_2)\to
  \psi(r_2,r_1)$ corresponds to $\psi_{rel}(\vr)\to \psi_{rel}(-\vr)$ with
  $\vr=r_1-r_2$ in the relative part of the WF; $\vr\to -\vr$ is half the way
  of going around zero). Thus exchanging two particles with $s$ attached
  vortices, the wavefunction acquires phase $\pi s$. For two {\em fermions}
  with $s$ attached vortices, it is $\pi(s+1)$. Thus the wavefunction changes
  sign at exchange of two particles when $s$ is even and does not change the
  sign when $s$ is odd.

\bibitem{comm:ch02-04}
The magnetic field described by the vector potential in (\ref{eq-ch02-58}) is
  proportional to electron density, $\Psi^\dag(\vek{r}_1)\Psi(\vek{r}_1)$. In
  other words: the magnetic field felt by an electron at $\vek{r}$ is only
  non-zero if $\vek{r}=\vek{r}_1$, or, an electron at $\vek{r}$ sees magnetic
  field consisting of delta--functions located at positions of other electrons.
  However, these points in space are inaccessible to the electron by virtue of
  the Pauli principle.

\bibitem{comm:ch02-05}
In Fig. \ref{fig-ch03-50}b, energies of low--lying states at $\nu=\tt$ are
  plotted against the aspect ratio $a:b$, assuming the short--range interaction
  (\ref{eq-ch02-14}). The ground state at $\nu=\ot$ has zero energy (cf.
  Subsec. \ref{pos-ch02-09}) for any value of $a:b$ and hence the energy of the
  $\nu=\tt$ ground state is equal to the energy of the completely filled lowest
  Landau level, $E_f$, multiplied by $(m-2n)/m$. The energy $E_f$ is not
  completely independent on $a:b$. As long as the $a\times b$ rectangle is
  still large enough to be a good description of a $\nu=1$ 2DEG, $E_f$ is
  constant. As soon as $b$ ($<a$) becomes too small (a reliable indication is
  that the density of the state $\ket{1}$ becomes markedly inhomogeneous) $E_f$
  starts to change (this is the case for $a:b>4$ in Fig. \ref{fig-ch03-50}b).

\bibitem{comm:ch03-09}
Consider the action of $S^-$ (the lowering operator for the $z$-component of
  spin) on the $\nu=2$ (or $\nu_{CF}=2$) ground state $\ket{\Psi, S_z=0}$ at
  zero Zeeman energy ($0\up$ and $0\dn$ LLs are filled). On one hand, the state
  $S^-\ket{\Psi, S_z=0}$ may not contain any particles in higher LLs (up to
  Zeeman energy, it should have the same energy as $\ket{\Psi, S_z=0}$). On the
  other hand, there is no room for an extra spin down in the lowest LL which is
  completely filled and therefore flipping a spin $\up\to\dn$ (as contained in
  $S^-$) must annihilate the state. Finally, $S^- \ket{\Psi, S_z=0}=0$ implies
  that $\ket{\Psi, S_z=0}$ is a $S^2=0$ state.

\bibitem{comm:ch03-01}
If we were able to distinguish two particles in a state described by a single
  Slater determinant, even such a state would also be entangled: consider a
  state $\Psi(1,2)=\psi_L(1)\psi_R(2)- \psi_R(1)\psi_L(2)$ and say the states
  $L$ and $R$ are localized on the left and on the right, respectively. There
  is an uncertainity in whether it is the particle 1 or the particle 2 which is
  on the right. If we measure one particle on the right, we automatically know
  that it is the {\em other} particle which is on the left. Owing to the
  indistinguishability of indentical particles, however, we cannot say whether
  it was the particle 1 or the particle 2 which we have just caught on the
  right. If the particles 1 and 2 have different spin, the indistinguishability
  requirement is lifted and the state $\Psi$ is entangled. In that case, $\Psi$
  is the spin--singlet state.

\bibitem{comm:ch03-03}
Several lowest of these polynomials are: $L_0^1(x)=1$, $L_1^1(x)=2-x$,
  $L_2^1(x)=\frac{1}{2}(6-6x+x^2)$, $L_3^1(x)=\frac{1}{6}(24-36x+12x^2-x^3)$.

\bibitem{comm:ch03-05}
In fact, there are {\em some} analytical results. Very appealing schemes how to
  evaluate energy and correlation functions were suggested by Girvin
  \cite{girvin:07:1984} Takano and Isihara \cite{takano:07:1986}. Interesting
  extension of the former work was presented by G\"orbig (Subsec. 1.2.2. in
  \cite{goerbig:2004}). All these schemes however present closed formulae
  neither for energy nor for correlation functions.

\bibitem{comm:ch03-06}
If we completely fill the lowest Landau level with spin up electrons and with
  spin down electrons (imagine $\nu=2$ and zero Zeeman energy), then spin up
  and spin down electrons are uncorrelated, $g_{\up\dn}(\vek r)=1$. It is {\em
  not} a claim of composite fermion theories that the same is true if we do the
  same with CF Landau levels. The attachment of flux quanta introduces
  correlations between the originally uncorrelated $(n=0,\up)$ and $(n=0,\dn)$
  levels: spin up CFs do not feel the spin down CFs (owing to LL mixing
  neglect) but they do feel fluxes attached to the spin down CFs.

\bibitem{allesch:01:2004}
M.~Allesch, E.~Schwegler, F.~Gygi, and G.~Galli.
\newblock {\em cond-mat}, page 0401267, 2004.

\bibitem{apalkov:02:2001}
V.M. Apalkov, T.~Chakraborty, P.~{Pietil\"ainen}, and K.~{Niemel\"a}.
\newblock {\em Phys. Rev. Lett.}, 86:1311, 2001.

\bibitem{bertotti:1998}
G.~Bertotti.
\newblock {\em Hysteresis in magnetism : for physicists, materials scientists
  and engineers}.
\newblock Acad. Press, San Diego, California, 1998.

\bibitem{bonsall:02:1977}
L.~Bonsall and A.~Maradudin.
\newblock {\em Phys. Rev. B.}, 15:1959, 1977.

\bibitem{chakraborty:07:2000}
T.~Chakraborty.
\newblock {\em Adv. Phys.}, 49:959, 2000.

\bibitem{chakraborty:1995}
T.~Chakraborty and {Pietil\"ainen, P.}
\newblock {\em The Quantum Hall Effects}.
\newblock Springer, Berlin, second edition, 1995.

\bibitem{cho:09:1998}
H.~Cho, J.B. Young, W.~Kang, K.L. Campman, A.C. Gossard, M.~Bichler, and
  W.~Wegscheider.
\newblock {\em Phys. Rev. Lett.}, 81:2522, 1998.

\bibitem{clark:03:1989}
R.G. Clark, S.R. Haynes, A.M. Suckling, J.R. Mallett, P.A. Wright, J.J. Harris,
  and C.T. Foxon.
\newblock {\em Phys. Rev. Lett.}, 62:1536, 1989.

\bibitem{sarma:1997}
S.~Das~Sarma and A.~Pinczuk.
\newblock {\em Perspectives in Quantum Hall Effects}.
\newblock Wiley, New York, 1997.

\bibitem{dePoortere:??:2000}
E.P. de~Poortere, E.~Tutuc, S.J. Papadakis, and M.~Shayegan.
\newblock {\em Science}, 290:1546, 2000.

\bibitem{dietsche:??:2001}
W.~Dietsche and S.~{Kronm\"uller}.
\newblock {\em Physica E}, 10:71, 2001.

\bibitem{dunford:07:2002}
R.B. Dunford, M.R. Gates, V.W. Rampton, C.J. Mellor, O.~Stern, W.~Dietsche,
  W.~Wegscheider, and M.~Bichler.
\newblock Absence of the huge longitudinal resistance maxima in surface
  acoustic wave measurements of narrow quantum wells.
\newblock In {\em Proceedings of the 26th International Conference on the
  Physics of Semiconductors}, page P143, Edinburgh, UK, July 2002. IOP.

\bibitem{eisenstein:03:1989}
J.P. Eisenstein, H.L. Stormer, L.~Pfeiffer, and K.W. West.
\newblock {\em Phys. Rev. Lett.}, 62:1540, 1989.

\bibitem{eom:09:2000}
J.~Eom, H.~Cho, W.~Kang, K.L. Campman, A.C. Gossard, M.~Bichler, and
  W.~Wegscheider.
\newblock {\em Science}, 289:2320, 2000.

\bibitem{fano:08:1986}
G.~Fano, F.~Ortolani, and E.~Colombo.
\newblock {\em Phys. Rev. B.}, 34:2670, 1986.

\bibitem{freytag:09:2001}
N.~Freytag, Y.~Tokunaga, M.~{Horvati\'c}, C.~Berthier, M.~Shayegan, and L.P.
  {L\'evy}.
\newblock {\em Phys. Rev. Lett.}, 87:136801, 2001.

\bibitem{girvin:07:1984}
S.M. Girvin.
\newblock {\em Phys. Rev. B.}, 30:558, 1984.

\bibitem{girvin:07:1999}
S.M. Girvin.
\newblock {\em cond-mat}, page 9907002, 1999.

\bibitem{girvin:02:1985}
S.M. Girvin, A.H. MacDonald, and P.M. Platzman.
\newblock {\em Phys. Rev. Lett.}, 54(6):581, 1985.

\bibitem{goerbig:2004}
M.O. Goerbig.
\newblock {\em Etude th\'eorique des phases de densit\'e inhomoge\`ene dans les
  syste\`emes \`a effet Hall quantique}.
\newblock PhD thesis, Universit\'e de Fribourg, Switzerland, 2004.

\bibitem{gradshteyn:1980}
I.S. Gradshteyn and I.M. Ryzhik.
\newblock {\em Tables of Integrals, Series, and Products}.
\newblock Academic Press, New York, any edition, 1980.

\bibitem{haldane:08:1983}
F.D.M. Haldane.
\newblock {\em Phys. Rev. Lett.}, 51:605, 1983.

\bibitem{haldane:11:1985}
F.D.M. Haldane.
\newblock {\em Phys. Rev. Lett.}, 55(20):2095, 1985.

\bibitem{haldane:02:1985}
F.D.M. Haldane and E.H. Rezayi.
\newblock {\em Phys. Rev. B.}, 31:2529R, 1985.

\bibitem{haldane:01:1985}
F.D.M. Haldane and E.H. Rezayi.
\newblock {\em Phys. Rev. Lett.}, 54:237, 1985.

\bibitem{haldane:03:1988}
F.D.M. Haldane and E.H. Rezayi.
\newblock {\em Phys. Rev. Lett.}, 60:956, 1988.

\bibitem{halperin:??:1983}
B.I. Halperin.
\newblock {\em Helv. Phys. Acta}, 56:75, 1983.

\bibitem{hashimoto:04:2002}
K.~Hashimoto, K.~Muraki, T.~Saku, and Y.~Hirayama.
\newblock {\em Phys. Rev. Lett.}, 88:176601, 2002.

\bibitem{helias:2003}
M.~Helias.
\newblock Quasihole tunneling in the fractional quantum hall regime.
\newblock Master's thesis, University of Hamburg, Germany, 2003.

\bibitem{jain:07:1989}
J.K. Jain.
\newblock {\em Phys. Rev. Lett.}, 63:199, 1989.

\bibitem{jain:11:1994}
J.K. Jain.
\newblock {\em Science}, 266:1199, 1994.

\bibitem{jungwirth:12:2000}
T.~Jungwirth and A.H. MacDonald.
\newblock {\em Phys. Rev. B.}, 63:035305, 2000.

\bibitem{jungwirth:11:2001}
T.~Jungwirth and A.H. MacDonald.
\newblock {\em Phys. Rev. Lett.}, 87:216801, 2001.

\bibitem{kamilla:11:1997}
R.K. Kamilla, J.K. Jain, and S.M. Girvin.
\newblock {\em Phys. Rev. B.}, 56:12411, 1997.

\bibitem{kraus:12:2002}
S.~Kraus, O.~Stern, J.G.S. Lok, W.~Dietsche, K.~von Klitzing, M.~Bichler,
  D.~Schuh, and W.~Wegscheider.
\newblock {\em Phys. Rev. Lett.}, 89:266801, 2002.

\bibitem{kronmuller:05:1999}
S.~{Kronm\"uller}, W.~Dietsche, K.~von Klitzing, G.~Denninger, W.~Wegscheider,
  and M.~Bichler.
\newblock {\em Phys. Rev. Lett.}, 82:4070, 1999.

\bibitem{kronmuller:09:1998}
S.~{Kronm\"uller}, W.~Dietsche, J.~Weis, K.~von Klitzing, W.~Wegscheider, and
  M.~Bichler.
\newblock {\em Phys. Rev. Lett.}, 81:2526, 1998.

\bibitem{kukushkin:05:1999}
I.V. Kukushkin, K.v. Klitzing, and K.~Eberl.
\newblock {\em Phys. Rev. Lett.}, 82:3665, 1999.

\bibitem{lam:07:1984}
P.K. Lam and S.M. Girvin.
\newblock {\em Phys. Rev. B.}, 30:473, 1984.

\bibitem{laughlin:05:1983}
R.~Laughlin.
\newblock {\em Phys. Rev. Lett.}, 50:1395, 1983.

\bibitem{laughlin:03:1983}
R.~Laughlin.
\newblock {\em Phys. Rev. B.}, 27:3383, 1983.

\bibitem{leadley:11:1997}
D.R. Leadley, R.J. Nicholas, D.K. Maude, A.N. Utjuzh, J.C. Portal, J.J. Harris,
  and C.T. Foxon.
\newblock {\em Phys. Rev. Lett.}, 79:4246, 1997.

\bibitem{mariani:12:2002}
E.~Mariani, N.~Magnoli, F.~Napoli, M.~Sassetti, and B.~Kramer.
\newblock {\em Phys. Rev. B.}, 66:241303, 2002.

\bibitem{morf:08:2002}
R.H. Morf, N.~{d'Ambrumenil}, and S.~Das~Sarma.
\newblock {\em Phys. Rev. B.}, 66:075408, 2002.

\bibitem{mueller:2005}
Ch. {M\"uller}.
\newblock {\em Disorder and vortex detachment in fractional quantum Hall
  liquids}.
\newblock PhD thesis, University of Hamburg, Germany, 2005.

\bibitem{murthy:01:2000}
G.~Murthy.
\newblock {\em Phys. Rev. Lett.}, 84:350, 2000.

\bibitem{murthy:10:2001}
G.~Murthy.
\newblock {\em Phys. Rev. Lett.}, 87:179701, 2001.
\newblock comment.

\bibitem{murthy:10:2003}
G.~Murthy and R.~Shankar.
\newblock {\em Rev. Mod. Phys.}, 75:1101, 2003.

\bibitem{niemela:xx:2000}
K.~Niemel\"a, P.~Pietil\"ainen, and T.~Chakraborty.
\newblock {\em Physica B}, 284:1717, 2000.

\bibitem{nomura:2003}
K.~Nomura.
\newblock {\em Various broken symmetries in two--component quantum Hall
  systems}.
\newblock PhD thesis, Department of Basic Science, University of Tokyo, Japan,
  2003.

\bibitem{pan:01:2003}
W.~Pan, H.L. Stormer, D.C. Tsui, L.N. Pfeiffer, K.W. Baldwin, and K.W. West.
\newblock {\em Phys. Rev. Lett.}, 90:016801, 2003.

\bibitem{prange:1987}
R.~E. Prange and S.~M. Girvin.
\newblock {\em The Quantum Hall Effect}.
\newblock Springer, Berlin, 1987.

\bibitem{rezayi:11:1985}
E.H. Rezayi and F.D.M. Haldane.
\newblock {\em Phys. Rev. B.}, 32:6924, 1985.

\bibitem{rezayi:12:1994}
E.H. Rezayi and F.D.M. Haldane.
\newblock {\em Phys. Rev. B.}, 50:17199, 1994.

\bibitem{rezayi:05:2003}
E.H. Rezayi, T.~Jungwirth, A.H. MacDonald, and F.D.M. Haldane.
\newblock {\em Phys. Rev. B.}, 67:201305(R), 2003.

\bibitem{schollwock:09:2004}
U.~{Schollw\"ock}.
\newblock {\em cond-mat}, page 0409292, 2004.

\bibitem{schollwoeck:04:2005}
U.~{Schollw\"ock}.
\newblock {\em Rev. Mod. Phys.}, 77:259, 2005.

\bibitem{shibata:06:2001}
N.~Shibata and D.~Yoshioka.
\newblock {\em Phys. Rev. Lett.}, 86:5755, 2001.

\bibitem{shibata:03:2003}
N.~Shibata and D.~Yoshioka.
\newblock {\em J. Phys. Soc. Jpn.}, 72:664, 2003.

\bibitem{shibata:08:2003}
N.~Shibata and D.~Yoshioka.
\newblock {\em cond-mat}, page 0308122, 2003.

\bibitem{shibata:03:2004}
N.~Shibata and D.~Yoshioka.
\newblock {\em cond-mat}, page 0403493, 2004.

\bibitem{smet:01:2002}
J.H. Smet, R.A. Deutschmann, F.~Ertl, W.~Wegscheider, G.~Abstreiter, and K.~von
  Klitzing.
\newblock {\em Nature}, 415:281, 2002.

\bibitem{smet:03:2001}
J.H. Smet, R.A. Deutschmann, W.~Wegscheider, G.~Abstreiter, and K.~von
  Klitzing.
\newblock {\em Phys. Rev. Lett.}, 86:2412, 2001.

\bibitem{takano:07:1986}
K.~Takano and A.~Isihara.
\newblock {\em Phys. Rev. B.}, 34:1399, 1986.

\bibitem{tao:03:1986}
R.~Tao and F.D.M. Haldane.
\newblock {\em Phys. Rev. B.}, 33:3844, 1986.

\bibitem{trugman:04:1985}
S.A. Trugman and S.~Kivelson.
\newblock {\em Phys. Rev. B.}, 31:5280, 1985.

\bibitem{wojs:02:2000}
A.~{W\'ojs} and J.J. Quinn.
\newblock {\em Phil. Mag. B}, 80:1405, 2000.

\bibitem{wojs:07:2002}
A.~{W\'ojs} and J.J. Quinn.
\newblock {\em Phys. Rev. B.}, 66:045323, 2002.

\bibitem{wu:07:1993}
X.G. Wu, G.~Dev, and J.K. Jain.
\newblock {\em Phys. Rev. B.}, 71:153, 1993.

\bibitem{yoshioka:06:1984}
D.~Yoshioka.
\newblock {\em Phys. Rev. B.}, 29:6833, 1984.

\bibitem{yoshioka:2002}
D.~Yoshioka.
\newblock {\em The Quantum Hall Effect}.
\newblock Springer, Berlin, 2002.

\bibitem{yoshioka:04:1983}
D.~Yoshioka, B.I. Halperin, and P.A. Lee.
\newblock {\em Phys. Rev. Lett.}, 50:1219, 1983.

\bibitem{yoshioka:xx:1984}
D.~Yoshioka, B.I. Halperin, and P.A. Lee.
\newblock {\em Surf. Sci.}, 142:155, 1984.

\bibitem{yoshioka:xx:2002}
D.~Yoshioka and N.~Shibata.
\newblock {\em Physica E}, 12:43, 2002.

\bibitem{zak:06:1964}
J.~Zak.
\newblock {\em Phys. Rev.}, 134:A1602, 1964.

\bibitem{zak:06-2:1964}
J.~Zak.
\newblock {\em Phys. Rev.}, 134:A1607, 1964.

\bibitem{zhang:12:1984}
F.C. Zhang and T.~Chakraborty.
\newblock {\em Phys. Rev. B.}, 30:7320, 1984.

\bibitem{zhang:11:1985}
F.C. Zhang, V.Z. Vulovic, Y.~Guo, and S.~Das~Sarma.
\newblock {\em Phys. Rev. B.}, 32:6920, 1985.

\end{thebibliography}
\bibliographystyle{plain}
\addtocontents{toc}{\protect\pagebreak}


\end{document}